Understanding learning within a commercial video game: A case study

Allan Fowler
B.Bus, MBA, M.Ed.

a thesis submitted to the faculty of design and creative technologies
AUT University
in partial fulfilment of the
requirements for the degree of
doctor of philosophy

School of Computing and Mathematical Sciences

Auckland, New Zealand
2014

## DECLARATION

I hereby declare that this submission is my own work and that, to the best of my knowledge and belief, it contains no material previously published or written by another person nor material which to a substantial extent has been accepted for the qualification of any other degree or diploma of a University or other institution of higher learning, except where due acknowledgement is made in the acknowledgements.

...........................

# ACKNOWLEDGMENTS

This thesis has been undertaken at the Faculty of Design and Creative Technologies of AUT University in Auckland, New Zealand. I would like to take this opportunity to thank all the people who have given me invaluable feedback and support throughout the development, research, and completion of the research project and the production of this thesis. I received considerable support from many colleagues, peers, and leading academics at the various conferences I have been fortunate enough to attend. Thank you. I apologise if I fail to mention all the people who helped me, but the list would fill several pages.

First and foremost, I would like to thank my supervisor, Dr. Brian Cusack for your support, feedback, and guidance over the last five years. Without your support, I am very sure this research would not have been possible. I also thank you for the opportunity and support to go to Boston, this was an incredible experience, and I could not have managed it without the support of AUT.

I would like to thank my wife, Qiao Hao and my children, Ciaran and Annah. You have been my inspiration and motivation. Thank you for your understanding, guidance and constant support.

I also want to thank the many people at Northeastern University for their help and guidance: Associate Professor Magy Sief El-Nasir, Associate Professor Rupal Patel, Professor Susan Gold and Kali Gold, thank you for hosting me. I want to also acknowledge the team of incredible people at Northeastern that helped me: Associate Professor Alessandro Canossa, Dr. Jeremy B. Badler, Dr. Joshua Gross, Dr. Kate Connaghan, Dr. Sree Durga, Ms. Dara-Lynn Pelechatz, and all the staff and students in the lab, thank you for looking after/putting up with me.

I also would like to acknowledge some of the people that I met at the Games for Learning and Society Conference: Professor James Paul Gee, Professor Rick Halverson and his wife Associate Professor Erica Halverson, Professor Kurt Squire and his wife, Associate Professor Constance Steinkuehler. Although I did not get to meet Professor Valerie Shute, your paper was a source of inspiration for this study, thank you. I would like to thank Dr. Alex Games and Microsoft Corporation for making it possible for me to attend this valuable and important conference.

I also want to thank the team at SMI USA: Mr. Mark Mento, Mr. Jonathan Hyland, and Ms. Lisa Richardson. Your generosity, support and advice was beyond amazing. Without your support, this research would not have been achievable. To the team at 2D Boy, thank you for making such an awesome game and for letting me use images from your game in the production of this thesis

I want to thank the co-authors and the many reviewers that helped me by providing the valuable feedback and support needed to get my conference and journal papers published, thank you.

I would like to dedicate this thesis to my late Mother, Margaret Fowler. You instilled in me the value and importance of continuous improvement. You will always be in my heart.

# ABSTRACT

There has been considerable interest and debate about the value of video games. There are researchers that suggest that video games are useful tools for engaging learners in engaging and authentic learning experiences (Gee, 2003; McGonigal, 2012; Shaffer, Squire, Halverson, & Gee, 2005; Squire, 2008; Steinkuehler, 2005). However, there are researchers that suggest that video games are harmful, addictive, or exploitive (Chan & Rabinowitz, 2006; Fisher, 1994; Spain & Vega, 2005; Phillips, Rolls, Rouse, & Griffiths, 1995) and disruptive to the learning experience (Kirriemuir & McFarlane, 2003). There has been considerable research on the educational value of educational video games, and the findings have found that the educational video games are very engaging and interesting. However, as the organisations that make the educational games lack the production budgets results in the video games not being widely adopted by the target consumer. Furthermore, many of these video games incorporate what Bruckman (1999) referred to as 'chocolate covered broccoli' approach. This is when a video game is presented as a reward for completing a learning outcome (Bruckman, 1999). Through a review of the literature, gaps in the literature on the educational value of commercial video games were identified. Therefore, this research sought to undertake an empirical study into what types of learning transpires while using a commercial video game.

As the process of learning is typically an internal process, the task of identifying if and when learning transpired is sometimes challenging. Although traditional methods of assessments and verbal reflection provide some indicators that learning may have transpired, this research sought to obtain additional evidence through using observational methods. This additional method was possible through advances in video based eye tracking cameras. The availability of high-resolution desk mounted video based eye tracking camera made it possible for this research to obtain quantitative data on potential indicators of cognition or problem solving; endogenous eye blinks and fixations that lasted longer than 600 milliseconds (ms). Ponder & Kenned (1927) suggest that the endogenous (having an internal cause or origin) blink is an indicator of human cognition. This theory has been supported by more recent researchers (Stern, Walrath, & Goldstein, 1984; Orchard & Stern, 1991; Tanaka & Yamaoka, 1993). The other indicator or cognition is when the eye fixates on a specific object or stimuli. The duration the eyes are fixed on a particular object can potentially indicate the amount of processing that is taking place (Just & Carpenter, 1976; Just & Carpenter, 1980).

This study used a desktop video-based eye tracking camera to monitor and record the endogenous eye blinks and eye fixations.

While the studies into the value of educational video games have provided a valuable contribution to knowledge, very few investigated the transferability of the learning that took place within the game to an external context. To address this limitation, the study conducted in this paper measured pre and post exposure to the learning principles embedded in the treatment. Furthermore, this study observed any improvements in the embedded learning concepts through a physical exercise that replicated the in-game learning concepts to an out-of-game test. The research questions are:

RQ 1: What learning takes place when playing the video game World of Goo?

RQ 2: Does problem solving ability improve through playing video games?

RQ 3: Do the participants that played the video game World of Goo learn tower construction from playing the game?

This study identified that conceptual learning did transpire through exposure to a commercial video game. The adult participants exhibited this learning through in-game performance and how to play the game. The children exhibited this learning through demonstrating an advanced understanding of the embedded learning concepts within the game to the out-of-game tests. Furthermore, the children exhibited improvements in in-game performance and reductions in cognition and cognitive problem solving after they were exposed to additional treatment.

These findings will be valuable to educators who find it challenging to educate twenty-first-century learners. These learners have grown up with Internet-connected multimedia devices and the pedagogical concepts embedded in traditional media (books, video) may not provide a complete learning experience that they are familiar with. Further, these findings will be valuable for video game developers as the method employed could be used by developers to help identify the parts of the game that the user is struggling with. These findings could also be used by developers or regulators in helping identify the most appropriate target age group for the game. These findings could also be used in future research as this research validated a method for collecting empirical data on individual cognition.

# PUBLICATIONS

| | |
|---|---|
| *"The evolution and significance of the Global Games Jam"* | Fowler, A., Khosmood, F., & Arya, A. (2013). The evolution and significance of the Global Games Jam, In *Proceedings of Inaugural workshop of the Global Game Jam, collocated with the Foundations of Digital Game Conference (FDG 2013)*, Chania, Crete, Greece. |
| *"The Global Game Jam as a venue for teaching and learning"* | Fowler, A., Khosmood, F., Arya, A., & Lai, G. (2013). The Global Game Jam as a venue for teaching and learning, In M. Lopez & M. Verhaart (Eds.) *Proceedings of the 4th Annual Conference of Computing and Information Technology, Education and Research in New Zealand (incorporating 26th Annual NACCQ)*, Hamilton, New Zealand. |
| *"Kodu Game Lab: A programming environment"* | Fowler A., Fristoe, T., & MacLaurin, M. (2012). Kodu Game Lab: A programming environment, *The Computer Games Journal: Whitsun 2012*, TuDocs Ltd, 17-22 |
| *"Enriching student learning programming through using Kodu"* | Fowler, A. (2012b). Enriching student learning programming through using Kodu, In M. Verhaart & M. Lopez, (Eds.), *Proceedings of the 3rd Annual Conference of Computing and Information Technology Education and Research in New Zealand (incorporating 25th Annual NACCQ)*, Christchurch, New Zealand. 33-39. |
| *"Enhancing introductory programming with Kodu Game Lab in a high school classroom"* | Fowler, A. (2012a). Enhancing introductory programming with Kodu Game Lab in a high school classroom, In K. Squire, C. Martin, & A. Ochsner (Eds.), *Proceedings of the Games, Learning, and Society Conference*: *3*. Pittsburgh PA: ETC Press. 561-564 |

# TABLE OF CONTENTS

# List of Tables

## List of Figures

## List of Abbreviations

| | |
|---|---|
| AVG | Action Video Game |
| CTA | Concurrent Think Aloud |
| CVG | Commercial Video Game |
| EVG | Educational Video Game |
| FPS | First Person Shooter |
| Hz | Hertz |
| IRB | Internal Review Board |
| MMORPG | Massively Multiplayer Online Role Playing Game |
| ms | Millisecond |
| NFL | National Football League |
| OCD | Obsessive Completion Distinction |
| PC | Personal Computer |
| RTA | Retrospective Think Aloud |
| SMI | SensoMotoric Instruments |
| TAM | Think Aloud Methods |
| WoW | World of Warcraft |

## 1.0    INTRODUCTION

As a student, educator, and researcher I have long been interested in the learning opportunity that video games represent a corollary of a pedagogic awareness of the considerable benefit of applied and practical learning experiences. However, when I look at the textbooks that students use in today's classrooms, I see that they have not changed much since I was a high school student. Apart from the advent of colour printing and the introduction of supplementary materials, the words on the page have not changed very much at all. To me, education and learning are fascinating and interactive experiences. However, it would be challenging to identify this in many introductory textbooks. From my own observations and experience, I feel that with the advent of such powerful electronics and computing devices, there must be a better way. The potential of video games as an educational tool needs to be considered. Video games offer the potential to learn in so many ways that a traditional textbook cannot. Video games offer a rich and engaging learning opportunity. Video games also offer the opportunity to experiment and learn from mistakes. With the increased improvement of the capabilities of portable computing devices such as tablet computers and smart phones, video games are also highly portable and offer the user the opportunity to engage with these devices anywhere, and at any time. The primary motivation of this thesis is to qualify and quantify the benefits and the educational potential that is inherent in a video game. The research questions are:

RQ 1: What learning takes place when playing the video game World of Goo?

RQ 2: Does problem solving ability improve through playing video games?

RQ 3: Do the participants that played the video game World of Goo learn tower construction from playing the game?

The purpose of this chapter is to provide the motivations for this thesis and provide an outline of the research findings and the structure of the thesis. In Section 1.1, the motivation behind this research will be discussed. The outline of the thesis will be provided in Section 1.2. Section 1.3 provides a summary of this chapter.

## 1.1    MOTIVATION

This researcher first experienced the use of computer or video games as an educational tool as a final year student at RMIT University in Melbourne Australia. The experience helped the researcher and his fellow students put together the many theories into practical and applied ways that helped facilitate a deeper understanding of the many implications to the theories that had been learnt. The next experience of computer games was a more personal reason. In 1998, the researcher joined an enthusiast car club and through this club, the members were able to drive their cars on a professional race track. As the researcher had no prior experience in racing cars, he did not do very well on the first race day. However, the researcher saw an opportunity to learn how to improve these skills by buying a video console and car racing game. This particular game was then played every night until the next race day. The result of this learning resulted in the researcher finishing third in the first race, and first in the next. This experience would not be forgotten. I started teaching martial arts in my mid-to-late twenties, and quickly learnt the educational value of using games to help young students learn and develop. One of the games I used was called crab soccer. This game involved the students crawling in an inverted crab position (with the body facing up and the arms and legs on the floor) while playing soccer with a small boxing glove. The purpose of this exercise was to improve upper body strength (especially in the arms) and flexibility. The game element of this exercise was to encourage team work and improve flexibility.

When I changed my career to become a fulltime educator, I wanted to use what I had learnt as a student and as a martial art instructor in the classroom (although I would not use crab soccer). It was to my surprise that I met some resistance from established educators. In 2010, I watched a demonstration of the beta release of the Microsoft game development tool called Kodu Game Lab which used the process of making video games to engage young students in learning some fundamental principles of programming. From this insight, I wanted to see what impact this tool would have in the classroom. This research led to the opportunity to publish four conference papers, which resulted in meeting several of the foundation authors in this area. Although the research into the use of Kodu Game Lab concluded that this tool could be an effective tool to engage students in learning computer science, I wanted to look deeper into the broader use of video games within an educational context, and see what impact commercial video games could have in the classroom. While there are considerable studies that have investigated the use of video games in education, the results of the literature review did not identify any studies that have measured the change in cognitive processes in participants. The intention of this thesis is to extend the conversation by implementing a mixed- methods approach to testing this link.

## 1.2 PREVIOUS STUDIES

The following section provides a summary of the research identified in Section 1.1. This initial research provided the experience needed to identify the foundation literature and methods which lead to the development of the experimental design undertaking in this thesis.

### 1.2.1 Kodu Game Lab: A programming environment

Through exposure to the game development tool, Kodu Game Lab, the author published a journal paper to provide a foundation for the use of this tool in formal education (Fowler, Fristoe, & MacLaurin, 2012). The intention of this journal paper was to provide an overview of the capabilities of a game development tool from Microsoft called Kodu game Lab. Microsoft developed this game development tool to help engage students in learning programming through making and playing games.

### 1.2.2 The results of the first study

To get a deeper understanding of the effectiveness of Kodu Game Lab, a study was undertaken at a New Zealand Intermediate (the students were between 11 to 12 years old) (Fowler & Cusack, 2011a). With the consent of the host school, the children, and their parents, it was possible to get some valuable data on the validity of the proposed benefits of this software. The School was selected because it was relatively close to where the researcher worked, and represented a typical regional Intermediate school in New Zealand. The grade level was selected on the basis that these students had not been formally introduced to programming concepts, and this would ensure that the benefits of any prior knowledge would not distort the results of this study.

### 1.2.3 The results of the second study

The information collected in the first study provided a basis for continuing the study. The feedback from the reviewers was generally positive, however, to be able to generalise these findings, more studies would be needed. Therefore, with the help of Microsoft Fuse Labs, it was possible to recruit students from another Intermediate school (middle school) in New York. Furthermore, while the results of the first study were very positive, the researcher wanted to see how effective these technologies would be at in a secondary school environment. Therefore, with the assistance of Microsoft Fuse Labs, it was possible to recruit a High School in New Zealand and a High School in the United Kingdom (Fowler, 2012b).

These studies provided the researcher the opportunity to further develop, refine and polish some research skills. Through the assistance of Microsoft and AUT, the researcher was able to meet many of the foundation authors on this field of study. This was an invaluable experience. This research while concluding that the tool (Kodu Game Lab) was beneficial, the study had some limitations. One of the biggest limitations is that the study did not include a control group. According to the reviews, without a control group, it was not possible to identify that these changes in perception and understanding would not have occurred without the treatment. However, although this data would have been interesting, the practicality of going into one classroom and telling them that they will be making video games and then going to the next classroom and telling them they will not be making video games, made this very difficult. Although, it may have been possible to undertake a longitudinal approach, and study the treatment group one term and the control group the following term, this still would have been problematic. A typical school term may have added confounding variables that could have possibly interfered with the result. Moreover, given the researcher had a fulltime teaching job, supporting a long-term study would have been very challenging to manage.

While the result of this study was a valuable learning experience, the challenges with implementing a control group and the difficulty finding more tangible indicators of cognition, the study of Kodu Game Lab was concluded.

## 1.3 THE STRUCTURE OF THE THESIS

This thesis is structured to provide the reader with the motivations for this research (Chapter 1). Chapter 2 reviews the literature and academic studies on the use of video games in education. Chapter 3 provides an overview of the methodologies used in these studies and concludes with the methods chosen for this study. Chapter 4 includes the results of the findings from the two studies that were conducted. Chapter 5 discusses the implications of these findings and reviews the contribution that this thesis has made. Chapter 6 summarises the research findings and answers the research questions. The avenues for future research will also be discussed in Chapter 6, which will be followed by a concluding statement.

<div align="center">**Chapter Two**</div>

<div align="center">**LITERATURE REVIEW**</div>

## 2.0    INTRODUCTION

Throughout the last century, researchers and educators have attempted to understand the process and potential of human learning. The ambit of the research to date has attempted to understand the learning process, establish learning theories, and understand the ideal conditions for learning.

One of the challenges with trying to understand human learning and knowledge acquisition is that these are ambiguous concepts and are, therefore, challenging to define (Hamilton & Ghatala, 1994). Another challenge with researching the activity of learning is that it can be difficult to quantify if any actual learning has occurred as much can be hidden and not demonstrated throughout the duration of the experimental phase. While it is possible to measure if a new skill has been learnt through performance, it is more difficult to measure changes in knowledge or attitudes (Hamilton & Ghatala, 1994). Bloom (1956) suggests that the purpose of the learning process is to acquire new skills, knowledge, and/or attitudes.

An accepted learning theory for games-and-learning research is what Gee (2003; 2004) referred to as situated cognition. Situated cognition describes human learning, thinking, and problem solving as being embodied within a context (Kirshner & Whitson, 1997). According to this theory, humans also learn through active social experiences and critical interpretation of experiences, and this is enhanced through personal reflection and interpersonal discussion (Gee, 1997; Gee, 2004b; Shute & Kim, 2012).

## 2.1    LEARNING THEORIES

To help understand the learning process, a number of theories have been developed. Although there is considerable debate as to the relevance of some of these theories, they still provide a basis for helping the understanding of the many possible explanations of how humans learn.

There are four approaches to categorising learning theories: behaviourist, cognitivist, humanist, and social/situational (Greeno, Collins, & Resnivk, 1996; Merriam and Caffarella, 1991).

### 2.1.1   The behaviourist view of learning

The behaviourist view is based on the belief that knowing or knowledge is an organised collection of associations and components of skill (Greeno, Collins, & Resnivk, 1996). This view is based on the understanding that learning is a process where associations and skills are acquired through stimulation and reinforcement. Further, transfer occurs when a behaviour that has been learnt in one context can be used in another context (Greeno, Collins, & Resnivk, 1996).

The behaviourist approach views learners as passive recipients of knowledge (Skinner, 1968). Skinner (1968) believed that learning is a result of positive reinforcement. Moreover, negative reinforcement can also strengthen behaviour when the negative condition is avoided as a consequence of that behaviour. Skinner's (1968) work, although initially based on rats, identified that this theory could also be applied to humans. Behaviourism is underpinned by three main assumptions: first, learning is evidenced by changes in behaviour, second behaviour is shaped by the environment, and third, reinforcement is central to the learning process.

### 2.1.2   The cognitive view of learning

The cognitive view of learning emphasises interaction as a way of developing a general understanding of the domain (Greeno, Collins, & Resnivk, 1996). The cognitive theorists believe in the importance of active assimilation of learners through the accommodation of new information into existing cognitive structures (Piaget, 1970).

The Gestalt cognitive approaches are based on the idea of grouping of stimuli cause humans to structure or interpret a problem in a certain way (Kohler, 1947; Koffka, 1935). The three core principles of Gestalt theory are that learners need to be encouraged to discover the nature of a topic or problem, gaps, incongruities are an important stimulus for learning and instruction should be founded on the laws of organisation (proximity, closure, similarity, and simplicity) (Wertheimer, 1924). What is also pertinent is the concept that the learning has different requirements at different times and may have individual and unique interpretations in different contexts.

The constructivist view of learning is as an active process which is unique to each individual. Learning is a process of constructing conceptual relationships and meaning from information and experiences that have already been experienced. Piaget (1970) suggested that as children develop, their capacity to assimilate and accommodate more advanced theories and concepts also increases. Accordingly, Piaget (1970) children are born with basic mental structure on which all subsequent learning and knowledge are based. The three basic components of Piaget's theory (1952) are schemas, adaption processes, and the stages of development. According to Piaget, (1952) a schema is a cohesive, repeatable action that is tightly interconnected and governed by core meaning. Piaget (1952) believed that the process of intellectual growth is a process where the learner adapts to the environment. The process of adaption as viewed by Piaget (1952) is through a process of assimilation, accommodation, and/or equilibration. Piaget (1952) suggests that the process of assimilation involves using existing schema to deal with a new object or situation. Accommodation is when the existing schema could not assimilate the new object or situation. Equilibration is the internal force that motivates the learning process (Piaget, 1952).

Another school of thought in the cognitive domain is the theory of learning by experience. Based on the work of Lewin (1935), Kolb (1984) provided a descriptive model of the adult learning process. The suggestion is that adult learners develop concrete experience and then through the process of reflecting on this experience; the abstract concepts are then conceptualised, and through active experimentation, these concepts form concrete experience. Although disputed (Jarvis, 2006; Seaman, 2008), Kolb's theory still has supporters (Honey & Mumford, 1992; Merriam, Caffarella, & Baumgartner, 2007; McCarthy, 1987).

### 2.1.3 The humanist view of learning

The humanist perception of a human being is that humans behave intentionally, and this behaviour is based on core values (Kurtz, 2000). This view which was heavily influenced by psychology (Maslow, 1987; Rogers, 1980). Maslow (1987) was interested in the personality of the individual learner and how this affected their motivation in achieving self-actualisation.

### 2.1.4 The social/situational view of learning

Social/situational learning is based on the premise that people learn through observation and through interacting with others (Merriam & Caffarella, 1991; Vygotsky, 1978). The social/situational view considers knowledge as an idea or concept that is distributed amongst people and the environment they work or live in. One of the recent adaptations of the social view of learning is situated learning (Lave, 1993; Lave & Wenger, 1991). Situated learning is an "attempt to infuse careful case studies with concepts of cognitive science, while contrasting learning in the field with thought processes taught and measured in Western schools." (Gardner, 1985, p. 256). This view originates from an anthropologist perspective of how communities and societies passed on knowledge and understanding onto new members or future generations.

Within the context of learning languages Gee (2004b) asserts that humans frame an understanding and derive meaning of words based on past experience. According to Gee (2004b) these meanings can evolve over time based on experience or interactions with "more advanced peers and adults" (p. 55). Although, it is not clear if the interactions necessarily need to be with more advanced peers, these interactions help frame meaning in the use and new uses of a word. Thus according to Gee (2004b) the meaning (of language) is derived from a context that is dependent on the situation.

According to Greeno et al. (1998) the situated perspective "can provide a synthesis that subsumes the cognitive and behaviourist theorists." (p. 5). Greeno et al. (1998) posit that the situated perspective focuses on systems that are "larger than behavior and cognitive processes of an individual agent" (p. 6). The systems that Greeno et al., (1998) suggest that are important to human learning include the behavioral and cognitive processes **and** the "social, material and informational environments as contexts in which the behavior occurs." (p. 6).

Wilson (2002) suggests that there are (at least) six interpretations of the term situated cognition: cognition is situated, cognition is time pressured, humans off-load work to the environment, the environment is part of the cognitive system, and off-line cognition is body based. One of the foundation views of embodied cognition is that cognition is situated (Clark, 1997). This view suggests that cognition is a situated activity. Situated cognition "takes place in the context of task-relevant inputs and outputs. That is, while a cognitive process is being carried out, perceptual information continues to come in that affects processing, and motor activity is executed that affects the environment in task relevant ways." (Wilson, 2002, p. 626). The limitation of this view is that it appears to exclude any cognitive activity that takes place in the absence of a task. This would exclude the cognitive process of visualisation (which could be situated). Another view of situated cognition is that cognition is time pressured (Verplanken, 1993). The view that the pressure (perceived or induced) of time to complete a task is the determinant in cognitive architecture (Wilson, 2002). However, the limitation of this view is that while some cognition is dependent on real-time situated cognition (driving a car), not all cognition is. Furthermore, not many tasks need to be completed 'on-time' and the completion time is entirely at the discretion of the cognisor (although the task may be late). Although, the exception may be playing sport, driving in peak hour traffic, or playing video games. Another interpretation of situated cognition is that we off-load cognitive work onto the environment. This interpretation suggests to deal with situations where the cognisor is overloaded, one solution is to off-load information or data that is not currently required to the environment (books, notepads, and so on) to avoid memorising that information at that particular time (or ever - if possible). Another interpretation of situated cognition is that the environment is part of the cognitive system. An extreme interpretation of this view is that cognition is not an activity of the mind, but also includes the mind, the body, and the environment (Greeno & Moore, 1993). This view suggests that the "forces that drive cognitive activity do not reside solely in the head of an individual, but instead are distributed across the head of the individual and the situations as they interact." (Wilson, 2002, p. 630). Another interpretation of situated cognition is that cognition is for action. This interpretation is based on the view that considers cognitive mechanisms in terms of their function in serving adaptive activity. This view appears to incorporate more of an embodied cognitive view than the previous interpretations listed. The final interpretation is that, off-line cognition is body based. This broader view of cognition recognises that many cognitive activities make use of sensor-motor functions.

While several of these theorists disagree about which learning theory is correct, these views provide valuable contributions to the understanding of how humans learn.

These broad definitions of the term 'learning' make any discussion on the topic more challenging due to the different interpretations and use of the term. However, this paper will adopt the interpretation of Gee (1997; 2004b) that humans learn through active social experiences and critical interpretation of experiences, and this is enhanced through personal reflection and interpersonal interaction.

## 2.2    MEASURING LEARNING

According to Thorndike (1931), most humans require multiple exposures to a new concept before new knowledge can be retained. Furthermore, the acquisition of this knowledge can also depend on the frequency and duration of each exposure as well as the delay between each period of time (Thorndike, 1931). For example, when an adult learns a new language, it may take several attempts to acquire new vocabulary. The frequency and duration of exposures to new knowledge will generally depend on the individual, the method of instruction, and the environment where the new knowledge and practice took place (Thorndike, 1931). Furthermore, these variables potentially influence the length of time the new knowledge will be retained. However, if this vocabulary is not used, or not used very frequently after it has been acquired, then it may be forgotten. Siang and Rao (2003) assert that many theorists believe that the information that is stored in long-term memory is never lost. While these authors do not identify the theorists that they mention, it is an assertion that needs more evidence. Jaeggi, Buschkuehl, Jonides and Shah (2011) provided evidence that after a six-month break from training, the knowledge was retained. There is uncertainty as to the frequency, and the duration of the exposures to a concept for the knowledge to be semi-permanent or permanent. The debate about what, how much, and how long information is retained in the human brain is beyond the scope of this paper. However, from the literature reviewed, it is clear that most humans require more than a single exposure to a concept to learn it. Further, acquiring expert levels of performance requires extended practice and/or preparation (Cianciolo et al., 2006).

Video games incorporate both cognitive and physical activity, it is therefore important to understand the; skill learning, perceptual learning, motor learning, conceptual learning, and procedural learning that may result in playing them. According to Ackerman (1988), skill learning is the process of learning a task to give accuracy, speed and performance after a high degree of practice. Such skills may be perceptual, cognitive, motor or combination of any these skills. Perceptual learning is the process of learning through changes in human perception (Gibson & Pick, 2000). Humans perceive through their senses (auditory, visual, touch, smell, and taste) (Fielder, 1993). Motor learning is the process of learning through physical actions required in performing a task (Schmidt & Wrisberg, 2004). Conceptual learning within a video game consists of learning how to play the game (the explicit and implicit rules/game mechanics) and the implicit (and explicit) concepts (Klabbers, 2009; Squire, Barnett, Grant, & Higginbotham, 2004). Procedural learning is the process of learning a concept or skill through repetition (Beaunieux et al., 2006).

## 2.3    THE POTENTIAL OF DIGITAL LEARNING

The advent of low-cost computing has increased ownership of personal computers in the last twenty years (Harchaoui & Tarkhani, 2004). A result of these decreasing costs has seen a significant increase in household ownership of personal computing devices and video game consoles for entertainment, education and enterprise. The sale of video games represented $40,000,000,000 in annual revenue in 2011, which is almost double, the revenue of the movie industry ($27,000,000,000) (Entertainment Software Association, 2011; Global movie ticket sales hit record high, 2008).

The increased ownership of personal computing equipment has seen a significant increase of ownership and use of video games, which is boosted by the innovations of video game console technology (Morris, 2007). The increasing interest and use of video games for entertainment has motivated researchers and educators to evaluate the potential of using video games as tools for learning and education.

Prensky (2001) refers to the current generation of learners as "Digital Natives" (p. 65), as in most cases (especially in developed economies) these learners have grown up with easy and almost immediate access to a range of digital video game devices. This generation of learners have played video games frequently and have potentially acquired many skills from playing and watching many hours of gameplay. The skills that these learners may potentially have include (Akilli, 2007):

•      dealing with considerable amounts of data in a relatively short timeframe;

- finding and using multiple ways and means to acquire information;
- finding answers to their own problems through various communication networks;
- persistence when faced with new situations; and,
- learning through trial and error and/or self-discovery.

According to Bransford, Brown, and Cocking, (1999) technology (and therefore video games) can be used to: bring real-world problems into the classroom, provide effective scaffolding, increase the opportunity for a learner to receive feedback, and to build both local and global communities of learners.

## 2.4    DEFINING GAMES, VIDEO GAMES AND SIMULATIONS

Although there does not appear to be full agreement on the definition of what is a video game, the common elements are:

- an element of fun;
- the opportunity to play;
- established rules;
- a quantifiable outcome;
- the chance of winning or losing; and
- an element of competition.

The literature uses a number of terms to explore the concept of electronic games that are played on a range of computing and communication devices. The terminology used includes – videogame or video game (Gee, 2003; Squire, 2008), digital games (Prensky, 2001), and computer games (Malone, 1980; Prensky, 2001). In the context of this paper, the term video game will be used to describe an electronic form of a game that is played on personal computers, portable devices, communications devices, or video game consoles. This includes video games that are played using a computing or communications device and video games that are hosted on private and public computer networks (including the Internet).

It is also beneficial to differentiate video games from computer-based simulations. A simulation is an interactive abstraction of some real life activity or construct (Heinch, Molenda, Russell, & Smaldino, 2002). Although simulations incorporate many elements of games, they rarely integrate all of them (Heinch et al., 2002).

## 2.5    THE IMPORTANCE OF PLAY

Play is a powerful mediator for learning and socialization (Blanchard & Cheska, 1985; Csikszentmihalyi, 1990; Huizinga, 1950; Provost, 1990). Play is not only a preparatory exercise for physical development (Groos, 1901), but it can also be seen as an assimilation of reality (Piaget, 1970). Moreover, pretend or symbolic play contributes to learning by supporting the development of metacognitive or self-regulatory skills, which are essential to the development of problem solving and creativity (Whitehead, 2004). Vygotsky (1962) also understood the importance of play and its effect on developing a sense of control and self-regulation. During play, learners construct their own level of challenge. Vygotsky (1962) posited that through language children make sense of the world around them. Language serves the purpose of regulation over cognitive processes such as memory and thought.

There have been several attempts by academics to describe what play is. Scales, Almy, Nicolopulou, and Ervin-Tripp (1991) described play as an absorbing activity which people engage in with enthusiasm. Whereas Csikszentmihalyi (1981, p. 14) describes play as "a subset of life, an arrangement in which one can practice behaviour without dreading its consequences." Although there does not appear to be any absolute agreement on the meaning of the word 'play,' the general consensus is that it is generally voluntary, is intrinsically motivating (usually through an embedded recognition or reward system) and includes a form of invented or distorted reality (Blanchard & Cheska, 1985; Csikszentmihalyi, 1990; Huizinga, 1950).

## 2.6    VIDEO GAMES AS TOOLS FOR LEARNING

Based on existing learning theory, video games have the potential to provide valuable learning experiences. Video games provide an effective medium for learning through difficulty of tasks, motivation and arousal, feedback, variability, and the support of failure (Gee, 2003). Learning from a video game can involve learning the game mechanics (the rules intended to produce gameplay) (Adams, 2010), as well as learning other aspects external to the game (problem solving, spelling, geography, mathematics, languages, and so on). Moreover, through playing video games it is possible to improve perceptual-motor skills including: manual dexterity, hand-eye coordination, reaction time, and fine motor ability (Drew & Waters, 1986; Green, Li, & Bavelier, 2010; Staiano, & Calvert, 2011).

Furthermore, repeated or prolonged exposure to action video games can lead to improvements in decision-making and the allocation of cognitive resources (Green, Pouget & Bavelier, 2012).

### 2.6.1 Task difficulty

Most video games utilise the principle of introducing small incremental increases in task difficulty. As players progress through each level of the game, they are usually required to learn new skills that will assist the player in the later stages of the game (Gee, 2003; Green & Bavelier, 2008). This progressive and staged teaching technique is not unique to video games. However, what is unique to video games is that the user can proceed at their own pace and only when they have mastered the skill they are required to learn, then they can proceed to the next stage. It has also been argued that accurate and regular practice is essential for the development of a given task (Gibson & Gibson, 1955; Gee, 2003).

### 2.6.2 Motivation

Another consideration that may influence learning is motivation (Ackerman & Cianciolo, 2000; Prensky, 2002). Hull (1943) defined motivation as the "initiation of learned, or habitual, patterns of movement or behavior" (p. 226). Hull (1943) postulated that instinctive behaviours usually satisfied primary needs and that learning occurred when these behaviours were ineffective. Further, learning also occurred when secondary reinforcers were present (for example, working to earn a financial reward). When combined with a primary need (food) acquired secondary reinforcing power (money buys food).

An implicit feature of the zone of proximal development (ZPD) (Vygotsky, 1978) is motivation. If a learner is within the ZPD, (Vygotsky, 1978) they will be more motivated. Tasks that are too easy or too difficult can lead to lower levels of motivation, and this in turn can lead to lower levels of learning.

In many video games, the user usually selects the level of difficulty they are comfortable with. The level of difficulty can be increased or decreased by the user through the user options when starting the game, or by the player choosing a particular character or guild/clan in a role-playing game (RPG). The difficulty level can also be increased (or decreased) by the game system which is based on a player's ability (or inability) to play the game.

Video games offer positive reinforcement for achievement (Skinner, 1968) through reward points, bonus levels, and high scores, which may have an influence on the player's motivation (Gee, 2003). Although Kohn (1999) argues that there are minimal long-term benefits of rewards, the short term benefits of the rewards provided in video games appear to have an influence on player motivation (Gee, 2003).

### 2.6.3 Feedback

There is some debate as to the value and importance of feedback in the learning process. Some research has indicated that feedback is necessary for learning (Herzog & Fahle, 1997; Seitz, Nanez, Holloway, Tsushima & Watanabe 2006). Ball & Sekuler (1987) and Karni & Sagi (1991) suggest that feedback is not necessary. However, Bransford, Brown, and Cocking (1999) assert the value and importance of providing feedback to learners. Feedback is important for enabling learners to focus on understanding. Feedback is also an important element in facilitating transfer (Bransford, Brown, and Cocking, 1999).

Video games provide the player with feedback that can be presented in different ways. One common type of feedback used in video games is implemented through rewarding the player (based on Skinner, 1968). The utility or value of the reward is not universal. What one person may value as being extremely beneficial may not appeal to another. Some players value in-game status symbols that are only obtainable through effective gameplay (like high scores or avatar enhancements), whereas other players may value being part of a particular guild or increasing their social status within that guild (Steinkuehler & Williams, 2006). Another common type of feedback used in video games is the implementation of negative feedback. This provides information to the player that they are not controlling the avatar correctly, either through visual or auditory feedback or through the termination of the game (Juul, 2009). Although traditional methods of instruction provide feedback, one of the benefits of video games is that the feedback is immediate. Anderson (1985) notes the importance of immediate feedback. In a traditional classroom setting, it is often difficult (if not near impossible) to provide every student with immediate feedback. The delay in providing students with feedback could result in students learning something incorrectly (Pellegrino, & Hilton, 2013). According to Roscoe & Chi (2007) feedback that provides an indicator as to why the practice is incorrect is more valuable than feedback that just highlights the errors (Pellegrino, & Hilton, 2013; Shute, 2008). Video games have the ability to provide direct and immediate feedback if the user has made either the correct choice or the incorrect choice.

### 2.6.4 Variability

Video games provide variability in task and input. If the skill to be learnt is presented in a highly variable context then participants learn to recognize these tasks in a more flexible way (Brady & Kersten, 2003). Moreover, through presenting the task in a variety of ways, this should potentially reduce learner boredom or frustration.

Video games frequently present the player with the same repetitive task. However, through presenting this task in a different context, player boredom is reduced, and engagement is potentially increased.

### 2.6.5 Supporting failure

In video games, there is an opportunity to try (and try) again without any serious repercussion (Gee, 2003; Juul, 2009). They offer the opportunity to learn through trial and error in a 'safe' or 'low risk,' environment (Gee, 2003; Grammenos, 2008), also permitting the user to practice in safety (Prensky, 2003). While Prensky's (2003) attention was primarily on the health and safety of the user, video games also provide an equally effective opportunity for users to 'fail' in a safe environment. Video games provide the user the opportunity to fail without any negative personal or social impact. Moreover, massively multiplayer online role-playing games (MMORPG) users get to form social relationships, or take on roles that they may feel uncomfortable with or incapable of forming in the real world (Steinkuehler & Williams, 2006).

In video games, the concept of failure is multifaceted (Juul, 2009). Failure at playing video games can be as straightforward as not controlling the avatar correctly, or as serious as losing the game (Juul, 2009). However, any loss is temporary as the user can restart or replay the game (or level) at any time.

### 2.6.6 Good video games

One of the key motivations for using video games as tools for learning is the potential for making learning fun (Gee, 2003; Malone, 1980; Prensky, 2001). Some of the essential elements of a good video game are (Malone, 1980; Malone & Lepper, 1987) the challenge, elements of fantasy and player curiosity. For a video game to be challenging, it needs to "provide a goal whose attainment is uncertain" (Malone, 1980, p. 162). For a challenge to be effective, the goal needs to be clear and consistent. An uncertain outcome provides a measure of unpredictability to a video game through (Malone, 1980):

• variability in the level of difficulty;

• having several categories of goals (and rewards);

• hidden information; and

• a level of randomness or unpredictability.

Fantasy can lift us from the milieu of everyday life and good video games based on fantasy can allow users the opportunity to escape from their reality (McGonigal, 2012). Further, good video games can stimulate the player's interest by providing a simulated environment that has an optimal level of complexity (Gee, 2003; Piaget, 1952). Curiosity is one of the factors that motivates people to learn (Johnson & Johnson, 1985).

### 2.6.7 The challenges of using video games as tools for learning

There are some concerns that video games are addictive (Chan & Rabinowitz, 2006; Fisher, 1994; Spain & Vega, 2005; Phillips et al., 1995) and that the content of the games teaches anti-social behaviour and reduces the threshold for violence (Huesmann, Moise, Podolski, & Eron, 2003). However, these studies fail to recognise the clinical definition of the term addiction. According to O'Brien (2008, p. 3278), addiction is a "compulsive drug-seeking behaviour." Although the subjects studied by Fisher (1994), Phillips et al., (1995) and Chan & Rabinowitz (2006) reported some of the symptoms of addiction, these subjects did not report many (if any at all) of the genuine symptoms of withdrawal.

While it is possible that many video games are extremely compelling and in some cases the user may have played a video game instead of attending to other tasks (going to work, spending time with family, and so on), the number of reported cases of excessive (or addictive) video game use (Spain & Vega, 2005; Yee, 2002) represents a relatively small minority when compared to the total number of people who play video games.

Another challenge is that video games may be seen as a disruption to the learning process (Kirriemuir & McFarlane, 2003). Unlike a regular structured classroom setting, when playing a video game the user can terminate or pause gameplay at any time. While this potentially breaks the player's immersion and the gameplay experience, it does give the player the opportunity to leave the game temporally, recover from a major mistake, consider an alternative solution, or play a different game (Adams, 2010).

With the advent of the Internet, it has become easier to find solutions to in-game challenges. These walkthroughs or cheats provide the user with a detailed guide on how to solve a particular challenge. Just like study guides or abridged versions of classic texts, these walkthroughs provide the user a quick or easy solution through the many challenges in a video game. It is possible that this method of study may not afford the user a full experience.

Finally, not everyone likes playing video games (Holt, Guram, Smith, & Skinner, 1992). If video games are introduced into a formal learning environment, this issue needs to be considered.

Clark (1994) asserts that electronic media fails to influence learning and is not directly responsible for motivating learning. While this view is contrary to the results of empirical research, it is still a view that needs to be considered.

### 2.6.8   Transfer of learning

The ability to use skills acquired in one context in another context indicates that the learning was transferred (Woodworth and Thorndike, 1901). According to Perkins and Salomon (1992), "near transfer refers to transfer between very similar contexts" (p. 3). Further, far transfer "refers to transfer between contexts that, on appearance seem remote and alien to one another" (p. 3). There has been considerable interest and debate into what skills are transferred from a video game to external environment (Bavelier et al., 2011). Part of the challenge has been finding a consistent understanding or interpretation of what constitutes a video game. As there are a vast array of different video game genre and gaming devices, finding a common definition is not easy. Furthermore, there are several different categories of video game. For example, there are commercial, educational, therapeutic video games, and video games that combine elements of these categories (for example, video games that are both entertaining and educational). However, while there is some consensus that video games have an effect on the user, another challenge is whether this effect is positive or negative. According to (Bavelier et al., 2011) "the many games that are effective teachers of perceptual and cognitive skills can also be harnessed to produce maladaptive effects on brain and behaviour." (p. 764). According to Lieberman, Beily, Thai and Peinado (2014), video games are effective tools to teach transfer because "video games can adapt to the player's changing abilities and can adjust the difficulty level in response to player success or failure… so gameplay does not become too easy and boring or too difficult and frustrating." (p. 192). This assertion (which was based on a review of the literature) implies that video games have the potential to retain the player in the ZPD (Vygotsky, 1978).

While it is possible that some positive elements can be transferred from a video game, there is also the potential for some negative elements that can be transferred. However, Bavelier et al. (2011) suggest that "violent video games alone are unlikely to turn a child with no other risk factors into a maniacal killer. However, in children with many risk factors, the size of the effect may be sufficient to have practical negative consequences." (p. 764).

## 2.7    STUDIES OF VIDEO GAMES AS TOOLS FOR LEARNING

Despite some of the noted challenges of researching the educational benefits of video games, there have been a number of studies that have investigated this topic and generally found a positive result.

### 2.7.1   The perceptual and cognitive benefits of playing a video game

One of the potential benefits of playing video games is improved spatial and sensory-motor skills (Greenfield, De Winstanley, Kilpatrick, & Kaye, 1994; Orosy-Fildes & Allan, 1989). Fitts (1964, p. 283) states that "skilled performance is dependent on discrete or quantized processes." Therefore to understand discrete perceptual motor responses, the measurement of reaction time, movement time and response accuracy are needed. Furthermore, Fitts (1964) states that these measures will contribute to gaining "an understanding of serial and continuous communication and control skills on one hand and to an understanding of organization of thinking decision making, and verbal behavior on the other hand." (p. 283). Orosy-Fildes & Allan (1989) suggest that a measurement of visuo-motor skills is the time to react to various stimuli. Orosy-Fildes and Allan (1989) conducted a study where the participants who played video games had faster reaction times than those who did not.

Another potential benefit of playing video games is the ability to gather and manipulate spatial information (Feng, Spence, & Pratt 2007). McClurg and Chaille (1987) concluded that children who were trained on video games performed better than those who had not. Moreover, the younger children that played the video game performed better (at the Mental Rotation test) than older children without the same experience.

It is also possible that playing video games can improve visual attention (Greenfield et al., 1994). Visual attention is the ability to focus on a specific object and ignore others still in the field of view. Greenfield et al., (1994) demonstrated that the participants that played video games were more efficient and more effective at finding and focusing on a given object.

However, given the premium that video games put on spatial and sensory motor skills, these findings are probably no surprise. Moreover, in many video games, to advance to higher levels, the acquisition and improvement of spatial and sensory skills is usually a prerequisite condition. These findings are important and have established that video games can be useful for improving spatial and sensory motor skills. In professions such as surgery, the military, or motor racing where these skills are extremely valuable, the use of video games to advance them could be very useful.

### 2.7.2   The benefits of video games in formal education

There has been considerable interest in the benefits of using video games in education. The following Table 2.1 summarises the current literature on the application of video games used in a formal educational context. Table 2.1 identifies if the game was a commercial video game (CVG), or an educational video game (EVG). A CVG is a video game made for a commercial (non-educational) market and an EVG is a video game specifically made for educational purposes. The following table identifies if the evidence of the research indicated if game supported learning. Support for learning was operationalised as having either cognitive, physical, or emotional benefits.

| Author(s) | Type | Domain | Support Learning | Results |
|-----------|------|--------|------------------|---------|
| Abbey (1993) | EVG | Problem solving | Yes | Video games provide for the transfer of learning significantly more than traditional instruction. |
| Adams (1998) | CVG | Urban geography | Yes | Video games increase motivation and teach students about the role of urban planners. |
| Alkan & Cagiltay (2007) | CVG | Problem solving | Yes | Video games promote trial-and-error strategies for problem solving. |
| Anderson (2005) | CVG | Business management | Yes | Team dynamics influenced students' game playing performance and their affect toward game. |
| Bai, Pan, Hirumi, & Kebritchi (2012) | EVG | Mathematics | Yes | The video game had a significant effect on student academic performance. |
| Barab et al. (2007) | CVG | Science education | Yes | The video game provided significant gains in student performance. |
| Barab et al. (2009) | EVG | Environmental Science | Yes | Participants working in pairs and used the video game did significantly better than the other participants. |
| Barab, Pettyjohn, Gresalfi, Volk, & Solomou (2012) | EVG | Writing | Yes | The students in the game-based class demonstrated significant learning gains and reported higher levels of engagement. |
| Barker, Brinkman, & Deardorff (1995) | EVG | Divorce adjustment | Yes | The video game produced significant improvements in subject knowledge and participants reported positive behaviour improvements. |
| Bartholomew et al. (2000) | EVG | Health education | Yes | Video games increased subject knowledge for older children. |
| Barzilai, & Blau (2014) | EVG | Finance/ Mathematics | No | Significant gains in problem-solving were not found through playing a video game. |
| Beale, Kato, Marin-Bowling, Guthrie, & Cole (2007) | EVG | Health education | Yes | Video games improved cancer-related knowledge. |
| Becker (2001) | n/a | Programming | Yes | Games are found to be more effective and motivating than traditional teaching. |
| Bensen et al. (1999) | EVG | Sex education | Yes | Video games are motivating and can increase knowledge related to sex education. |
| Ben-Zvi (2007) | EVG | Business functions | Yes | Affective response was favourable. |
| Betz (1995) | CVG | Engineering | Yes | Video games increase motivation and learning among students. |

| Boot, Kramer, Simons, Fabiani, & Gratton (2008) | CVG | Problem Solving | No | More experience in playing video games may result in better basic cognitive abilities, but does not result in substantive improvements in most cognitive tasks. |
|---|---|---|---|---|
| Brown et al. (1997) | EVG | Diabetes | Yes | The study finds that children can learn about diabetes from playing video games and help change their daily habits. |
| Chen, Hwang, Wu, Huang, & Hsueh (2011) | EVG | The impact of digital game-based learning | Yes | Students held firm intentions to use a game embedded in a social network. Enjoyment was considered to be a key factor. |
| Conati & Zhao (2004 ) | EVG | Mathematics | No | Students learned little from the video game without any external guidance. |
| Dede (2009) | EVG | Environmental science | Yes | Students obtained substantial knowledge and skills. |
| Dickey (2011) | EVG | Literacy | Yes | The game narrative had a positive impact on intrinsic learner motivation, curiosity, plausibility and transfer into pre-writing activities. |
| Dowey (1987) | EVG | Dental health | Yes/No | Children learn best from a combination of teaching and video games. Although they learn about dental hygiene, this does not transfer into change of everyday practice. |

| Forsyth & Lancy (1987) | EVG | Geography | Yes | The adventure game resulted in children learning geographic locations with strong recall. |
|---|---|---|---|---|
| Foss & Eikaas (2006) | EVG | Engineering | Yes | Affective feedback to using the video game was encouraging. |
| Gander (2000) | EVG | Programming | Yes | Video games are effective for teaching specific knowledge. |
| Giannakos (2013) | EVG | Mathematics | Yes | Attitudinal factors affect knowledge acquisition gained by a video game. |
| Holmes (2011) | EVG | Literacy | Yes | The students believed they had learnt some useful spelling techniques and strategies through playing the game. Student engagement increased. |
| Inal & Cagiltay (2007) | CVG | Social skill development | Not measured | Gender and challenge level in the game influenced students' flow experiences and game-playing behaviours. |
| Kambouri, Thomas, & Mellar (2006) | EVG | Literacy | Yes | The video game was engaging, and learners made significant literacy gains. |
| Khalili et al. (2011) | n/a | Immunology & Game Design | Yes | The game design goal provided the opportunity for the participants to explicitly express their understanding. |
| Lim, Nonis, & Hedberg (2006) | EVG | Science education | Yes | Through using a video game, there was a significant subject knowledge gain. |
| Miller & Robertson (2011) | EVG | Mathematics | Yes | The students that played the video game improved at a significantly faster rate than the control group. |
| Nilsson & Jakobsson (2011) | CVG | Environmental Science | Yes | Video games like SimCity 4 may contribute to improved reflection and problem solving methods. |
| Noble, Best, Sidwell, & Strang (2000) | EVG | Drug education | Yes | Students taught through video games find the experience motivating and want to play the video game again. |
| Pannese & Carlesi (2007) | EVG | Business functions | Yes | Affective feedback from using the video game was unusually high. |
| Piper, O'Brien, Morris, & Winograd (2006) | EVG | Social skill development | Yes | The video game provided an engaging experience. |
| Rai, Wong, & Cole (2006) | n/a | Computer programming | Yes | Video game construction promoted active engagement with the content and increased enthusiasm. |

| Ricci, Salas, & Cannon-Bowers (1996) | EVG | Chemical, biological, and radiological defence training | Yes | Video games increase knowledge retention. |
|---|---|---|---|---|
| Robertson & Miller (2009) | EVG | Brain training | Yes | Video games have a positive effect on mental computational skills. |
| Rosas et al. (2003) | EVG | Reading and mathematics | Yes | Video games increase motivation, and there is a transfer of competence in technology from using them. |
| Schmitz, Czauderna, Klemke, & Specht (2011) | EVG | Information technology knowledge | Yes | The game motivated students to engage with information technology. |
| Shute & Kim (2012) | CVG | Assessment | Yes/No | The potential to learn from playing video games depends on how the player interacts with the game. |
| Squire & Barab (2004 ) | CVG | History, geography, and political science | Yes | The video game can be a powerful tool for engaging learners. |
| Squire, Barnett, Grant, & Higginbotham (2004) | EVG | Physics | Yes | Students using the simulation game performed better in comparison to the control group. |
| Squire & Jan (2007) | EVG | Environmental Science | Yes | Location-based augmented video games have the capacity to engage participants in scientific argumentation. |
| Sung & Hwang (2013) | EVG | Environmental Science | Yes | Participants improved attitudes and learning motivation, but also improves their learning achievement and self-efficacy. |
| Tanes & Cemalcilar (2010) | CVG | Social skills | Yes | Participants who played the video game changed their perceptions of the city they live in. |
| Tao, Cheng, & Sun (2009) | EVG | Business and management education | Yes | The playfulness of the simulation has an important influence on the students continued use of the game. |
| Turnin et al. (2000) | EVG | Nutrition | Yes | Video games can teach students about nutrition and lead to a significant change in daily habits. |
| Tuzun (2007) | EVG | Science education | Yes | Thirteen categories of motivational elements to play the video game emerged. |
| Vogel, Greenwood-Ericksen, Cannon-Bowers, & Bowers (2006) | EVG | Mathematics and language arts | No | The video game did not promote learning for deaf children. |
| White (1984) | EVG | Physics | Yes | Playing the game improves students' problem-solving expertise in physics in relation to how forces influence motion. |
| Whitehall & McDonald (1993) | EVG | Electronics | Yes | Motivation plays a substantial role in game-based learning materials. |

| Wiebe & Martin (1994) | EVG | Geography | No | There is no difference in learning geography facts and attitudes, between video games and teaching activities not on a computer. |
| --- | --- | --- | --- | --- |
| Wrzesien & Raya (2010) | EVG | Science | No | Playing the video game increased student engagement and enjoyment. However, there were no differences in student learning. |
| Yip & Kwan (2006) | EVG | English vocabulary | Yes | The video game has a significant positive effect on learning. |

*Table 2.1 An overview of studies on the effectiveness video games have on education.*

The majority of the cited examples in Table 2.1 found that video games generally had a positive impact on student engagement and motivation. Less than 11% of the cited examples found that video games did not promote learning or were not as effective as more traditional methods, such as using textbooks (although very few directly compared the two methods). Of the cited examples 19% were commercial video games. The studies that investigated commercial video games, the majority (over 85%) included studies of semi-educational games like Sims. The remainder included studies of Massively Multiplayer Online Role Playing Games (MMORPG) like Lineage (Steinkuehler, 2005).

## 2.8   EDUCATIONAL AND COMMERCIAL VIDEO GAMES

The majority of the literature reviewed above has primarily focused on video games that have been made for, or tailored to educational purposes. These educational or serious video games potentially provide educators the ability to use video game technologies developed for specific learning outcomes.

However, while there has been some improvement in the quality of these educational video games, many still lack the production values that commercial video games have (Ryan, Costello, & Stapleton, 2012). This is probably no surprise given the resources and investment made in the mainstream commercial market (Ryan, Costello, & Stapleton, 2012). In video game titles such as Halo 4 (343 Industries, 2012), publishers have spent multi-million dollar budgets and several years developing these titles (Leone, 2012). The results of these research projects are appealing in terms of how these basic video games have engaged and motivated the learner. As the developers of these educational games did not have the budgets or development teams that the leading game studios had, the production quality of these games is generally not as high as the types of games typically played by their target audience. Therefore, while the researchers discovered that the subjects found these educational video games potentially more engaging or motivating than the tools typically used in the classroom, they also identified some residual dissatisfaction with their perceived lack of quality and poor production values. Further, another challenge with some of the educational video games is that the developers have taken what Bruckman (1999) referred to as a 'chocolate covered broccoli' approach. This is when a video game is presented as a reward for completing the learning outcome. According to Green (2014), a lot of educational video games are structured on the basis that practise makes perfect and thus stripping the game of its full potential.

Although there has been some professionally developed educational software, for example, Citizen Science (Gaydos & Squire, 2010), there is still a gap in the quality and perceived value of these two different products. Given the majority had considerable exposure to professionally designed AAA-rated games, which is a high quality game that is among the year's bestsellers (CA Made Games, n.d) that most of the subjects had played, it is worth considering what learning can take place in a CVG. Gee (2005) suggests that the massively multiplayer online role-playing game (MMORPG) World of Warcraft (Blizzard Entertainment, 2004) provides twenty-first-century skills, such as individual specialisation in cross-functional teams that work together to achieve mutually beneficial goals. In this virtual world "people save for homes, build relationships of status and solidarity, and worry about crime" (Steinkuehler, 2008, p. 614). In this and other similar video games, players are making complex decisions, solving problems (social, numeric, and in physics), making strategic decisions (Gee, 2005; Steinkuehler, 2008), and in some cases managing voluntary teams of other players, that if they existed in the real world would be equal in size to many commercial enterprises. It is therefore possible that through playing these video games, participants not only learn to play the game and improve spatial and sensory motor skills, but also improve their problem solving, leadership, communication, teamwork, and social skills.

There has been an increase in the interest of the effects of playing commercial action video games (see Green & Bavelier, 2003; Green & Bavelier, 2008; Bavelier, Green, Pouget, & Schrater, 2012). One of the main reasons for this interest is that playing action video games (AVG) can lead to an improvement in probabilistic inference (Green, Pouget, & Bavelier, 2010). Repeated exposure to AVG can lead to improvements in decision-making and the allocation of cognitive resources. Repeated and prolonged exposure to playing AVG can improve a general capacity to control top-down attention and learning a new task (Bavelier, Green, Pouget, & Schrater, 2012). The results of the studies that have compared habitual players of AVG with non-AVG player suggest that the long-term players of AVG achieve enhancements in many visual-perceptual skills including: contrast sensitivity, improvements in peripheral vision, improvements in divided attention (for example, Posner cueing, flanker effects, multiple object tracking, enumeration), visual search, change detection, spatial cognition, and executive function (task switching, distractor suppression) (Green and Bavelier, 2003; Green and Bavelier, 2006; Green and Bavelier, 2007; Li, Polat, Makous, and Bavelier, 2009; Posner, Snyder, and Davidson, 1980). Green (2014) suggests that "in most commercial (video) games, the information to be learned is used in many contexts and domains" (p. 39).

### 2.8.1   Ensuring deeper learning through video games

While it is clear that some video games are effective tools for learning, it is also evident that not all video games are effective for learning (Gee, 2004; Green, 2014). According to Mayer (2009), there are a number of principles that are recommended for ensuring deeper learning in multimedia design. To reduce extraneous processing, Mayer provides the following guidelines: ensuring coherence, including signalling, reducing redundancy, providing spatial contiguity, and temporal contiguity. Mayer further provides guidelines for managing essential processing. These guidelines include: segmenting information, pre-training, and using multiple modalities. Mayer also suggests that the following principles for managing generative processing: using multimedia, personalisation, and using a human voice. Table 2.2, provides a further explanation of this theory. Furthermore, while the National

Research Council (2010) found that there is no compelling evidence that that games are effective tools for learning if the learner is left to freely explore unassisted, **through providing initial guidance** and **scaffolding** deep learning is possible (Azevedo & Aleven, 2010; de Jong, 2005).

| Empirical Results | Practical Applications |
|---|---|
| **Multimedia Principle**: Students learn better from words and pictures than from words alone. | On screen animation, slide shows, and narratives should involve both written or oral text, and still or moving pictures. Simple blocks of text or auditory only links are less effect than when this text or narration is coupled with visual images. |
| **Spatial Contiguity Principle**: Students learn better when corresponding words and pictures are presented near rather than far from each other on the page or screen. | When presenting coupled text and images, the text should be close to or embedded within the images. Placing text under an image (i.e., a caption) is sufficient, but placing the text within the image is more effective. |
| **Temporal Contiguity Principle**: Students learn better when corresponding words and pictures are presented simultaneously rather than successively. | When presenting coupled text and images, the text and images should be presented simultaneously. When animation and narration are both used, the animation and narration should coincide meaningfully. |
| **Coherence Principle**: Students learn better when extraneous words, pictures, and sounds are excluded rather than included. | Multimedia presentations should focus on clear and concise presentations. Presentations that add "bells and whistles" or extraneous information that impede student learning. |
| **Modality Principle**: Students learn better from animation and narration than from animation and onscreen text. | Multimedia presentations involving both words and pictures should be created using auditory or spoken words, rather than written text to accompany the pictures. |
| **Redundancy Principle**: Student learn better from animation and narration than from animation, narration, and on-screen text. | Multimedia presentations involving both words and pictures should present text either in written form, or in auditory form, but not in both. |
| **Individual Differences Principles**: Design effects are stronger for low-knowledge learners than for high knowledge learns and for high spatial learners rather than from low spatial learners. | The aforementioned strategies are most effective for novices (e.g., low-knowledge learners) and visual learners (e.g., high-spatial learners). Well-structured multimedia presentations should be created for those they are most likely to help. |

*Table 2.2 The nature and effects of multimedia presentations on human learning (Doolittle, 2002, pp. 2-3).*

## 2.9   LEARNING WHILE PLAYING VIDEO GAMES

Sternberg (1998) asserts that playing video games is a form of interactive learning where participants demonstrate key skills: metacognitive, learning, and thinking. Some of these demonstrations require proxy variables (for example, examinations, interviews, puzzles and so on), to quantify. Klabbers (2009) goes a step further and categorises the types of knowledge a game player gains and has to acquire for successful gameplay, as being explicit and tactical. Further, as playing a video game incorporates both cognition and behaviour, it is also important to consider the perceptual-motor skills including manual dexterity, hand-eye coordination, reaction time, and fine motor ability (Drew & Waters, 1986; Green, Li, & Bavelier, 2010; Staiano, & Calvert, 2011). As noted in section 2.8, playing AVG can lead to improvements in decision-making and the allocation of cognitive resources (Green and Bavelier, 2003; Green and Bavelier, 2006; Green and Bavelier, 2007).

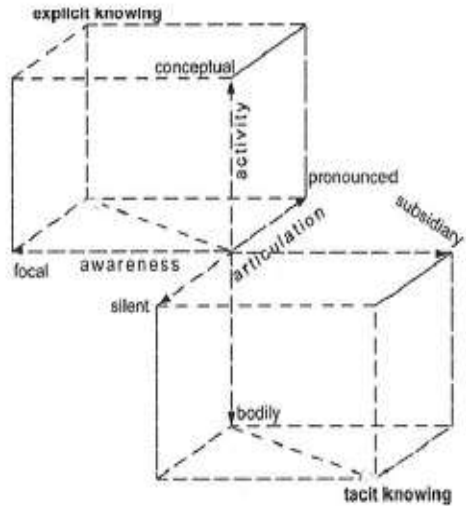

*Figure 2.1 Game learning dimensions (Klabbers, 2009, p. 71)*

In Figure 2.1, (Klabbers, 2009 p. 71) the complexity and multi-dimensional aspect of learning is expressed as a series of interrelating properties. The visualisation is multi-dimensional and is an attempt to represent the complexity of what is learned in gameplay. Klabbers (2009) categorizes human knowledge in three dimensions: awareness, articulation, and activity. Tacit knowledge is difficult to articulate; awareness of this knowledge is subsidiary (Polanyi, 1966) and is closely associated with bodily experience (Engel, 2008). Whereas explicit knowledge is conceptual, awareness is focal and is easier to articulate (Fisher et al., 2006).

Klabbers (2009) goes on to say that gameplay is "experience in action" that involves many concurrent activities. Figure 2.2 (from p. 86) captures the forms of knowing in relation to action and the potential for formulating categories for what may be learned. Each group, therefore, has the potential to develop a set of indicators so that performance can be measured.

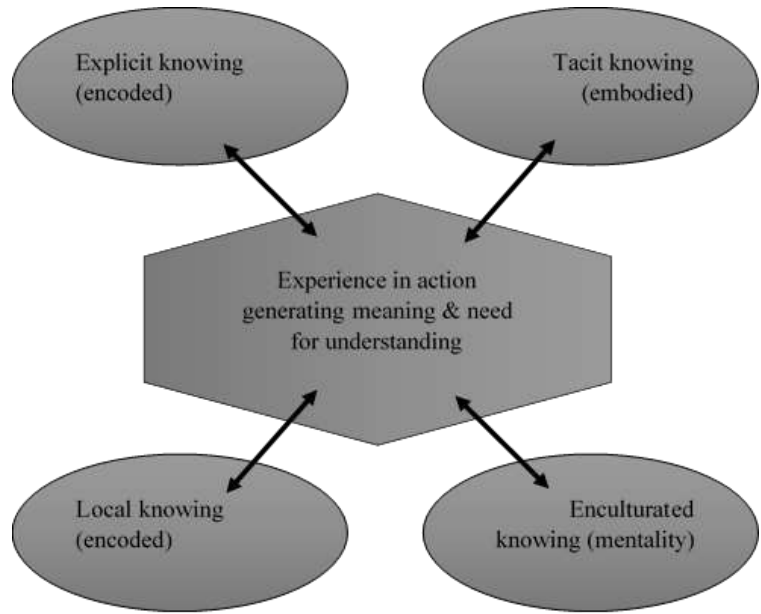

*Figure 2.2 Knowledge learning (Klabbers, 2009, p. 86)*

Learning in these senses embodies a complexity of categories that are present in action, but may not necessarily be recognizable by the action. Learning and action are, therefore, complete units in this approach that may be observed by a completed action. In Figure 2.3 (Klabbers, 2009, p. 92), the variables of capability (dependent) and time (independent) are added to the conceptualisation of game learning to capture the notion that the game players learn multi-faceted skills through interaction (experience in action). The notion that players have skill capability and can improve that capability (skill level) over time suggests that there is potential for beneficial effects. The concept of learning in action is not dissimilar to the concept of situated cognition (Clarke, 1997; Gee, 1997; Kirshner & Whitson, 1997).

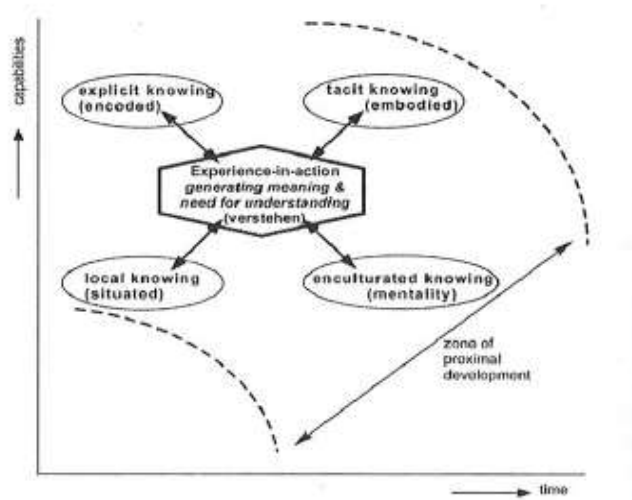

*Figure 2.3 The taxonomy of game learning (Klabbers, 2009, p. 92)*

### 2.9.1 Explicit and tacit knowing

Explicit knowing is the knowledge we can explain (Fisher, Drosopoulos, Tsen, & Born, 2006). It is the knowledge that can be encoded and articulated and expressed in formal language (Klabbers, 2006). It is worth observing that Klabbers (2006) used the term 'knowing' versus the term 'knowledge.' The use of the noun 'knowing' is to indicate an action rather than the physically passive state of knowledge (Klabbers, 2006).

Tacit knowing is the understanding of concepts that are difficult to transfer to another person by means of communication (Polanyi, 1966). Engel (2008) cites examples of activities that involve tacit knowledge such as, riding a bike, playing the piano, or driving a car.

In video games, explicit knowledge is typically articulated in the formal and system rules (Klabbers, 2006). Tacit knowing is typically embodied in the user's mental, social, and physical organisation and cooperation with other players (Gee, 2004a). Furthermore, when playing video games, users have particular preferences for the type and look of an avatar or style of play, and yet they may not be able to articulate the rationale for this preference completely.

### 2.9.2 Local and enculturated knowing

Local knowing refers to the knowledge of the environment around us (Klabbers, 2006; Nasir, 2005). In video games, local knowing is manifest through the user's knowledge of the virtual environment. However, the physical environment may also have an impact on the players' actions (or inaction). Nacke & Drachen (2011) posit that the player's experience is dependent on the in-game (or virtual) experience and also the physical experience which consists of the technical experience, mental state, and the context (Figure 2.4).

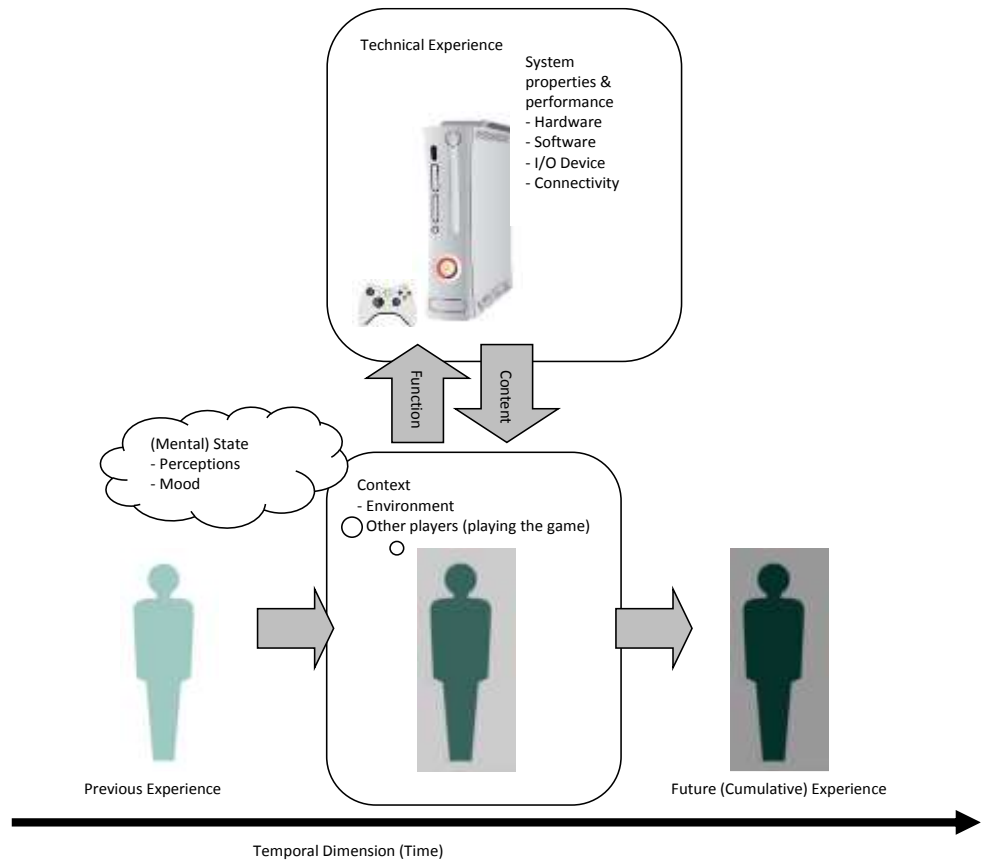

*Figure 2.4 Player experience (adapted from Nacke & Drachen, 2011, p. 6)*

Users can play a video game in a diverse range of physical environments. For example at home on their own, with friends (physically), and with friends, acquaintances, or strangers in a virtual environment. Furthermore, they can play in a public space (with or without friends). Therefore, the knowledge of this environment will vary depending on that user's experience. Furthermore, technical issues such as Internet connectivity, the input/output device, the gaming platform used, or the performance of the software, may have an impact on the playing experience and thus influence the cognitive process of the user.

Enculturated knowledge according to Klabbers (2006) has a direct impact on experience in action. This enculturated knowledge may also have a significant influence on tacit knowing. Vygotsky (1978) stated that much of the cognitive process is culturally situated. The knowledge obtained from cultural influences significantly influences what we know (Lave, 1991; Sfard, 1998). Lave and Wenger (1991) refer to a community of practice, in which communities of like-minded individuals can assist and facilitate learning (Lave, 1991). In video games, cultural knowing is typically manifested through the shared and hidden codes of a group or community of players (Klabbers, 2006).

MMORPGs represent global communities of players that share a common language, behaviour, ethos, and goals (Steinkuehler & Williams, 2006). One MMORPG that has been studied by researchers is World of Warcraft (Blizzard Entertainment, 2004) (see, for example, Steinkuehler & Williams, 2006; Thurau & Bauckhage, 2010). World of Warcraft (WoW) (2004) is currently one of the more popular MMORPGs. According to statista (2012) during the third quarter of 2012, WoW had over 10,000,000 active subscribers (the number of people paying the monthly subscription fee). Although actively discouraged by the developers of the game, some players participate in the game for commercial reasons (a process called farming or gold farming) (Steinkuehler & Williams, 2006; Thurau & Bauckhage, 2010). Although banned by the developers of the game, these professional players play the game for the sole purpose of selling the assets they accumulate (Steinkuehler, 2006). These commercial transactions take place in the real world and will enable the buyer to fast track the acquisition of a particular asset. This practice has caused many participants to avoid interacting with other players from a particular geographic region or demographic because of the wide-spread practice of farming within this community (Steinkuehler, 2006).

### 2.9.3   Communities of practice or affinity spaces

The communities formed within the global virtual worlds of MMORPGs could be described as a community of practice (Lave & Wenger, 1991; Lave, 1991). Wenger and Snyder (2002) provide valuable guidance to what they considered as a community of practice. "Communities of practice are groups of people who share a concern or a passion for something they do and learn how to do it better as they interact regularly" (p. 1). The authors also provide guidance on the key attributes of what they consider the three characteristics of communities of practice: the identity is defined a **shared domain of interest**, members **interact and learn** from each other, the membership only includes active practitioners that participate in **shared practice**. While not all participants that play MMORPGs are part of a community of practice, the participants of specific guilds, leagues, or clans meet the criteria as defined by Wegner and Snyder (2002). However, two siblings, friends, or class mates, or a parent and a child that play a video game together, may not meet the strict interpretation of this definition. Gee (2004) suggests that a more appropriate term for these types of collaborative learning environments is affinity spaces. Gee (2004) asserts that these affinity spaces provide the opportunity to interact and learn in a collaborative domain (or space), but may not necessarily incorporate the formality or shared traditions of a community of practice. The example provided of two siblings, friends, or classmates playing together on the same or shared video game is what Gee refers to an affinity space. In the case where an elder sibling or a parent plays or assists the sibling or child in playing the video game, this provides the opportunity to interact and learn within this virtual or physical space. Gerber, Cavallo, & Marek (2001) concluded that the family is an important part of a child's formal and informal learning.

### 2.9.4   Experience in action

According to Klabbers (2006), explicit knowing, tacit knowing, enculturated knowing, and local knowing, all influence and are influenced by experience in action. Experience in action according to Klabbers (2006) is the production of meaning and the need for understanding. Learning through experience (action learning) is a valuable and meaningful experience for both the individual and the group (Revans, 1980). Experience in action appears to be similar to what other authors refer to as situated cognition (Clarke, 1997; Barab et al., 2007; Gee, 1997; Gee, 2003; Kirshner & Whitson, 1997; Shaffer, Squire, Halverson, & Gee., 2005).

Video games provide for the process of learning through action (and interaction) and reflection (Gee, 2003; Klabbers, 2006; Prensky, 2001). Users are presented with solvable problems or challenges and are provided incentives to solve these challenges.

### 2.10 PROBLEM SOLVING

According to Newell and Simon (1972), "A person is confronted with a problem when he wants something or does not immediately know what series of actions he can perform to get it: (p. 72). Problem solving is a key element of the learning process (Murray, Olivier, & Human, 1998). Problems are generally situations where there is a good idea about what to do, but there is no clear idea of how to solve them (Prawat, 1989). Problem solving tools and techniques can provide intellectual challenges that can enhance cognitive development, promote a conceptual understanding, and foster the ability to reason (Hiebert & Wearne, 1993; Marcus & Fey, 2003; van de Walle, 2003). Although there does not appear to be any consensus on the optimum problem solving strategies or the number of steps in this process, there does appear to be some common agreement on the importance of search behaviour in this problem solving process (Chi, Glaser, & Rees, 1982).

It is also evident that there is a correlation between reductions of search behaviour and expertise (Chi et al., 1982). That is as expertise in a given subject matter or practice increases; the less search behaviour takes place. It is, therefore, plausible that once the learner has found a solution that works, then potentially less searching is needed. When the problem is encountered a second time, and each subsequent time, there should be less search behaviour. This suggests that with each successful attempt at solving the same problem, less searching for solutions is required. The participant, having learnt how to solve this problem, has, therefore, found the optimum solving strategy that works for them (Knoblich, Ohlsson, & Raney, 2001; Piaget & Blanchet, 1976).

In the study conducted by Beaunieux et al. (2006) the authors used the measures of number of moves and time to complete as an indicator of improvements of cognitive problem solving. Through using a disc transfer task (the Tower of Hanoi or the Tower of London puzzle game), the authors identified improvements in cognitive problem solving and procedural learning.

Frustrated with the inability of the existing behaviourist theories to explain complex perceptual and problem solving, Vygotsky (1978) sought to resolve the crisis in psychology and sought to develop a new theoretical basis based on a synthesis of existing views. According to Crawford (1996), Vygotsky focused on the relationship between people and the sociocultural context in which they act and interact in shared experiences. One of Vygotsky's contributions to the field of learning was the zone of proximal development (ZPD) (Vygotsky, 1978). According to Vygotsky (1978), "the distance between the actual developmental level as determined by independent problem solving and the level of potential development as determined through problem solving under adult guidance, or in collaboration with more capable peers" (p. 86). That is the difference between the perceived capability of the learner and the potential capability with guided instruction. Figure 2.5 provides a diagrammatic interpretation of Vygotsky's theory.

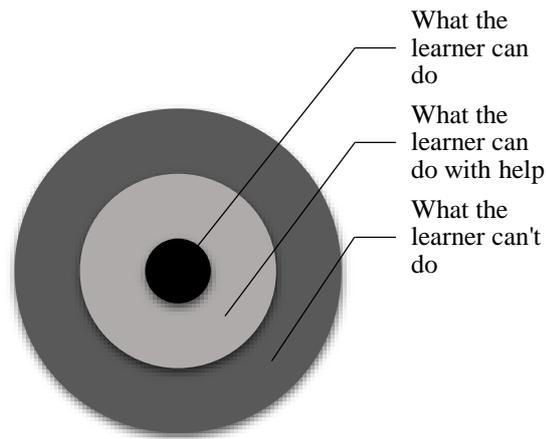

*Figure 2.5 Zone of proximal development (Vygotsky, 1978)*

Vygotsky (1978) asserted that the function of formal education is to provide learners with experiences which are in their ZPD, which will encourage and advance their learning. One of the important elements of Vygotsky's theory is that there is no indication of source of the help while a learner is in the ZPD. While, the assumption that this help would be provided through formal education, this does not suggest that formal education is the only source of help. Bransford, Brown and Cocking (1999), suggest that there is a social element to this theory. Bransford, Brown & Cocking (1999) posit that while a lot of learning that takes place that is self-directed, other people such as parents, coaches, caregivers, and other children also play an important part if the learning process.

Jonassen (1997) stated that there are two kinds of problems: well-structured problems and ill-structured problems. Table 2.3 summarises the main features of these two kinds of problems. Jonassen provides guidance on designing instructional materials for these different kinds of problem.

| Well-structured Problems | Ill-structured Problems |
|---|---|
| Present all elements of the problem | One or two elements of the problem are unknown |
| A probable solution is presented | The goals are ambiguous |
| The rules and principles of the problem are limited and are organized in a prescribed manner with well-defined (and constrained) parameters. | Include a number of solutions or no ideal or appropriate solution |
| Have correct and limited or constrained solutions | Provide multiple criteria for evaluating the solutions |
| Provide understandable solutions and "all problem states is known or probabilistic" (p. 68) | Include a limited number of manipulable parameters |
| Have defined solution processes. | Do not generally include an exemplar |
| | Provide limited information about which rules or concepts are necessary for the solution |
| | Include no specific means for determining the appropriate actions |
| | Involve the learner to decide how to approach the solution and evaluate these decisions |

*Table 2.2 A comparison between well-structured problems and ill-structured problems (Jonassen 1997, pp. 68-69).*

In video games, players are presented with problems that need solving. These problems are often key learning outcomes that need to be mastered before the player can go to the next stage or level in the game. However, most commercial video games do not have a set or constrained way of meeting the objectives of the level or the game. Furthermore, most commercial video games provide a lot of manipulable parameters. Therefore, the problems presented are neither well-structured problems nor ill-structured problems as per the definition posited by Jonassen (1997). The kinds of problems that are presented in video games appear to be semi-structured problems.

In many video games, challenges that are presented at the introductory or intermediate levels serve as preparatory training for the skills needed in the advanced levels of the game (Adams, 2010). The process of supporting the required skill acquisition needed for further advancement in the game is similar to what is referred to as 'scaffolding' (Sawyer, 2006). These scaffolds are gradually removed as the learner develops independent learning. The process of supporting the learning process to facilitate more advanced learning is an implicit feature of the ZPD (Vygotsky, 1978).

Through repeated exposure to the same problem in a video game, players acquire knowledge through repeated experience. It is through this repeated experience (or procedural learning) that advanced skills are obtained and maintained (Beaunieux et al., 2006; Norman, 1993).

## 2.11 INDICATORS OF PLAYER COGNITION

One of the challenges of measuring player cognition is that cognition is difficult to accurately measure. Instead proxy variables are used which provide potential indicators of cognition and learning. In some studies, the researchers have used either verbal protocols or performance measures. Verbal protocols (such as think-aloud methods) have been used to study personal insights (Kaplan & Simon, 1990), whereas performance measures have been used to record the time it takes the player to solve a particular problem (Knoblich, Ohlsson, Haider, & Rhenius, 1999). However, according to Newell and Simon (1972) performance measures do not typically measure the underlying thought processes.

To test the functionality of a video game, publishers such as Microsoft and Electronic Arts use a number of techniques (Kim et al., 2008; Pagulayan, Keeker, Wixon, Romero, & Fuller, 2003; Kobayashi & Ridlen, 2012). In most cases, a research subject is asked to play the video game and will either think-aloud while they play the video game or think-aloud while they review a video recording of their game play experience. Think-aloud methods (TAM) are processes which get participants in research studies to verbalise their thoughts while they are performing a given task (Guan, Lee, Cuddihy, & Ramey, 2006; Van den Haak et al., 2003) and are utilised in usability testing (Medlock, Wixon, Terrano, Romero, and Fulton, 2002). There are two types of TAM – concurrent think aloud (CTA) and retrospective think aloud (RTA) (Hyrskykari, Ovaska, Majaranta, Räihä, & Lehtinen, 2008). CTA is the process of the participants verbalising their thoughts or describing their experiences while they are performing a given task (Medlock et al., 2002). RTA is the process of the participants providing a description of their experience after each or all of the tasks are completed (Hyrskykari et al., 2008).

These methods are a relatively easy and cost effective way to identify usability issues (Medlock et al., 2002). Therefore, it is possible that these methods could also be used for identifying what learning has taken place while playing a video game.

Through recording video game play experience, video game researchers can also identify some design issues with the software. Through measuring user actions (or inaction) researchers can identify specific items in that the users struggle with. By using video game telemetry analysis of a group of users' mouse clicks, avatar action and inaction, time to complete specific tasks or challenges in the video game, researchers have been able to identify which areas of the video game need to be modified. Plot maps (also known as heat maps) can be produced using the telemetry data to provide a visual guide to these areas of the game that may need to be addressed (see Figure 2.6). This image plots the avatar deaths (in red). From the image, it is possible to identify a high number of avatar deaths in the centre of the map. Whereas, there are not many avatar deaths on the perimeter of the map. This suggests that the players that are in the middle of the map appear to have a strategic disadvantage. This image enabled the developer to identify usability issues and address these issues before the product was made available to the general population (Thompson, 2007).

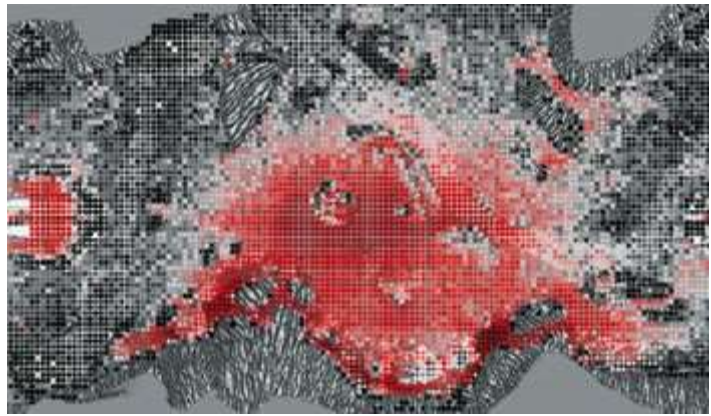

*Figure 2.6 Gameplay Metric Data from a level in Halo 3 (Thompson, 2007).*

Another potential measurement technique for identifying where the player is looking is through measuring the eye movements of the player. There is a close connection between what a player is looking at and what they are thinking about (Duchowski, 2007; Goldberg & Kotval, 1998; Just & Carpenter, 1976). Where users look on the screen, how long they look, and how many times they look at that particular object, can not only provide an indicator of what the user is thinking about, but it can also indicate where the user was having difficulty with that particular part of the game. The duration the eyes are fixed on a particular object can also potentially indicate the amount of processing that is taking place (Just & Carpenter, 1976; Just & Carpenter, 1980). In order to comprehend visual information, the eyes fixate on areas that are surprising, significant, important, and/or need further investigation (Just & Carpenter, 1976; Loftus & Mackworth, 1978). The eye-mind hypothesis (Just & Carpenter, 1984) suggests that there is a correlation between what is being looked at and what is being thought about. From the fixations or gaze points (the cumulative fixation time), it is possible to map these against the reference image (Figure 2.7) to ascertain the sequence of eye movements and areas of interest or attention (Rakoczi & Pohl, 2012).

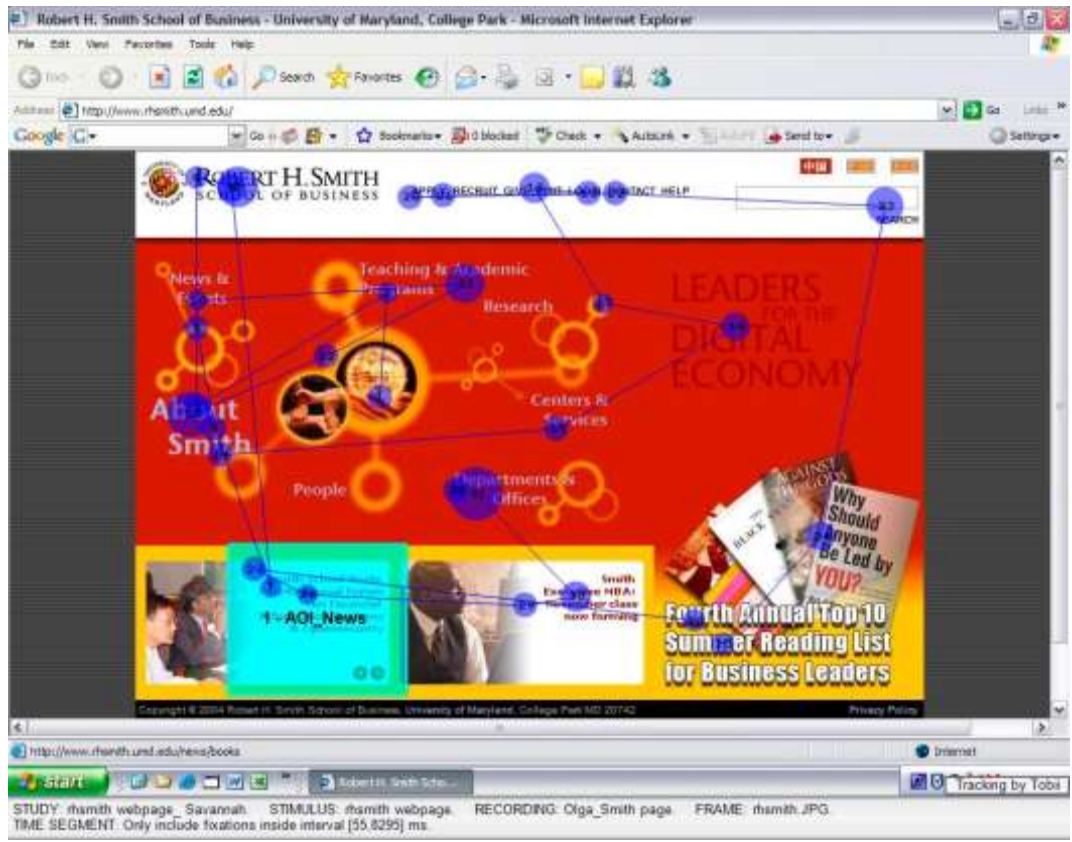

*Figure 2.7 Gaze plot data (Eye Tracking Lab, n.d.).*

The Gaze Plot Map (Figure 2.7) can also provide scan paths which are "defined by a saccade-fixate-sequence on a display" (Goldberg & Kotval, 1998, p.635). The scan path sequence and duration can provide insights into search behaviour (Goldberg & Kotval, 1998).

Saccades are defined as "rapid ballistic movements of the eye that abruptly change the point of fixation of the eye" (Purves, Augustine, & Fitzpatrick et al., 2001, p. 457). Saccades are one of four fundamental types of eye movements which also include, smooth pursuit movements, vergence movements, and vestibulo-ocular movements (Purves, Augustine, & Fitzpatrick et al., 2004). Saccades can be from very short in duration to relatively long in duration. The saccades that are very short typically occur during visually demanding events like reading or playing a video game. The long saccades, typically occur during tasks that are not as visually demanding, like looking around the room (Purves, Augustine, & Fitzpatrick et al., 2001). However, the duration of the saccade within these tasks can vary widely between individuals and the individual's familiarity of the subject matter being observed. Goldberg & Kotval (1998) posit that when presented with subject matter that they are familiar with, the participant tends spend more time focused on specific aspects of the object of focus. However, when the participant is not familiar with the subject matter, then more saccades (visual search) take place.

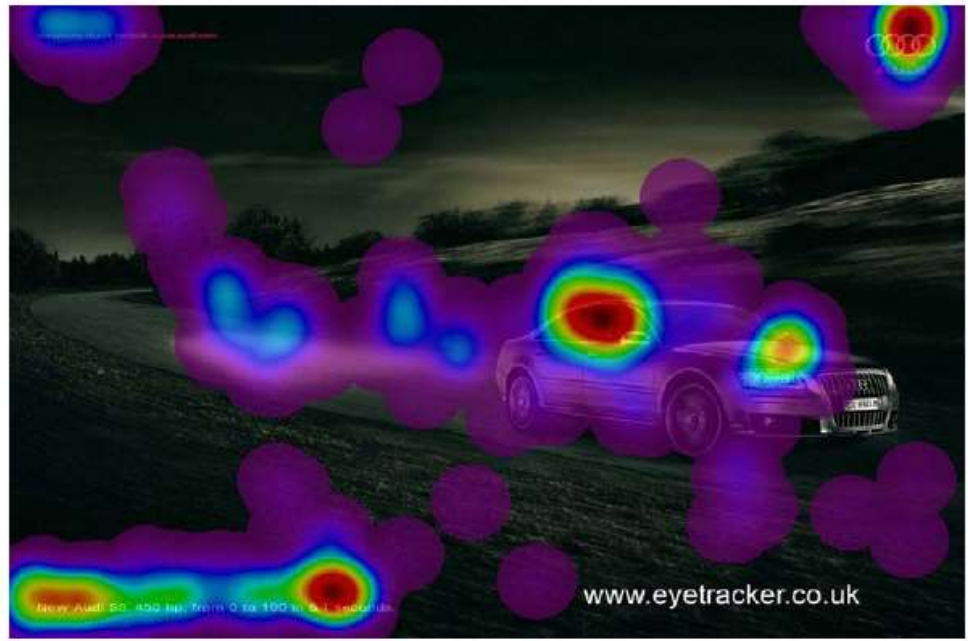

*Figure 2.8 Heat map produced by an eye tracker system (Janes, 2009.).*

The eye tracking heat map visualisation tools offer the view of the areas most visited of the image that are focused on by the viewer (or viewers) (Janes, 2009). In this heat map (Figure 2.8), it is possible to identify where on this image this participant focused their eyes (and attention) (Duchowski, 2007; Goldberg & Kotval, 1998; Just & Carpenter, 1976). The red areas on this image indicate the areas that garnered the most focus, and the purple areas identify the areas where that garnered the lowest attention (apart from those areas that have no colour, as these areas received no or limited focus).

Another indicator of human cognition is when the eye blinks (Orchard & Stern, 1991; Ponder & Kennedy, 1927; Stern, Walrath, & Goldstein, 1984; Tanaka & Yamaoka, 1993). The occlusion of vision lasts for 300 to 400 ms (Volkman, 1986). This lack of vision would interfere with behaviour if the blinks were not regulated. When a human blinks their eyes this could be a result of reflex or startle action (a response to something about to physically invade the eye, or the eye is dry), a voluntary action (resulting from a decision to blink), or an endogenous action (due to perception, a reaction, or information processing) (Fogerty & Stern, 1989; Orchard & Stern, 1991; Volkmann, 1986). The endogenous blink (also referred to as a spontaneous blink or involuntary blink) is a blink that is controlled or is a result from within an internal thought process (Doughty, 2001; Stern, Walrath, & Goldstein, 1984). The difference between the voluntary blink and an endogenous blink is the level of conscious thought processes. When a subject is directed to blink or consciously volunteers to blink, then this is considered a voluntary blink. Whereas, the endogenous blink is a blink that the subject does not consciously choose to perform. A blink interrupts vision and this presents potential problems for detecting visual stimuli and the integration of the information with behaviour. The endogenous blink should not be confused with an attentional blink (Raymond, Shapiro, and Arnell, 1992). According to Raymond, Shapiro, and Arnell, (1992) an attentional blink is when the human brain is unable to identify or recollect the presence of an additional object. This does not suggest that the eyes are actually closed (as in an eye blink), but instead the brain is not able to process or recall this additional information (as in a brain blink).

When humans engage in visually demanding tasks, such as playing a video game, they blink less frequently than we would when performing a non-visually demanding task (Goldstein, Bauer, & Stern, 1992; Kennedy, 1927; Orchard & Stern, 1991). Furthermore, humans will blink more frequently when they become excited or angry (Ponder & Kennedy, 1927). The research by Doughty (2001) suggests that there is a tri-modal distribution of endogenous eye blinks that are dependent on the task being performed. Doughty found that when engaged in conversation for endogenous eye blinks the mean blink rate per minute is 23. Whereas, the mean endogenous eye blink rate per minute for reading was 7.3 blinks per minute. The primary gaze activity resulted in a mean endogenous blink per minute rate of 15.3. Furthermore, according to Pivik and Dykman (2003), when engaging in activities that involve greater cognitive load, the endogenous blink rate increases. Fukuda (2001) suggests that the endogenous eye blink is a potential indicator of deception. This finding supports the research findings that specific blinks are a result of semiconscious and subconscious thought processes.

Endogenous blinks are typically the lowest in amplitude and the shortest in duration (Figure 2.8) (VanderWerf, Brassinga, Reits, Aramideh, & Ongerboer de Visser, 2003, p. 2790).

| Type of Blinking | Author | Comments | Down Phase | Up Phase |
|---|---|---|---|---|
| Spontaneous | Current study | | 92 ± 17 | 242 ± 55 |
| | Evinger et al. (1991) | | Slowest | Slowest |
| | Guitton et al. (1991) | | 85 | 200 |
| | Sun et al. (1997) | | 84.5 | 198.2 |
| | Stava et al. (1994) | Left | 104.7 ± 28 | |
| | Stava et al. (1994) | Right | 98.0 ± 26 | |
| Voluntary | Current study | | 88 ± 13 | 187 ± 34 |
| | Evinger et al. (1991) | | Middle | Similar to SO |
| | Collewijn et al. (1985) | | 100–150 | 300 |
| | Sun et al. (1997) | | 91.86 | 221.1 |
| SO stimulated | Current study | | 53 ± 6 | 152 ± 6 |
| | Snow and Frith (1989) | Early *R1* movement | | |
| | Evinger et al. (1991) | Early *R1* movement | Fastest | Similar to voluntary |
| | Evinger et al. (1991) | | | |
| | Bour et al. (2000) | | | |
| | Bour et al. (2000) | Early *R1* movement | | |
| Air puff–induced ipsilateral | Current study | | 87 ± 15 | 216 ± 32 |
| Air puff–induced contralateral | Current study | | 70 ± 11 | 173 ± 14 |

Values are means with SD in milliseconds, except for the amplitude, which is shown in degrees.

*Figure 2.9 Comparison of the results of the studies on eye blinks (VanderWerf et al., 2003, p. 2790).*

The study by VanderWerf et al. (2003) (Figure 2.9) presents the speed of each blink in milliseconds. The study that VanderWerf et al. (2003) conducted suggests that cumulative mean duration of an endogenous blink is approximately 334 ms (±67 ms) (although these authors do not provide any data to substantiate this claim). Guitton, Simard, and Codere (1991) found that the mean spontaneous blink was 285 ms. Whereas, Sun et al. (1997) found that the mean spontaneous blink was 283 ms. In contrast to this, Evinger, Manning, & Sibony (1991) found that the mean endogenous blink could be between 95 and 220 ms (Figure 2.10). One of the challenges in providing a standard duration of a blink is that the eyelid seldom returns to its original position, thus making the measurement of a complete blink a debatable topic (VanderWerf et al., 2003).

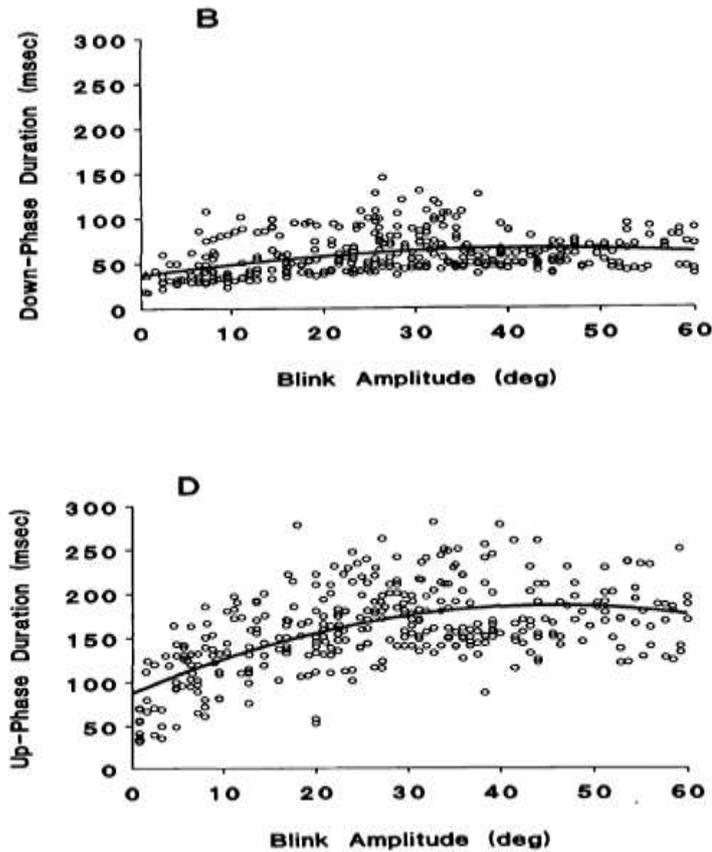

*Figure 2.10 Endogenous blinks (Evinger, Manning, & Sibony, 1991, p. 4).*

### 2.11.1  Eye fixations and endogenous blinks

The minds-eye theory states that where the eyes are focused is what the person is thinking about (Just & Carpenter, 2013). As noted, the eyes tend to fixate when an object (or event) is surprising, significant, important, and/or needs further investigation (Just & Carpenter, 1976; Loftus & Mackworth, 1978). The long fixations are objects that the person is investigating or trying to understand (or problem solving). Furthermore, the endogenous blink provides an indicator that perception, or information processing is taking place (Fogerty & Stern, 1989; Orchard & Stern, 1991; Volkmann, 1986). Ackerman (1986) demonstrated that information processing manages the relationship between specific abilities and performance during skill acquisition. The indicators of cognition (endogenous blinks and long fixations), could provide information about human problem solving. That is, decreases in the frequency of long eye fixations and endogenous blinks could indicate less problem solving has taken place. Furthermore, increases in the frequency of long fixations and endogenous blinks indicate that more problem solving has taken place. When combined with physical evidence of skill acquisition or performance improvement, then these measurements have the potential to verify this assertion.

## 2.12 REVIEW OF ISSUES AND PROBLEMS

A challenge with understanding the learning process is that it is generally a very private experience. In some cases, humans may not even be aware they are learning (Bunce, Bernat, Wong, & Shevrin., 1999). Subsequently, proxy variables can be used as indicators of learning. These proxy variables need to be used with extreme caution as it is possible to assume the research subject has learnt to solve a particular problem, when this may not have been the case. For example, when playing a video game it is possible to complete a particular challenge through pure chance (a fluke). Moreover, it is possible for a user to 'cheat' by obtaining a solution to the problem via a friend or through searching on the Internet.

While it is possible to hypothesise that people who play video games obtain resultant perceptual and cognitive benefits through playing them, an equally valid hypothesis could be that people who benefit from playing video games are more attracted to playing them than those who do not. In this case the underlying causality would potentially be experience and socialisation and not video gameplay.

The particular challenge with measuring eye movements and blinking is that this requires specific technology (eye tracking hardware) that may distract the research subject and thus bias the data. Furthermore, to minimize the external distractions, research subjects are typically asked to sit in a usability lab. These labs are usually considerably different from the research subjects 'natural' environment, and this may influence the data.

The challenges associated with using think-aloud methods in researching video games are as follows:

• Certain cognition is unconscious, and participants will generally not verbalise some or all of their thought processes (Eger, Ball, Stevens, & Dodd., 2007; Polanyi, 1966)

• Cognition is faster than verbal processes and therefore the participants could be thinking about a lot more than what they can verbalise (Eger et al., 2007)

• As the user needs to verbalise their thoughts as they play the game, concurrent thinking aloud can interrupt the game play experience. This process involves a greater cognitive workload (Van Den Haak, de Jong, & Schellens, 2004) and for most users this is not how they would naturally use the software or play a video game.

- The first impressions of an experience may bias the participant's experience in CTA, or the participant may forget the first impression in RTA (Van den Haak et al., 2004).

- Participants' sometimes forget to express their thoughts out loud (Pagulayan et al., 2003)

- RTA relies on participant's medium to long term memory (Bell, Bricker, Lee, Reeve, and Zimmerman, 2006) and participants may forget the key steps, or alternatively may fabricate (intentionally or unintentionally) the experience (Russo, 1979).

However, think aloud methods provide some clear insights into usability (Cooke, 2010) and could provide some valuable information.

Another challenge is the issue of transferability. While this study will look at what learning takes place while playing a commercial video game, it will not investigate if any of these skills are transferable to other situations that are beyond the scope of what is learnt in the game (or far transfer (Dede, 2009; Paas, 1992)). Although this is a worthy and intriguing question, it is beyond the scope of this study. This does not imply that what is learnt in a video game cannot be transferred to another dissimilar context. The work of Brown (Brown, 1989; Brown & Kane, 1988) shows that deeper learning (understanding causal relationships) can facilitate transfer. As video games potentially develop deep learning experiences (Gee, 2007) then there is a greater chance of learning being transferred through playing a video game.

Another challenge is that the process of observation can affect participants' performance, and they may change their behaviour to fit what they think the researcher expects (French, 1950).

As detailed in the research methodology chapter, a number of steps will be taken to ensure that these issues and challenges are addressed, which will be detailed in Chapter four.

## 2.13 CONCLUSION

In this chapter the process, potential and challenges of measuring human learning were identified. From the literature reviewed, video games represent the potential to support procedural and skill learning which is identifiable and transferable. While much of the existing literature has focused on educational video games, the latent dissatisfaction with the presentation or quality of these games has left many learners ultimately dissatisfied with the experience. However, the potential of commercial video games as tools for educational purposes represents the potential for further research. These video games have the content and presentation to keep many thousands of players engaged and motivated. The study of action video games provides an indication of the potential cognitive and educational benefits of commercial video games. However, the study of non-action video games has been largely ignored. The question of transfer from a commercial video game to an educational context is more speculative but can be addressed by discussion arising from the knowledge of what is learned in gameplay.

Chapter 3 includes a review of the different measurement tools used to understand learning that can be applied in a video game. This chapter will include a review of similar studies. The design of the study, the data collection methods, and the limitations of these methods can be found in Chapter 3.

**Chapter Three**

METHODOLOGY

## 3.0 INTRODUCTION

This chapter introduces the research approach used for this study. Section 3.1 provides a review of the research methodologies that have been utilised in undertaking research on video gaming. A review of similar studies is included in Section 3.2. An overview of the research questions and hypotheses is presented in Section 3.3. Section 3.4 presents the design of the study and Section 3.5 describes the collection methods that were used to obtain the data that will be analysed in this study Section 3.6 provides the details of the proposed study and finally, Section 3.7 reviews the limitations of the methods used.

This study utilised both quantitative and qualitative data to examine what learning and the types of learning that take place while playing video games. From the literature reviewed (Chapter 2), current research into learning from video games has used either qualitative or quantitative methods or a mixed-methods approach. The benefit of this approach is that by including a range of data types it will facilitate a diverse range of evidence that will provide justification for accepting or rejecting the hypothesis. Using a hybrid approach by utilising both qualitative and quantitative methods, which is also known as triangulation (Denzin, 1978), enables the researcher the opportunity to cross-validate the two different methods. The term triangulation originated in navigation where multiple reference points are used to locate an object's precise location (Smith, 1975). In social research the process of using more than one method (or reference points) to ensure that any variance that is identified is from the object or trait being studied and not from the method itself (Bouchard, 1976).

The arguments against using a mixed-methods approach cannot be ignored however. The first argument is that each methodology has its own epistemological commitment and, therefore, cannot be applied in another context (see, for example, Cook & Reichardt, 1979; Smith, 1983). The second argument is that the quantitative and qualitative research strategies are separate paradigms and incompatible (see, for example, Smith & Heshusius, 1986). Bryman (2006), states that it is necessary to understand and articulate the rationale for using a mixed-methods approach before embarking on a multi-strategy research project. Therefore, if the understanding is clear, then there is a reduced risk of data redundancy and research resources will not be wasted (Bryman, 2006).

The process of using a mixed-methods approach for video game user research is supported by Hazan (2013) and is used extensively by the video game industry (Pagulayan et al., 2003).

## 3.1 VIDEO GAME RESEARCH METHODOLOGIES

From the literature review (in Chapter 2), it is possible to identify a number of methodologies used to study the learning acquired while playing video games. While these methods have many advantages, they do also have some limitations.

### 3.1.2 Ethnographic studies

One method used to understand the learning and community found within a video game is through using an ethnographic study. Ethnographic studies allow the researcher to acquire a more comprehensive perspective than some other forms of social research – this is primarily due to the first-hand observation that takes place in a natural (or 'normal') setting (Wilson, 1977). Further, the observation of behaviour in a natural context enables contextualization (Wiersma, 1986). Ethnographic research does however have some limitations. It is dependent on the particular researcher's own observations or mind-set. As a result, it is almost impossible to eliminate some observer bias (Burns, 1994). Moreover, because the observations are based on a certain set of events within a specific period, it is extremely difficult to make generalisations from an ethnographic study. Furthermore, it is often difficult to replicate the findings of ethnographic research (Wiersma, 1986; Burns, 1994). From the literature reviewed, ethnographic studies have been primarily used to study massively multiplayer online role playing games (MMORPG) (see, for example, Nardi 2010; Steinkuehler, 2008; Galarneau, 2009). Due to the multiplayer aspects of the game and disparate nature of the playing community, channels of communication (text, voice, wiki's) within the game have provided a rich data set for scholars to make substantiated claims about the educational and learning potential within these games.

### 3.1.2   Think-aloud methods

Think-aloud methods provide the potential to get a better understanding of the thought process and problem solving methods through verbal expression of what Vygotsky (1962) referred to as inner speech. According to Olson, Duffy, and Mack (1984), think-aloud methods are one of the most effective ways to assess higher-level thinking processes. Higher-level thinking processes are those that involve working memory, and this working memory is where concurrent reasoning is processed into a verbal form (Ericsson and Simon, 1980).

One advantage of think-aloud methods is that they can provide researchers an efficient and effective way of finding out usability issues, which is one reason this technique is used by the video game industry (Medlock et al., 2002). To identify usability issues, organisations such as Electronic Arts, Mattel, and Microsoft use think-aloud methods to identify major usability issues with their products during and after the development process (Kim et al., 2008; Pagulayan et al., 2003; Kobayashi & Ridlen, 2012). However, while this method appears to be useful, it is not without its limitations.

One of the challenges of think-aloud methods is that they rely on the translation of abstract thought into inner speech, which is then subsequently verbalised. As a result, it may be difficult for someone else to understand or interpret this. Think-aloud methods depend on the recollection of thought processes. However, there are many thought processes that are not verbalised, or they are processed so quickly there is no time to verbalise them (especially those processes relating to tacit knowledge) (Davis & Bistodeau, 1993). Furthermore, there is a risk that when using think-aloud methods and the cognitive load is high, the subject may forget to express or recall precisely what they are thinking (Preece, 1994; Preece et al., 2002). Branch (2000) found that participants involved in think-aloud studies reported that they felt they were being observed and judged, and this influenced their performance.

From the literature reviewed, think-aloud methods have been used to understand learning acquired while playing video games. Shute & Kim (2012) used a think-aloud method to help understand the players' problem solving and causal reasoning while playing the video game World of Goo (2D Boy, 2008).

### 3.1.3 Eye tracking methods

The motivation for recording eye movements is based on the theory that humans move their eyes when they seek to focus their attention on an object (Duchowski, 2007; Palmer, 1999). To derive meaning from the environment, the human brain implements multiple processes to select, organise and interpret information from the senses. Vision consists of two components – perception and cognition (Palmer, 1999). When undertaking a task or activity, human visual perception combines low-level features into high-level representations that inform cognitive processes. If these tasks involve visual stimuli, these cognitive processes guide perception and focus the eyes (and other senses) onto the subject.

The action of focus on a particular object (or objects) is a cognitive process and therefore can provide clues as to what the observer found interesting, fascinating, disgusting, or what required further investigation (Duchowski, 2007).

Eye tracking systems enable researchers to record and measure movements and fixations of the eyes (Holmqvist et al., 2011; Sunderset et al., 2013). The video-based systems record infrared light, which is projected onto and reflected from the participants' cornea or retina (Holmqvist, Nyström, Andersson, & Dewhurst, 2011). This allows the measurement of fixations (the time the eye is fixated on a particular object), saccades (the rapid movement from one fixation point to another), and blinks (the duration the eye is temporally closed) (Holmqvist et al., 2011; Orchard & Stern, 1991). These events typically occur at very high speed and require specialist equipment to record and measure them. The duration of an eye fixation is typically for 200-300 milliseconds. However, the duration of a saccade typically takes 30-80 milliseconds, and a blink normally lasts for 100-400 milliseconds (Holmqvist et al., 2011; Stern et al., 1984). To measure these events, an eye tracking system is needed that has a high sampling frequency. In video-based eye tracking systems, sampling frequency is measured in hertz (Hz), this means that a 50 Hz eye tracker will record the eye gaze direction of the participant at 50 times per second (Holmqvist et al., 2011). If a researcher wants to measure saccade or eye blink data, then a 50 Hz eye tracking system will not capture this information.

As noted in Chapter 2, when someone blinks his or her eyes this could be a reflex or startle action (a response to something physically about to invade the eye or the eye is dry), a voluntary action (resulting from a decision to blink), or an endogenous action (due to perception, a reaction, or processing information). Each different type of blink varies in duration and amplitude (Thorsheim, Mensink & Strand, 2001). The endogenous blink is generally the lowest in amplitude and the shortest in duration (Orchard & Stern, 1991; Evinger et al., 1991; Guitton et al., 1991; Stava, Baker, Epstein, & Porter, 1994).

The current eye tracking systems offer a range of sampling speeds of 50 to 1000 Hz. The current systems that provide a sampling speed of over 500 Hz require the participant to wear a headset, which can be intrusive. The SMI Red500 operates at 500 Hz (Red500 Technical Specification, 2011) and, therefore, is capable of identifying eye fixations, saccades, and the quantity and frequency of endogenous eye blinks. Moreover, as this device can be mounted on the desktop or placed under the computer monitor it does not require the participant to wear a headset or special glasses, thus enabling a relatively natural engagement with the video game (see Figure 3.1).

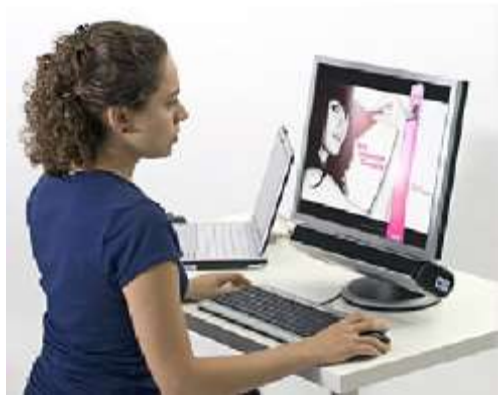

*Figure 3.1 The SMI Red500 (SensoMotoric Instruments, n.d.)*

### 3.1.4    Cognitive Load Measurement

Although not widely used to study learning within a video game, techniques that measure cognitive load (Paas & van Merrienboer, 1994) use: rating scales, mental efficiency, or physiological measures (heart rate, pupillary responses), that could be applied. The use of rating scales depends on the participant to accurately reflect on the mental load they experienced. While, this method has support (see Paas, Tuovinen, Tabblers, & Gerven, 2003), the methods that use self-reflection and/or self-ratings have been questioned (Branch, 2000; Davis & Bistodeau, 1993; Preece, 1994; Preece et al., 2002). Furthermore, the physiological measures used in cognitive load measurement (heart rate, or pupillary responses) are affected by cognitive load and other physiological behaviour that could potentially confound the results. Moreover, the literature review did not identify any studies that used self-rating scales, or measured the heart rate, or pupillary responses of young children, as a result, these measures were not applied.

## 3.2 REVIEW OF SIMILAR STUDIES

From the literature reviewed there does not appear to be any previous studies that have used a combination of the methods proposed in this study. The following is a review of methods used in video game user research.

### 3.2.1 Ethnographic studies

Ethonographic studies enable researchers to engage with the participants in their natural environment. Through ethnographic research, a more sympathetic understanding is possible (Fine, 1993). Although ethnography is typically associated with the study of society and cultures of the physical world, several authors (Nardi 2010; Steinkuehler, 2008; Galarneau, 2009) have explored online video games through using an ethnographic study.

Nardi (Nardi, 2010; Nardi & Harris, 2008) used an ethnographic study to understand the types of learning and the importance of collaborative play in the online video game World of Warcraft (WoW) (Blizzard Entertainment, 2004). In their study Nardi and Harris (2008) conducted an immersive ethnographic study to undertake participant observation, using semi-structured interviews, collecting chat logs and reading online resources. Nardi and Harris (2008) spent approximately two months playing WoW, as well as conducting twenty-six in-depth semi-structured interviews with a range of players. The researchers asked the participants questions about the length of time they had been playing WoW, the appealing attributes of the game, the quantity of characters each participant had created within WoW, and whom they played with (Nardi & Harris, 2008).

Steinkuehler (Steinkuehler & Williams, 2006; Steinkuehler, 2005) has undertaken ethnographic studies of Linage (NC Soft, 1998) and Linage II (NC Soft, 2003), and World of Warcraft (Blizzard Entertainment, 2004). The study of Linage (NC Soft, 1998; NC Soft 2003) included 24 months of participant observation of actual gameplay. This study also collected data from discussion boards, in-game player discussions (chat room transcripts), instant message conversations, and email communications between the players. The researcher also undertook unstructured and semi-structured interviews of 16 key informants. From these studies Steinkuehler (Steinkuehler & Williams, 2006; Steinkuehler, 2005) concluded that MMORPGs provide a valuable (virtual) social space for groups of users. Through these studies, both Nardi (2010) and Steinkuehler (2005) have been able to observe tacit and implicit knowledge acquisition within a virtual environment.

### 3.2.2   Think-aloud studies

Shute & Kim (2012) used a think-aloud method to help understand the player's problem solving and causal reasoning within the video game World of Goo (2D Boy, 2008). This exploratory study involved four adult participants who had not played the game before. The participants were provided with a warm-up exercise to familiarise them with the think-aloud procedure. The participants then played the game while their in-game actions were recorded using Fraps (Beepa, 2010) and the reactions to their experience were recorded. Through this exploratory study, the researchers were able to examine the acquisition of certain knowledge and skills during gameplay. The authors found that video games have great potential as a rich context for learning. However, learning depends on how the user interacts with the game. Reflection and failure (and possibly reflection on failure) were identified as key elements of the learning process.

### 3.2.3   Eye tracking studies

In their study of how the participants learn to play a new video game, Alkan and Cagiltay (2007), used the Eyegaze (LC Technologies, 2013) eye tracking system to help find what strategies are used to learn a new video game, the attentional differences at differing levels of a game and the usability issues. In this study, Alkan and Cagiltay (2007) asked 15 participants to play the video game Return of the Incredible Machine: Contraptions (Sierra Entertainment, 2000). Through this study, the authors specifically sought out a video game that the participants had probably not played before. Furthermore, the video game required decision-making, and problem-solving constructs that the authors believed were applicable in education. The research also found that the participants needed to use (or learn) strategies that were different from the other games they had played. The study found that the participants used both trial and error as well as external advice (asking a friend) as sources of information (although this study did not make it clear what information was sought). The eye tracking system data was found to be consistent with the descriptive data. Although more detail about this study would be beneficial, it did identify that the eye tracking system provided some information about the cognitive processes of the participants. This study also provides some useful information about the type of game played and the number of participants used.

Pretorius, Gelderblom, & Chimbo (2010) used an eye tracker to investigate the aspects of software users find easy or difficult to use. They also considered different patterns of learning behaviour used by both adults and children. Through using a Tobii 1750 eye tracker, Pretorius, Gelderblom, and Chimbo (2010) explored if and how users read the on-screen instructions of a video game. The sample size in this experiment was eight participants and included four adults and four children. The researchers used the eye tracking system to compare the total number of fixations on the on-screen instructions with the number of fixations on other parts of the screen. Through this research, it was found that the adults and the children differed in their use of the instructions. Further, although not a goal of this study, the results show that in this case, the children generally had fewer fixations in total than the adults.

In a study conducted by Kenny, Koesling, Delaney, McLoone, and Ward (2005), they investigated the participants playing a first-person shooter game (FPS). This study involved six participants playing a custom-made game for the purpose of furthering the understanding of psycho-perceptual experiences. Kenny et al. (2005) used an SR Research EyeLink 2, which is a head-mounted binocular video-based eye-tracking system that has a sampling rate of 500 Hz. This study compared the differences between two gameplay experiences. Through analysing the final score of the individual participants' gameplay experience, the authors were able to compare the level of success in the game (operationalised as the final score) with the percentage of fixations near the centre of the screen. The authors concluded that there may be a correlation between task proficiency and the duration of eye movement (and fixation). This study also provides some valuable guidelines for the analysis and reporting of fixation data.

The study by Evans and Saint-Aubin (2005) used an eye tracking system to measure participant behaviour and cognition. In this study, the authors observed 15 infants that were engaged in a shared reading experience as their parents read them a story. The participants were 48 to 61 months old and were read a contemporary storybook that included both text and graphic illustrations. The purpose of this study was to investigate whether children were looking at the text or the pictures during shared reading. This research used an SR Research EyeLink II, which consists of three cameras attached to a headband that is worn by the participant. This enabled the researchers to track both head and eye movement. The researchers found that the primary point of interest is on the graphic images and not on the text.

### 3.2.4   Telemetry data acquisition studies

Video game telemetry data is used in several aspects of the video game development lifecycle (Zoeller, 2010; Drachen, Seif El-Nasr, 2013; Canossa, 2013). This data can provide developers with valuable information to help track bugs, find design problems, detect balance issues, and evaluate player usability challenges (Zoeller, 2010; Medlock et al., 2002; Drachen, Seif El-Nasr, 2013). Drachen, Canossa, and Yannakais (2009) examined gameplay activity from Tomb Raider: Underworld (Eidos Interactive, 2008) to identify four unique types of players (or categories of player behaviour). Thurau and Bauckhage (2009) mined player statistics from World of Warcraft (Blizzard Entertainment, 2004) and discovered patterns in the evolution of guilds.

In their study of Tomb Raider Underworld (Eidos Interactive, 2008), Drachen et al. (2009) analysed an existing data set of 1365 players who had played Tomb Raider Underworld (Eidos Interactive, 2008) during November, 2008. In this study, the researchers used a data collection system that provided information on the participant activity while the game was played (using Xbox Live data). The data collected provided the researchers with information about the causes of avatar death. It was found that avatar deaths were caused by other players, a non-player controlled character, environmental factors, or the player's inability to master a particular challenge. Moreover, through analysing this data, the researchers were able to identify the total number of avatar deaths, the time to complete the game, and the number of times the player used the online help system.

### 3.3   RESEARCH QUESTIONS AND HYPOTHESIS

From the studies reviewed in the previous section, it is apparent that there are a number of ways to measure and understand what learning takes place while playing a video game. From the review of these and previous studies it is possible to conclude that while video games are engaging and provide motivational elements, the challenge has been quantifying and measuring what learning takes place within a commercial video game. Finding reliable proxy measures for learning has not been easy. Further, there does not appear to be any consensus on the frequency or duration of the learning that needs to take place before reliable measures can be taken.

The main question this research attempts to ascertain is what learning takes place within the context of playing a commercial video game. This question assumes that some learning does take place. Although the result of the research to date is not entirely conclusive, in most cases some learning does, in fact, take place within an educational video game. However, the type of learning that takes place from playing a commercial video game is still unclear. Although the learning may be as simple as learning to navigate the user interface, or as complex as solving a puzzle. If a novice user (a user that has not played that game or genre before) has successfully completed a level or part of a game then, it is plausible that while procedural learning (and possibly perceptual learning) has taken place, some conceptual learning has also transpired. By taking a mixed-methods approach, this research intends to establish a reliable method for measuring explicit and implicit learning gained by playing a commercial video game.

### 3.3.1   The research questions

The research questions are:

RQ 1: What learning takes place when playing the video game World of Goo?

RQ 2: Does problem solving ability improve through playing video games?

RQ 3: Do the participants that played the video game World of Goo learn tower construction from playing the game?

### 3.3.2   The hypothesis

This study tested two hypotheses. The first hypothesis is that playing a commercial video game positively affects user learning. The second hypothesis is that playing commercial video games positively affects problem solving ability.

### H1 - Playing a video game positively affects user learning

The null hypothesis is:

$H_0$ Playing video games does not affect user learning.

The alternate hypothesis is:

$H_1$ Playing video games positively affects user learning.

**H2 - Playing a video game positively affects problem solving ability**

The null hypothesis is:

$H_0$ Playing video games does not affect problem solving ability.

The alternate hypothesis is:

$H_1$ Playing video games positively affects problem solving ability.

### 3.4   DESIGN

As detailed in Section 2.0, the definition of learning adopted by Gee (2003; 2011) and the focus it provides, is a platform for investigating learning in a commercial video game. The limitations and exclusions discussed above help focus a study onto executable processes so that the research questions could be answered.

The study was undertaken in three phases: pilot testing, data acquisition, and data analysis as detailed in Table 3.1.

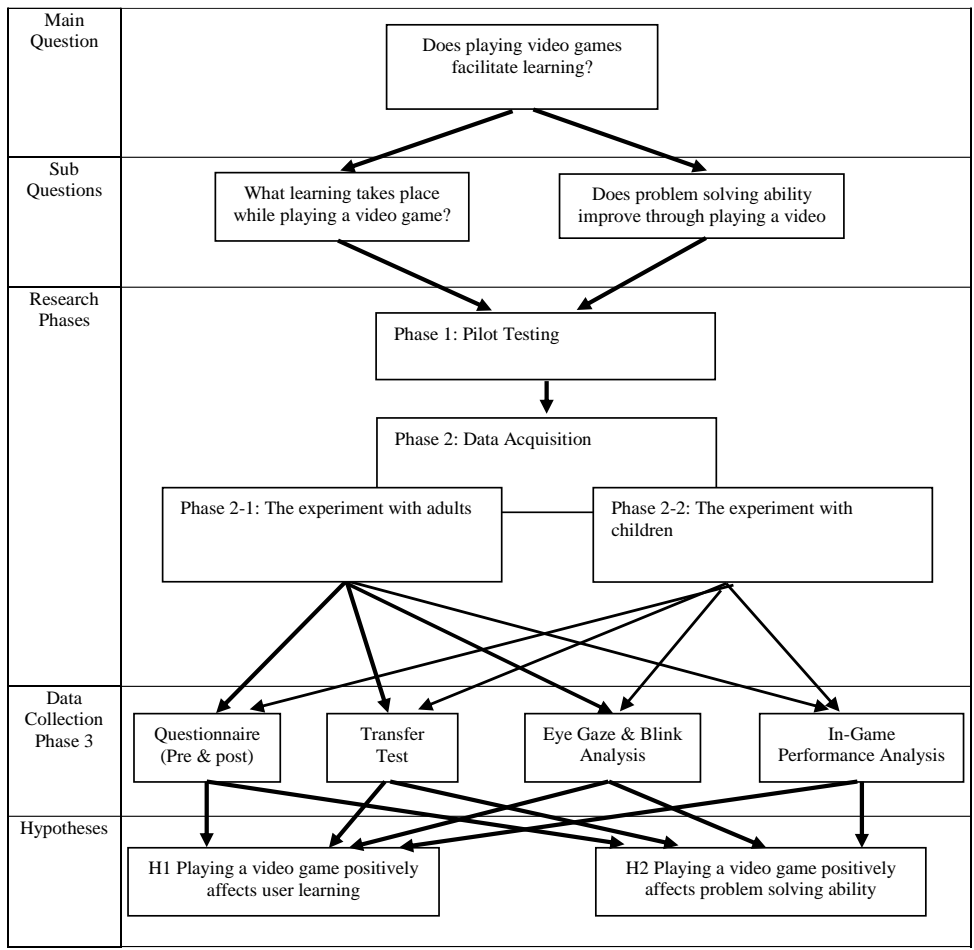

*Table 3.1 Data Map*

## 3.5   THE STUDY

In this study, participants in the treatment group were asked to play a level of the commercial video game, World of Goo (2D Boy, 2008).

The proposed study will involve two different experiments that will consist of two different participant groups: Experiment 1 – children aged six to eight years old and Experiment 2 – adult students enrolled at Northeastern University.

### 3.5.1   Experiment 1

To understand the learning process that takes place within a commercial video game this study involved getting a convenience sample of children aged six to eight years old. This particular group was chosen because the participants should have been able to understand how to play the video game and use the magnetic construction set. That is, they were at the pre-operational and concrete operational stages of development (Piaget, 1952). Furthermore, these participants have the necessary hand-eye coordination to play the video game, and based on the results of the initial pilot study, these participants have generally not been exposed to any formal instruction in the basics of tower construction (a core element of the treatment). This study was conducted in a laboratory to minimise external distractions. A sample size of 10 participants was chosen based on Pagulayan et al. (2003) who suggested that in game user research that most of the data can be obtained with 10 to 15 participants through using multi-method studies. Furthermore, previous studies have used a similar or smaller number of participants (see table 3.2).

| Author(s) | Sample Size |
| --- | --- |
| Alkan & Cagiltay (2007) | 15 |
| Barker, Brinkman, & Deardorff (1995) | 13 |
| Kenny et al. (2005) | 6 |
| Khalili et al. (2011) | 16 |
| Kenny, Koesling, Delaney, McLoone, & Ward (2005) | 6 |
| Pretorius, Gelderblom, and Chimbo (2010) | 8 |
| Shute & Kim (2012) | 4 |
| Squire & Barab (2004 ) | 18 |

*Table 3.2 Sample sizes of similar studies*

The children were invited to join the study through a recruitment drive using social media that was focused on community groups in close proximity to Northeastern University. Although anyone from outside of this area was more than welcome to join the study, these particular regions were selected due to the relative convenience of travelling to Northeastern University. Informed consent was obtained from the participants involved, and ethics approval was obtained from the Northeastern University Internal Review Board (IRB # 13/10/13). Northeastern University was chosen on the basis that there were fully functional laboratories, and Northeastern was accessible to SensoMotoric Instruments (SMI) which provided the eye tracker (SMI RED500) and the training necessary to conduct the research.

The children in the treatment group were asked to play one or two levels of the video game, World of Goo (2D Boy, 2008) on a Personal Computer (PC). The participants in the treatment group were exposed to one or two levels of the game twice (two exposures per level). This made it possible to measure any improvements in performance, as well as any changes in the cognitive measure that were being observed (eye fixations and endogenous blinks).

The participants in the control group played the video game, Bad Piggies (Rovio Entertainment, 2013) on a PC. Bad Piggies is another problem solving game that is non-violent and success in the game does not depend on advanced reaction speeds (or twitch). Game play in Bad Piggies is based on building vehicles, and successful gameplay does not rely on any knowledge of tower building construction principles. Participants in the control group played the game one of level one (the beginner's level) of Bad Piggies twice.

The participants were invited to join the study through posting messages on websites in the area (Appendix 2). The participants were offered a USD40 gift certificate from the iTunes Store. When the participants and their parents arrived at the lab, the study was explained to them, and informed consent was obtained (Appendix 4). Upon completion of the study, the parents were debriefed and told the purpose of the study (Appendix 5).

### 3.5.2   Experiment 2

The participants in the adult group included adult students enrolled at Northeastern University. The observations were conducted in a laboratory setting, and the sample size of the study consisted of 10-20 participants (based on Pagulayan et al., 2003). Furthermore, previous studies have used a similar or smaller number of participants (see table 3.2).

A convenience sampling method was used and selected from students enrolled in undergraduate programs at Northeastern University. This particular group was chosen because the participants should have had some exposure to playing video games.

This study was conducted in a laboratory to minimise any distractions and to ensure the best possible lighting conditions were available for the eye-tracker. The participants in this study were chosen on the basis that they are relatively close to the research laboratory, and it was comparatively easy to recruit them. The students were invited to join the study by email (Appendix 2), and informed consent (Appendix 4) was obtained from the participants involved. The Northeastern University Internal Review Board (IRB #13-9-10) provided ethics approval (Appendix 3).

The participants in the treatment group played one or two levels of the game, World of Goo (2D Boy, 2008) twice (two exposures per level). This made it possible to measure any improvements in performance, as well as any changes to the duration of fixations on any object (a possible indicator the participant is struggling with or is fascinated with a particular part of the game).

The treatment group and the control group played the video games on a tablet computer (an Apple iPad). The participants were offered a USD20 gift certificate from the iTunes Store. When each student arrived at the lab, the study was explained to the student, and informed consent was obtained. Upon completion of the study, each participant was debriefed and was told the purpose of this study (Appendix 5).

### 3.5.3   World of Goo

The game World of Goo (2D Boy, 2008) was chosen because it is not overly popular and, therefore, potentially few participants will have had prior experience in playing it. Furthermore, the game is non-violent (which was considered important when asking young children to play it), it does not require advanced reaction speeds, and there is the potential for participants to learn some basic principles of physics. According to Shute and Kim (2012), World of Goo incorporates the physics concept of static equilibrium. The other implicit physics concept was the importance of building sound structures (Davidson, 2009). Through the pilot study, the researcher also identified that the additional physics concept embedded in the game was the concept of force (gravity, wind, or buoyancy). Furthermore, success in the game also depended on learning the basic principles of construction (the importance of strong foundations, the importance of support structures, and/or the importance of level structures). Furthermore, there is an opportunity to improve analytical thinking skills (Anderson et al., 2009; Davidson, 2009; Shute & Kim, 2012).

World of Goo is a physics based puzzle/construction game (Davidson, 2009). The basic premise of World of Goo (depending on the level) is to construct a tower, bridge, or chain to enable the goo (or gooballs[1]) (the protagonists) to get close enough to the pipe (Figure 3.2). When the structure gets close to the pipe, the gooballs are extracted into an extraction pipe to join the liquefied goo. However, the forces of gravity and wind, spikes, bog (swamp), fire, cogs, or machines

---

[1] The individual goo are referred to as gooballs.

(antagonists) challenge the player to build something that is strong enough to resist the antagonist(s) but is also tall or long enough to reach the extraction pipe.

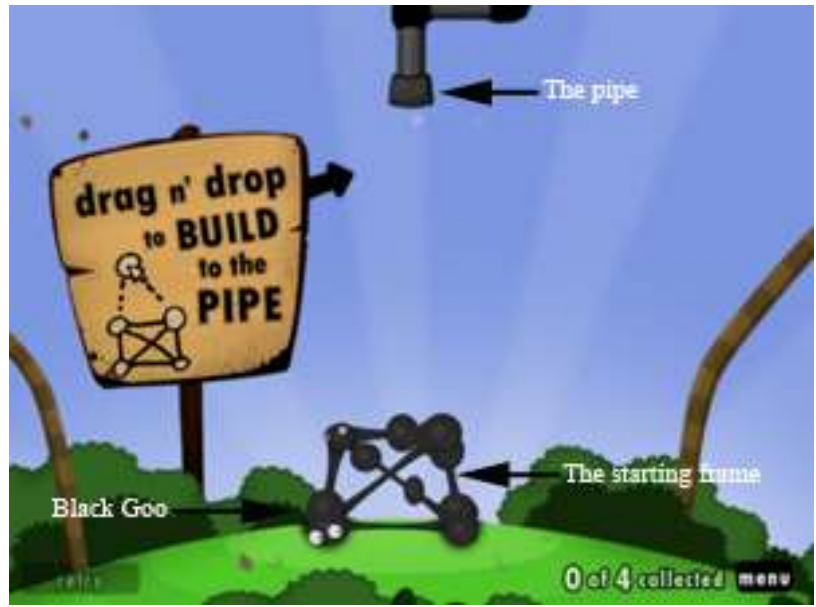

*Figure. 3.2. World of Goo. (2D Boy, 2008)*

Within the game, there are several species of goo. There are black goo, white goo, green goo, dark black goo, and an eyeball goo. Each species of gooball have different features and capabilities. The black goo are the most basic of the goo species, they are used for building. The black goo have two legs (or connectors) and once placed in a construction, cannot be reused. Unlike the black goo, the white goo has four legs (or connectors), and the length of the legs are shorter. However, like the black goo, once connected, they cannot be re-used. The green goo have three legs and can be re-used multiple times. The dark black goo is used only for collection, they have no legs and there cannot be connected to anything or used in construction. The eyeball goo is used for lifting (in the air) or anchoring an object (in water). Further, the eyeball goo has no legs.

The controls in the World of Goo are very simple. The user interacts with the game through selecting a gooball (with their finger or mouse) and then drag this (either by continuing to hold the mouse button or by pressing their finger) to the desired location. The game provides a visual cue through providing guidelines for where the legs (or connectors) of the gooballs will be located and attached (see Figure 3.3). If the user pulls the gooball beyond the reach of the legs, the white lines will disappear. If the user de-selects the gooball and the gooball is within the range of the gooballs legs, the gooball will automatically connect its legs with the adjoining structure, gooballs, or terrain. If the user de-selects the gooball and it is outside of the range of its legs, the gooball will fall.

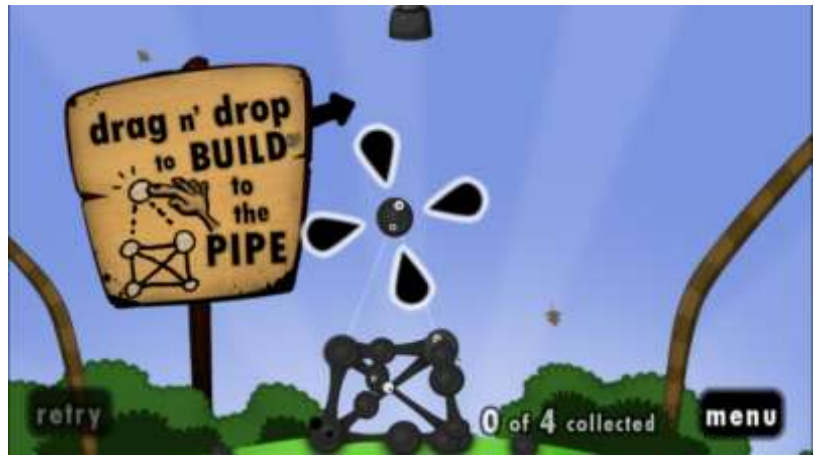

*Figure. 3.3. World of Goo – Level 1. (2D Boy, 2008)*

The goo have a personality. The live (or usable) goo have eyes and provide auditory feedback to the user. If a gooball is successfully placed, the gooball responds with a positive auditory acknowledgement (that almost sounds like woo who, and yippee). If a gooball is dropped or lost the gooball responds with a negative auditory protest (that almost sounds like bother, and oh bother). As noted the user is required to build an appropriate structure to enable the goo to get close enough for the extraction pipe. However, with each level, there is a limit as to the total number of goo and there are unique environmental factors to challenge the user. The difficulty of the game is to build a sound structure to save enough goo. However, if the structure is over engineered, not enough goo will be saved. If the structure is not built to resist the environmental factors within the level, it will fall and no (or not enough) goo will be saved.

There is no tutorial or help file within World of Goo, although the first level provides a simplified introduction (beginner's level) to the game. An ever-present, yet unseen character in World of Goo, the Sign Painter, provides 'guides' or directions throughout the game. The work of the Sign Painter can be seen in the first level (Figure 3.3). The Sign Painter provides the user with the directions and guidance to success in the level. However, the assistance of the Sign Painter is less direct in the more advanced levels. In the more advanced levels of the game there is an 'undo' feature. This is provided through time bugs, which when selected, take the user back one step or move in the game. The total number of time bugs is limited in each level, but the allocation of time bugs varies depending on the difficulty of each level.

The score is based on the total number of gooballs saved. If not enough gooballs are saved, the user cannot move to the next level. The user is provided with a constant visual indicator of the goal (how many gooballs need to be saved) and the status (how many have been saved). This information is provided on the lower-left of the screen (0 of 4 collected information on the screen in Figure 3.3). The difficulty level is self-selecting. It is up to the user to decide if they want to save more than the minimal number of required gooballs. Upon completing each level, the user is presented an audio confirmation (applause) and the results of the level will be displayed. The information provided includes the time taken to complete the level, the number of moves, the number gooballs saved, the number of extra gooballs saved, and the cumulative time it took to complete the level (Figure 3.4). This information can be shared on social media and websites of fan-based leader boards (on several versions of the game, there is an option to submit the results to the 'World of Goo Leaderboard of Excellence').

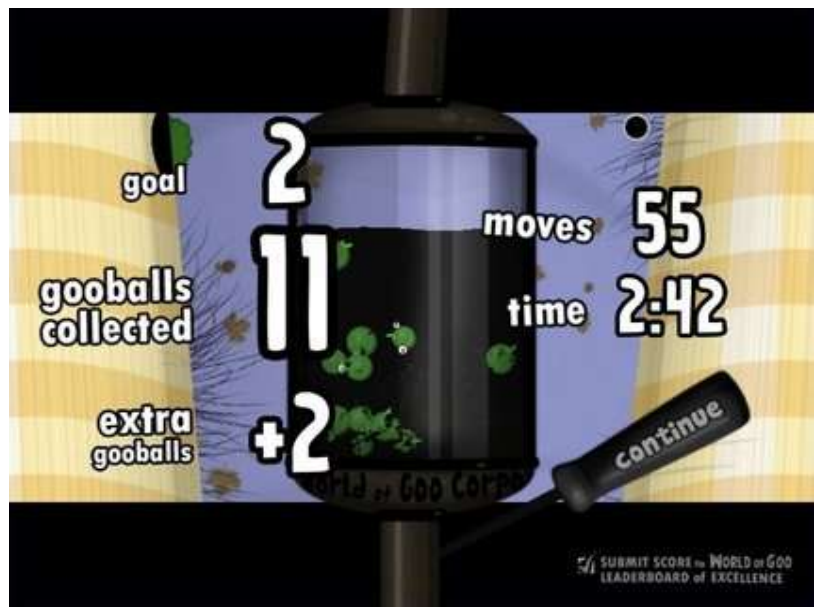

*Figure. 3.4. World of Goo – Level completion screen. (2D Boy, 2008)*

The only tangible rewards in the game are the results that are available upon the completion of each level. Further, the only way to progress through the various levels and chapters in the game are through saving the required number of gooballs. With each level, the user can pre-select or select OCD (Obsessive Completion Distinction). OCD is an optional achievement. Getting OCD requires one of three things: getting a required quantity of gooballs (more than double the normal requirement), completing the level within a certain time limit, or completing the level with a minimal amount of moves. Obtaining OCD status does not assist the user or provide any additional in-game benefits.

The audio in the game is presented through background music and sound effects (as noted auditory feedback to the user). The background music has a dark whimsical note to it. According to Gabler (2008) (the artist that wrote the music), the intention was for the theme to reflect the song Libertango (by Astor Piazzolla).

Within the game, when the user selects the menu option, the in-game menu is displayed (Figure. 3.5). The options displayed are; skip level, retry, quit level, OCD, and resume. This menu can also be used to pause the game.

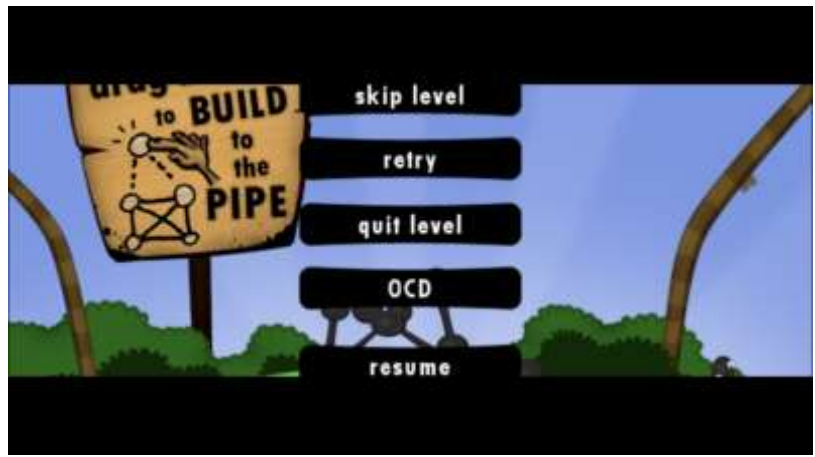

*Figure. 3.5 World of Goo – Menu option. (2D Boy, 2008)*

The visual presentation differs between each chapter. There are differences in art direction or style within each chapter. There are four chapters and one epilogue. The chapters are: Chapter 1 – The Goo Filled Hills, Chapter 2 – Little Miss World of Goo, Chapter 3 – Cog in the Machine, Chapter 4 – The Information Super Highway, Epilogue (Chapter 5) – The End of the World.

Unlike the puzzle game Tower of Hanoi, (which has been used by other researchers - see for example; Beaunieux et al. 2006; Piaget & Blanchet, 1976; Simon, 1975), in the game World of Goo, there is an unlimited number of options or choices where the participant can place the gooball.

### 3.5.3.1 Exposure to World of Goo

The first level all participants in the treatment group encountered was Chapter 1 – Going Up (Level 1) (Figure 3.6). In this level, the user is required to build a small tower and save four gooballs. To get OCD status, 11 gooballs need to be saved. This level was chosen because it is the introductory (or tutorial level) in the game and, therefore, the participants should be able to complete this level regardless of prior exposure to this video game. The level starts with the camera focused on the extraction pipe, the goo drop from the sky and the camera tilts to follow the goo as they fall to the ground. As this is an introductory level, the scene includes a starting frame with four black gooballs connected together to potentially provide a guide as to the construction process. In this level the gooballs that drop from the sky automatically roll onto the starting frame (however, in more advanced levels, the user is required to construct something to reach the additional goo). Further, the wooden sign provided by the Sign Painter provides a visual construction guide and the text 'Drag n' Drop to Build to the Pipe'. The sign also includes an arrow that points to the pipe. The camera focuses on the starting frame. As the extraction pipe is out of the view of the camera, a small semi-transparent triangle indicates the direction of the pipe (Figure 3.6).

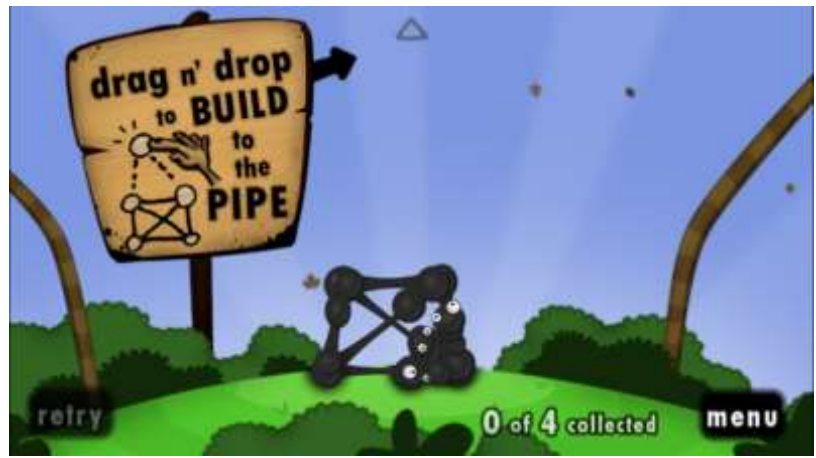

*Figure. 3.6. Level 1. (2D Boy, 2008)*

The second level the adult participants were asked to complete was Chapter 1 – Tower of Goo. This is a more advanced version of the Going Up level and required the participants to build a taller tower and save 25 gooballs. To get OCD status, 68 gooballs need to be saved.

The reason for choosing these levels was the possibility to replicate a similar tower construction process using a physical magnetic construction toy (Figure 3.7). Although, the magnetic construction toy may require three-dimensional construction principles, World of Goo utilises two-dimensional construction principles. Initial pilot testing demonstrated that learning was transferable from the video game to the magnetic construction set. Figure 3.7 is one of the magnetic towers that was constructed by a participant during the pilot test.

Another reason for choosing this video game is that it has been used in previous academic studies and literature (Anderson et al., 2009; Davidson, 2009; Shute & Kim, 2012).

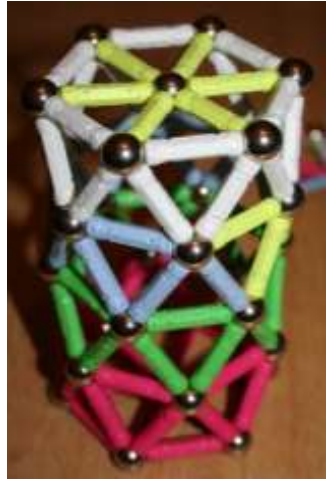

*Figure. 3.7 Magnetic construction set*

### 3.5.4 Bad Piggies

The video game, Bad Piggies (Rovio Entertainment, 2013) was chosen for the control group because although it is a problem solving game unlike World of Goo (2D Boy, 2008) it does not involve constructing towers.

The basic premise of Bad Piggies is to build a vehicle (a wooden car or aeroplane) to transport a pig (the protagonist) across a variety of terrain (an antagonist) to the end of a pre-defined path to a nest. In the more advanced levels, there are also characters from the game Angry Birds that will try (by throwing something) to stop the pig getting to the egg. If the user gets the pig to the nest egg, the end goal is achieved. However, additional rewards are available for: breaking/not breaking the vehicle, bringing the King Pig in the vehicle, not using a specific vehicle part, collecting star crates, getting the pig to the destination on time, carrying an egg in the vehicle, and/or carrying cakes or treats. There are 168 story levels within the game. Movement to the higher levels is dependent on obtaining two or three stars in each of the lower levels. If three stars is obtained, the user may also be rewarded with cake to feed to the King Pig, who will provide free power ups.

In the establishment (or opening) shot of each level, the camera (what the participant sees) pans from the objective (the nest) to the starting point of the game. This serves the purpose of informing the user the direction of game play and the level data (what obstacles) the player will face. The camera is consistent with the game play. While the car is being built, the camera is stationary. When the car moves, the camera pans to follow the car, ensuring that the vehicle is at the centre of the player's attention.

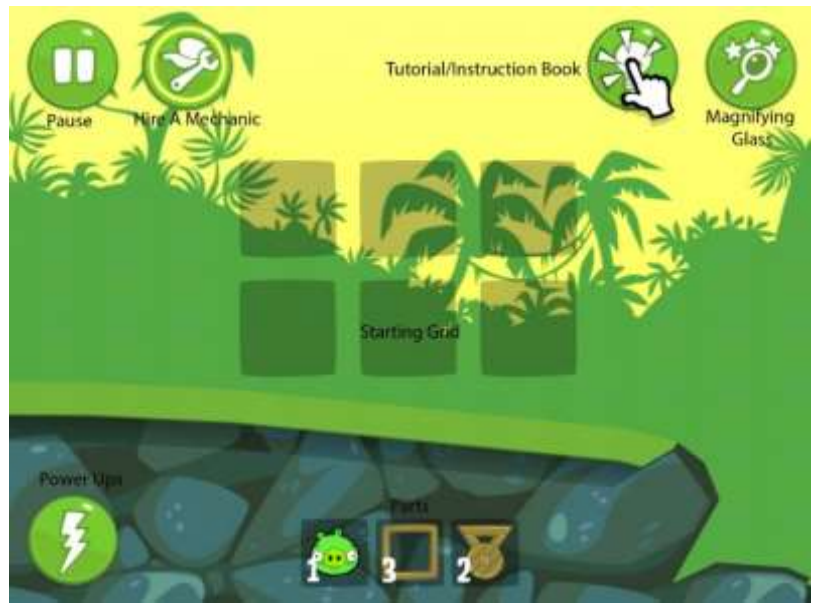

*Figure 3.8 Vehicle construction screen. (Rovio Entertainment, 2013).*

Several screens are available to the user. After the establishment shot, the first screen is the vehicle construction screen. The vehicle construction screen provides the user with a blank start grid (Figure 3.8). When the user moves a part onto the starting grid it will automatically snap into place on the grid. The screen provides the user with information about what parts are available (the images of the pig, box, and wheel) and how many are available (the numbers on the images). The user is also presented with on screen options that include: a pause icon, a Hire a Mechanic icon, a Tutorial/instruction Book icon, a Magnifying Glass icon, and a Power Ups Icon. The pause button enables the user to pause the game or to select another level. The Hire a Mechanic icon enables the user to pay a 'mechanic' to build a suitable vehicle that is capable of getting through the level.

       The (tutorial) instruction book icon provides the user with contextual visual cues on how to successfully build a vehicle. The image in Figure 3.9 shows the user that if the user places the pig in a well-balanced car (the left image), then there it

may not fall over (the right image). The optimal construction has a tick (or check mark) and the less-optimal construction has a red x mark. The less optimal construction image also suggests that the car is about to fall.

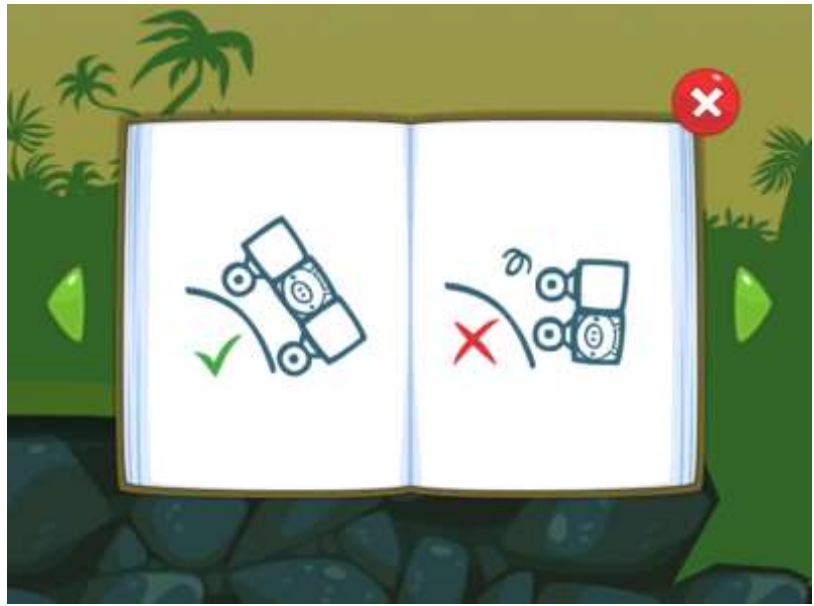

*Figure 3.9 The Instruction Book screen. (Rovio Entertainment, 2013).*

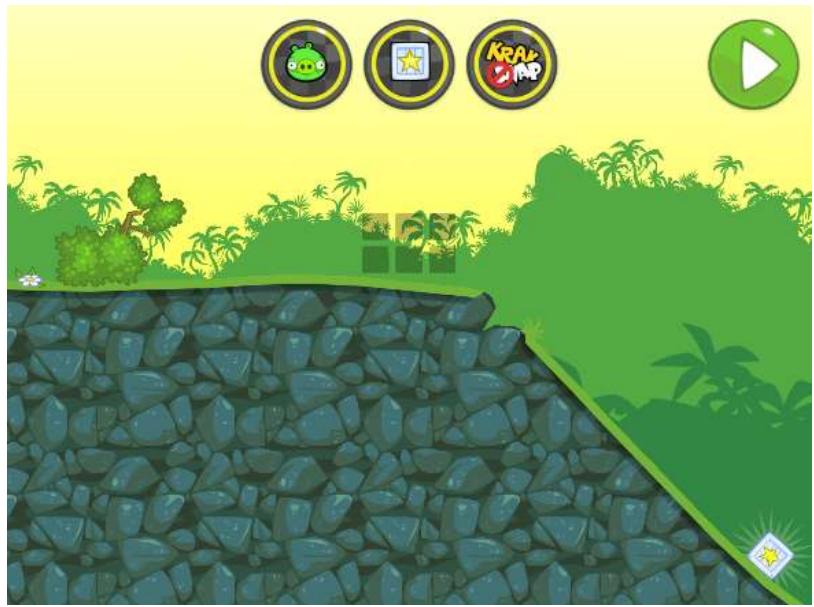

*Figure 3.10 The Magnifying Glass screen. (Rovio Entertainment, 2013).*

The magnifying glass icon provides the user a view of the level and the level objectives. Figure 3.10 shows the view of the starting grid and part of the terrain that will be encountered in this level. To view the remaining terrain, the user can pan or zoom the camera. As noted, the view also provides the user with the objectives of the level. The objectives of the level displayed in Figure 3.10 include: a pig icon, a blue star create icon, and a 'krack-snap' icon. The pig icon indicates that in this level the user is required to get the pig to the destination. To achieve two or three stars, the user also needs to collect a star create (the blue star create icon) and achieve 'Krak-snap' (the krak snap icon) the vehicle (which indicates that the user needs to inflict significant damage to the vehicle).

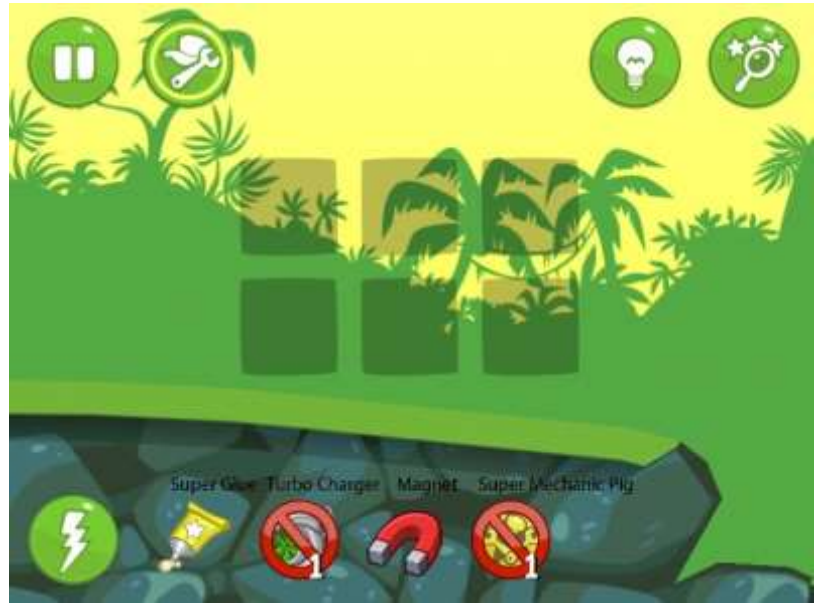

*Figure 3.11 – The Power ups screen. (Rovio Entertainment, 2013).*

The power ups screen (Figure 3.11), enables the user to add additional features (or power ups) to the vehicle. The power ups that may be available include: super glue, a turbo charger, a magnet, and a super mechanic pig (the circle with the line through it indicates that this power up is not currently available). Super glue makes the vehicle stronger. The turbo charger makes the vehicle faster. The magnet provides additional assistance to attract and collect items (for example, the blue star creates). The power ups are obtained through achieving three stars in a previous level, through feeding the King Pig cake, or through purchases of the power ups (in-app purchases).

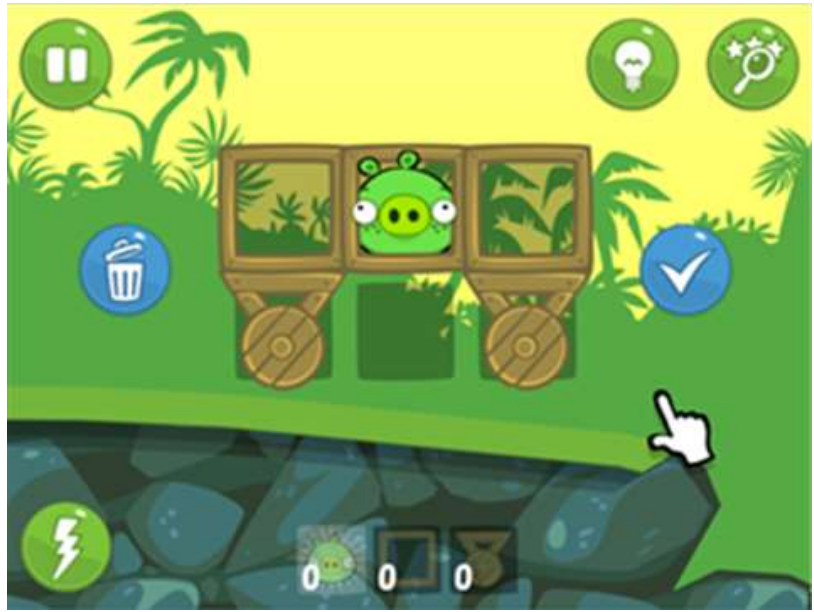

*Figure. 3.12. Vehicle completion screen. (Rovio Entertainment, 2013).*

When the user has built a vehicle that has the minimal part (a pig), the check (tick) icon will be displayed (Figure 3.12) which indicates that the vehicle is ready to go. While constructing the vehicle a trash can (rubbish bin) icon provides the user the option to delete the entire construction and start again. When the user selects the check icon, the starting grid disappears, the vehicle will descend to the terrain, the camera will dolly out (pull away from the scene) and the user will see the vehicle and a partial view of the terrain that will be encountered in the level. The camera will atomically track the movement of the vehicle. If the vehicle has any propulsion devices on-board, icons will be on the screen for the user to select when the user wants the propulsion device to be activated/deactivated.

The audio presentation is provided through background music and sound effects. The background music has an uplifting tempo with a cheerful melody. It invokes a feeling of fun. The sound effects include some oinks, and squeals from the pig as the vehicle proceeds through the level. There are also sound effects for each of the propulsion devices that are activated when the vehicle crashes or explodes.

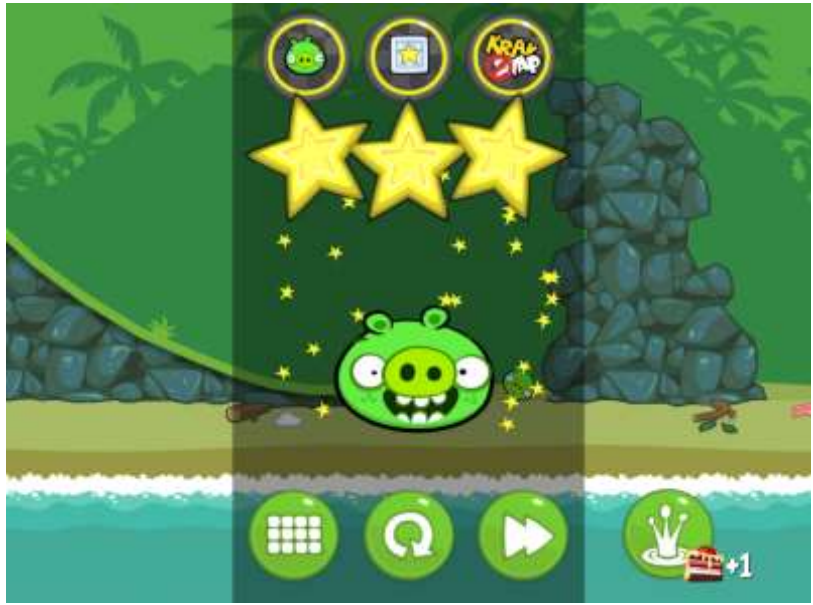

*Figure. 3.13 Level Completion screen. (Rovio Entertainment, 2013).*

Upon completion of the level, the user is also presented with the level completion screen (Figure 3.13). The level completion screen provides the user with: the number of stars awarded the objectives that have been achieved, and an animated image of the pig bouncing up and down. The level completion screen also includes the following icons: the grid icon, the replay icon, the fast forward icon, and a crown icon. The grid icon when selected enables to user access the level selection screen. The replay icon if selected will restart the same level. The fast forward icon when selected will advance the game to the next level. The crown icon when selected will initiate the feeding the King Pig sequence.

### 3.5.4.1 Exposure to Bad Piggies

All participants in the control group were exposed to the video game Bad Piggies (Rovio Entertainment, 2013). The participants were exposed to the beginning level of Bad Piggies, as it was the tutorial level in this game. This level starts with the camera (view) focused on the nest egg. The camera slowly pans from right to left, showing the user the terrain, and any achievements. When the camera has shown the level, the camera focuses on the starting grid. As this is the beginning level, an animation of a moving hand with the index finger indicating where the user needs to look (the pointing hand or 'pointy hand' icon) moves to the instruction book, indicating where the user can get help (Figure 3.14). If the user opens the instruction book, once the instruction book is closed, the pointy hand icon moves from a part icon to the starting grid, to help the user understand how and where to place the parts. Once the minimal part is on the starting grid, the pointy hand icon will move to the check icon, indicating that this is how to start. The pointy hand icon will continue the animation sequence until the user selects the relevant icon.

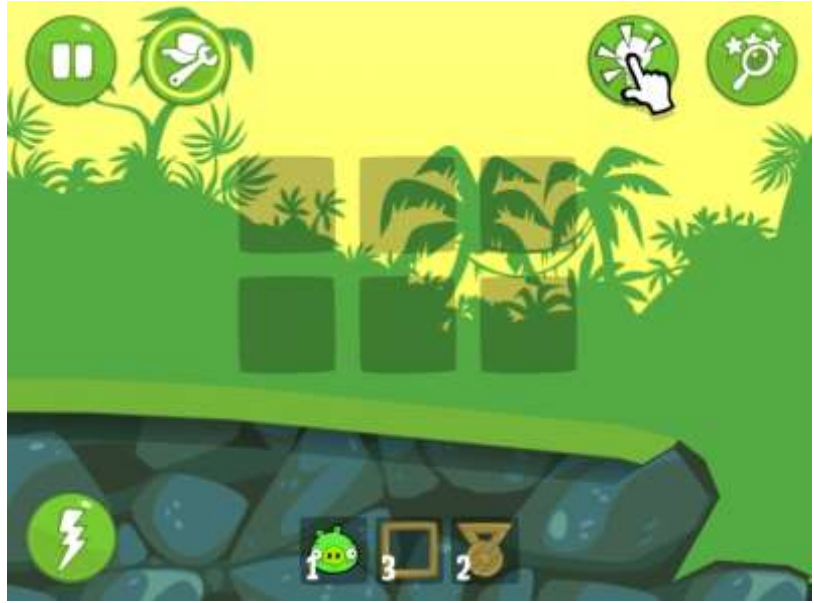

*Figure 3.14 Level 1. (Rovio Entertainment, 2013).*

The objective of this level is to get the pig to the nest from using: one to three boxes, one or two wheels, and the pig. To get three stars in this level, the user is required to get the pig to the egg nest and collect one blue star create, and Krack-snaps (breaks) the car.

### 3.5.5  Transfer

As noted in the literature review, there is some debate as to the transferability of skills that are learnt from playing a video game. The experimental design deliberately included one game that included implicit or explicit skills that would be tested outside of the game and one game that would not. The magnetic construction set was used to test what (if any) of the identified skills were transferable from World of Goo (2D Boy, 2008) to the magnetic construction set. As noted the potential construction skills that could be acquired in World of Goo include; the importance of strong foundations, the importance of support structures, and/or the importance of level structures. The control group was implemented to test if the transfer was a result of the treatment.

The video game Bad Piggies (Rovio Entertainment, 2013) did not incorporate the specific skills that would assist in the construction of the magnetic construction set. To succeed in the game, the skills required in Bad Piggies include an understanding: of the influence of force (gravity, wind, engines, compressed gas, and/or explosive devices), the importance of balance (putting the pig in the optimum position in the vehicle), the influence of momentum, and/or the importance of robust structures (the impact of an adhesive, or wood versus steel). Table 3.3 provides a comparison of the potential transferable skills that are embedded in the games. In the World of Goo, the objective was to build a tower. In Bad Piggies, the objective was to build a car.

| Embedded Concept | World of Goo | Bad Piggies |
|---|---|---|
| Static equilibrium | Yes | No |
| The concept of force: | | |
| - gravity | Yes | Yes |
| - wind[2] | Yes | No |
| - buoyancy[2] | Yes | No |
| Basic principles of tower construction: | | |
| - the importance of strong foundations | Yes | No |
| - the importance of support structures, | Yes | No |
| - the importance of level structures. | Yes | No |
| Construction objective | Build a tower | Assemble a car |

*Table 3.3 Comparison of the embedded concepts in each game*

Both games have the potential to develop analytical thinking skills (Anderson et al., 2009; Shute & Kim, 2012), the transfer test did not provide for the measurement of any advancement of analytical thinking skills. However, as noted in section 3.2.2, this study will investigate if the number of endogenous blinks and the frequency of long eye fixations are indicators of cognitive problem solving and/or information processing.

---

[2] These concepts were not in the levels played

### 3.6  DATA COLLECTION METHODS

The data was collected using a mixed-methods approach. Participants were given a pre-exposure questionnaire (Appendix 1) which sought data on gender, demographics, prior gameplay experience, experience in playing the games that were used in this study, and provide a baseline understanding of the participants existing knowledge of construction principles. The participants were then asked to build a tower with the magnetic toy construction set. Each participant was provided with a finite number of construction pieces (27 magnetic rods and 44 steel balls). Then the participants were asked to play the World of Goo (2D Boy, 2008) or Bad Piggies (Rovio Entertainment, 2013). When the level was completed, the participants were then asked to build a second tower with the magnetic toy construction set.

During gameplay, the participants' eyes were tracked, and their eye blinks were monitored using the eye tracker. The computing platforms used for each study varied slightly due to some challenges identified through using the iPad. Early testing found that it was difficult to keep the participants' eyes consistently elevated high enough for the eye tracker to keep a constant line of sight with the retina while the participant used the iPad. However, the results of these two studies were not compared. These two experiments should be viewed in isolation.

The data that is available at the completion of each level will be used to identify improvements in cognitive problem solving and procedural learning. This method is consistent with the method used by Beaunieux et al. (2006).

### 3.6.1  Experiment 1

In the the study of the children the participants used a personal computer. A personal computer with the game software and an eye tracking system was connected together to facilitate dual recording of the game play and capture the eye gaze behaviour (Figure 3.15). A web camera was connected to a second PC to facilitate confirmation and validation of when the participant blinked. To ensure the participants sat relatively still, they were asked to sit in a chair that could not swivel, but the height could be manually adjusted to allow for the different heights of the participants. This ensured that the eye tracking system maintained constant monitoring of the participants' retina. The web cameras were placed above and beside the computer monitor to ensure, they did not interfere with the eye tracker's line of sight.

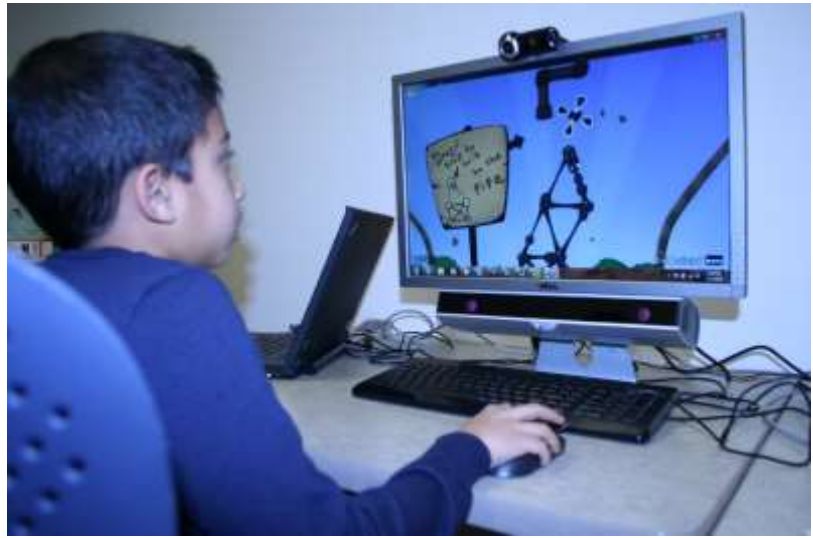

*Figure 3.15 Experiment 1 set up*

Calibration of the eye tracker was facilitated through the SMI calibration software. The software was set up to place an animated circle on the screen. The participant was asked to look at that point on the screen. Once the first calibration point was confirmed, the calibration system moved an animated dot to five separate points on the screen. This was repeated until the unit was calibrated with the participants' eyes and was confirmed by the operator. The SMI software also includes a tracking monitor and provides a visual indicator of the placement of the retina. This provides a visual warning if the participant is sitting too close or too far away (Figure 3.16). This allows the operator to ask the participant to adjust their position.

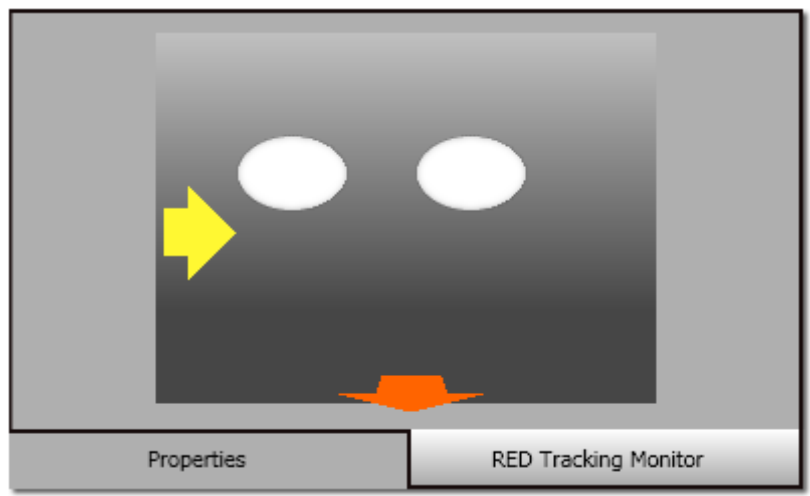

*Figure 3.16 SMI RED tracking monitor*

### 3.6.2   Experiment 2

In the study of the adults participants used an Apple iPad (iPad). A web camera that was mounted above the screen (Figure 3.17) recorded the in-game choices and gameplay. This enabled the participant's hand gestures and on screen decisions to be recorded. The camera was connected to the computer that had the eye-gaze recording software. The eye-gaze recording software recorded real-time video of the user's actions and decisions. When the participant finished the experiment, each data set was collected and analysed.

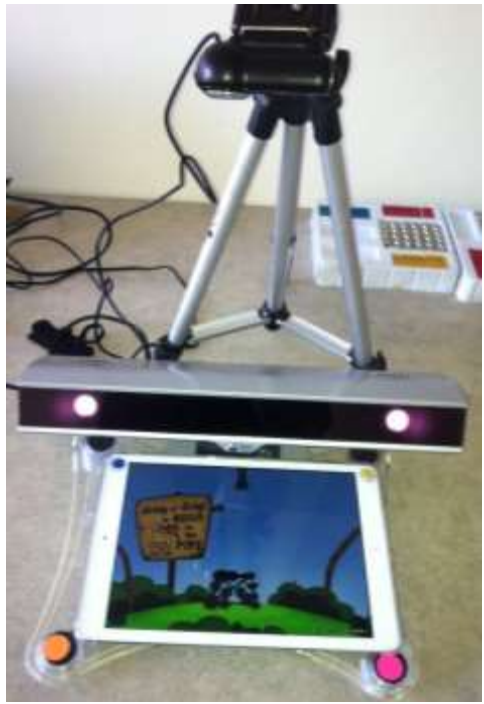

*Figure 3.17 Experiment 2 set-up*

The SMI RED500 provides a five-point calibration system. Four coloured dots were placed at each corner of the mobile device stand to facilitate this calibration (Figure 3.17). The participant was asked to place a finger on each coloured dot, and then look at it while the system was calibrated for each point. The fifth reference point for calibration used an image near the centre of the iPad Screen (the game centre icon).

The iPad was secured to the mobile device stand to ensure a clear and consistent line of sight between the eye tracker and the participants' eyes. Although the iPad can be used on the participants' lap or held in the user's hand, this would provide a challenge for the eye tracker to maintain a clear line of sight of the participants' eyes. The mobile device stand and the eye tracker was fastened to the desk to ensure the system did not move after the unit had been calibrated. The mobile device stand was built-for-purpose, based on guidelines provided by Trenchard-Seys (2011).

### 3.6.3   Pilot test

A small exploratory study was conducted to test the process and hypothesis. The researcher first played the video game, World of Goo (2D Boy, 2008), which is a video game the researcher had not played before. The researcher explored the core game mechanics and corroborated the conclusions of Shute and Kim (2012). Shute and Kim found that the world of Goo was an appropriate game for both young children and adults, as it was not violent and was a game that would appeal to any age group.

The eye tracker was not available for the pilot test and therefore, the researcher acquired a high-speed camera and put this in front of the screen to record eye-gaze behaviour. World of Goo (2D Boy, 2008) provides the player with the results of the game play including the number of moves, the number of gooballs collected, and the time taken to complete the level. This information provided the capacity to make comparisons between each attempt at the game.

The children of both the researcher and his friends (n=5) tested the system and to ensure that the game was playable by children and suitable for young children (it was not too violent). Further, it was possible to test the comparative performance of the magnetic tower set and measure any associated improvements of the construction process with each gameplay experience.

### 3.6.4   Pre- and post-exposure questionnaire

The data collected from the pre-exposure questionnaire primarily collected demographic data, data about previous game play experience, and prior exposure to the video game World of Goo (2D Boy, 2008) (Appendix 1). However, the pre-exposure questionnaire also asked the participant a question which garners a baseline understanding of basic construction principles through asking the question, "What shape do you think is the strongest for building a really tall tower?" Participants were asked to answer with one of the following options: a Square, a Circle, a Triangle, or Not Sure (the correct answer was a triangle). The post-exposure questionnaire asked the same question and, therefore, comparisons can be made between the answers to the two questions. Furthermore, in the post-exposure questionnaire, participants were asked the following questions:

- Did you have fun?

- What do you think was the most fun part?

- What do you think is important when building a tower?

- Why did the tower break or fall down? (This question will only be asked if the tower fell)

The last two questions were open-ended, thus enabling the participant to express their opinions in their own words. These answers were coded for comparative analysis. The questions about fun are dependent on the participants' interpretation of what fun is. Although the concept of fun is subjective, based on the literature review undertaken (in Chapter 2) it is still an important attribute to consider when adopting a video game into an educational context.

The categorical data obtained from the pre-exposure structured questionnaire was used to obtain an understanding of the demographics, game- play experience, and obtained a baseline understanding of basic construction principles. Both the pre-exposure and the post-exposure questionnaire asked the same closed-answer question, "What shape do you think is the strongest for building a really tall tower?" Comparisons were made between the individual answers from the pre- and post-questionnaires. This data was categorised and then used to make comparisons. Furthermore, the post-exposure questionnaire was used to seek data to support this question through an open question, "What do you think is important when building a tower?" and "Why did the tower break or fall down?" The answers to these questions were then analysed using text analysis and keyword identification. The data analysis was used to identify keywords related to construction principles. From this analysis, descriptive statistics was used to report on the frequency of these keywords. The post-exposure questionnaire asked qualitative questions. The first question – a closed-ended question, "Did you have fun?" was analysed through using descriptive statistics. The second question was open-ended and sought feedback through asking, "What do you think was the most fun part?" Although, not directly related to this research project, answers to this question was used to validate any positive answers to the closed-ended question.

### 3.6.5   Magnetic tower construction

Before the participants were exposed to the video game, a baseline measurement was obtained by asking them to build a tower with a magnetic construction set. To ensure a baseline of the participant's construction ability, product packaging and construction guides was removed to eliminate any possible guides for ideal construction.

Upon completion of the building of the magnetic tower, measurements of the type and number of shapes used in the construction process was observed and recorded. After each gameplay experience, the participants were asked to build another tower with the magnetic construction set. Comparisons were then made between each magnetic tower. Where possible, the magnetic tower construction process was video recorded. This enabled detailed analysis of the time it took to complete each tower, as well as the number of attempts to build the tower. Further analysis was performed by comparing the results of the treatment group and the control group.

In this phase of the study the dependent variable (DV) was the instructional treatment and the independent variable (IV) was the construction methods used in building the magnetic towers.

### 3.6.6   Eye gaze and blink data analysis

To monitor the participants eye-gaze, saccade, and endogenous eye blinks, this research used an SMI Red500 eye tracker (SensoMotoric Instruments, n.d.). The SMI Red500 eye tracking system operates at 500 MHz and, therefore, was capable of tracking eye-gaze behaviour and measuring the quantity and frequency of endogenous eye blinks. This device does not require the participant to wear a headset or special glasses and, therefore, it facilitated relatively natural engagement with the video game. Moreover, through using the SMI RED500 eye tracker (SensoMotoric Instruments, n.d.), it was possible to observe and measure the following (Duchowski, 2007, p. 173):

•      Eye fixation

•      Fixation duration (gaze)

•      Fixation rate (overall)

•      Fixation duration mean (overall)

•      Number of fixations (overall)

•      Eye blinks

The data from the SMI Red500 eye tracker made it possible to compare the differences in the metrics collected from each individual replay experience.

Recent improvements in eye trackers have made the calibration process a lot better for the user (Duchowski, 2007; Jacob & Karn, 2003), through allowing constrained head movement. The latest eye trackers have rendered the chin rest of previous systems unnecessary. Although these systems are still susceptible to the issues of interference from eyelashes or the rims of certain glasses, they have made the process of calibration a lot more user-friendly. Furthermore, the latest systems provide software to monitor and therefore maintain the correct distance between the user and the eye tracker (Duchowski, 2007).

Every measure was taken to ensure a consistent and clear line of sight was maintained between the eye tracker and the participants' cornea. Should the participant move too far away from the eye tracker, they were informed before the study of the optimal position and posture, and were reminded during the study when the participant deviated too far from the desired position. As noted, extremely long eyelashes and the rims of certain glasses can interfere with the data collection process. Therefore, if any objects interfered with the data collection process, that particular data was excluded from the analysis.

The data obtained from the eye tracking system software for each participant was mapped against each level of the video game played. According to Duchowski (2007, p. 168), eye movement data is generally parametric "because related metrics can be represented by a uniform interval/ratio scale." The dependent variable (DV) in this phase of the study was attentional quality. Attentional quality will be operationalised as the number of long eye fixations ($> 600$ ms) and endogenous blinks. The independent variable (IV) is each replay experience. A two tail t-test was performed on the pair of means test for statistical significance in the difference of the means. According to Neill (2008) "significance tests conducted with low power can be misleading" (para. 2) and therefore the effect size will be reported. A measure of the standardised mean effect size (Cohen's $d$) was used to understand what was the effect of the intervention.

### 3.6.7   In-game performance

Video games enable the user to make choices, decisions, mistakes, and in some games – in-game purchases. The variables that are derived from changes to the game state are the result of the interaction of the player with the game (Canossa, 2013). These changes in the game state can be captured from within the source of the game or by using external data acquisition systems (Canossa, 2013). Analysis of user-decisions and actions within video games have been used for a variety of purposes. It can help the developer refine and polish the game, and it can be used to get an understanding of the user decisions and actions while playing the game (Drachen et al., 2013; Canossa, 2013). These points of data can then be analysed to help the video game developer make informed business decisions about the product during and after the product has been developed. There are several ways that this data can be obtained. Game telemetry data acquisition is the process of getting data from a distance (Drachen et al., 2013). This data can be obtained by remote monitoring, analysis of video game servers, in-game purchase transaction analysis, conversion tracking, or production data (Distimo, 2013; Drachen et al., 2013). This raw data can then be transformed into game metrics that can be used as quantifiable measures such as average completion time, income per day, the number of active users, in-game purchases, and so on (Distimo, 2013; Drachen et al., 2013).

While the adult participants played the video game, their choices and selections were video recorded. The rationale for recording the hand gestures is that when the participant is using an iPad, it is not possible to record mouse clicks (as the iPad does not have a mouse). Further, the participant may move their finger to a particular part of the screen (without touching the screen) and hesitate. The application software would not record this hesitation, as the participant had not selected anything. However, this hesitation could be a result of the user thinking about how to solve that particular problem (amongst other reasons) and is, therefore, potentially valuable data that needs to be collected.

In the video game World of Goo (2007), data on the number of moves made, how much time is taken to complete the level, and the number of 'gooballs' saved is available at the completion of each level. Comparisons were made between the replay experiences of each individual user. Furthermore, comparisons of each completed level were made to analyse any changes in tower construction between each level.

## 3.7 LIMITATIONS

The limitations of this study are that laboratory work is restricted to two commercial video games, and the definition of learning is that adopted by Gee (2003). Although it is possible to identify transfer of learning from the previous video games the participants have played, the transfer of the findings will be limited by the case study methodology and no causal links can be made to any other video games.

Another limitation is that through the experimental design it will not be possible to identify if the participant was guessing and/or completed the game by chance. However, the mechanics of the video game in the treatment require the player to repeat a similar task multiple times (move gooballs), if the outcome of the decision was a guess or by chance, this would be evident in their next move. However, there was the possibility that the participant could guess the answers to the questions in the structured questionnaire. The results of the magnetic construction task would also corroborate these answers. Thus, if the participant guessed the answer to the question in the structured questionnaire, it would be unlikely that this would be supported by the construction method used to build the magnetic tower.

The other limitation is that the physical measurements and observations made are proxy indicators of learning. As noted in Chapter Two, it is possible to conclude from the evidence that the research subject has learnt to solve a particular problem, when this may not have been the case. For example, the subject may have solved the problem by chance (a fluke). It also could be argued that if the participant sought the solution through seeking it outside of the game (by asking a friend or finding one on an online forum), then it is possible that the participant may have only found the solution but may not have thoroughly learnt how to solve the problem. Furthermore, this study is based on the basis that delays in solving problems can be a result of deficiencies in learning. However, this may not necessarily be the case. A delay in solving the problem could be a result of other influences. For example, the participant could have been thinking about something else (what they will have for lunch, daydreaming, or any other thought process). Furthermore, the participant could have known the solution but may have wanted to experiment or play with other elements of the game. The experiments conducted gave the participants one or two attempts to play the video game before the measurement of the transfer of learning will take place.

Another limitation is the effect of the repeated exposure to the treatment. Newell and Rosenbloom (1981), demonstrate that repeated practice results in improvements in performance (the law of practice). In the experimental design, the participants were subjected to two exposures of the video game. Any short-term improvements in performance could have been due to learning from the embedded pedagogy in the game or through the repeated exposure. However, the experimental design used a number of measures to test what learning transpired. The treatment did not include any explicit instruction on tower construction or identification as to what shape was the strongest shape for tower construction.

Finally, another potential limitation of this study was that the temporal delay between the exposure to the video game and the transfer test was minimal (five to ten minutes only). This study did not investigate the impact of an extended temporal delay between the exposure to the video game and the tower construction process. While it would be appealing to get a better understanding of the antecedents of retained learning, this is outside the scope of this study but will be seriously considered for future research.

## 3.8 CONCLUSION

In this chapter, the methodology was developed for this empirical research. A mixed-methods approach was adopted to alleviate some of the challenges with the research methods used in game user research. This research triangulated data from a variety of sources so that reliability and validity can be assured. Five different data collection sources were proposed, to ensure that the measurement of any performance improvements were reliably measured. Two decidedly different types of participants were chosen to provide an understanding of the potential and benefit of transference from prior experience and existing knowledge. Half of the participants were exposed to the treatment, and the other half were exposed to a placebo (play a video game that does not involve tower construction). Although, it was possible to observe and measure performance improvement in both groups, the study also measured the transfer of the in-game learning to an external non-game context. This phase of the study measured any transference from the concepts learnt in the video game and an external non-game context.

As many of the concepts under investigation were intangible, a wide variety of measures has been included to ensure that the necessary data is available to test the research questions and hypotheses fully.

This chapter also acknowledges some of the limitations of the methods used in this study. While each method in isolation has its own limits, this proposed research uses a mixed-methods approach to mitigating these limitations.

The outcome of this research could provide a quantitative and qualitative basis for understanding the potential for learning from a commercial video game. Furthermore, the method could also be adopted to provide the game development community a method for testing game usability, which could be used to obtain metrics that may help the general public get a better understanding of what age groups are suitable for a particular educational or commercial video game.

<div align="center">

**Chapter Four**

**FINDINGS**

</div>

## 4.0 INTRODUCTION

The method outlined in Chapter Three was implemented through running two studies at Northeastern University during the Spring and Winter term of the 2013-2014 academic year. The time of year was especially challenging to recruit participants due to the two major public holidays (Thanksgiving and Christmas). It was, however, possible to recruit adult students that were enrolled in courses at Northeastern University, the week prior to and after the Thanksgiving holiday. The University term break made the recruitment of the students a lot more difficult, but because this coincided with the Primary School holiday, it became easier to recruit the children.

This chapter will report the findings of the two studies through using the relevant frameworks used in previous studies – these were identified in Chapter Three. The findings from the study of adults will be reported in Section 4.1.1, and the findings from the study of children will follow in Section 4.1.2. The implications of the findings will be further discussed in Chapter Five.

## 4.1 THE STUDIES

The study of the adults took place in the PlaIT Laboratory, which is located on campus at Northeastern University's main campus in Boston. The Northeastern University Internal Review Board (ethics committee) provided ethics approval (IRB #13-9-10). The laboratory consisted of a number of meeting rooms, a user experience lab, and offices. The availability of an unused office made it possible to dedicate a room to this study. As a result, the equipment could be set up and left on site (Figure 4.1). This room was on the top level of the building and had reasonable access and sufficient lighting.

The iPad was fixed to the mounting, which was then affixed to the desk to ensure that the iPad did not move (Figure 4.2). This was to ensure that the webcam had a consistent line of sight with the iPad. However, to use the iPad when it was fixed to the desk, the angle of the participants' head was aimed directly at the iPad, which was frequently too acute to get a consistent line of sight between the eye tracker and the participants' retina. From this initial study, it was found that the gaze of each participant would frequently fall outside the range of the eye tracker. Therefore, after consultation with representatives from SMI, it was decided to study children while they used a Personal Computer. Through displaying the game on a computer monitor, this would ensure that the participants' head position (and, therefore, their eyes) would be within the range of the eye tracker.

The study of the children took place in the Speech Language Pathology and Audiology Laboratory at Northeastern University. The Northeastern University Internal Review Board (IRB 13-10-13) provided the ethics approval. The lab is used extensively for the research of children, and this facilitated the identification of potential participants and recruitment channels. The purpose built laboratory made it possible to set up the measurement equipment and leave it in place (Figure 4.3). Once the equipment was set up and calibrated, it did not have to be reconfigured for each experiment. From the pilot testing, it was found that the children would move excessively in the office chair (this would result in the loss of eye gaze data). Therefore, a fixed seat was used that ensured the participant did not swivel in the chair. Further, this chair had an adjustable seat which would enable the height to be adjusted for each participant. The laboratory also had a frosted glass window (Figure 4.4), which enabled parents and siblings to be separated from the child if needed. Both configurations allowed the researcher to sit relatively close to the participants which enabled immediate communication and direct observation.

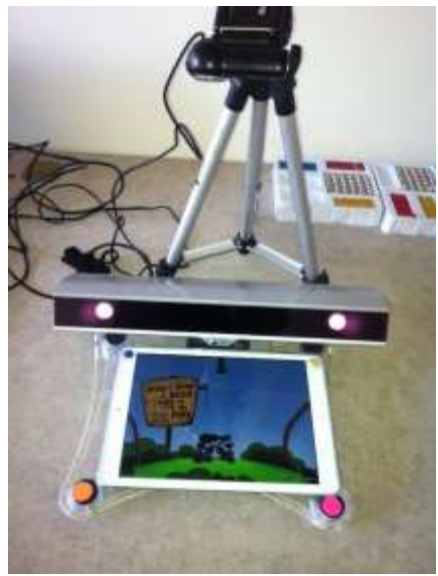

*Figure 4.1 Configuration of the adult study*

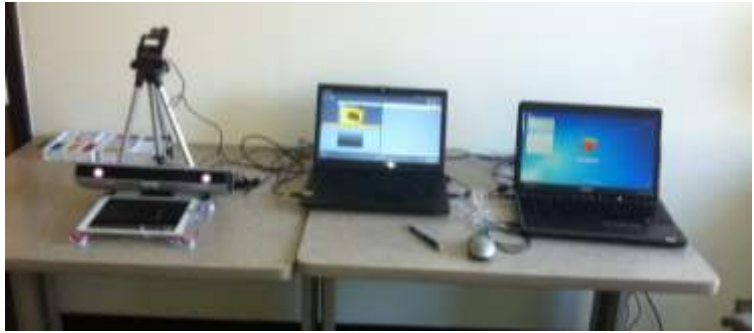

*Figure 4.2 Setup of the adult study*

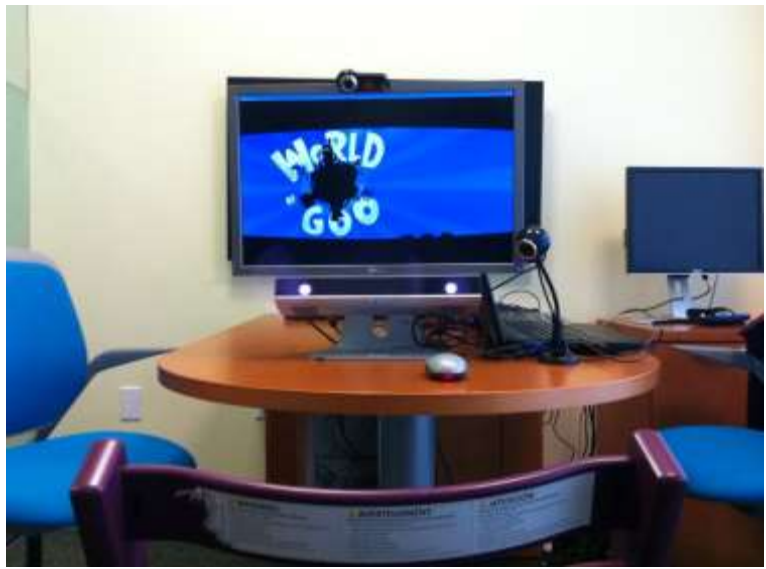

*Figure 4.3 Configuration of the study of the children*

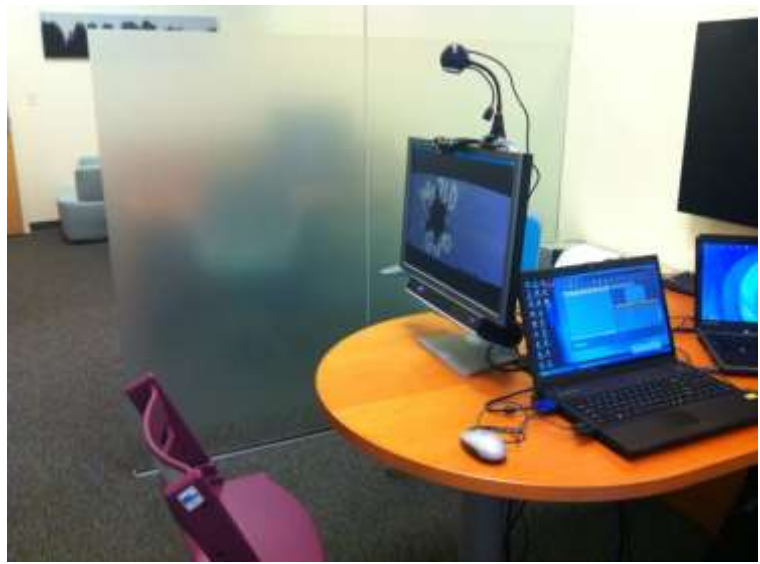

*Figure 4.4 Setup of the study of the children*

### 4.1.1 The study of the adults

Due to the availability of the adult student population, this study took place first. The adult study consisted of 20 participants. All of these participants were aged between 18-29 years old. Eighty-six percent of the participants were male, and 14 percent were female. All of the participants reported that they played video games. Forty-five percent of the participants indicated that they played video games five to six times a week, whereas 25 percent said that they played video games three to four times a week. The participants reported that they played video games on a Personal Computer (PC) (95%), console (Xbox, PlayStation, or Wii) (85%), cellular connected phone or tablet device (80%), and handheld devices (PSP, Nintendo DS, Game Boy, or iPod) (45%). The participants were studying a variety of disciplines that included: Computer Science, Engineering, Economics, Architecture, and Pharmacy. However, one participant had not declared their major. Prior exposure to playing World of Goo was higher than anticipated, with 50 percent of the respondents reporting that they had played the game before. However, the amount of prior exposure was low as only four participants reported that they had played the game more than four times. Ten percent of the participants reported prior exposure to playing Bad Piggies.

As detailed in Chapter Three, all the participants participated in playing video games. The treatment and control group played a different game. Due to the nature of the experiment, the first and second attempts at the game were sequential. The two conditions were undertaken in the same order.

Table 4.1 shows the answers to the question, "What shape do you think is the strongest for building a really tall tower?" The correct answer was a triangle.

| Participant | Group | Pre-exposure answer | Post-exposure answer |
|---|---|---|---|
| A-03 | Control | Triangle | Square |
| A-04 | Control | Triangle | Triangle |
| A-05 | Control | Circle | Triangle |
| A-07 | Control | Triangle | Triangle |
| A-09 | Control | Square | Triangle |
| A-13 | Control | Square | Square |
| A-15 | Control | Triangle | Triangle |
| A-17 | Control | Triangle | Triangle |

| A-18 | Control | Triangle | Triangle |
|------|---------|----------|----------|
| A-01 | Treatment | Triangle | Triangle |
| A-02 | Treatment | Triangle | Triangle |
| A-06 | Treatment | Triangle | Triangle |
| A-08 | Treatment | Triangle | Triangle |
| A-10 | Treatment | Triangle | Triangle |
| A-11 | Treatment | Triangle | Triangle |
| A-12 | Treatment | Square | Triangle |
| A-14 | Treatment | Triangle | Triangle |
| A-16 | Treatment | Triangle | Triangle |
| A-19 | Treatment | Triangle | Triangle |
| A-20 | Treatment | Triangle | Triangle |

*Table 4.1 Answers to the tower construction question*

The participants were then asked to build a tall tower with the magnetic construction set after they played the video game. The methods used in the constructed process were noted by using structured observation. Table 4.2 presents the construction method of each participant.

| Participant | Group | Pre-exposure construction | Post-exposure construction |
|---|---|---|---|
| A-03 | Control | Triangles, Diamonds, Stick | Triangles, Stick |
| A-04 | Control | Triangles, Squares | Triangles, Stick |
| A-05 | Control | Triangles, Diamonds, Stick | Triangles, Squares, Stick |
| A-07 | Control | Squares, Triangle | Triangles, Squares |
| A-09 | Control | Squares | Squares |
| A-13 | Control | Squares | Triangles |
| A-15 | Control | Triangles, Squares | Triangles, Squares |
| A-17 | Control | Triangles, Squares, Stick | Triangles, Squares, Stick |
| A-18 | Control | Squares, Triangles, Squares | Squares, Triangles, Squares |
| A-01 | Treatment | Squares | Triangles, Squares |
| A-02 | Treatment | Triangles | Triangles, Squares |
| A-06 | Treatment | Triangles, Squares | Triangles, Stick |
| A-08 | Treatment | Triangles, Diamonds | Triangles, Squares |
| A-10 | Treatment | Triangles | Triangles, Squares |
| A-11 | Treatment | Triangles, Squares | Triangles, Squares |
| A-12 | Treatment | Triangles, Squares | Triangles |
| A-14 | Treatment | Triangles, Squares, Stick | Triangles |
| A-16 | Treatment | Triangles, Squares | Triangles, Stick |
| A-19 | Treatment | Triangles, Squares, Stick | Triangles, Stick |
| A-20 | Treatment | Triangles, Squares | Triangles, Squares |

*Table 4.2 Magnetic Tower construction methods*

The participants were asked two open-ended questions relating to tower construction after the gameplay had concluded. The results of the answers are presented in Table 4.3.

| Participant | Group | What is important when building a really tall tower? | Why did the tower break? |
|---|---|---|---|
| A-03 | Control | Trying to attend to outside forces | Not sure |
| A-04 | Control | Base strong, not thin. Did not support it properly | Bottom fragile |
| A-05 | Control | The foundations are stable | Not stable |
| A-07 | Control | Stability | Because I focused on too much height, not enough stability |
| A-09 | Control | Looks, stability, precision needed | Construction was weak. Used squares that are not as strong as a triangle, and an uneven distribution of weight |
| A-013 | Control | Using a few pipes to have a strong support | Built it the wrong way around, instead of building it forward |
| A-015 | Control | Trying to keep the structural integrity. Strong shapes like triangles or maybe an arch | Too top heavy, nothing to fasten it to |
| A-017 | Control | Worry whether it is structurally sound. Can take out support where it is not needed | N/A - Did not fall |
| A-018 | Control | Structural stability on the base | I tried to go straight up, but there is more room for error |
| A-01 | Treatment | Foundation | Foundation was not strong enough |
| A-02 | Treatment | Strong base | Weight not distributed evenly |
| A-06 | Treatment | Foundation | Shaky hands |
| A-08 | Treatment | Make sure the foundations are stable | Too much focus on height and structure |
| A-010 | Treatment | Foundations | The foundation was not strong |
| A-011 | Treatment | Keeping it balanced. Rigid shapes | Because it was not strong enough. Too heavy on one side |
| A-012 | Treatment | Good structure | N/A - Towers did not fall |
| A-014 | Treatment | Timing | It swayed unexpectedly, not perfectly balanced |

| A-016 | Treatment | Foundation | Foundation was not strong enough |
| A-019 | Treatment | Depends. A strong support in the base | Because maybe the wind; it was flexible; tool tips recommended going faster |
| A-020 | Treatment | Foundation | Foundation was not strong enough |

*Table 4.3 Answers to the open-ended questions*

Table 4.4 presents the video game performance of the treatment group. This table presents the number of gooballs collected, the number of moves taken to complete the tower construction, and the time taken to complete the level for both attempts one and two. The variance between each metric is also included which exhibits the additional balls collected, the number of additional moves, and the time saved. If the participant took more moves to complete the level in the second attempt at the game, this is presented as a positive number. If the participant took more time to complete the second attempt at the game, this is present as a negative number (in brackets). The delay between each attempt at the game was limited by the time the participant took to build the second magnetic tower. The delay was between five to ten minutes.

| | Attempt 1 | | | Attempt 2 | | | Variance | | |
| ID | Gooballs collected | Number of moves | Time taken | Gooballs collected | Number of moves | Time taken | Additional gooballs collected | Additional moves | Time saved |
|---|---|---|---|---|---|---|---|---|---|
| A-01 | 7 | 7 | 0:39 | 10 | 6 | 0:31 | 3 | -1 | 0:08 |
| A-02 | 8 | 9 | 0:40 | 11 | 5 | 0:30 | 3 | -4 | 0:10 |
| A-06 | 7 | 6 | 0:35 | 11 | 3 | 0:30 | 4 | -3 | 0:05 |
| A-08 | 9 | 5 | 0:29 | 10 | 3 | 0:20 | 1 | -2 | 0:09 |
| A-010 | 7 | 7 | 0:22 | 9 | 5 | 0:23 | 2 | -2 | (0:01) |
| A-011 | 8 | 11 | 1:00 | 7 | 7 | 0:16 | -1 | -4 | 0:44 |
| A-012 | 7 | 8 | 0:24 | 9 | 7 | 0:26 | 2 | -1 | (0:02) |
| A-014 | 7 | 8 | 0:34 | 10 | 6 | 0:21 | 3 | -2 | 0:13 |
| A-016 | 5 | 7 | 0:39 | 8 | 5 | 0:27 | 3 | -2 | 0:12 |
| A-019 | 6 | 8 | 0:37 | 8 | 6 | 0:15 | 2 | -2 | 0:22 |
| A-020 | 10 | 4 | 0:34 | 6 | 7 | 0:49 | -4 | +3 | (0:15) |

*Table 4.4 Treatment group game performance*

### 4.1.1.1 Endogenous blinks

At the conclusion of the experiments, the SMI BeGaze™ eye tracking software (BeGaze Analysis Software, n.d.) was used to identify the blinks of each participant. The data file was exported to a tab-delimited file, which was then used for analysis. Although there is some debate as to the duration of an endogenous blink, the endogenous blink was operationalised as being between 100 and 300 milliseconds (ms), which is within the minimum and maximum range identified in the literature (Evinger et al., 1991; Guitton et al., 1991; Stava et al., 1994; VanderWerf et al., 2003). As the SMI RED500 is a bi-ocular device, the system identified the closing of the eyelid in either or both eyes. As this study was focused on blinking (the closure of both eyelids) and not winking (the closure of one eyelid), when the software identified that the participant had blinked with both eyes, the data was included in the calculation. Where the software identified that the participant had blinked with one eye, this data was omitted. Furthermore, where the eye tracker identified a loss of signal as a blink that was outside of the range of an eye blink (Evinger et al., 1991; Guitton et al., 1991; Stava et al., 1994; VanderWerf et al., 2003), this data was also excluded. Table 4.5 and 4.6 present the frequency of blinks for each participant for the first and second attempts. The area of interest was those blinks that lasted between 100 and 300 ms. However, the blinks that were below and above this range have also been included for comparison. As a result of the acute angle between the participants' retina and the eye tracker, it was not possible to capture some data, therefore, the blink data from participants A-01, and A-13 is not included. The eye gaze data files from the experiments for participants A-07, A-05, A-06, A-07, A-09, A-10 and A-11 were corrupted or incomplete (due to signal loss) and could not be included.

| Blink duration (ms) | A-02 | A-03 | A-04 | A-12 | A-14 | A-15 | A-16 | A-17 | A-18 | A-19 | A-20 |
|---|---|---|---|---|---|---|---|---|---|---|---|
| 0-49 | 61 | 18 | 40 | 7 | 7 | 13 | 5 | 21 | 39 | 13 | 9 |
| 50-100 | 12 | 20 | 22 | 16 | 11 | 15 | 3 | 13 | 21 | 7 | 1 |
| 101-150 | 16 | 18 | 20 | 13 | 6 | 21 | 2 | 6 | 23 | 8 | 2 |
| 151-200 | 30 | 15 | 18 | 5 | 4 | 14 | 2 | 14 | 21 | 7 | 4 |
| 201-250 | 6 | 7 | 14 | 5 | 4 | 20 | 1 | 6 | 9 | 2 | 0 |
| 251-300 | 4 | 8 | 8 | 2 | 0 | 6 | 0 | 7 | 3 | 2 | 0 |
| 301-350 | 2 | 15 | 17 | 3 | 2 | 6 | 0 | 3 | 3 | 3 | 0 |
| 351-400 | 7 | 12 | 15 | 1 | 2 | 8 | 1 | 6 | 12 | 3 | 1 |

*Table 4.5 Eye blinks between 0 and 400 ms (Attempt 1)*

| Blink duration (ms) | A-02 | A-03 | A-04 | A-12 | A-14 | A-15 | A-16 | A-17 | A-18 | A-19 | A-20 |
|---|---|---|---|---|---|---|---|---|---|---|---|
| 0-49 | 62 | 13 | 40 | 2 | 32 | 31 | 4 | 30 | 14 | 19 | 0 |
| 50-100 | 24 | 17 | 17 | 20 | 18 | 16 | 2 | 8 | 11 | 12 | 1 |
| 101-150 | 5 | 13 | 17 | 11 | 10 | 12 | 1 | 5 | 6 | 7 | 0 |
| 151-200 | 4 | 13 | 20 | 6 | 8 | 11 | 1 | 2 | 7 | 4 | 1 |
| 201-250 | 3 | 6 | 8 | 1 | 5 | 6 | 0 | 2 | 4 | 2 | 0 |
| 251-300 | 4 | 6 | 9 | 7 | 9 | 4 | 1 | 1 | 4 | 1 | 0 |
| 301-350 | 3 | 8 | 9 | 7 | 7 | 3 | 0 | 3 | 2 | 3 | 0 |
| 351-400 | 6 | 11 | 13 | 5 | 4 | 12 | 0 | 2 | 6 | 1 | 1 |

*Table 4.6 Eye blinks between 0 and 400 ms (Attempt 2)*

A box-and-whisker plot (Tukey, 1977) was created to help understand the descriptive statistics and any extreme values. Figure 4.5 and Figure 4.6 and Figure 4.6 present a box-and whisker plot for the blinks that were between 50 and 300 ms in the first and then second attempt at the game.

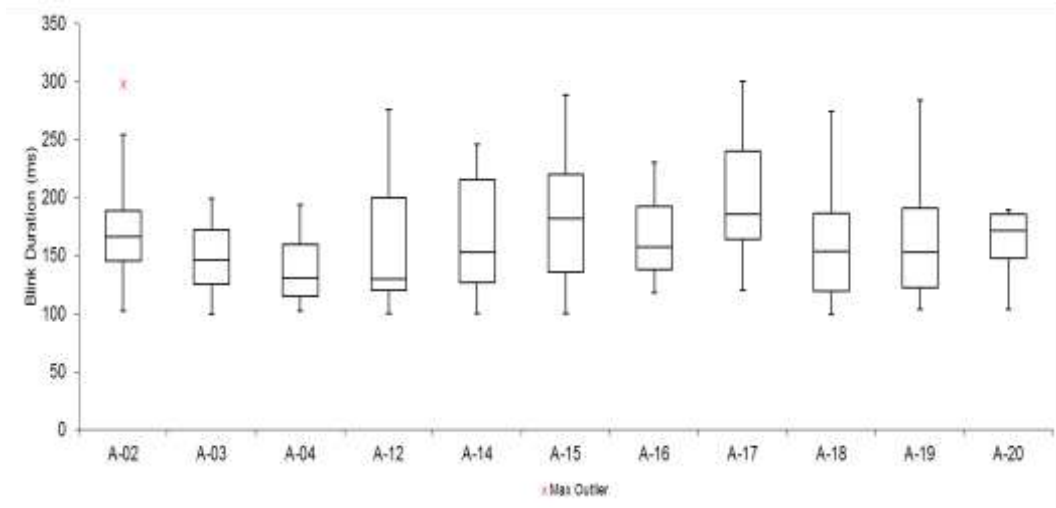

*Figure 4.5– Eye blinks between 100 and 300 ms (Attempt 1)*

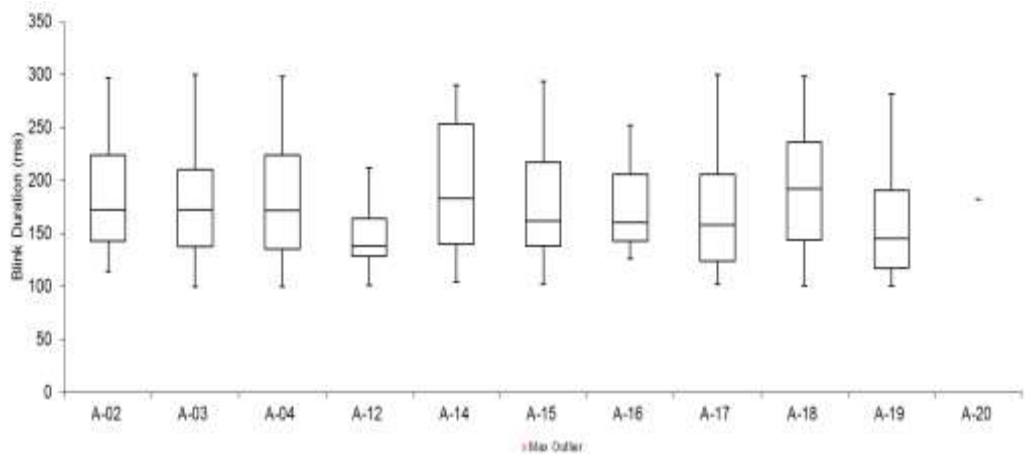

*Figure 4.6– Eye blinks between 100 and 300 ms (Attempt 2)*

For the blinks between 100 and 300 ms in the first attempt ($M = 34.82$, $SD = 22.12$) and the second attempt ($M = 22.45$, $SD = 15.98$). The effect size for this analysis ($d = 0.67$) was found to be above Cohen's (1988) convention for a medium effect ($d = .50$).

However, as the duration of each gameplay varied, these results may be misleading. The percentage of the endogenous blinks for each participant is presented in the following Tables (4.7 and 4.8) to provide a normalised comparison between each game play experience.

| Blink duration (ms) | A-02 | A-03 | A-04 | A-12 | A-14 | A-15 | A-16 | A-17 | A-18 | A-19 | A-20 |
|---|---|---|---|---|---|---|---|---|---|---|---|
| 0-49 | 44.2 | 15.9 | 26.0 | 13.5 | 19.4 | 12.6 | 35.7 | 27.6 | 29.8 | 28.9 | 52.9 |
| 50-100 | 8.7 | 17.7 | 14.3 | 30.8 | 30.6 | 14.6 | 21.4 | 17.1 | 16.0 | 15.6 | 5.9 |
| 101-150 | 11.6 | 15.9 | 13.0 | 25.0 | 16.7 | 20.4 | 14.3 | 7.9 | 17.6 | 17.8 | 11.8 |
| 151-200 | 21.7 | 13.3 | 11.7 | 9.6 | 11.1 | 13.6 | 14.3 | 18.4 | 16.0 | 15.6 | 23.5 |
| 201-250 | 4.3 | 6.2 | 9.1 | 9.6 | 11.1 | 19.4 | 7.1 | 7.9 | 6.9 | 4.4 | 0.0 |
| 251-300 | 2.9 | 7.1 | 5.2 | 3.8 | 0.0 | 5.8 | 0.0 | 9.2 | 2.3 | 4.4 | 0.0 |
| 301-350 | 1.4 | 13.3 | 11.0 | 5.8 | 5.6 | 5.8 | 0.0 | 3.9 | 2.3 | 6.7 | 0.0 |
| 351-400 | 5.1 | 10.6 | 9.7 | 1.9 | 5.6 | 7.8 | 7.1 | 7.9 | 9.2 | 6.7 | 5.9 |

*Table 4.7 Percentage of blink occurrence within each range (Attempt 1)*

| Blink Duration (ms) | A-02 | A-03 | A-04 | A-12 | A-14 | A-15 | A-16 | A-17 | A-18 | A-19 | A-20 |
|---|---|---|---|---|---|---|---|---|---|---|---|
| 0-49 | 55.9 | 14.9 | 30.1 | 3.4 | 34.4 | 32.6 | 44.4 | 56.6 | 25.9 | 38.8 | 0.0 |
| 50-100 | 21.6 | 19.5 | 12.8 | 33.9 | 19.4 | 16.8 | 22.2 | 15.1 | 20.4 | 24.5 | 33.3 |
| 101-150 | 4.5 | 14.9 | 12.8 | 18.6 | 10.8 | 12.6 | 11.1 | 9.4 | 11.1 | 14.3 | 0.0 |
| 151-200 | 3.6 | 14.9 | 15.0 | 10.2 | 8.6 | 11.6 | 11.1 | 3.8 | 13.0 | 8.2 | 33.3 |
| 201-250 | 2.7 | 6.9 | 6.0 | 1.7 | 5.4 | 6.3 | 0.0 | 3.8 | 7.4 | 4.1 | 0.0 |
| 251-300 | 3.6 | 6.9 | 6.8 | 11.9 | 9.7 | 4.2 | 11.1 | 1.9 | 7.4 | 2.0 | 0.0 |
| 301-350 | 2.7 | 9.2 | 6.8 | 11.9 | 7.5 | 3.2 | 0.0 | 5.7 | 3.7 | 6.1 | 0.0 |
| 351-400 | 5.4 | 12.6 | 9.8 | 8.5 | 4.3 | 12.6 | 0.0 | 3.8 | 11.1 | 2.0 | 33.3 |

*Table 4.8 Percentage of blink occurrence within each range (Attempt 2)*

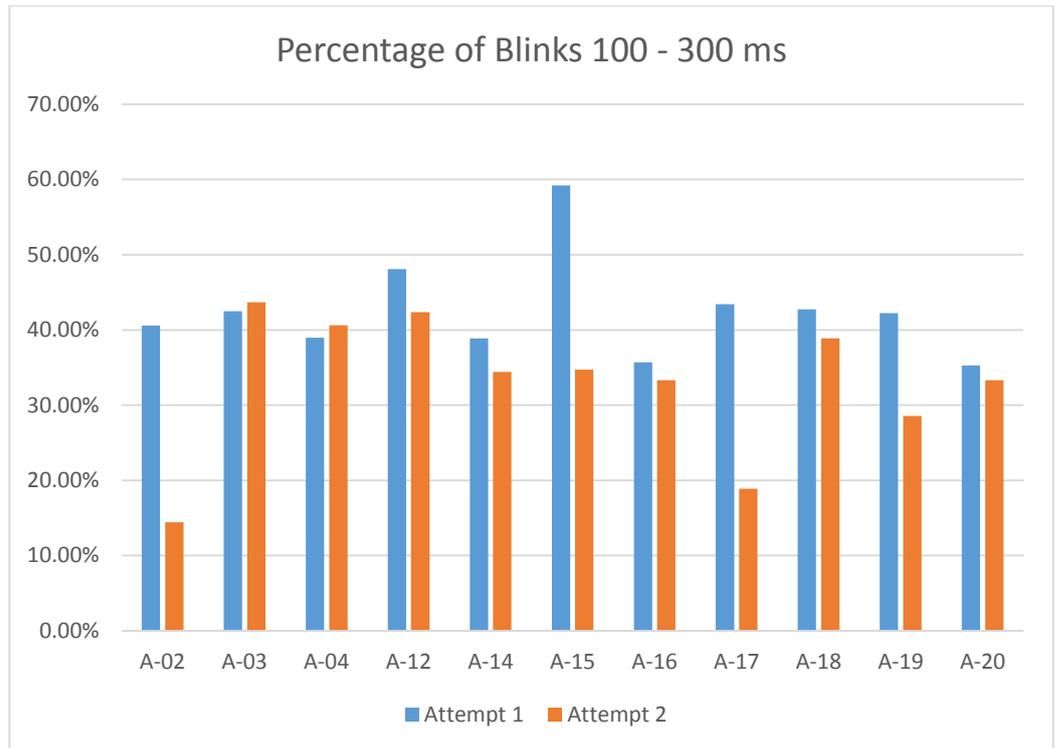

*Figure 4.7 The percentage of blinks between 100 and 300 ms*

Figure 4.7 presents the percentage of eye blinks that were between 100 and 300 ms for each participant for the first and second attempts at the game.

For the percentage of blinks between 100 and 300 ms in the first attempt ($M = 0.43$, $SD = 0.07$) and the second attempt ($M = 0.33$, $SD = 0.09$), the effect size for this analysis ($d = 1.30$) was found to be above Cohen's (1988) convention for a large effect ($d = .80$).

These results suggest that the frequency of blinks that were between 100 and 300 ms in the first attempt at the game may have a strong effect on the frequency of blinks that were between 100 and 300 ms in the second attempt at the game.

### 4.1.1.2 Fixation data

The SMI BeGaze™ eye tracking software (BeGaze Analysis Software, n.d.) was used to identify the individual fixations of each participant. Kenny et al., (2005) presented the results of the fixation data in a table that identified the number of fixations under 300 ms, the number of fixations between 300 ms and 600 ms, and the number of fixations over 600 ms. Using the framework provided by Kenny et al., (2005), Table 4.9 presents the fixation data for each participant for the first attempt at the video game. Table 4.10 presents the fixation data for the second attempt at the game. The table also identifies fixations from 600 to 899 ms, 900 to 1199 ms, and 1200 ms plus. The rationale for providing the additional data is based on the interest in longer fixations, which are considered to be indicators of cognition (Evinger et al., 1991; Guitton et al., 1991; Orchard & Stern, 1991; Stava et al., 1994). Reporting this fine detail may provide a greater understanding of these fixations which potentially indicate that the participant was struggling with understanding how to play the game, or make a decision. The data from participants A-01, A-05, A-06, A-07, A-08, A-09, A-10, A-11, and A-13 have been excluded as the data was incomplete or corrupt.

| Fixation duration freq. (ms) | A-02 | A-03 | A-04 | A-12 | A-14 | A-15 | A-16 | A-17 | A-18 | A-19 | A-20 |
|---|---|---|---|---|---|---|---|---|---|---|---|
| 0-299 | 44 | 56 | 81 | 42 | 11 | 53 | 5 | 142 | 41 | 4 | 3 |
| 300-599 | 5 | 28 | 53 | 3 | 7 | 15 | 4 | 44 | 8 | 4 | 5 |
| 600-899 | 8 | 7 | 15 | 1 | 1 | 3 | 3 | 19 | 3 | 2 | 0 |
| 900-1199 | 0 | 4 | 5 | 0 | 1 | 2 | 2 | 9 | 2 | 0 | 0 |
| 1200 + | 0 | 0 | 0 | 0 | 0 | 0 | 4 | 15 | 7 | 0 | 0 |

*Table 4.9 Fixation duration (Attempt 1)*

| Fixation duration freq. (ms) | A-02 | A-03 | A-04 | A-12 | A-14 | A-15 | A-16 | A-17 | A-18 | A-19 | A-20 |
|---|---|---|---|---|---|---|---|---|---|---|---|
| 0-299 | 34 | 56 | 97 | 26 | 3 | 32 | 0 | 88 | 8 | 7 | 2 |
| 300-599 | 8 | 11 | 27 | 1 | 4 | 6 | 2 | 18 | 1 | 0 | 0 |
| 600-899 | 3 | 4 | 2 | 0 | 1 | 3 | 2 | 11 | 1 | 0 | 0 |
| 900-1199 | 1 | 0 | 0 | 0 | 0 | 0 | 1 | 6 | 0 | 0 | 0 |
| 1200 + | 0 | 0 | 0 | 0 | 0 | 1 | 0 | 0 | 3 | 0 | 0 |

*Table 4.10 Fixation duration (Attempt 2)*

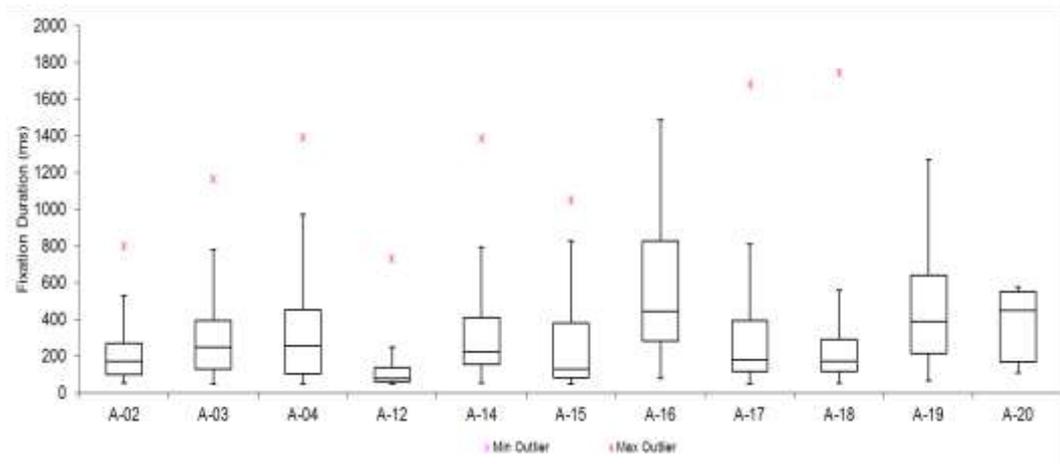

*Figure 4.8 Cumulative fixation (Attempt 1)*

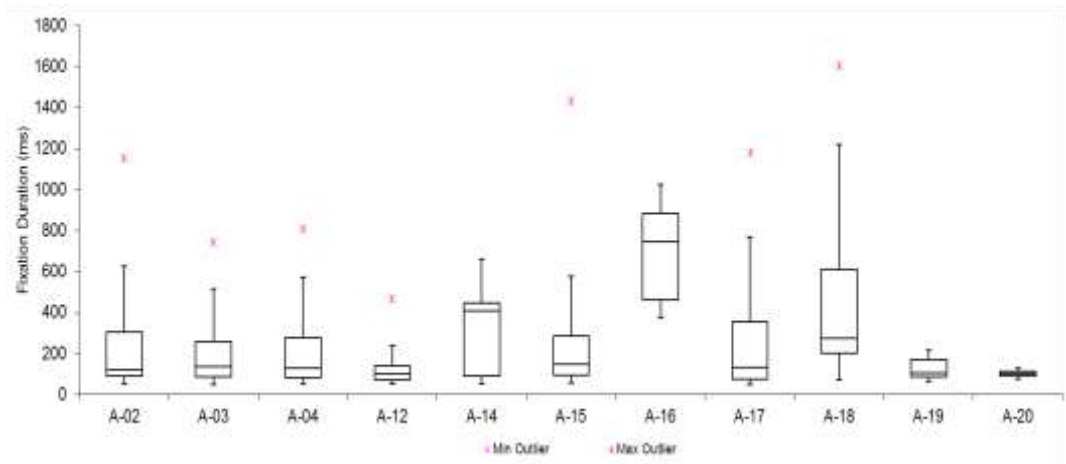

*Figure 4.9 Cumulative fixation (Attempt 2)*

Figure 4.8 presents a box-and-whisker plot (Tukey, 1977) of the cumulative fixations of the first attempt at the game for each participant. Figure 4.9 presents a box-and-whisker plot for the cumulative fixations for the second attempt at the game for each participant.

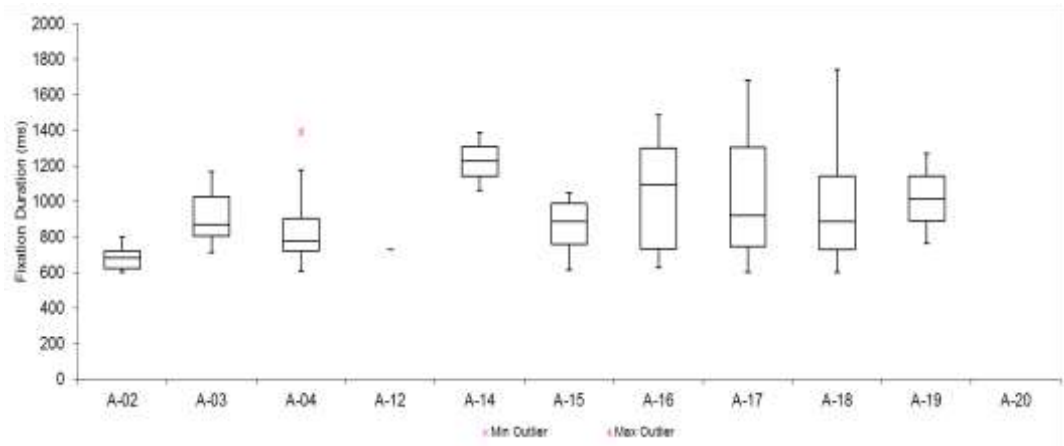

*Figure 4.10 Fixations above 600 ms (Attempt 1)*

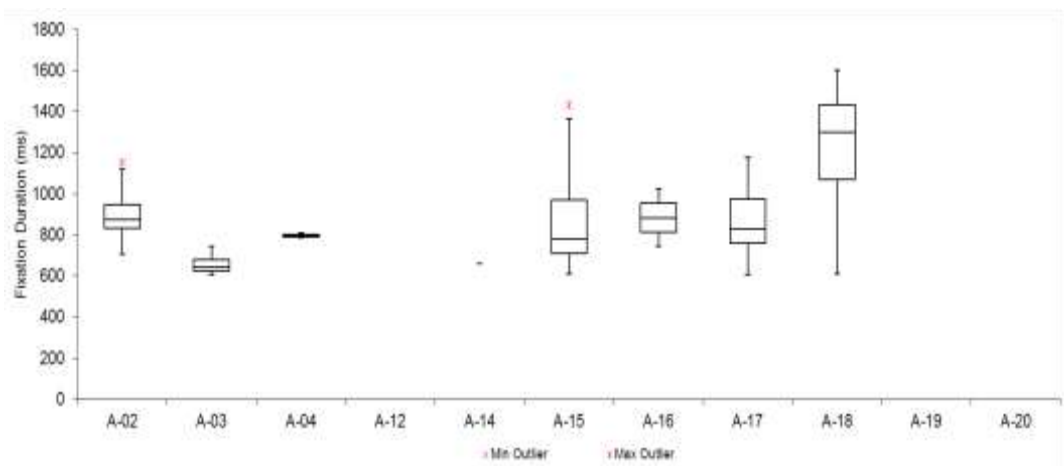

*Figure 4.11–Fixations above 600 ms (Attempt 2)*

Figure 4.10 and 4.11 exhibits the fixations that lasted more than 600 ms for each participant at each attempt of the game.

For the fixations greater than 600 ms in the first attempt ($M = 7.18$, $SD = 7.55$) and the second attempt ($M = 3.55$, $SD = 4.78$); $d =0.61$. The effect size for this analysis ($d = 0.61$) was found to be above Cohen's (1988) convention for a medium effect ($d = .50$). These results suggest that the number of fixations above 600 ms in the first attempt of the game has an effect on the number of fixations above 600 ms in the second attempt at the game.

As noted in the previous section, the comparisons between the first and second attempts may not provide an accurate comparison as the duration of gameplay varied in each attempt at the game. To provide a more standardised comparison of the data, the percentage of the fixation data for each participant is presented in Table 4.11 and Table 4.12.

| Fixation duration freq. (ms) | A-02 | A-03 | A-04 | A-12 | A-14 | A-15 | A-16 | A-17 | A-18 | A-19 | A-20 |
|---|---|---|---|---|---|---|---|---|---|---|---|
| 0-299 | 77.2 | 58.4 | 53.3 | 91.3 | 58.3 | 67.6 | 31.3 | 69.3 | 75.0 | 28.6 | 37.5 |
| 300-599 | 8.8 | 33.7 | 32.2 | 6.5 | 25.0 | 26.5 | 25.0 | 15.3 | 14.6 | 42.9 | 62.5 |
| 600-899 | 14.0 | 4.5 | 10.5 | 2.2 | 0.0 | 2.9 | 18.8 | 7.3 | 6.3 | 14.3 | 0.0 |
| 900-1199 | 0.0 | 3.4 | 2.6 | 0.0 | 8.3 | 2.9 | 12.5 | 2.9 | 2.1 | 0.0 | 0.0 |
| 1200 + | 0.0 | 0.0 | 1.3 | 0.0 | 8.3 | 0.0 | 12.5 | 5.1 | 2.1 | 14.3 | 0.0 |

*Table 4.11 Fixation duration % (Attempt 1)*

| Fixation duration freq. (ms) | A-02 | A-03 | A-04 | A-12 | A-14 | A-15 | A-16 | A-17 | A-18 | A-19 | A-20 |
|---|---|---|---|---|---|---|---|---|---|---|---|
| 0-299 | 73.9 | 78.9 | 77.0 | 96.3 | 37.5 | 76.2 | 0.0 | 71.5 | 61.5 | 100.0 | 100.0 |
| 300-599 | 17.4 | 15.5 | 21.4 | 3.7 | 50.0 | 14.3 | 40.0 | 14.6 | 7.7 | 0.0 | 0.0 |
| 600-899 | 6.5 | 5.6 | 1.6 | 0.0 | 12.5 | 7.1 | 40.0 | 8.9 | 7.7 | 0.0 | 0.0 |
| 900-1199 | 2.2 | 0.0 | 0.0 | 0.0 | 0.0 | 0.0 | 20.0 | 4.9 | 0.0 | 0.0 | 0.0 |
| 1200 + | 0.0 | 0.0 | 0.0 | 0.0 | 0.0 | 2.4 | 0.0 | 0.0 | 23.1 | 0.0 | 0.0 |

*Table 4.12 Fixation duration % (Attempt 2)*

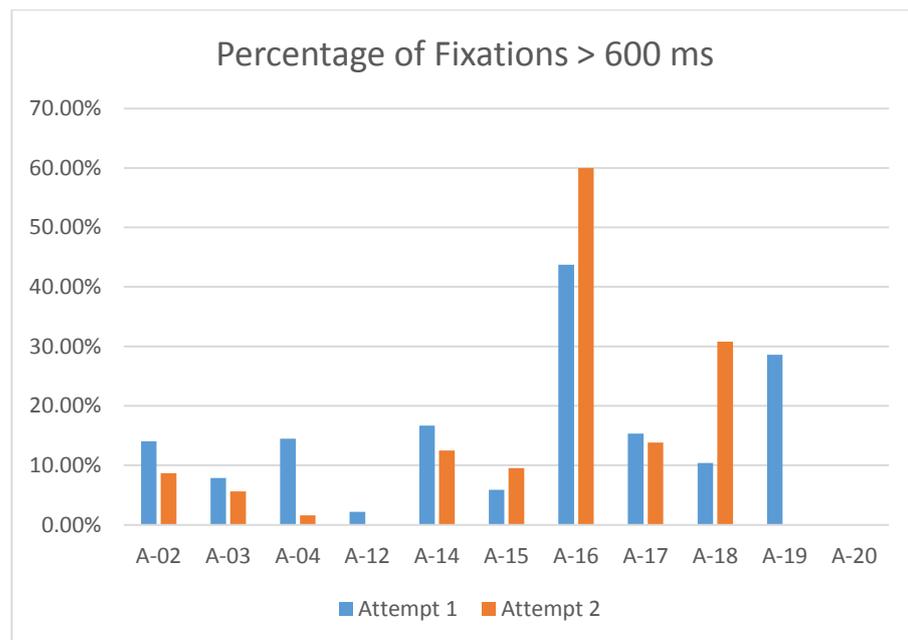

*Figure 4.12 The percentage of fixations of each participant greater than 600 ms*

Figure 4.12 presents the percentage of fixations that were greater than 600 ms for each participant.

For the fixations greater than 600 ms in the first attempt ($M = 0.14$, $SD = 0.12$) and the second attempt ($M = 0.13$, $SD = 0.18$); $d = 0.07$. The effect size for this analysis ($d = 0.07$) was found to be below Cohen's (1988) convention for a small effect ($d = .20$). These results suggest that the percentage of fixations greater than 600 ms in the first attempt of the game has some detectable effect on the number of fixations greater than 600 ms in the second attempt at the game.

### 4.1.1.3 Video Evidence

The gameplay was captured from the web camera and the eye tracking to further the understanding of what learning transpired during the game play. The data presented in the previous sections suggests that several participants did not demonstrate a reduction in the number of endogenous eye blinks and or the number of fixations that were longer than 600 ms. This section provides the specific evidence from the videos that were recorded. However, given the format of this document could not support video footage, screenshots of the video footage were taken. The video recordings were played on a personal computer and an image of what was on the screen was taken every second. These images were then scaled to fit onto the page and rotated 180 degrees to provide the reader with the same perspective as a participant. The highlights of the video evidence from participants in the treatment group, A-16, and A-20 is included below. Although the complete images of the video recording are available for these participants, presenting this would have added 20 additional pages to this document. The key images are presented to minimise the burden on the reader. To enable to reader to view these the complete screenshots of the video, a URL is provided in the description that links to this evidence.

The rationale for including this evidence will be discussed in detail in Chapter 5.

| 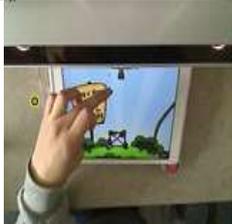 | 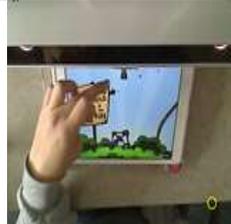 | 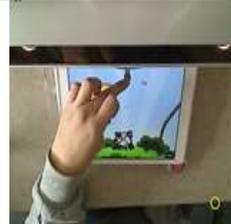 | 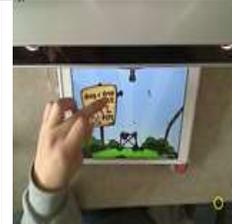 |
|---|---|---|---|
| Frame 8 | Frame 9 | Frame 10 | Frame 11 |
| In frames 1-5 (not shown), the participant watched the starting video sequence. <br> In frames 6-10, the participant moved a gooball to the pipe. Indicating he did not comprehend how to play the game | | | |

| 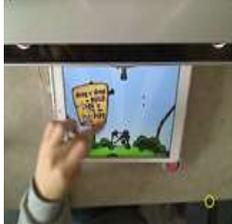 | 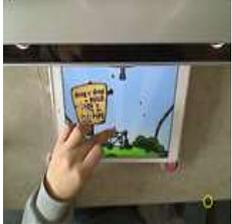 | 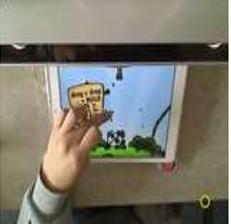 | 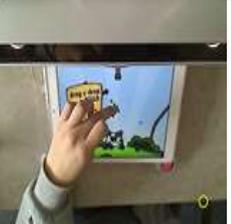 |
|---|---|---|---|
| Frame 12 | Frame 13 | Frame 14 | Frame 15 |

In frames 11-15, the participant moved another gooball to the pipe. Indicating he did not learn from the first experience.

| 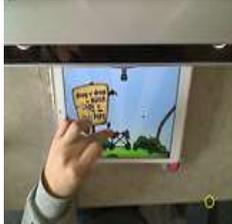 | 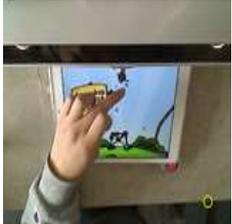 | 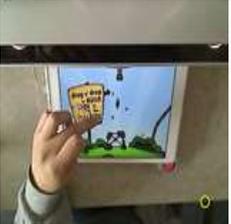 | 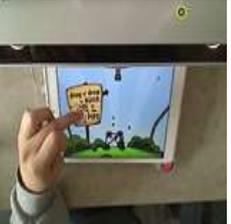 |
|---|---|---|---|
| Frame 16 | Frame 17 | Frame 18 | Frame 19 |

In frames 16-17, the participant moved another gooball to the pipe.
In frames 18-19, the participant watched the gooball fall from the pipe.

| 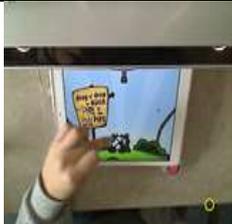 | 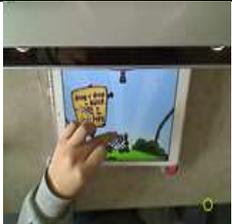 | 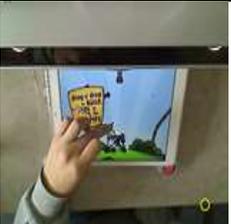 | 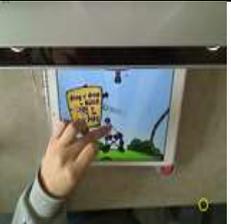 |
|---|---|---|---|
| Frame 20 | Frame 21 | Frame 22 | Frame 23 |

In frame 20, the participant selected a gooball.
In frames 21-25, the participant moves a gooball and starts to build a tower.
Thus suggesting he had worked out how to play the game.

| 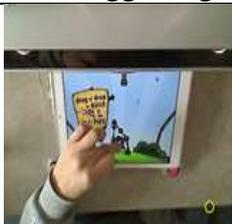 | 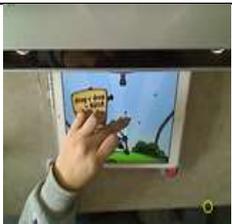 | 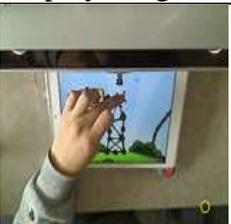 | 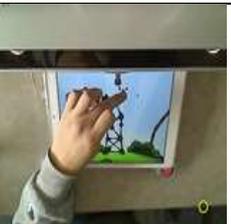 |
|---|---|---|---|
| Frame 24 | Frame 25 | Frame 31 | Frame 32 |

In frames 26-30 (not shown), the participant continued to build the tower.
In frames 31-35, the participant continued to build the tower with precision and speed. This indicated the required skills had been learnt

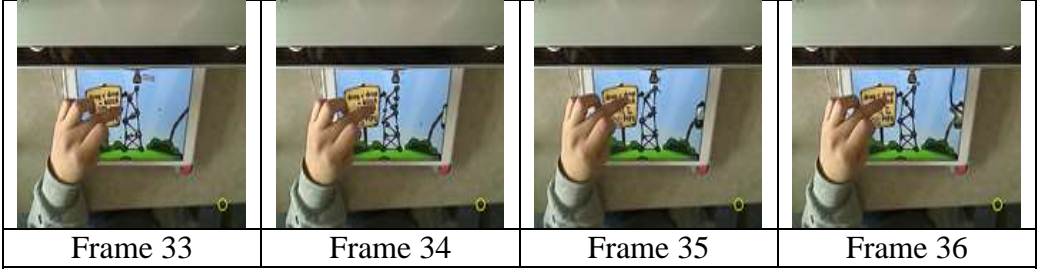

| Frame 33 | Frame 34 | Frame 35 | Frame 36 |

In frames 36, the participant watched the end of game video sequence
In frames 37-48 (not shown), the participant watched the end of game video sequence.

*Figure 4.13 Participant A-16 Attempt 1([Link](Link))*

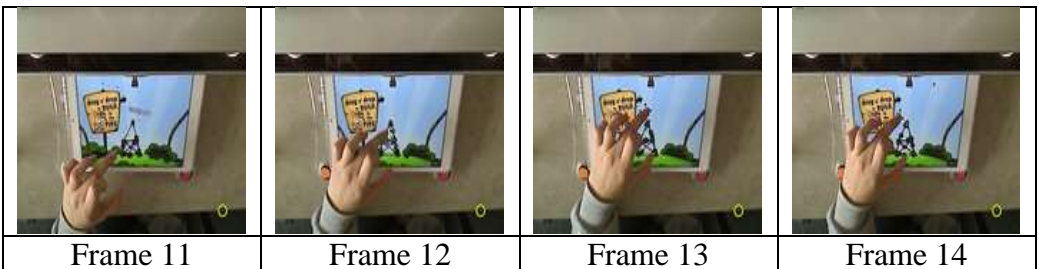

| Frame 11 | Frame 12 | Frame 13 | Frame 14 |

In frames 1-10 (not shown), the participant started moving the gooballs to form the basic structure that is shown in Frame 11.
In frames 11-14, the participant moved a gooball on the screen.
In frames 16-20, the participant starts to make the tower, suggesting that he now understood how to play the game

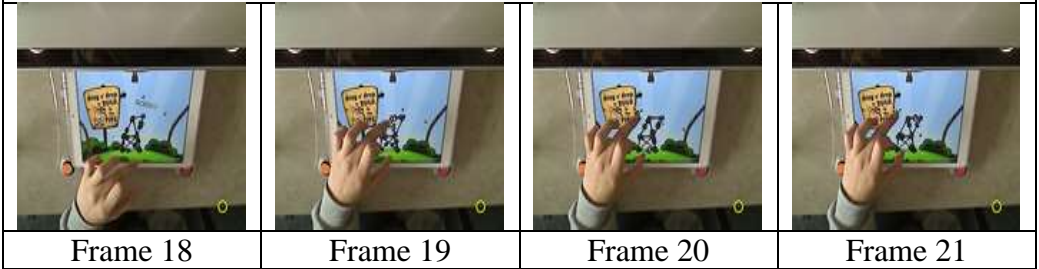

| Frame 18 | Frame 19 | Frame 20 | Frame 21 |

In frames 18-21, the participant continued to move gooballs on the screen

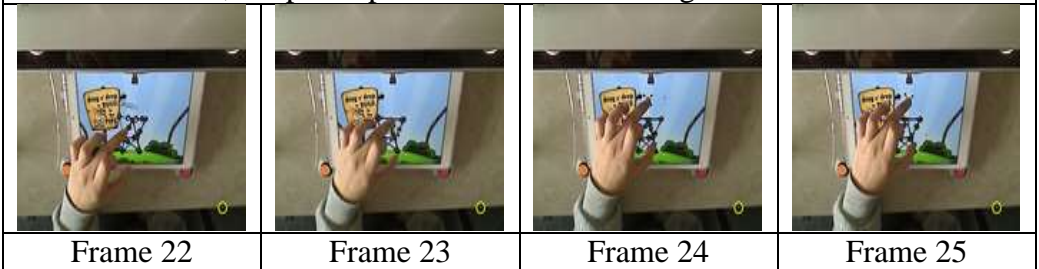

| Frame 22 | Frame 23 | Frame 24 | Frame 25 |

In frames 21-25, the participant continued to move gooballs on the screen.

| 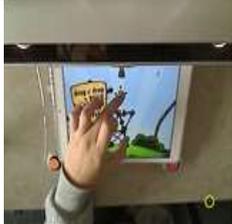 | 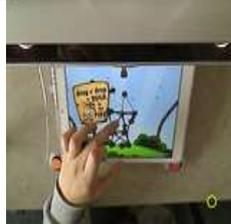 | 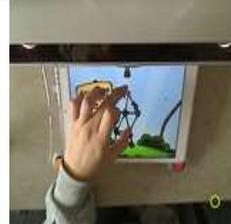 | 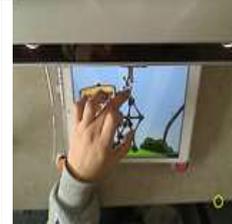 |
|---|---|---|---|
| Frame 26 | Frame 27 | Frame 28 | Frame 29 |

In frames 26-28 the participant continued to construct the tower.
In frame 29, the participant moved the gooball directly to the pipe. Which indicates, that while he had learnt how to build the tower, he had not learnt to build the tower so it is close to the pipe.
In frames 31-35 (not shown), the participant continued to construct the tower.

| 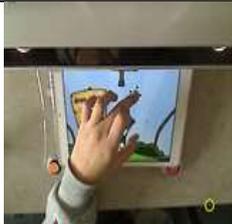 | 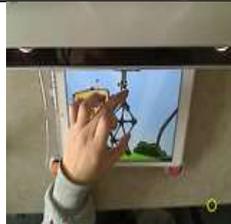 | 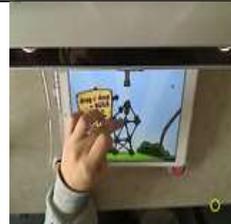 | 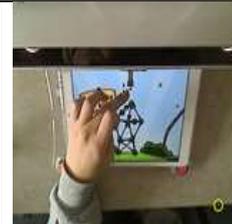 |
|---|---|---|---|
| Frame 36 | Frame 37 | Frame 38 | Frame 39 |

In frames 36-40, the participant made two attempts to place the gooball directly in the pipe, which suggests that the participant still did not realise that he needed to build a high tower.

| 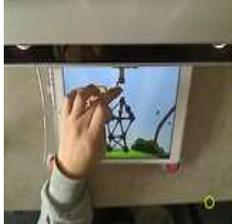 | 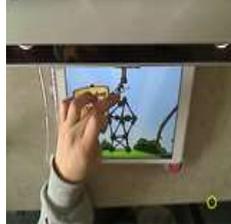 | 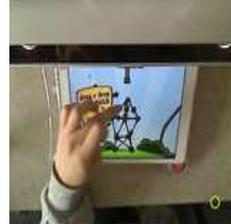 | 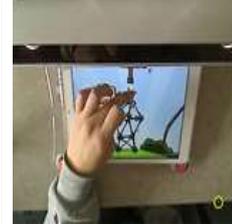 |
|---|---|---|---|
| Frame 40 | Frame 41 | Frame 42 | Frame 43 |

In frames 41-45, the participant made two more attempts to place the gooball directly in the pipe. It is clear from the multiple attempts he still had not learnt to build a tall tower.

| 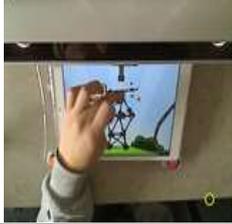 | 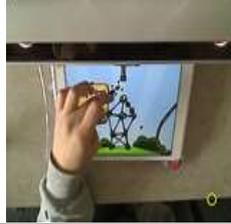 | 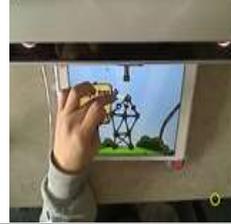 | 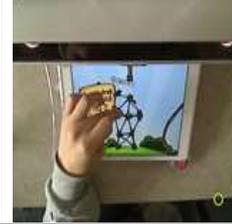 |
|---|---|---|---|
| Frame 44 | Frame 45 | Frame 46 | Frame 47 |

In frame 46-47, the participant continued to construct the tower.
In frame 48, the participant attempted to place the gooball directly in the pipe.

| 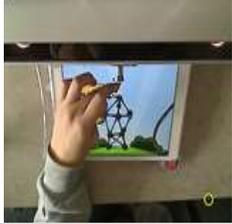 | 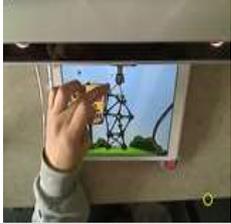 | |
|---|---|---|
| Frame 48 | Frame 49 | |

In frame 49, the participant successfully built the tower high enough to reach the tower and then completes the level.
In frame 50-63 (not shown), the participant watched the end of game video sequence.

*Figure 4.14 Participant A-16 Attempt 2 ([Link](#))*

| 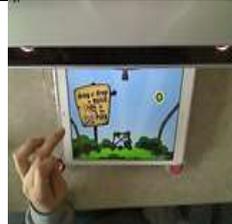 | 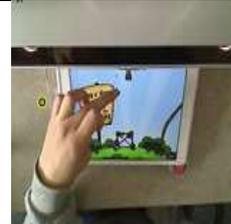 | 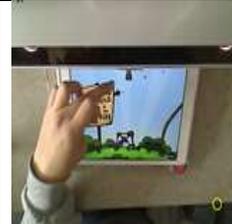 | 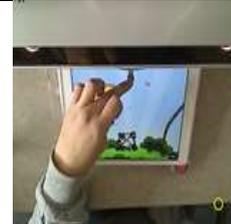 |
|---|---|---|---|
| Frame 6 | Frame 7 | Frame 8 | Frame 9 |

In frames 1-5, the participant watched the beginning video sequence.
In frames 6-9, the participant selected a gooball and moved it directly to the pipe. This suggests that the participant had not seen or understood the visual guide on the signpost.

| 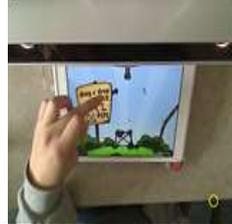 | 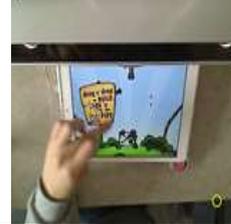 | 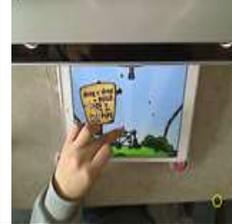 | 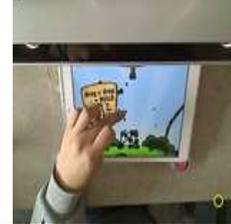 |
|---|---|---|---|
| Frame 11 | Frame 12 | Frame 13 | Frame 14 |

In frames 11-15, the participant selected and moved another gooball directly to the pipe. This suggests that the participant had not learnt from the first experience or had read/understood the visual guide on the signpost.

| 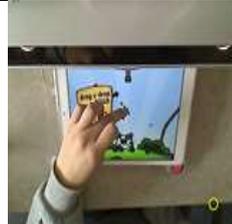 | 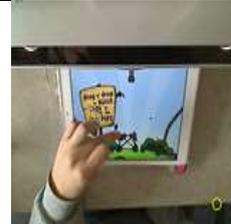 | 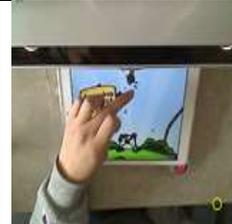 | 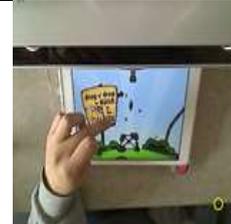 |
|---|---|---|---|
| Frame 15 | Frame 16 | Frame 17 | Frame 18 |

In frames 16-18, the participant selected and moved another gooball directly to the pipe. This suggests that the participant had not learnt from the first experience or had read/understood the visual guide on the signpost.

| 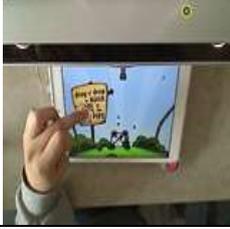 | 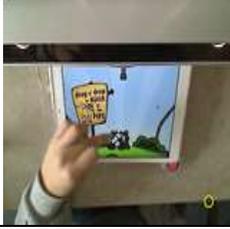 | 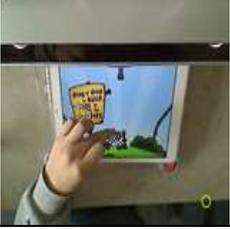 | 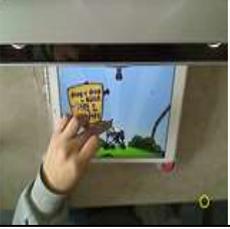 |
|---|---|---|---|
| Frame 19 | Frame 20 | Frame 21 | Frame 22 |

In frames 16-17, the participant selected and moved another gooball directly to the pipe. This suggests that the participant had not learnt from the first experience or had read/understood the visual guide on the signpost.
In frame 20, the participant selected another gooball and moved it around the starting frame.
In frame 21-22, the participant selects another gooball and moves it off the frame

| 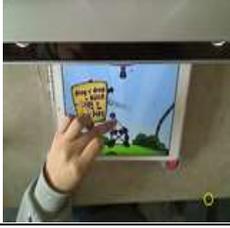 | 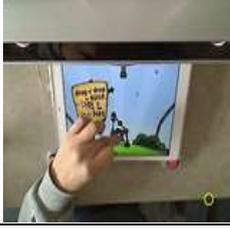 | 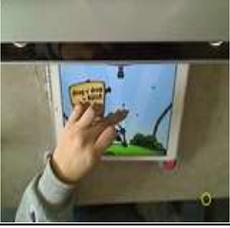 | 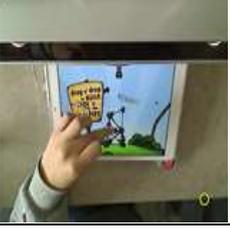 |
|---|---|---|---|
| Frame 23 | Frame 24 | Frame 25 | Frame 26 |

In frames 21-25, the participant selects another gooball and appears to have learnt the process of constructing the tower. .

| 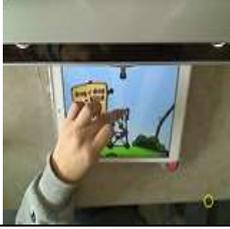 | 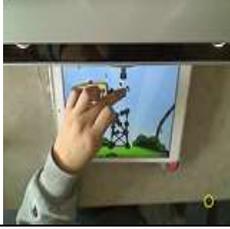 | 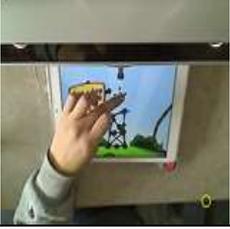 | 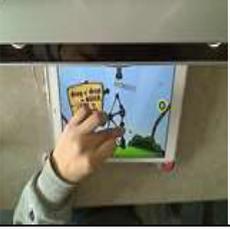 |
|---|---|---|---|
| Frame 27 | Frame 28 | Frame 29 | Frame 30 |

In frames 26-30, the participant continued to construct a tower.

| 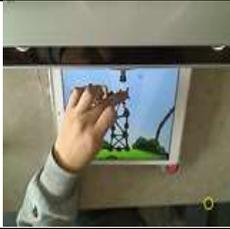 | 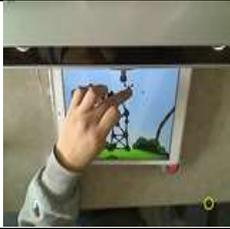 | 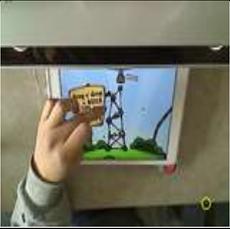 | 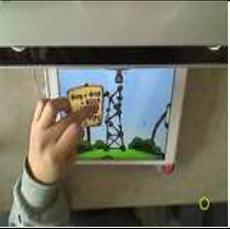 |
|---|---|---|---|
| Frame 31 | Frame 32 | Frame 33 | Frame 34 |

In frames 31-35, the participant continued to construct a tower. The participant appears to construct the tower with both speed and precision, this suggests the participant understood how to play this game

| 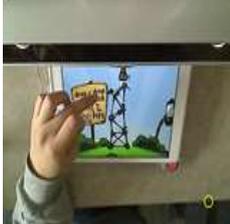 | 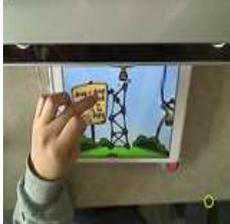 | |
|---|---|---|
| Frame 35 | Frame 36 | |

| In frame 36, the participant completed the level. |
| In frames 37-48 (not shown), the participant continued to watch the end of game video sequence. |

*Figure 4.15 Participant A-20 - Attempt 1 (<u>Link</u>)*

| 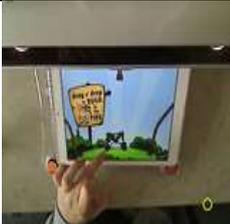 | 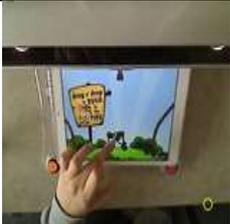 | 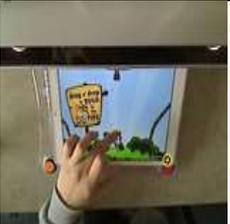 | 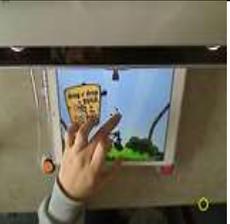 |
|---|---|---|---|
| Frame 6 | Frame 7 | Frame 8 | Frame 9 |

| In frames 1-5 (not shown), the participant watched the opening video sequence. |
| In frames 6-10, the participant started constructing the tower. |

| 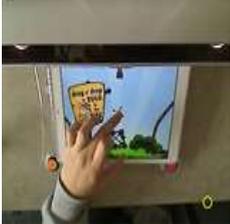 | 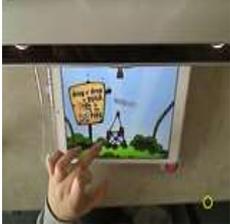 | 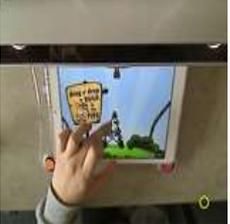 | 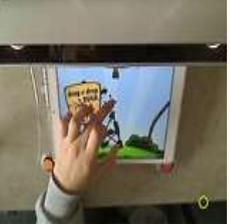 |
|---|---|---|---|
| Frame 10 | Frame 11 | Frame 12 | Frame 13 |

| In frames 11-15, the participant continued to construct the tower |

| 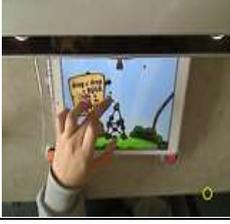 | 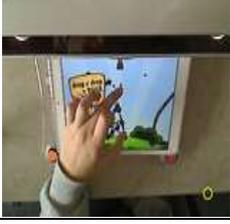 | 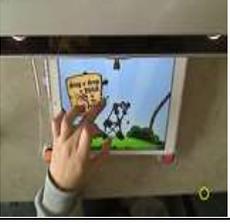 | 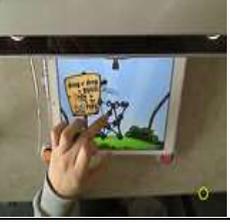 |
|---|---|---|---|
| Frame 14 | Frame 15 | Frame 21 | Frame 22 |

| In frames 16-20 (not shown), the participant continued to construct the tower. |

| 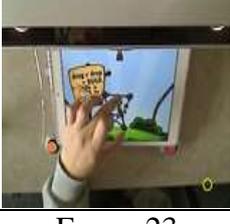 | 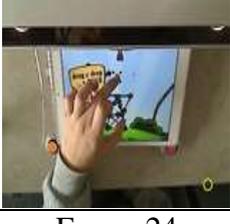 | 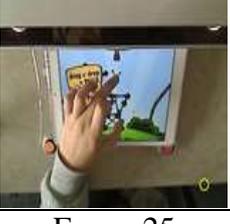 | 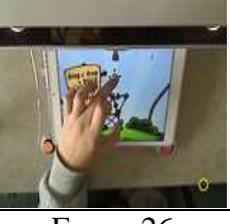 |
|---|---|---|---|
| Frame 23 | Frame 24 | Frame 25 | Frame 26 |

| In frames 21-23, the participant continued to construct the tower. |
| In frames 23-25, the participant moved a gooball directly to the pipe. This suggests that the participant did not see or understand the image on the sign post |

| | | | |
|---|---|---|---|
| 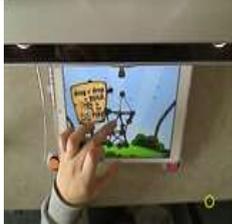 | 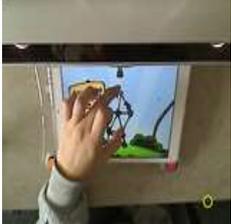 | 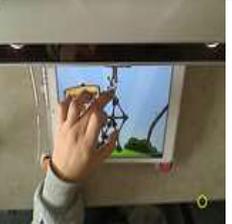 | 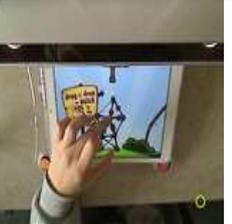 |
| Frame 27 | Frame 28 | Frame 29 | Frame 30 |

In frames 26-28, the participant continued to construct the tower.
In frame 29, the participant moved a gooball directly to the pipe.
In frame 30, the participant selected a gooball.

| | | | |
|---|---|---|---|
| 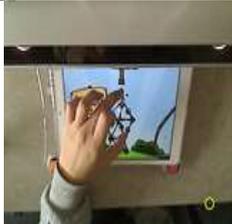 | 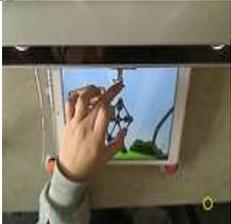 | 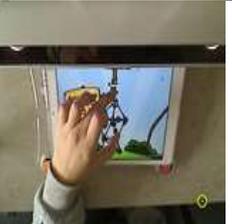 | 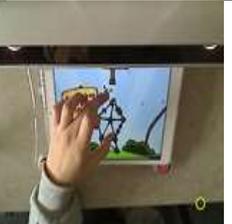 |
| Frame 31 | Frame 32 | Frame 33 | Frame 34 |
| 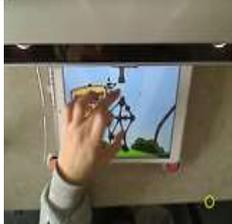 | 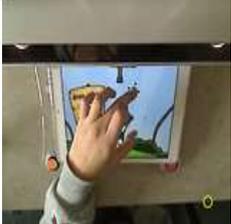 | 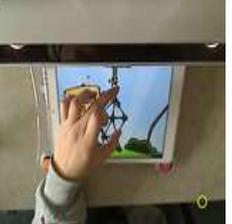 | 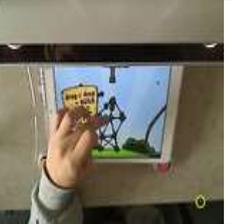 |
| Frame 35 | Frame 36 | Frame 37 | Frame 38 |

In frames 31-35, the participant moved the gooball directly to the pipe (twice). This suggests that the participant had not seen or understood the visual guide on the sign post

| | | | |
|---|---|---|---|
| 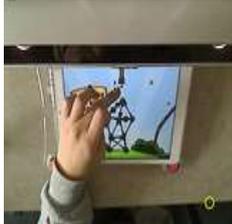 | 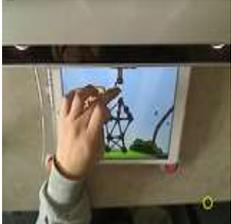 | 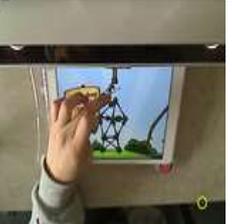 | 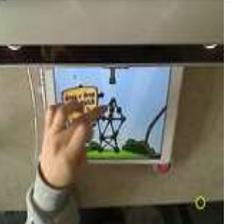 |
| Frame 39 | Frame 40 | Frame 41 | Frame 42 |

In frames 36-40, the participant moved the gooball directly to the pipe (twice). This suggests that the participant had not seen or understood the visual guide on the sign post

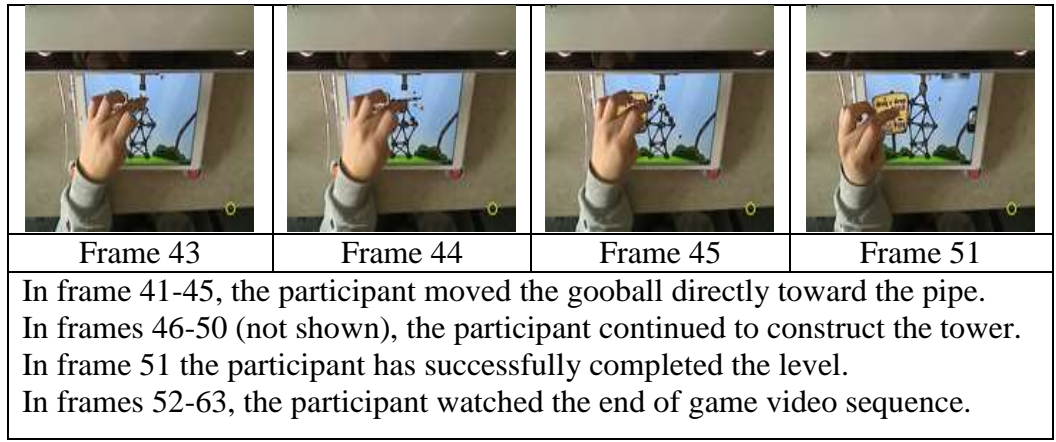

| Frame 43 | Frame 44 | Frame 45 | Frame 51 |

In frame 41-45, the participant moved the gooball directly toward the pipe.
In frames 46-50 (not shown), the participant continued to construct the tower.
In frame 51 the participant has successfully completed the level.
In frames 52-63, the participant watched the end of game video sequence.

*Figure 4.16 Participant A-20 - Attempt 2 (Link)*

### 4.1.2 The study of the children

In the study of the children, the recruitment programme resulted in 12 participants. The age of the 12 participants ranged between 6-8 years old (Mean 7.08, SD 0.9, Mode 8). Sixty-seven percent of the participants were male, and thirty-three percent were female. The frequency of video game play varied amongst the participants from never (16%), to more than six times a week (33%). Eighty-five percent of the participants that played video games primarily used a Personal Computer (PC). Seventy-five percent of the participants used a tablet or phone device and/or gaming console (Xbox, PlayStation, or Wii). The games the participants reported playing included: Minecraft, Lego Batman, NFL Madden, Angry Birds, Fast Math, and Fun Brain.

One participant had prior exposure to playing World of Goo (2D Boy, 2008), and two participants had previously played Bad Piggies (Rovio Entertainment (2013).

As detailed in the study of the adults, the children were asked to undergo the two attempts at the game sequentially. The delay between the two attempts was limited by the construction of the magnetic tower. The range of this delay was between 5-12 minutes.

Table 4.15 shows the results of the answer to the question, "What shape do you think is the strongest for building a really tall tower?" The preferred (or more correct) answer to the question was a triangle.

| Participant alias | Group | Pre-exposure answer | Post-exposure answer |
|---|---|---|---|
| K-04 | Control | Triangle | Not Sure |
| K-05 | Control | Square | Not Sure |
| K-06 | Control | Square | Square |
| K-07 | Control | Circle | Not Sure |
| K-09 | Control | Not sure | Not Sure |
| K-12 | Control | Square | Square |
| K-01 | Treatment | Square | Triangle |
| K-02 | Treatment | Square | Circle |
| K-03 | Treatment | Circle | Triangle |
| K-08 | Treatment | Square | Triangle |
| K-10 | Treatment | Square | Triangle |
| K-11 | Treatment | Square | Triangle |

*Table 4.15 Answers to the tower construction question*

The participants were also asked to build a really tall tower with the magnetic construction set before and after they played the video game. The shapes used in the constructed process were noted for each tower. The pilot tested identified that a triangle provided the strongest construction and used the least components.

| Participant alias | Group | Pre-exposure construction | Post-exposure construction |
|---|---|---|---|
| K-04 | Control | Flat sticks | Flat sticks |
| K-05 | Control | Flat sticks | Flat sticks |
| K-06 | Control | Stick (held up) | Sticks (held up) |
| K-07 | Control | Flat sticks | Flat sticks |
| K-09 | Control | Flat Sticks | Flat Sticks |
| K-12 | Control | Flat sticks | Flat sticks |
| K-01 | Treatment | Triangles & Squares | Triangles & Squares |
| K-02 | Treatment | Flat sticks | Triangles |
| K-03 | Treatment | Triangles | Triangles |
| K-08 | Treatment | Sticks (held up) | Triangles |
| K-10 | Treatment | Flat sticks | Flat sticks |
| K-11 | Treatment | Flat sticks | Triangles |

*Table 4.16 The results of the magnetic tower construction*

From the data presented in Table 4.15, it is possible to identify positive changes in the understanding of one of the construction principles in the treatment group through their answers to the question, "What shape do you think is the strongest for building a really tall tower?" This is further supported through the changes in the construction methods used in the out-of-game experience with the magnetic construction set (Table 4.16). The majority of the children placed the magnetic sticks flat on the ground during the pre-exposure test (Figure 4.17). However, in the post-exposure task, 83% of the children in the treatment group attempted to make a three-dimensional tower. The children did not receive any additional intervention or outside assistance; they were provided the same equipment and asked to achieve the same output. The only delay between the two construction tests was the time it took to play the game.

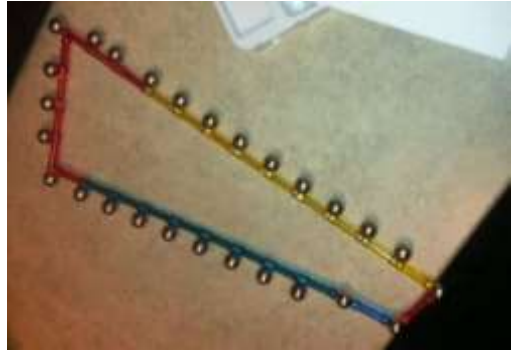

*Figure 4.17 Participant K-10 Pre-exposure tower*

Table 4.17 presents the results of the gameplay for the treatment group. Participant K-01 did not manage to save enough gooballs in either attempts at the game. The participant did not complete (DNC) the game.

| | Attempt 1 | | | Attempt 2 | | | Variance | | |
|---|---|---|---|---|---|---|---|---|---|
| ID | Gooballs collected | Number of moves | Time taken | Gooballs collected | Number of moves | Time taken | Additional gooballs collected | Additional moves | Time saved |
| K-01 | DNC | | 4:35 | DNC | | 1:52 | 0 | | 2:43 |
| K-02 | 4 | 9 | 5:30 | 7 | 7 | 0:44 | 3 | -2 | 4:46 |
| K-03 | 5 | 9 | 0:48 | 5 | 8 | 0:46 | 0 | -1 | 0:02 |
| K-08 | 9 | 5 | 1:33 | 10 | 4 | 1:29 | 1 | -1 | 0:04 |
| K-10 | 8 | 6 | 3:20 | 7 | 7 | 2:34 | -1 | 1 | 0:46 |
| K-11 | 4 | 8 | 2:42 | 10 | 4 | 0:46 | 6 | -4 | 1:56 |

*Table 4.17 Treatment group game performance*

### 4.2.2.1 Endogenous blinks

Tables 4.18 and 4.19 present the frequency of blinks for each participant for both the first and second attempts. The area of interest was those blinks that lasted between 100 and 300 ms. However, the blinks that were below and above this range have also been included for comparison. As identified during the study of the adults, the eye gaze software identified a loss of signal as a blink. Where the software identified a blink that was beyond the accepted range of a blink (Evinger et al., 1991; Guitton et al., 1991; Stava et al., 1994), the data was excluded. The data from participant K-05 has been excluded as this participant was extremely uncooperative and, as a result, did not replay the same game twice.

| Blink duration (ms) freq. | K-01 | K-02 | K-03 | K-04 | K-06 | K-07 | K-08 | K-09 | K-10 | K-11 |
|---|---|---|---|---|---|---|---|---|---|---|
| 0-49 | 2 | 17 | 11 | 9 | 11 | 1 | 118 | 17 | 13 | 19 |
| 50-100 | 0 | 13 | 5 | 4 | 4 | 2 | 41 | 13 | 13 | 22 |
| 101-150 | 1 | 15 | 2 | 7 | 5 | 2 | 36 | 15 | 15 | 20 |
| 151-200 | 1 | 15 | 2 | 6 | 6 | 1 | 199 | 15 | 21 | 26 |
| 201-250 | 0 | 22 | 1 | 2 | 4 | 4 | 40 | 22 | 16 | 19 |
| 251-300 | 2 | 19 | 1 | 1 | 3 | 2 | 8 | 19 | 12 | 11 |
| 301-350 | 0 | 12 | 2 | 5 | 2 | 3 | 6 | 12 | 5 | 6 |
| 351-400 | 0 | 4 | 0 | 3 | 9 | 1 | 6 | 4 | 4 | 7 |
| Mean | 0.8 | 14.6 | 3.0 | 4.6 | 5.5 | 2.0 | 56.8 | 14.6 | 12.4 | 16.3 |

*Table 4.18 Eye blinks between 0 and 400 ms (Attempt 1)*

| Blink duration (ms) freq. | K-01 | K-02 | K-03 | K-04 | K-06 | K-07 | K-08 | K-09 | K-10 | K-11 |
|---|---|---|---|---|---|---|---|---|---|---|
| 0-49 | 27 | 8 | 30 | 15 | 20 | 1 | 48 | 16 | 1 | 55 |
| 50-100 | 28 | 5 | 4 | 6 | 7 | 1 | 43 | 20 | 4 | 38 |
| 101-150 | 9 | 2 | 2 | 2 | 3 | 0 | 45 | 8 | 3 | 25 |
| 151-200 | 6 | 0 | 1 | 3 | 5 | 1 | 32 | 15 | 9 | 16 |
| 201-250 | 4 | 3 | 1 | 4 | 6 | 1 | 19 | 10 | 7 | 13 |
| 251-300 | 2 | 1 | 1 | 3 | 4 | 1 | 8 | 7 | 1 | 19 |
| 301-350 | 2 | 0 | 1 | 11 | 12 | 1 | 8 | 5 | 2 | 6 |
| 351-400 | 4 | 8 | 5 | 6 | 6 | 4 | 15 | 88 | 3 | 13 |
| Mean | 10 | 3 | 6 | 6 | 8 | 1 | 27 | 21 | 4 | 23 |

*Table 4.19 Eye blinks between 0 and 400 ms (Attempt 2)*

A box-and-whisker plot (Tukey, 1977) was created to help understand the descriptive statistics and any extreme values. Figures 4.18 and 4.19 present a box-and-whisker plot (Tukey, 1977) of the blink data collected in the first and second attempts at the game.

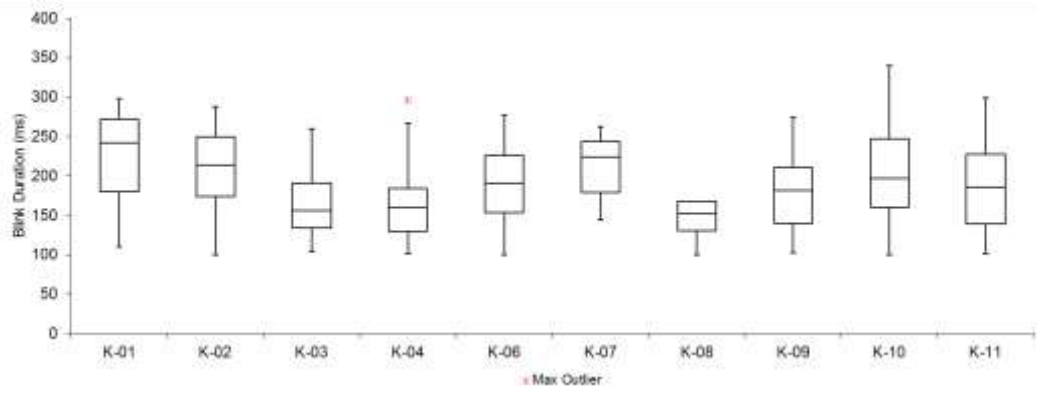

*Figure 4.18 Eye blinks between 100 and 300 ms (Attempt 1)*

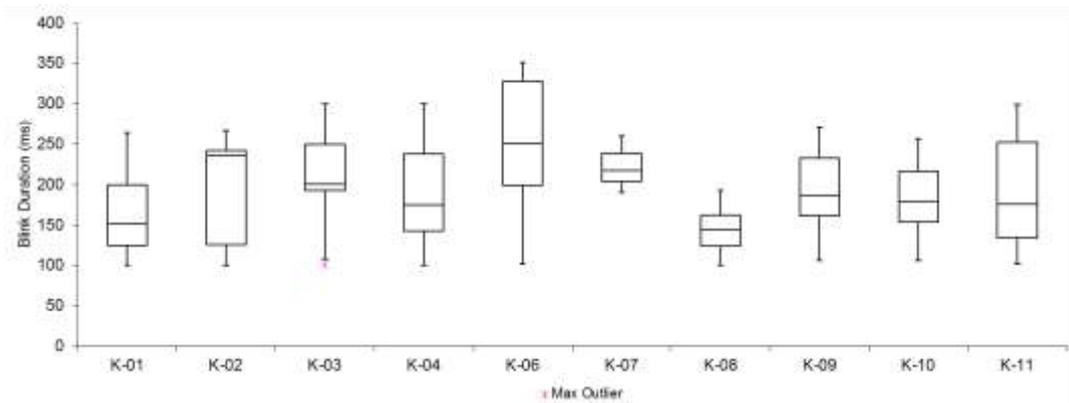

*Figure 4.19 Eye blinks between 100 and 300 ms (Attempt 2)*

For the frequency of blinks between 100 and 300 ms in the first attempt ($M = 73.50$, $SD = 95.30$) and the second attempt ($M = 45.80$, $SD = 47.96$); $d = 1.50$. The effect size for this analysis ($d = 0.39$) was found to be above Cohen's (1988) convention for a small effect ($d = .10$). These results suggest that the number of blinks in the first attempt at the game may have a small effect on the number of blinks in the second attempt at the game.

Although these statistics provide an insight into the frequency of blinks, as the duration of each attempt varied, this data may be misleading. To provide a more standardised comparison of the data, the percentage of blinks for each participant is presented below (Table 4.20 and Table 4.21). This data is represented graphically in Figure 4.20. The data from participant K-05 has been excluded as this participant was extremely uncooperative and, as a result, did not replay the same game twice.

| Blink duration (ms) % | K-01 | K-02 | K-03 | K-04 | K-06 | K-07 | K-08 | K-09 | K-10 | K-11 |
|---|---|---|---|---|---|---|---|---|---|---|
| 0-49 | 33 | 15 | 46 | 24 | 25 | 6 | 26 | 15 | 13 | 15 |
| 50-100 | 0.0 | 11 | 21 | 11 | 9 | 13 | 9 | 11 | 13 | 17 |
| 101-150 | 16.7 | 13 | 8 | 19 | 11 | 13 | 8 | 13 | 15 | 15 |
| 151-200 | 16.7 | 13 | 8 | 16 | 14 | 6 | 44 | 13 | 21 | 20 |
| 201-250 | 0.0 | 19 | 4 | 5 | 9 | 25 | 9 | 19 | 16 | 15 |
| 251-300 | 33.3 | 16 | 4 | 3 | 7 | 13 | 2 | 16 | 12 | 8 |
| 301-350 | 0 | 10 | 8 | 14 | 5 | 19 | 1 | 10 | 5 | 5 |
| 351-400 | 0 | 3 | 0 | 8 | 20 | 6 | 1 | 3 | 4 | 5 |

*Table 4.20 Percentage of blink occurrence within each range (Attempt 1)*

| Blink duration (ms) % | K-01 | K-02 | K-03 | K-04 | K-06 | K-07 | K-08 | K-09 | K-10 | K-11 |
|---|---|---|---|---|---|---|---|---|---|---|
| 0-49 | 33 | 30 | 67 | 30 | 32 | 10 | 22 | 9 | 3 | 30 |
| 50-100 | 34 | 19 | 9 | 12 | 11 | 10 | 20 | 12 | 13 | 21 |
| 101-150 | 11 | 7 | 4 | 4 | 5 | 0 | 21 | 5 | 10 | 14 |
| 151-200 | 7 | 0 | 2 | 6 | 8 | 10 | 15 | 9 | 30 | 9 |
| 201-250 | 5 | 11 | 2 | 8 | 10 | 10 | 9 | 6 | 23 | 7 |
| 251-300 | 2 | 4 | 2 | 6 | 6 | 10 | 4 | 4 | 3 | 10 |
| 301-350 | 2 | 0 | 2 | 22 | 19 | 10 | 4 | 3 | 7 | 3 |
| 351-400 | 5 | 30 | 11 | 12 | 10 | 40 | 7 | 52 | 10 | 7 |

*Table 4.21 Percentage of blink occurrence within each range (Attempt 2)*

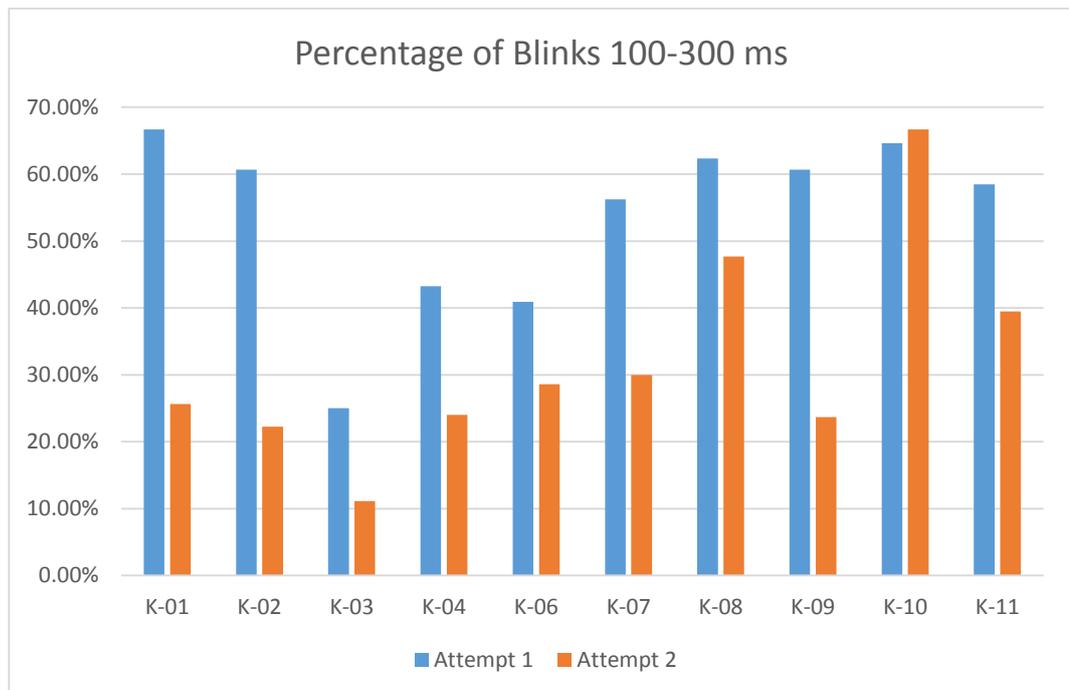

*Figure 4.20 The percentage of endogenous blinks (Both attempts)*

For the percentage of blinks between 100 and 300 ms in the first attempt ($M = 0.65$, $SD = 0.11$) and the second attempt ($M = 0.48$, $SD = 0.18$); $d = 0.39$. The effect size for this analysis ($d = 0.39$) was found to be below Cohen's (1988) convention for a medium effect ($d = .50$). These results suggest that the percentage of blinks in the first attempt at the game do have an effect on the percentage of blinks in the second attempt at the game.

### 4.2.2.2 Fixation data

Tables 4.22 and 4.23 present the cumulative fixation for each participant. As discussed in Section 4.1.1.2, the format used by Kenny et al., (2005) has been adopted and modified to include more detail on the frequency of fixations that lasted longer than 600 ms. The data from participant K-05 has been excluded as this participant was extremely uncooperative and, as a result, did not replay the same game twice. Figure 4.21 presents this data as a bar chart.

| Fixation duration (ms). | K-01 | K-02 | K-03 | K-04 | K-06 | K-07 | K-08 | K-09 | K-10 | K-11 |
|---|---|---|---|---|---|---|---|---|---|---|
| 0-299 | 33 | 442 | 170 | 73 | 77 | 24 | 24 | 31 | 309 | 463 |
| 300-599 | 16 | 120 | 36 | 7 | 33 | 3 | 3 | 4 | 57 | 142 |
| 600-899 | 5 | 41 | 7 | 1 | 6 | 2 | 2 | 2 | 12 | 12 |
| 900-1199 | 4 | 13 | 3 | 0 | 0 | 1 | 1 | 0 | 0 | 6 |
| 1200 + | 2 | 15 | 0 | 0 | 4 | 0 | 0 | 0 | 1 | 3 |
| Mean | 12 | 126.2 | 43.2 | 16.2 | 24 | 6 | 7.4 | 75.8 | 125.2 | 86 |

*Table 4.22 Fixation duration (Attempt 1)*

| Fixation duration freq. | K-01 | K-02 | K-03 | K-04 | K-06 | K-07 | K-08 | K-09 | K-10 | K-11 |
|---|---|---|---|---|---|---|---|---|---|---|
| 0-299 | 224 | 113 | 448 | 342 | 193 | 43 | 43 | 374 | 579 | 174 |
| 300-599 | 21 | 21 | 68 | 28 | 19 | 1 | 1 | 67 | 106 | 53 |
| 600-899 | 4 | 10 | 16 | 3 | 2 | 0 | 0 | 10 | 4 | 1 |
| 900-1199 | 1 | 3 | 5 | 0 | 1 | 1 | 1 | 4 | 2 | 0 |
| 1200 + | 0 | 1 | 6 | 0 | 0 | 1 | 1 | 2 | 1 | 0 |
| Mean | 50 | 29.6 | 108.6 | 74.6 | 43 | 9.2 | 91.4 | 138.4 | 45.6 | 108.6 |

*Table 4.23 Fixation duration (Attempt 2)*

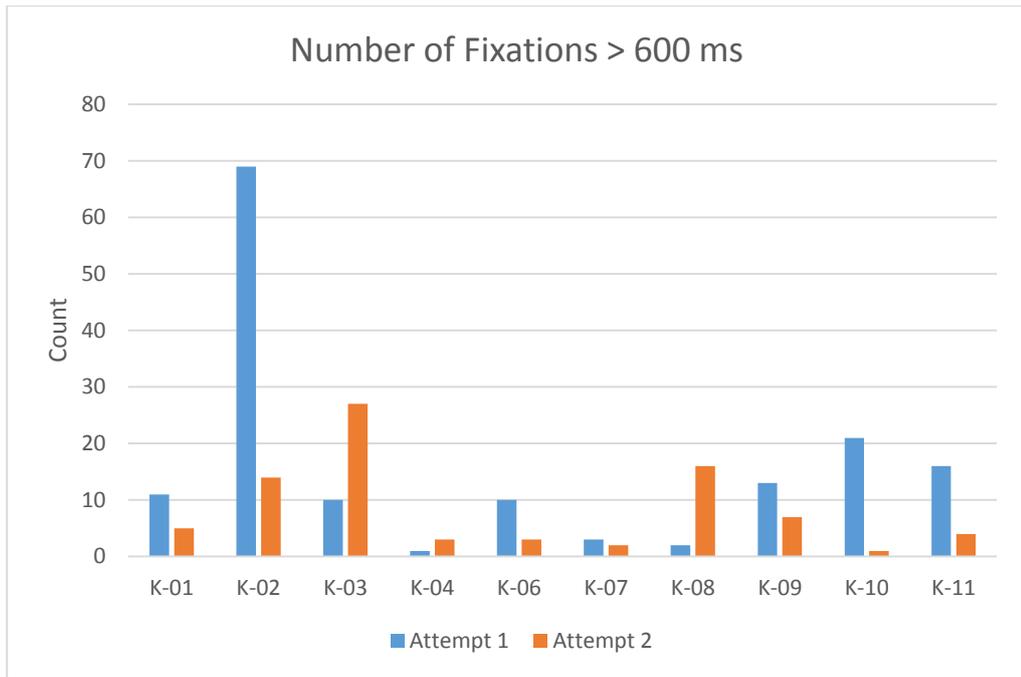

*Figure 4.21 Cumulative Fixations greater than 600 ms*

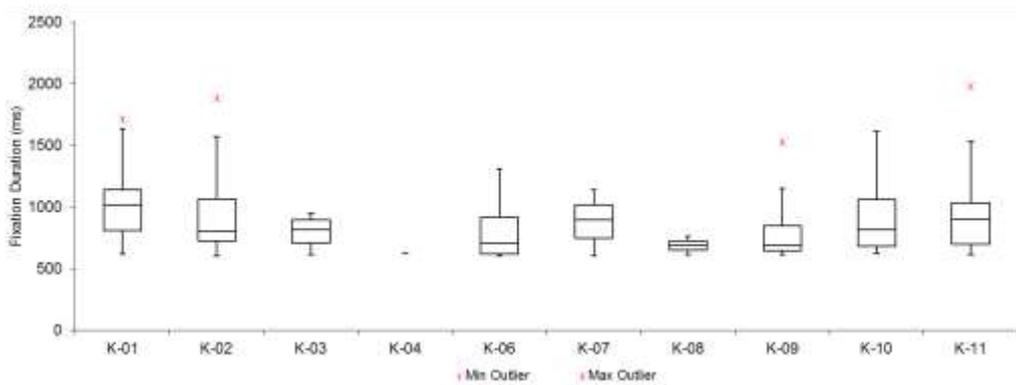

*Figure 4.22 –Fixations above 600 ms (Attempt 1)*

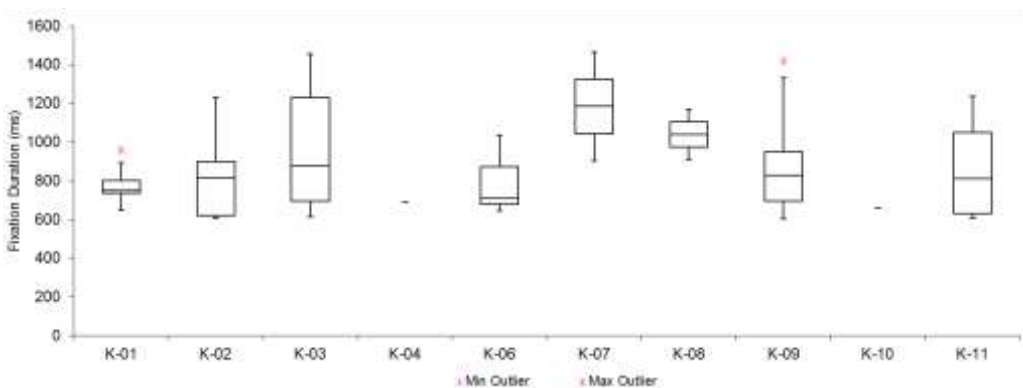

*Figure 4.23 –Fixations above 600 ms (Attempt 2)*

Figure 4.22 and 4.23 display a box-and-whisker plot (Tukey, 1977) of the fixations that were above 600 ms for the first and second attempt at the game. The results appear consistent with the results reported in Kenny et al., (2005), as there was a wide variance in fixation behaviour between each participant.

For the fixations greater than 600 ms in the first attempt ($M = 15.6$, $SD = 19.80$) and the second attempt ($M = 8.20$, $SD = 8.31$); $d = 0.44$. The effect size for this analysis ($d = 0.84$) was found to be above Cohen's (1988) convention for a medium effect ($d = .50$). These results suggest that the number of fixations above 600 ms in the first attempt at the game may have an effect on the number of fixations above 600 ms in the second attempt at the game.

While Kenny et al., (2005) provide a valuable template for presenting fixation data, they did not consider the duration of the gameplay, as each participant played the game for very different lengths of time. Therefore, the comparisons between the different fixations of each participant, or between each participants' first and second attempts, may not provide an accurate comparison. To provide a more standardised comparison of the data, the percentage of the fixation data for each participant is presented below (Table 4.24 and Table 4.25). This data is presented graphically in Figure 4.24.

| Fixation duration freq. | K-01 | K-02 | K-03 | K-04 | K-06 | K-07 | K-08 | K-09 | K-10 | K-11 |
|---|---|---|---|---|---|---|---|---|---|---|
| 0-299 | 55 | 70.05 | 78.7 | 90.12 | 64.17 | 80 | 80 | 83.78 | 81.53 | 73.96 |
| 300-599 | 26.67 | 19.02 | 16.67 | 8.64 | 27.5 | 10 | 10 | 10.81 | 15.04 | 22.68 |
| 600-899 | 8.33 | 6.5 | 3.24 | 1.23 | 5 | 6.67 | 6.67 | 5.41 | 3.17 | 1.92 |
| 900-1199 | 6.67 | 2.06 | 1.39 | 0 | 0 | 3.33 | 3.33 | 0 | 0 | 0.96 |
| 1200 + | 3.33 | 2.38 | 0 | 0 | 3.33 | 0 | 0 | 0 | 0.26 | 0.48 |

*Table 4.24 Fixation percentage (Attempt 1)*

| Fixation duration freq. | K-01 | K-02 | K-03 | K-04 | K-06 | K-07 | K-08 | K-09 | K-10 | K-11 |
|---|---|---|---|---|---|---|---|---|---|---|
| 0-299 | 89.60 | 76.35 | 82.50 | 91.69 | 89.77 | 93.4 | 93.48 | 81.84 | 83.67 | 76.32 |
| 300-599 | 8.40 | 14.19 | 12.52 | 7.51 | 8.84 | 2.17 | 2.17 | 14.66 | 15.32 | 23.25 |
| 600-899 | 1.60 | 6.76 | 2.95 | 0.80 | 0.93 | 0.00 | 0.00 | 2.19 | 0.58 | 0.44 |
| 900-1199 | 0.40 | 2.03 | 0.92 | 0.00 | 0.47 | 2.17 | 2.17 | 0.88 | 0.29 | 0.00 |
| 1200 + | 0.00 | 0.68 | 1.10 | 0.00 | 0.00 | 2.17 | 2.17 | 0.44 | 0.14 | 0.00 |

*Table 4.25 Fixation percentage (Attempt 2)*

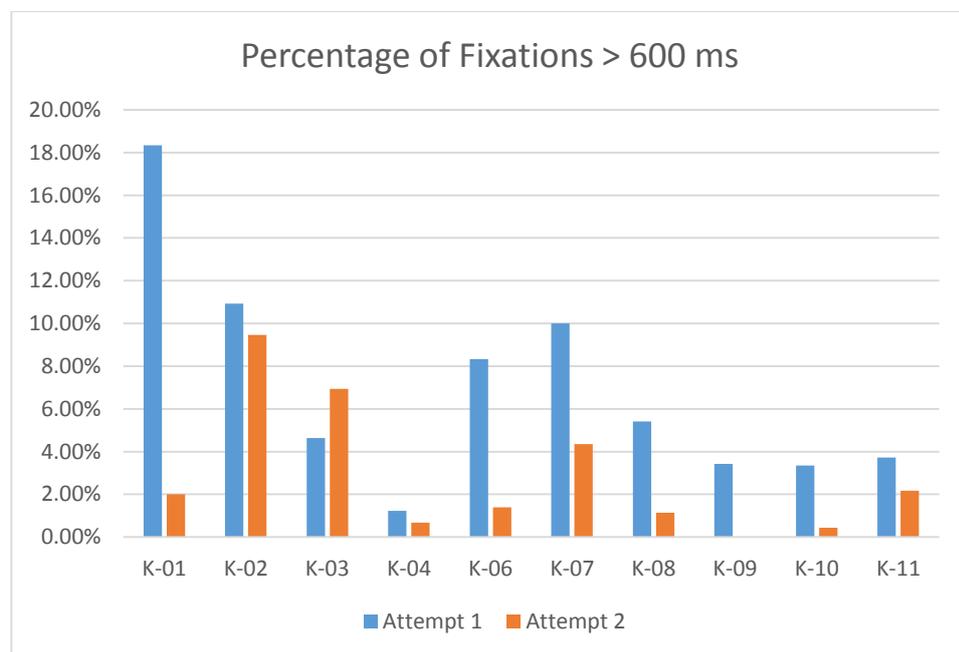

*Figure 4.24 Percentage of fixations of each participant greater than 600 ms*

For the fixations greater than 600 ms in the first attempt ($M = 0.07$, $SD = 0.05$) and the second attempt ($M = 0.03$, $SD = 0.03$); $d = 0.63$. The effect size for this analysis ($d = 0.63$) was found to be above Cohen's (1988) convention for a medium effect ($d = .50$). These results suggest that the percentage of fixations above 600 ms in the first attempt of the game has an effect on the percentage of fixations above 600 ms in the second attempt at the game.

### 4.2.2.3 Video Evidence

To help facilitate an understanding of the individual game play experience a video recording of the game play was obtained by recording what was on the computer screen. This facility was a function of the eye tracking software. This Section provides the evidence from the videos that were recorded. The video recordings were played on a personal computer, and an image of what was on the screen was taken every second. These images were then scaled to fit onto the page. The highlights of the video evidence from participants K-01, K-03, and K-08 are included below. Although the complete images of the video recording are available for these participants, presenting this would have added 200 additional pages to this document. The key images are presented to minimise the additional burden on the reader. The rationale for including this specific evidence will be discussed in detail in Chapter 5.

| 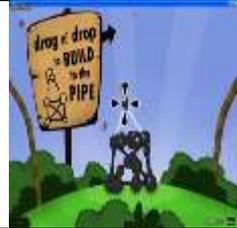 | 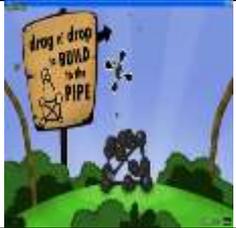 | 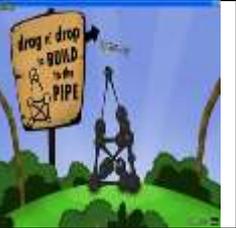 | 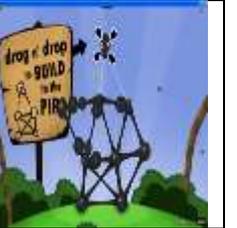 |
|---|---|---|---|
| Frame 65 | Frame 66 | Frame 67 | Frame 117 |

In the initial part of the game (frames 1-64), the participant moved the gooballs around the starting frame and appears to have been figuring out how to play the game. It was not until frame 65, that the participant moved the gooball off the frame. However, the participant pulled the gooball too far away from the frame (Frame 66). In frames 67, the participant has successfully placed the gooball correctly.
In frames 68 to 116, the participant continued to construct the tower. This suggests that the participant had learnt how to place the gooball.

| 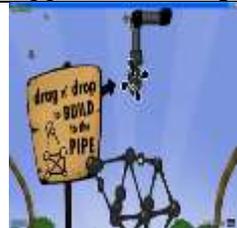 | 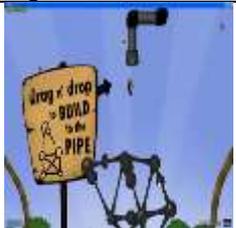 | 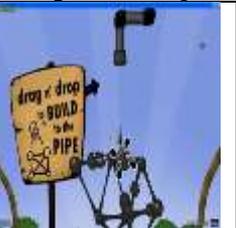 | 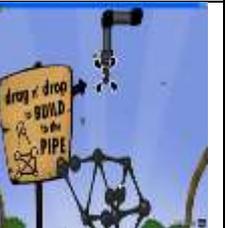 |
|---|---|---|---|
| Frame 118 | Frame 119 | Frame 120 | Frame 130 |

In frame 118, the participant read the sign and moved the gooball directly onto the pipe. Note that the pipe is now visible to the participant.
In frames 119-120, the participant watched the gooball that they had unsucessfully placed on the pipe fall.
In frame 130, the participant tried to place the gooball directly in the pipe.

| 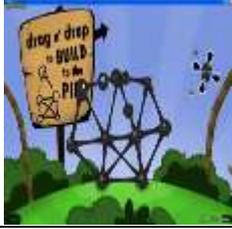 | 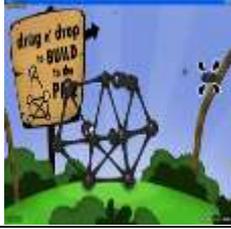 | 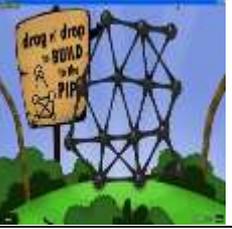 | 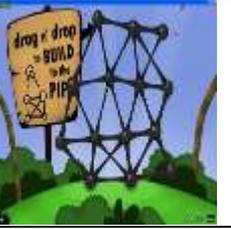 |
|---|---|---|---|
| Frame 165 | Frame 166 | Frame 267 | Frame 268 |

In frames 131-164 (not shown), the participant, continued to construct a tower.
In frames 165-166, the participant moved the gooball to the tree. This suggests that they had possibly mistaken the tree for a pipe. The first two attempts at moving the gooball to the pipe above the tower did not work, so the participant tried to place the gooball on the other object that looked like a pipe. In frames 167-266 (not shown), the participant continued to construct the tower.
In frames 267-268, the participant had used all of the gooballs and the game was restarted. The resulting construction suggests that this participant used a lot of trial and error to get the desired result. The participant did not reach the end goal, but there was no strong evidence that this was the result. Although the status at the bottom of the screen shows that '0 of 4 collected', it is not clear to the participant what needed to be collected and as this is the only confirmation that the end-goal had not been achieved, the user may have not actually realised this.

*Figure 4.25 Participant K-01 - Attempt 1 ([Link](Link))*

| 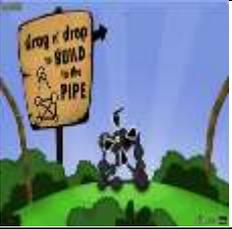 | 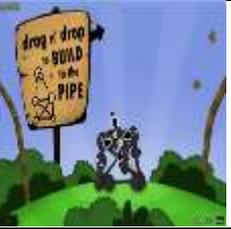 | 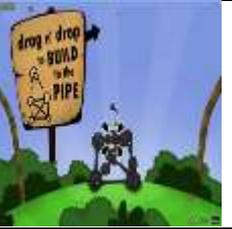 | 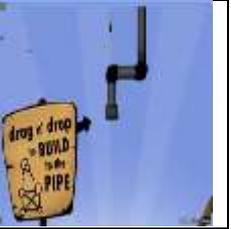 |
|---|---|---|---|
| Frame 1 | Frame 2 | Frame 5 | Frame 6 |

In frames 1-2 the participant moved the gooball around the starting frame.
In frames 3 to 5 (not shown), the participant continued to move the gooballs around the starting frame
In frames 5-6, the participant selected and moved a gooball to the pipe, which suggests that the participant did not read or understand the image on the sign.

| 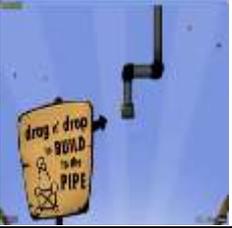 | 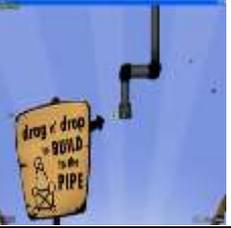 | 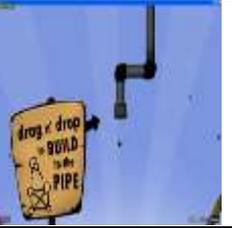 | 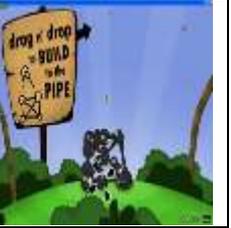 |
|---|---|---|---|
| Frame 7 | Frame 8 | Frame 9 | Frame 10 |

In frames 7-9, the participant released the gooball and watched it fall.
In frame 10, the participant moved the gooballs around the starting frame. This suggests that the participant still had not seen or understood the image on the sign.
In frames 11-14 (not shown), the participant continued to move the gooballs around the beginning frame.

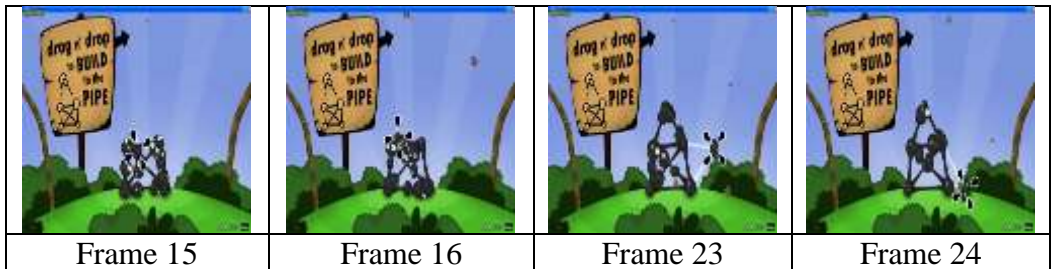

| Frame 15 | Frame 16 | Frame 23 | Frame 24 |

In frames 15-16, the participant continued to move the gooballs around the starting frame.

In frames 17-22 (not shown), the participant continued to move the gooball around the starting frame

In frames 23-24, the participant dragged a gooball around (without releasing it).

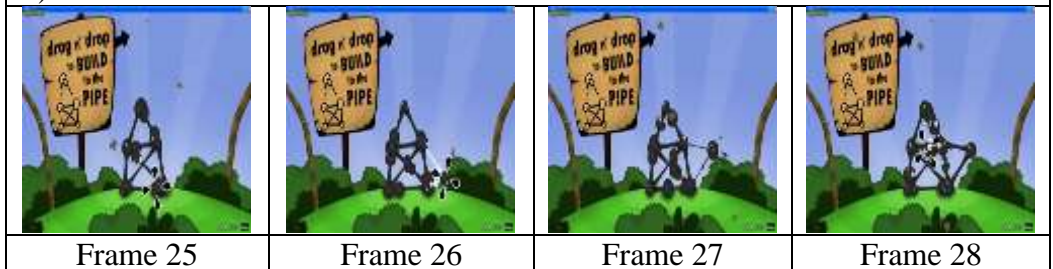

| Frame 25 | Frame 26 | Frame 27 | Frame 28 |

In frames 25, the participant continued to move and released the gooball.

In frames 26-28, the participant moved and released another gooball, this time to the side. This suggests that the participant had not seen or understood the image on the sign.

In frames 29-30, (not shown) the participant continued to move and release another gooball.

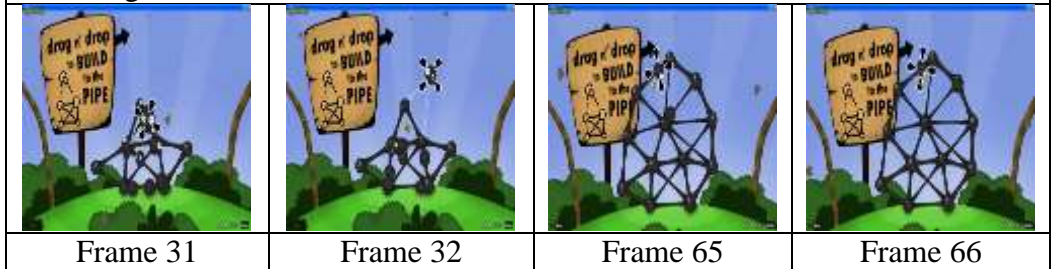

| Frame 31 | Frame 32 | Frame 65 | Frame 66 |

In frames 31-32, the participant continued to select and move another gooball. Once again, this looks accidental rather than deliberate.

In frames 33-64 (not shown), the participant select and release gooballs

In frames 65-66, the participant continued to select and release gooballs. The design appears haphazard.

In frames 67-74 (not shown), the participant continued to select and release gooballs.

| | | | |
|---|---|---|---|
| 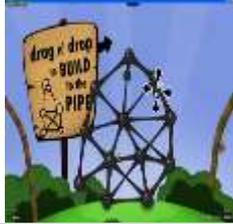 | 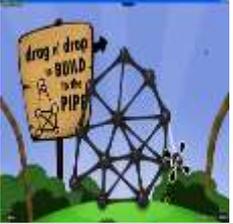 | 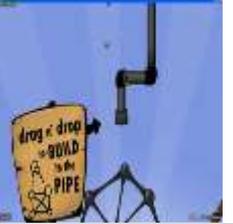 | 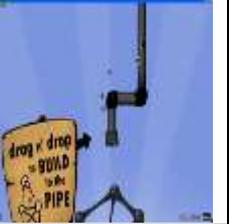 |
| Frame 75 | Frame 76 | Frame 93 | Frame 94 |

In frames 75-76, the participant continued to select and release gooballs.
In frames 77-92 (not shown), the participant continued to select and release gooballs.
In frames 93-94, the participant moved the gooball directly to the pipe.

| | | | |
|---|---|---|---|
| 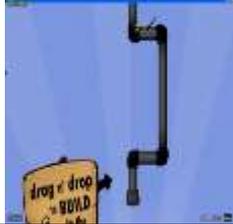 | 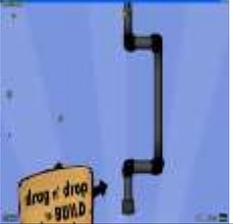 | 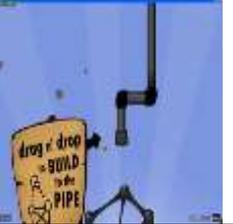 | 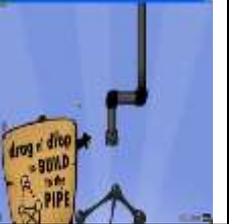 |
| Frame 95 | Frame 96 | Frame 105 | Frame 106 |

In frames 95-96, the participant continued to place the gooball directly on the pipe.
In frames 97-104 (not shown), the participant continued to place the gooball directly on the pipe.
In frames 105-106, the participant continued to place the gooball directly on the pipe.

| | |
|---|---|
| 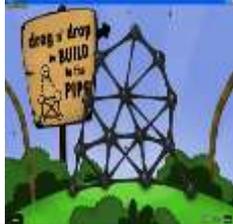 | 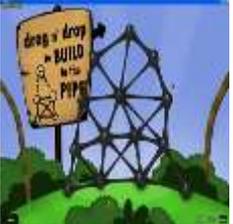 |
| Frame 113 | Frame 114 |

In frames 113-114, the participant had used all of the available gooballs and the game was terminated.

*Figure 4.26 Participant K-01 - Attempt 2 (Link)*

In frames 1-2 (not shown), the participant watched the opening video sequence. In frames 3-12 (not shown), the participant observed the scene without interacting with the game.
In frame 13-14, the participant selects and moves a gooball directly to the pipe.

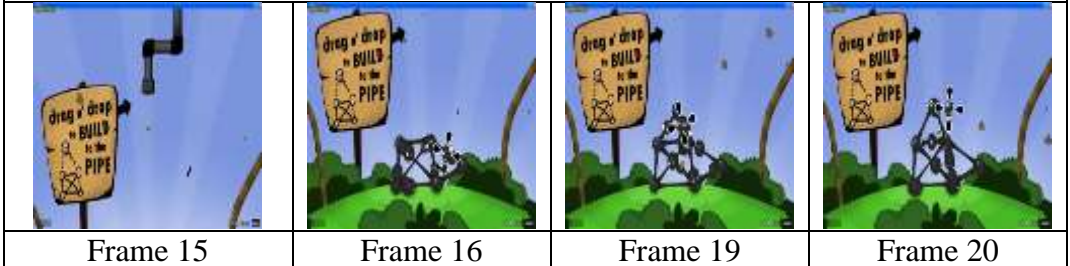

| Frame 15 | Frame 16 | Frame 19 | Frame 20 |

In frame 15-16, the participant watched the gooball fall and then proceeded to select and move another gooball and placed it to the left of the base frame.
In frames 17-18 (not shown), the participant continued to move the gooball around.
In frame 19, the participant has placed a gooball
In frame 20, the participant selected another gooball.
In frames 21-34, the participant started to build the tower.

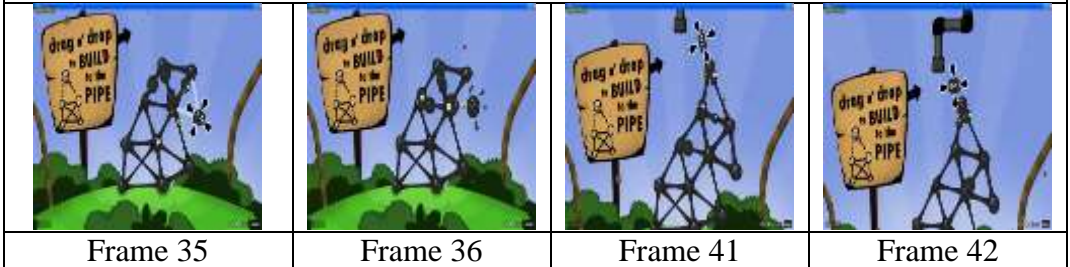

| Frame 35 | Frame 36 | Frame 41 | Frame 42 |

In frames 35-36, the participant continued to drag and drop gooballs This design appears very haphazard.
It appears that the participant has not seen or understood the image on the sign.
In frames 37-40 (not shown), the participant continued to drag and drop gooballs.
In frames 41-42, the participant moved the gooball directly to the pipe. As the tower was not close enough to the pipe, the gooball fell.
In frames 43-44 (not shown), the participant continued to attempt to move the gooball directly to the pipe.

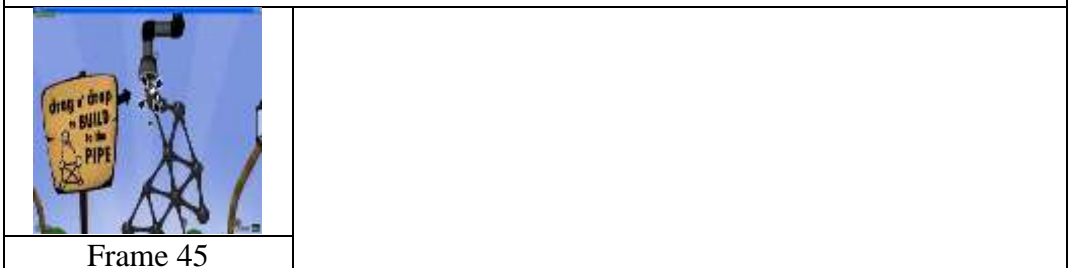

| Frame 45 |

In frame 45, the participant completes the level.
In frame 46-60 (not shown), the participant continued to watch the end of game video sequence.

*Figure 4.27 Participant K-03 - Attempt 1 (Link)*

| 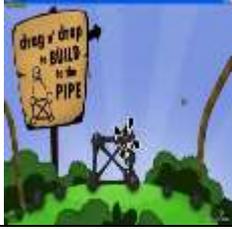 | 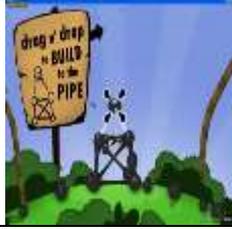 | 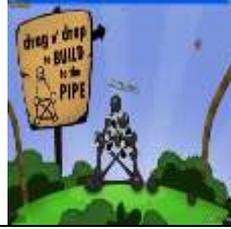 | 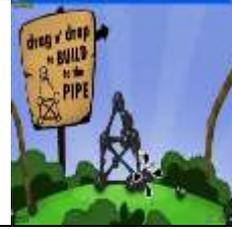 |
|---|---|---|---|
| Frame 7 | Frame 8 | Frame 9 | Frame 10 |

In frames 1-6 (not shown), the participant continued to watch the beginning sequence and then start to move the gooball

In frames 7-8, the participant selected and moved the gooball.

In frames 9-15 (not shown), the participant continued to select and move gooballs.

| 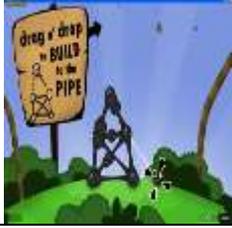 | 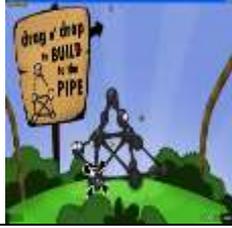 | 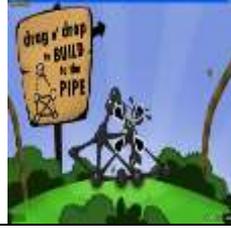 | 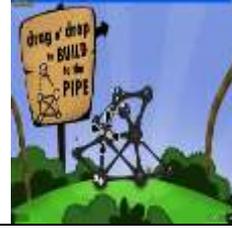 |
|---|---|---|---|
| Frame 16 | Frame 21 | Frame 26 | Frame 27 |

In frame 16, the participant selects another gooball and places it to the side of the tower. This suggests that the first placement may have been accidental and the participant had not read or understood the instructions on the sign.

In frames 21, 26, and 28, the participant selects more gooballs and places them to either side of the tower. The participant clearly has no idea what he should do.

| 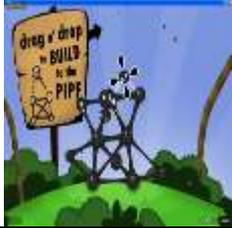 | 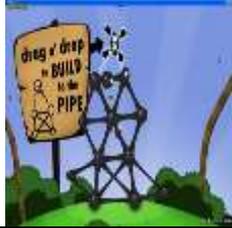 | 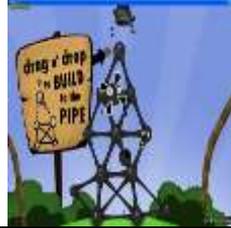 | 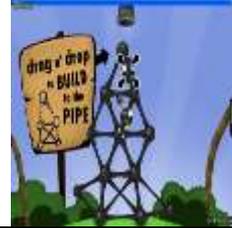 |
|---|---|---|---|
| Frame 32 | Frame 40 | Frame 43 | Frame 44 |

In frames 28-31 (not shown), the participant continued to select and move gooballs and placing them to the side of the tower.

In frame 32, the participant moves a gooball vertically. This is repeated in frames 33-41. In frame 40, it is clear that this participant has worked out the objective of the game and continues to build the tower toward the pipe.

| 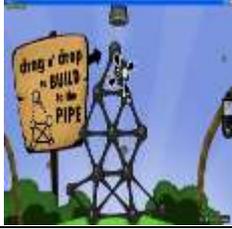 | 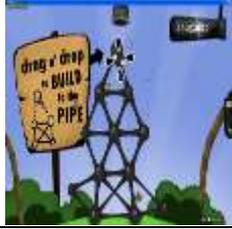 | | |
|---|---|---|---|
| Frame 45 | Frame 46 | | |

In frames, 43-46-the participant moved had built a tower close enough to the pipe and the level was complete.

In frame 49-60, the participant watched the end of game video sequence.

*Figure 4.28 Participant K-03 - Attempt 2 (Link)*

| Frame 13 | Frame 14 | Frame 15 | Frame 16 |
|---|---|---|---|

In frames 1-16, the participant started the game and watch the camera pan through the level. The participant looked at the starting grid, the instruction book and the magnifying glass. The participant eventually selected and looked at one page of the instruction book (Frames 17-27).

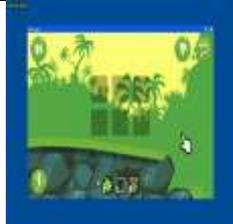 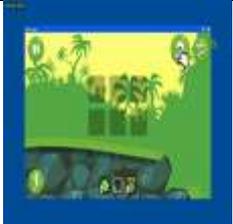 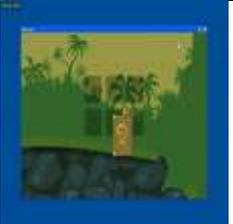 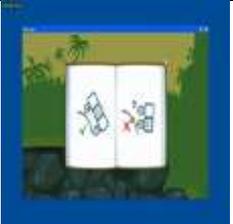

| Frame 17 | Frame 18 | Frame 19 | Frame 20 |
|---|---|---|---|

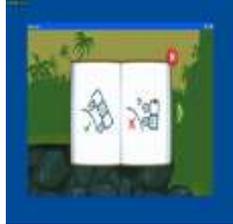 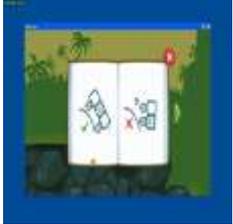 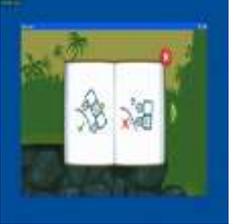 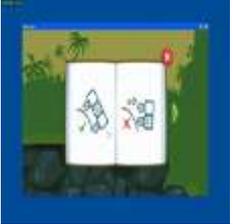

| Frame 21 | Frame 22 | Frame 23 | Frame 24 |
|---|---|---|---|

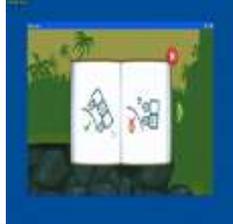 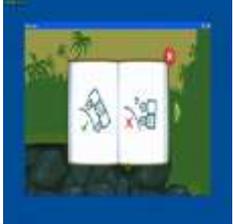 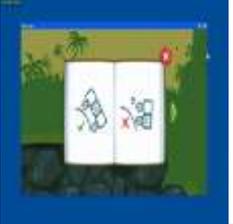 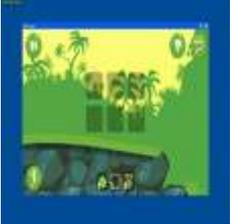

| Frame 25 | Frame 26 | Frame 27 | Frame 28 |
|---|---|---|---|

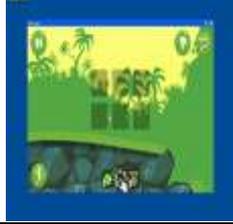 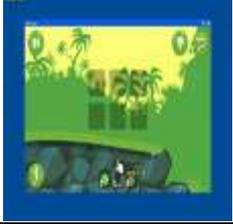 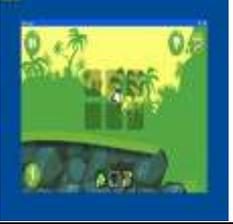 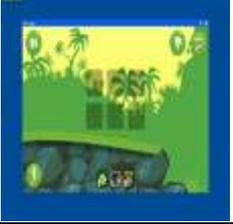

| Frame 29 | Frame 30 | Frame 31 | Frame 32 |
|---|---|---|---|

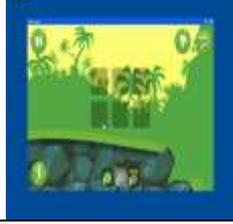 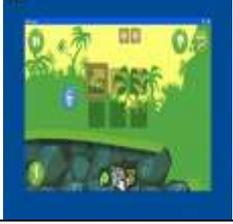 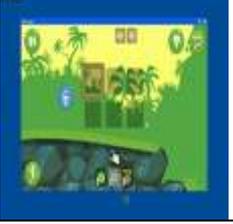 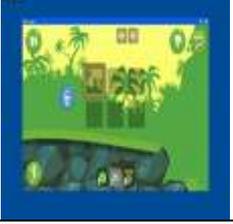

| Frame 33 | Frame 34 | Frame 35 | Frame 36 |
|---|---|---|---|

In frames 33-38, the participant explored the starting grid and selected and place a box on the starting grid. This suggests that through looking at the instruction book, the participant had learnt the basic knowledge of how and where to place the box.

| 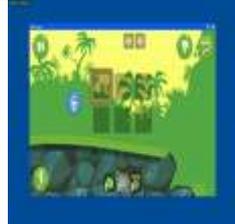 | 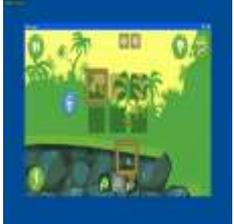 | 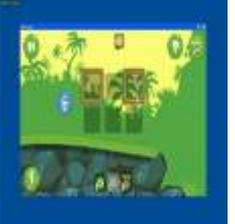 | 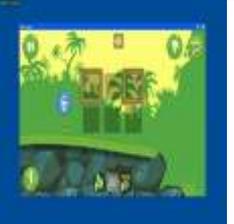 |
|---|---|---|---|
| Frame 37 | Frame 38 | Frame 39 | Frame 40 |

In frames 38-40, the participant selected another box and placed it on the starting grid. As the assembly process is identical to the example provided in the instruction book, this further suggests that the participant had learnt the basics principles of the car assembly process in this game.

| 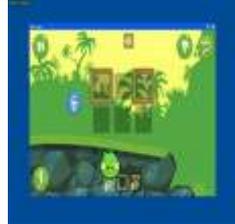 | 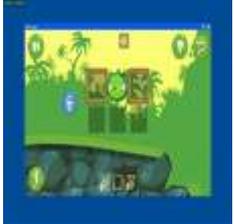 | 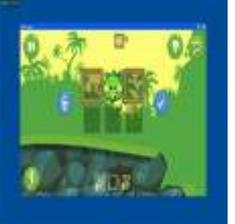 | 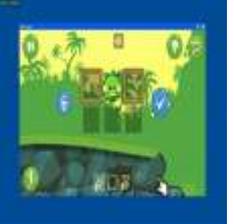 |
|---|---|---|---|
| Frame 41 | Frame 42 | Frame 43 | Frame 44 |

In frames 41-44, the participant selected and placed the pig on the grid. Note that the check icon and rubbish bin (trash can) is now displayed on the screen. Fom this assembly process, it is clear that the participant had not completely learnt the car assembly process. Although the car has the minimal components, it does not have any wheels and clearly wont go very far.

| 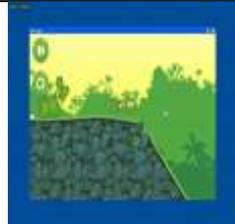 | 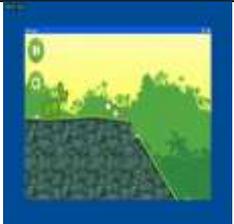 | 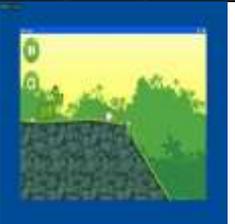 | 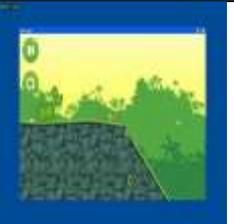 |
|---|---|---|---|
| Frame 45 | Frame 46 | Frame 47 | Frame 48 |

| 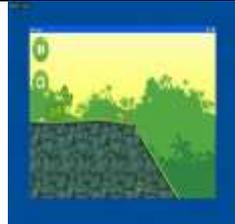 | 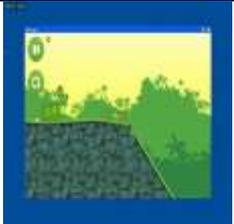 | 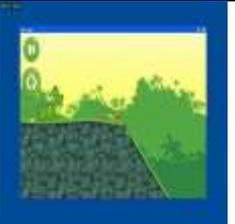 | 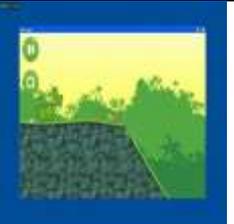 |
|---|---|---|---|
| Frame 49 | Frame 50 | Frame 51 | Frame 52 |

Frames 49-52 show the car sitting stationary on the top of the hill.

In frame 53 (not shown), the participant restarted the level.

Frames 54-56 (not shown), the participant looked at the screen.

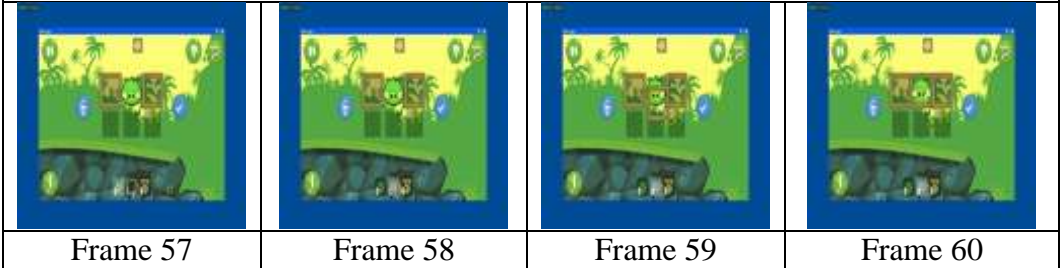

| Frame 57 | Frame 58 | Frame 59 | Frame 60 |

In frames 57-62, the participant selects another box and places on the grid. This demonstrates that the participant may have remembered this from the instruction book.

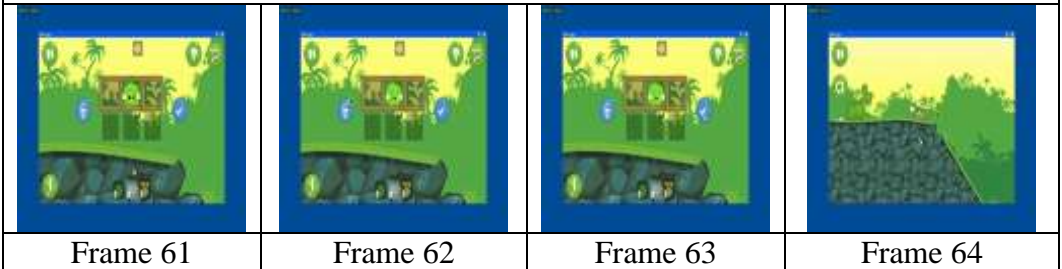

| Frame 61 | Frame 62 | Frame 63 | Frame 64 |

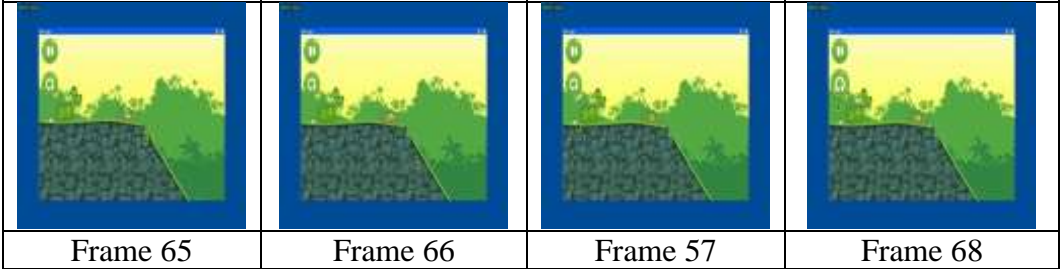

| Frame 65 | Frame 66 | Frame 57 | Frame 68 |

Frames 61-68, the participant selects the check icon and the car once again does not go anywhere.

In frames 69-72, the participant restarted the level and tried again.

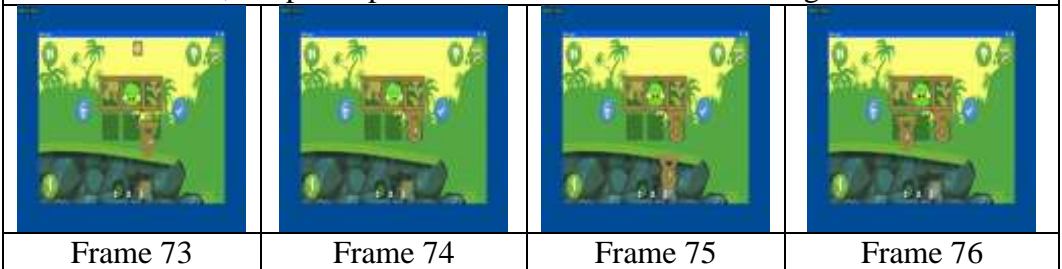

| Frame 73 | Frame 74 | Frame 75 | Frame 76 |

In frames 73-78, the participant sucessfully adds the two wheels. The optimum placement of the wheels suggets that the participant had connected the images in the instruction book with the task.

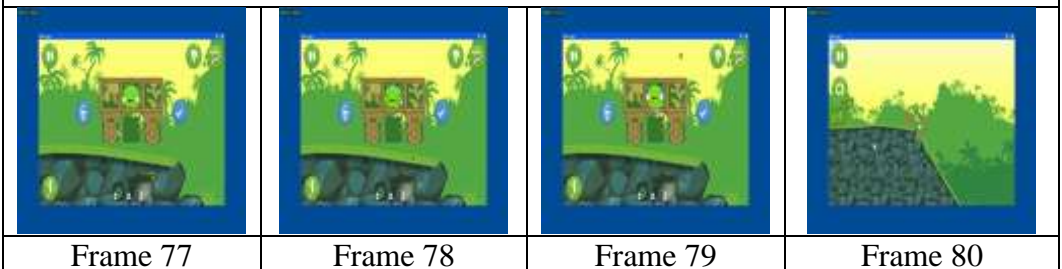

| Frame 77 | Frame 78 | Frame 79 | Frame 80 |

In frame 79, the participant selected the check icon.

| 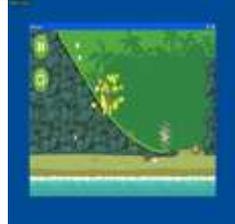 | 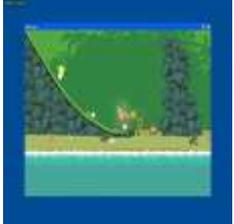 | 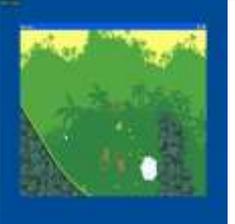 | 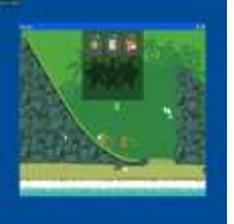 |
|---|---|---|---|
| Frame 81 | Frame 82 | Frame 83 | Frame 84 |

In frames 81-84, the car proceeds down the hill. The car makes it to the nest egg (in-tact), but crashes and breaks into the wall (meeting all three objectives of this level).

In frames 85-93 (not shown), the participant watched the confirmation of results sequence.

*Figure 4.29 Participant K-04 Attempt 1*

| 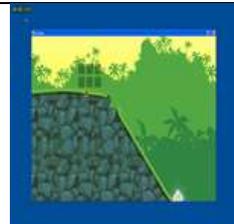 | 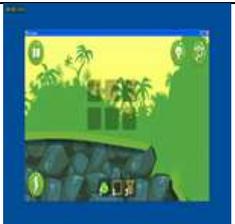 | 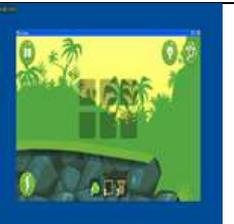 | 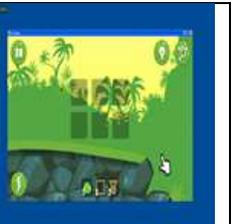 |
|---|---|---|---|
| Frame 1 | Frame 2 | Frame 3 | Frame 4 |

In frames 1-5, the participant looked around the screen.

| 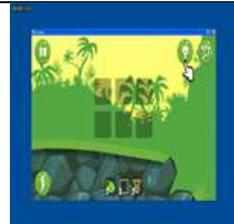 | 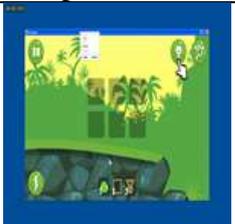 | 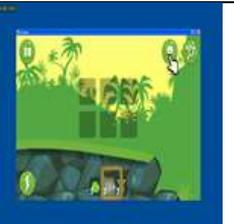 | 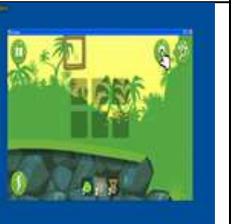 |
|---|---|---|---|
| Frame 5 | Frame 6 | Frame 7 | Frame 8 |

In frame 6-9, the participant slected a box and placed it on the grid.

| 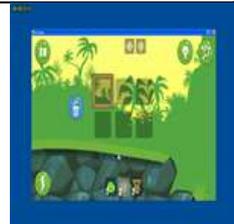 | 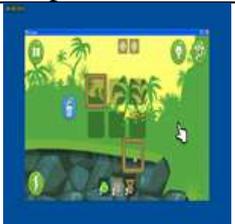 | 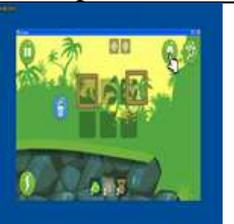 | 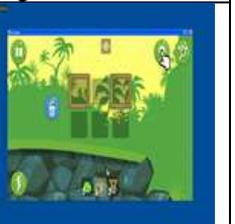 |
|---|---|---|---|
| Frame 9 | Frame 10 | Frame 11 | Frame 12 |

In frames 10-12, the participant selected the second box and placed this on the grid.

| 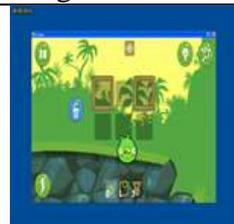 | 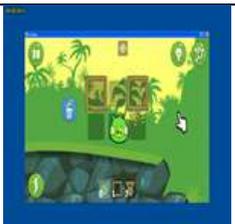 | 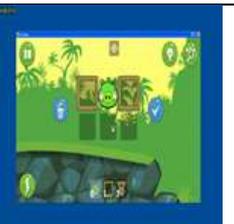 | 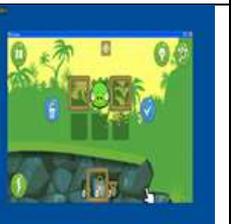 |
|---|---|---|---|
| Frame 13 | Frame 14 | Frame 15 | Frame 16 |

In frames 13-15, the participant selected and placed the pig on the grid.

| 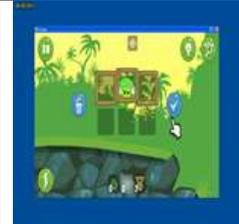 | 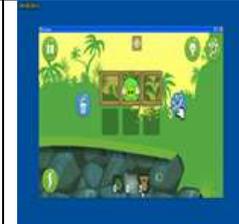 | 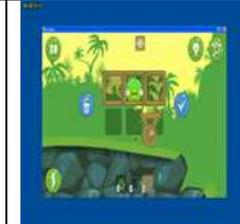 | 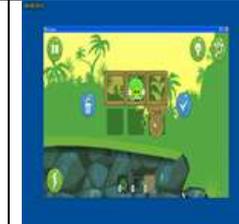 |
|---|---|---|---|
| Frame 17 | Frame 18 | Frame 19 | Frame 20 |

In frames 17 and 18, the participant selected and placed the third box. This process is identical to the first attempt. The participant placed the pig first and then the box. It is possible that the participant thought this was the correct procedure, given that this was successful last time.

In frames 19 and 20, the participant put the first wheel on the car.

| 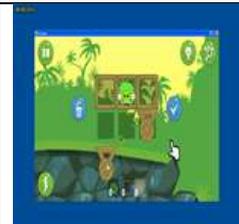 | 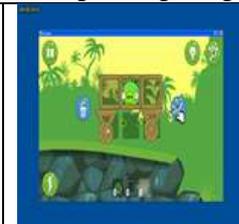 | 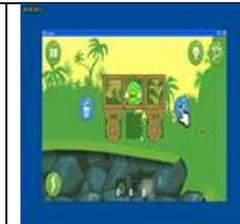 | 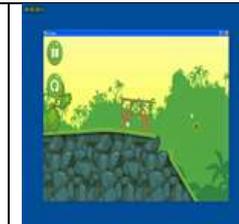 |
|---|---|---|---|
| Frame 21 | Frame 22 | Frame 23 | Frame 24 |

In frames 21 and 22, the participant selected and placed the second wheel on the car.

In frame 23, the participant selected the check icon.

| 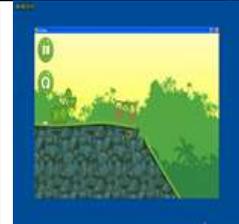 | 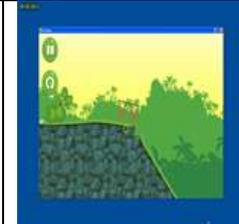 | 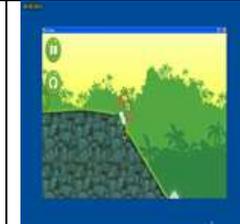 | 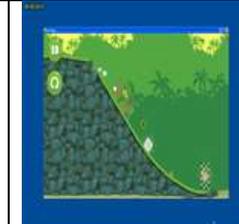 |
|---|---|---|---|
| Frame 55 | Frame 26 | Frame 27 | Frame 28 |

| 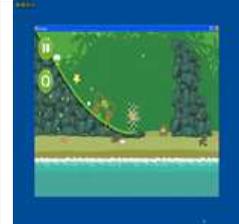 | 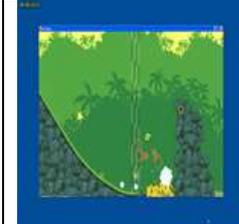 | 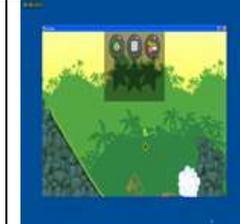 | 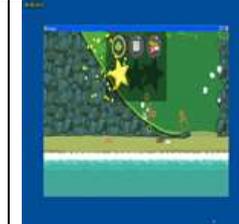 |
|---|---|---|---|
| Frame 29 | Frame 30 | Frame 31 | Frame 32 |

In frames 24-32, the participant watched the car proceed to the finish line.

In this attempt, the participant had achieved the requirements to get three stars.

*Figure 4.30 Participant K-04 Attempt 2*

| 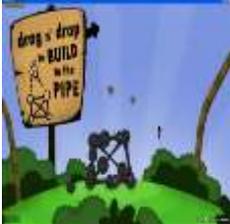 | 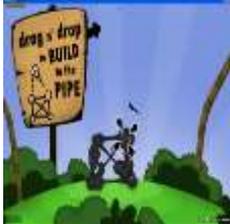 | 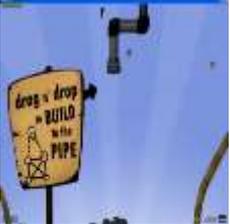 | 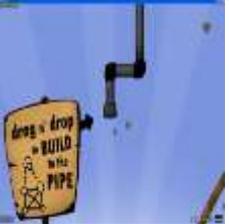 |
|---|---|---|---|
| Frame 31 | Frame 32 | Frame 33 | Frame 34 |

In frames 1-4 (not shown), the participant continued to watch the beginning video sequence.

In frames 5-30, the participant moved the gooball around the beginning frame.

In frame 31-32, the participant moved the gooball toward the tree. It appears that the participant may have confused the tree with a pipe.

In frames 33-34, the participant moved the gooball directly toward the pipe.

In frames 35-40 (not shown), the participant continued to move the gooball toward the pipe.

| 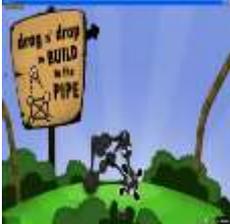 | 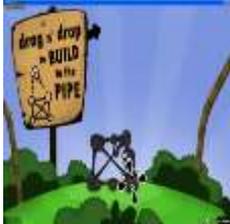 | 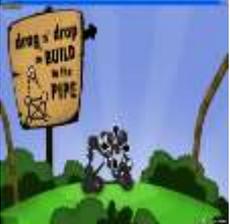 | 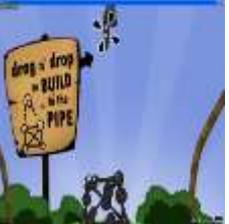 |
|---|---|---|---|
| Frame 41 | Frame 42 | Frame 43 | Frame 44 |

In frame 41-42, the participant moved the gooball around the base structure.

In frames 43-44, the participant selected another gooball and moved it to the pipe.

In frames 45-50 (not shown), the participant moved the gooball around the base frame.

It is clear from the actions that the participant had not completely understood or had seen the image on the sign.

| 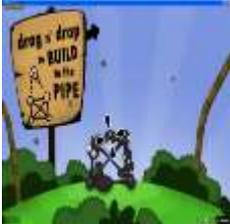 | 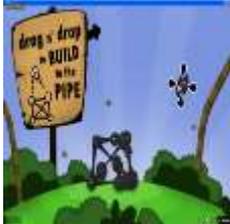 | 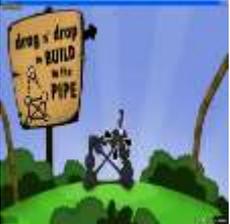 | 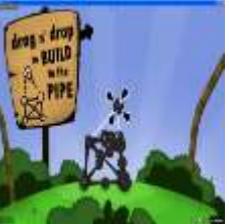 |
|---|---|---|---|
| Frame 51 | Frame 52 | Frame 57 | Frame 58 |

In frames 51-52, the participant moved the gooball to the tree.

In frames 53-56 (not shown), the participant continued to move the gooball around the base frame

In frame 57-58, the participant selected a gooball and dragged it away from the frame.

| 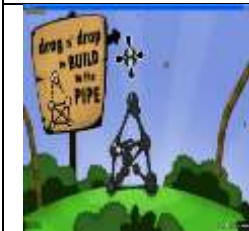 | 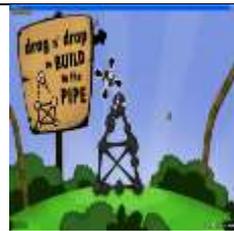 | 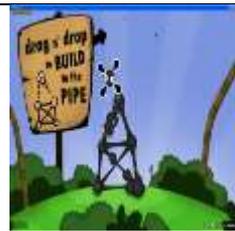 | 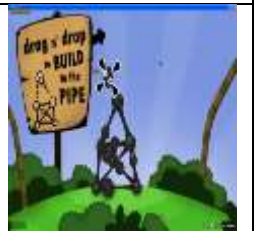 |
|---|---|---|---|
| Frame 59 | Frame 60 | Frame 61 | Frame 62 |

In frames 59-60, the participant continued to move the selected gooball.
In frame 61, the participant released the gooball.
In frame 62, the participant selected another gooball.

| 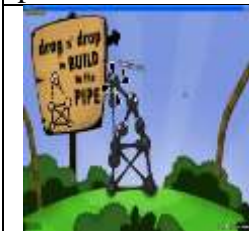 | 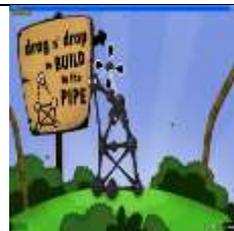 | 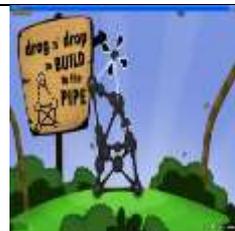 | 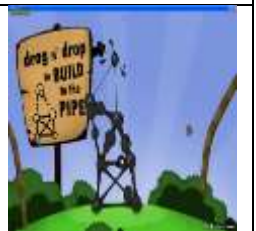 |
|---|---|---|---|
| Frame 65 | Frame 66 | Frame 67 | Frame 68 |

In frame 65-66, the participant dragged another gooball. The gooball was initially pulled too far away from the structure (frame 65), but the participant moved it back within range (frame 66).
In frame 67-68, The participant continued to select and move the gooball.
In frames 69-72 (not shown), the participant eventually placed the gooball and proceeds to build a tower

| 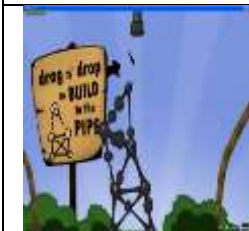 | 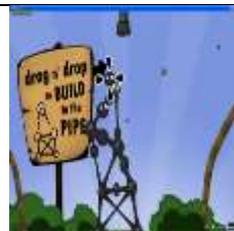 | 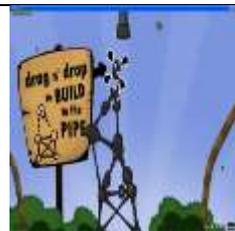 | 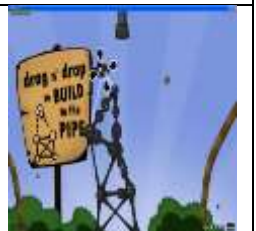 |
|---|---|---|---|
| Frame 69 | Frame 70 | Frame 71 | Frame 72 |

In frames 69-72, the participant appears to have understood how to correctly construct the tower.

| | | | |
|---|---|---|---|
| Frame 73 | Frame 74 | Frame 77 | Frame 78 |

In frame 73, the participant moved the gooball directly onto the pipe.
In frame 74, the participant selects another gooball and continues to build the tower.
In frame 77, the participant dragged the gooball too far away from the supporting structure.
In frame 78, the participant, moved the gooball back in range of the supporting structure.
In frames 79-82 (not shown), the participant continued to build the tower.

| | | |
|---|---|---|
| 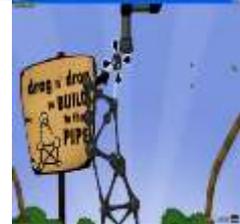 | 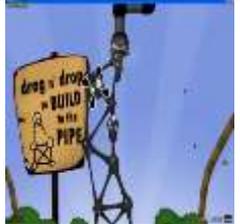 | |
| Frame 83 | Frame 84 | |

In frames 83-84, the participant successfully completed the level.
In frames 85-105 (not shown), the participant continued to watch the ending video sequence.

*Figure 4.30 Participant K-08 Attempt 1*

| | | | |
|---|---|---|---|
| 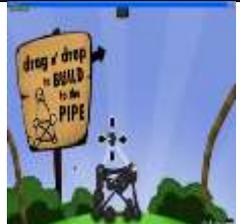 | 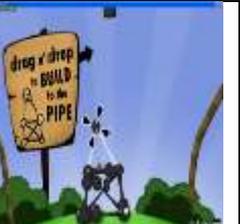 | 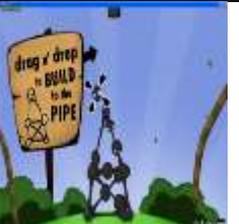 | 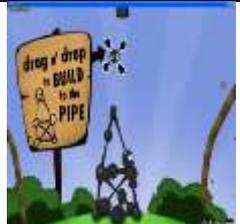 |
| Frame 43 | Frame 44 | Frame 49 | Frame 50 |

In frames 1-42 (not shown), the participant moved the gooball around the starting frame.
In frames 43-44, the participant selected and moved the gooball off the starting frame.
In frames 45-48 (not shown), the participant eventually dropped the gooball.
In frames 49-50, the participant moved the gooball directly to the pipe.

| | | | |
|---|---|---|---|
| 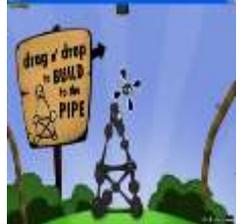 | 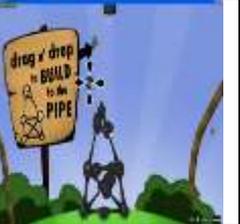 | 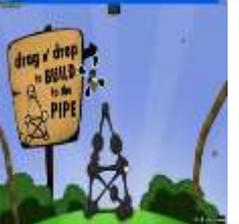 | 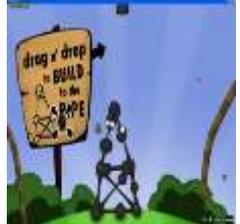 |
| Frame 51 | Frame 52 | Frame 53 | Frame 54 |

In frames 51-52, the participant moved the gooball. However, the ball was too far away from the connecting points for it to be successfully placed.
In frames 53-54, the participant moved the gooball to the sign.

| | | | |
|---|---|---|---|
| 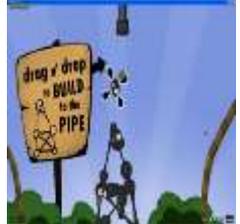 | 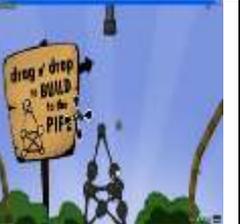 | 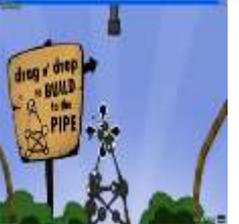 | 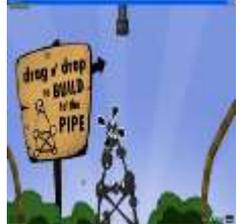 |
| Frame 55 | Frame 56 | Frame 61 | Frame 62 |

In frames 55-56, the participant moved the gooball to the pipe and the sign. This indicates that the participant did not completely understand the purpose of the game.
In frames 61-62, the participant continued to move the gooball.
In frames 63-76 (not shown), the participant continued to move the gooball.

| | | |
|---|---|---|
| 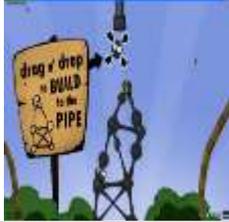 | 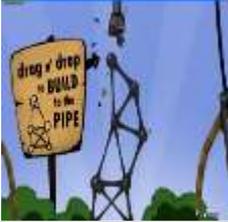 | |
| Frame 77 | Frame 83 | |

In frame 77, the participant moved the gooball directly to the pipe. This suggests that they did not completely understand the purpose of building the tower.
In frames 79-82 (not shown), the participant continued to construct the tower.
In frame 83, the participant successfully completes the level.
In frames 84-102 (not shown), the participant watched the end of game video sequence.

*Figure 4.31 Participant K-08 Attempt 2*

## 4.2.2.4 Enjoyment

While not of primary importance to the research questions, the majority of participants expressed that they enjoyed playing the game. Table 4.26 presents the answers from the adults and Table 4.27 displays the answers to the questions from the children.

| Participant | Answer |
|---|---|
| A-01 | Trying to make sure the structure does not fall down |
| A-02 | Real world physics |
| A-03 | The pigs |
| A-04 | Trying to figure out how to make it go faster |
| A-05 | Build the vehicles |
| A-06 | Liked the game, it was whimsical, cute gooballs |
| A-07 | Learning how to play |
| A-08 | Have to think quickly, which is fun |
| A-09 | Silly animations, gameplay mechanics |
| A-10 | The gooballs were cute; the challenge was really interesting |
| A-11 | Trying to figure the strongest thing to build fast enough, so it did not fall |
| A-12 | Trying to figure out the best structure and have the most at the end |
| A-13 | The silly pigs |
| A-14 | The challenge, the playfulness of the gooballs |
| A-15 | I liked how you could customize the car; it was more constrained |
| A-16 | Not sure |
| A-17 | The game made me think in a more practical sense |
| A-18 | I liked being able to build the vehicle, building with limited parts |
| A-19 | It was a challenge when I figured out the strategy to support it |
| A-20 | Physics & structures into a game |

*Table 4.26 The adult answers to the question, "What was the most fun part of the game?"*

| Participant | Answer |
|---|---|
| K-01 | Build cool towers |
| K-02 | Build the tower |
| K-03 | Making the tower |
| K-04 | Clicking the buttons |
| K-05 | Making the cars |
| K-06 | Build the cars |
| K-07 | Making the car go |
| K-08 | The sign said "they are stronger than you think" |
| K-09 | All of it |
| K-10 | Playing the game |
| K-11 | Playing with the magnets |
| K-12 | The pigs |

*Table 4.27 The childrens' answers to the question, "What was the most fun part of the game?"*

## 4.4    CONCLUSION

The data presented in this chapter suggests there is evidence of learning acquired by playing a commercial video game. Furthermore, there is evidence that there have been changes in the cognitive processing of each individual for each attempt at the game. These matters will be discussed further in Chapter Five.

**Chapter Five**

**DISCUSSION**

## 5.0 INTRODUCTION

The primary research question focuses on what learning is acquired by playing video games. An extensive review of the literature was performed, and this was presented in Chapter Two. In Chapter Three, the research questions were presented, and the methods for testing these questions were evaluated and the proposed method defined. The research methodology was applied, and the results of this research were presented in Chapter Four.

This chapter contains four main sections. Section 5.1 will discuss and analyse the findings from the study of adults. Section 5.2 will discuss and analyse the findings from the study of children. Section 5.3 will discuss these findings and the relationship to the research questions and the contribution to new knowledge. Section 5.4 will conclude this chapter.

## 5.1 THE STUDY OF ADULTS

The study of adults identified that the first hypothesis is supported; playing video games does positively affect user learning. The study of the adults did identify some support for the second hypothesis: playing video games positively affects problem solving ability. The answers to the research questions will follow.

The results that support the hypotheses will be presented in three sections. The first Section (5.1.1) will discuss the performance of the group. The second Section (5.1.2) will discuss the individual performance. The final Section (5.1.3) will discuss the implications of these findings.

### 5.1.1    The combined adult performance

As detailed in Section 4.1.1, there was only one participant in the treatment group that changed their incorrect answer to the question, "What shape do you think is the strongest for building a really tall tower?" into a correct answer (a triangle) (see Table 5.1). However, there were two participants in the control group that changed their answer from an incorrect answer to a correct one. This suggests that the treatment has no effect on the understanding of tower construction principles.

| Participant | Group | Pre-exposure answer | Post-exposure answer |
|---|---|---|---|
| A-03 | Control | Triangle | Square |
| A-05 | Control | Circle | Triangle |
| A-012 | Treatment | Square | Triangle |

*Table 5.1 Answers to the tower construction question*

Furthermore, there is limited evidence that the treatment had any affect on the magnetic tower construction process. Table 5.2 presents the results of the participants that changed the construction process from a structurally unsound method to a more structurally sound one. One participant in the treatment group and one participant in the control group changed from using squares to using triangles. This suggests that the treatment had no effect on the out-of-game experiment. Furthermore, these results suggest that the treatment did not have an effect on the understanding of the construction principles involved in building the magnetic tower.

| Participant | Group | Pre-exposure construction | Post-exposure construction |
|---|---|---|---|
| A-01 | Treatment | Squares | Triangles, Squares |
| A-013 | Control | Squares | Triangles |

*Table 5.2 Magnetic tower construction methods*

However, the answer to the question, "Why did the tower break?" indicated that the participants in the treatment group had an understanding of what was important in building a tall tower before the experiment. After the game-play, they were able to reflect on the reasons why the tower broke.

| Participant | Why did the tower break |
|---|---|
| A-01 | Foundation was not strong enough |
| A-02 | Weight not distributed evenly |
| A-06 | Shaky hands |
| A-08 | Too much focus on height and structure |
| A-10 | The foundation was not strong |
| A-11 | Because it was not strong enough. Too heavy on one side |
| A-12 | N/A - Towers did not fall |
| A-14 | It swayed unexpectedly, not perfect balance |
| A-16 | Foundation was not strong enough |
| A-19 | Because maybe the wind; it was flexible; tool tips recommended going faster |
| A-20 | Foundation was not strong enough |

*Table 5.3 Why did the tower break?*

Although this does not clearly demonstrate that transferable conceptual learning transpired during gameplay, it possibly suggests that the participants were cognisant of what went wrong.

The gameplay data provides evidence of skill and proceedural learning that is more apparent. In Table 5.4, 82% of the participants improved on the number of gooballs collected. Furthermore, 90% of the participants reduced the number of moves to meet this goal. Seventy-two percent of the participants reduced the amount of time it took to complete the level. Of the three participants that did not reduce the amount of time to complete the level, two of them took one or two additional seconds.

| | Attempt 1 | | | Attempt 2 | | | Variance | | |
|---|---|---|---|---|---|---|---|---|---|
| ID | Gooballs collected | Number of moves | Time taken | Gooballs collected | Number of moves | Time taken | Additional gooballs collected | Additional moves | Time saved |
| A-01 | 7 | 7 | 0:39 | 10 | 6 | 0:31 | 3 | -1 | 0:08 |
| A-02 | 8 | 9 | 0:40 | 11 | 5 | 0:30 | 3 | -4 | 0:10 |
| A-06 | 7 | 6 | 0:35 | 11 | 3 | 0:30 | 4 | -3 | 0:05 |
| A-08 | 9 | 5 | 0:29 | 10 | 3 | 0:20 | 1 | -2 | 0:09 |
| A-010 | 7 | 7 | 0:22 | 9 | 5 | 0:23 | 2 | -2 | (0:01) |
| A-011 | 8 | 11 | 1:00 | 7 | 7 | 0:16 | -1 | -4 | 0:44 |
| A-012 | 7 | 8 | 0:24 | 9 | 7 | 0:26 | 2 | -1 | (0:02) |
| A-014 | 7 | 8 | 0:34 | 10 | 6 | 0:21 | 3 | -2 | 0:13 |
| A-016 | 5 | 7 | 0:39 | 8 | 5 | 0:27 | 3 | -2 | 0:12 |
| A-019 | 6 | 8 | 0:37 | 8 | 6 | 0:15 | 2 | -2 | 0:22 |
| A-020 | 10 | 4 | 0:34 | 6 | 7 | 0:49 | -4 | +3 | (0:15) |

*Table 5.4 Gameplay data*

Table 5.4 indicates that the majority of participants (95%) learnt to play the game after two attempts. This is evident through improvements in one (or more) of the in-game performance metrics (the number of gooballs collected, the number of moves, and/or the time taken to complete the game). The only exception was participant A-20.

The potential of the eye fixations and blinks came from a discussion at the Games for Learning and Society Conference in Maddison, Wisconsin and the additional literature that was reviewed as a result of this conversation (Green & Bavelier, 2008; Just & Carpenter, 1976; Just & Carpenter, 1980; Just & Carpenter, 1984; Shute & Kim, 2012). Figure 5.1 presents frequency of the endogenous blinks for both attempts at the game.

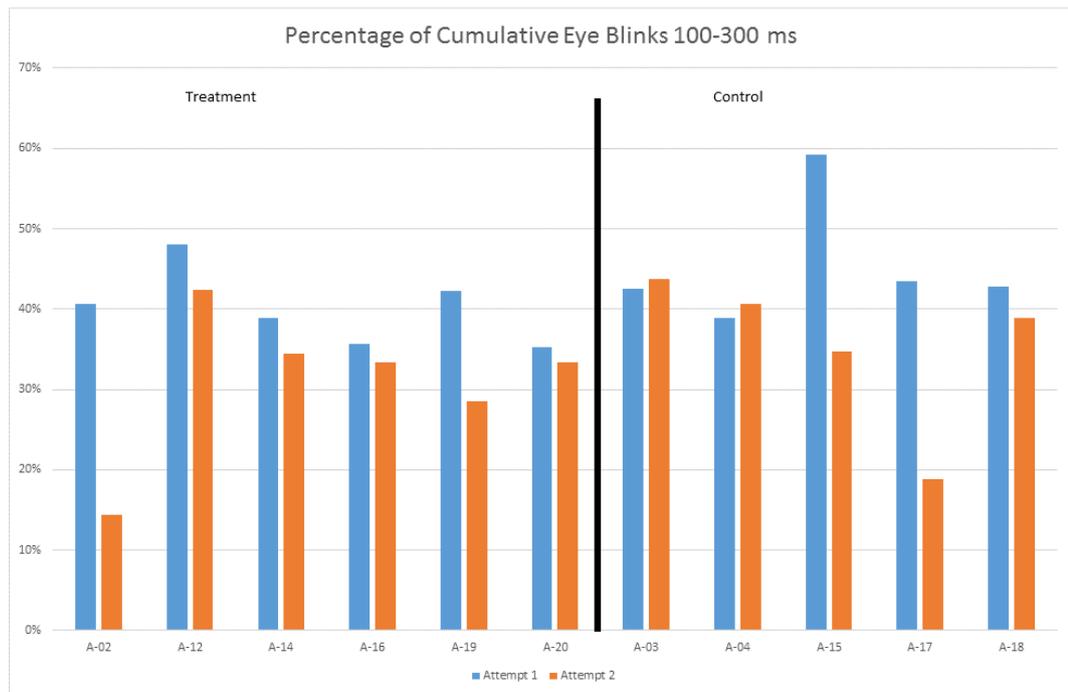

*Figure 5.1 Percentage of cumulative endogenous blinks*

From this data, it is possible to identify that the participants A-02, A-12, A-15, A-14, A-15, A-16, A-17, A-18, A-19 and A-20 did not perform as many endogenous blinks in the second attempt at the game as they did in the first attempt. This suggests that there was less information processing in the second attempt than there was in the first attempt at the game. Participant A-03 and A-04 performed slightly more endogenous blinks in the second attempt at the game than they did in the first. The possible reasons for this will be discussed in Section 5.1.2.

The result of the analysis of the number of endogenous eye blinks suggests that these variables provide an indicator of the amount of information processing that takes place while playing a commercial video game. The data presented in Figure 5.1 indicates that the majority of participants required less information processing (or cognition) through the repeated exposure to the game mechanics (creating a tower or assembling a car). The difference between the treatment and the control group is minimal. In both cases the participants have been presented with a concept and difference between the frequency of endogenous eye blinks between the first and second exposure indicate that majority have acquired or demonstrated the skills (perceptual or cognitive) required to complete the level. Further, the second attempt indicates the benefits of the repeated practice (Ackerman, 1988). While the data does not provide evidence of any changes in motor learning, the general decrease in endogenous blinks does indicate that the amount of information processing (or cognition) did decrease.

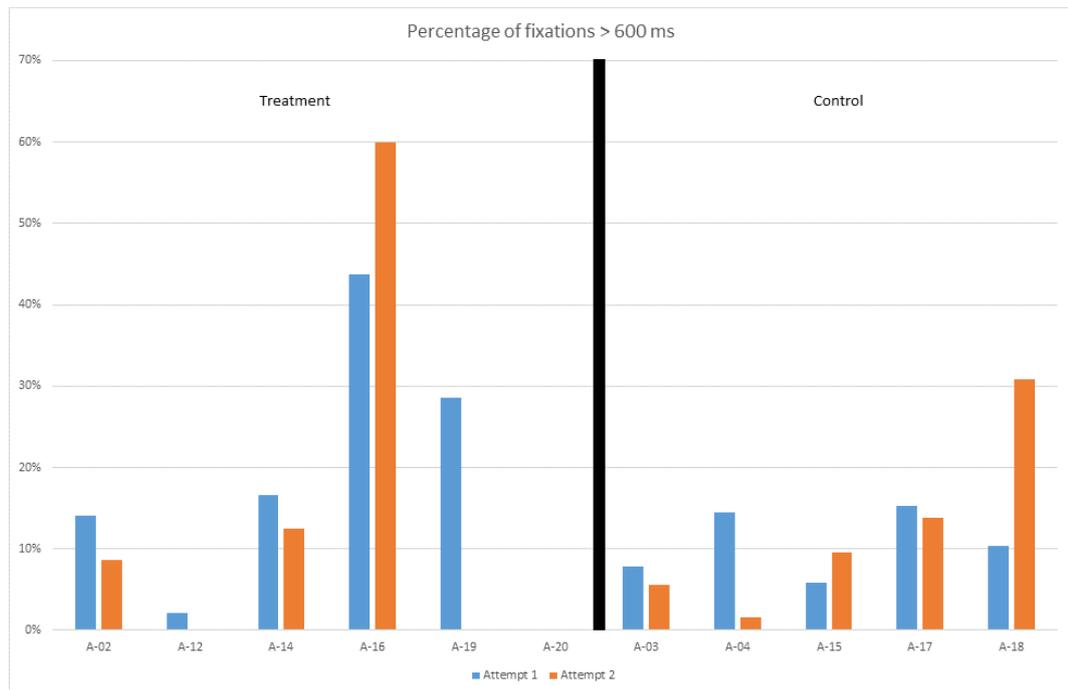

*Figure 5.2 The percentage of fixations above 600 ms*

In Figure 5.2, a change in the percentage of the fixations that were longer than 600 ms can be identified. There has been a reduction in the percentage of long fixations for participants A-02, A-03, A-04, A-12, A-14, A-17, and A-19. This suggests that there was less cognition (or problem solving) during the second attempt at the game for these participants. Participants A-15, A-16, and A-18 demonstrated an increase in the percentage of fixations that were above 600 ms and participant A-20 demonstrated no change.

The effect size for this analysis ($d = 0.7$) was found to exceed Cohen's (1988) convention for a medium effect ($d = .50$). The result of the analysis of the number of fixations that were above 600 ms suggests that these variables provide an indicator of the amount of problem solving that takes place while playing a commercial video game. The data presented in Figure 5.2 indicates that the majority of participants spent less time fixated on specific aspects of the game through the repeated exposure to the game mechanics (creating a tower or assembling a car). The difference between the treatment and the control group is minimal. In both cases the participants have been presented with a concept and the difference in the frequency of long fixations between the first and second exposure indicate that majority have acquired or demonstrated the skills (perceptual or cognitive) required to complete the level. Further, the second attempt indicates the benefits of the repeated practice (Ackerman, 1988). The general decrease in long fixations does indicate that the amount of cognitive problem solving did decrease.

### 5.1.2   The individual adult performance

In the literature review in Section 2.9.2, an adaptation of Nacke & Drachen's (2011) model was presented, that suggests that a player's experience is dependent on the in-game (or virtual) experience and also the physical experience which consists of the technical experience, mental state, and context. The experience that Nacke & Drachen (2011) refer to is dependent on the player's prior exposure to playing the game and/or games of a similar genre.

As identified in Section 5.1.1, there was one participant (A-14) that performed more endogenous blinks in the second attempt at the game. Furthermore, participant A-15 demonstrated a notable reduction in the percentage of endogenous eye blinks. Moreover, participant A-20 performed no fixations that were above 600 ms in both attempts at the game. Prior gameplay experience and the in-game experience of each of these participants will be discussed to help understand the underlying causes.

### 5.1.2.1 Participant A-15

Participant A-15 was a computer engineering major student who reported that they played video games three to four times a week. The participant was in the control group and had not played the video game Bad Piggies (Rovio Entertainment, 2013) before. In the first attempt at the game, the participant blinked 61 endogenous blinks, which is above the mean of the group (35). In the second attempt at the game, participant A-15 blinked 33 endogenous blinks, which is above the mean of the group (22). However, this still represents a decrease in endogenous blinks. Furthermore, the percentage of endogenous blinks decreased by 93.96%, which was considerably higher than the other participants ($M = 21\%$). The participant also demonstrated an increase in the percentage of fixations that were above 600 ms in the second attempt at the game (62%). From the observations it is possible to conclude that the participant did not know how to play the game in the first attempt. In that first attempt, participant A-15 selected the wrong options and then read through the instruction book. Then the participant tried to build a vehicle, but did not seem to understand how to do this. The participant then re-read the instruction book. The participant then proceeded to work out how to build the vehicle through a process of trial and error, and eventually made a workable solution. In the second attempt, the participant proceeded to make the vehicle. The participant did not read the instruction book; they did not select the wrong options, and they selected the right options for the intended purpose. It is clear from this evidence that the participant did learn how to play this game. The variation in the quantity and the percentage of endogenous blinks supports this. The increase in the percentage of fixations that were above 600 ms suggest that the percentage of problem solving increased in the second attempt. However, the fixation data suggests that the total amount of information process (or problem solving) did not change. From the video evidence, it is possible to conclude that there was a reduction in problem solving behaviour during the second attempt at the game. Although the total number of fixations did not increase, the percentage of fixations that lasted longer than 600 ms decreased, this suggests that the participant spent more time problem solving in the first attempt than in the second attempt.

### 5.1.2.2 Participant A-16

Participant A-16 was a Mechanical Engineering major who reported playing video games once or twice a week. The participant was in the treatment group and had not played the video game World of Goo (2D Boy, 2008) previously. In the first attempt, participant A-16 performed 7 endogenous blinks, which was equal to the mean (7). In the second attempt participant A-16 performed 3 endogenous blinks, which was considerably below the mean of the group (3.55). This variation represents a 43% decrease in the number of endogenous blinks when compared to the first attempt. However, the participant demonstrated a 37% increase in the percentage of fixations that were above 600 ms. Participant A-16 collected three additional gooballs and reduced the number of moves from seven to five. The participant reduced the amount of time playing the level by 12 seconds.

From the video evidence presented (Figure 4.13 and Figure 4.14), it is clear that the participant learnt to play the game. In the first attempt at the game, the participant tried to move the gooball directly to the pipe (Figure 4.13, Frames 7 to 10). The participant tries to repeat this in Frames 14 to 17. The participant then proceeds with building a tower and successfully completes the level (Frames 21 to 38). In the second attempt at the game (Figure 4.14), the participant proceeds to build a tower and does not attempt to move the gooballs directly to the pipe. Although the construction of the tower is not as refined as the first attempt, it is clear from the evidence that the participant did learn that they needed to build a tower to get the gooballs to the pipe.

Although the number and percentage of endogenous blinks decreased in the second attempt at the game, the percentage of fixations above 600 ms increased by 37 %. The variation in the percentage of fixations that were above 600 ms could be explained by the participant thinking about how to build the tower successfully. The variation in construction methods used could be a result of of of the increased percentage of problem solving in the second attempt at the game. The construction method in the first attempt could have been unintentional. Alternatively, the participant may have wanted to explore an alternative construction method in the second attempt.

### 5.1.2.3 Participant A-20

Participant A-20 was a Pharmacy major who reported playing video games three to four times a week. The participant was in the treatment group and had not played the video game World of Goo (2D Boy, 2008) previously. In the first attempt, participant A-20 performed 6 endogenous blinks, which was below the mean (35). In the second attempt, the participant A-20 performed 1 endogenous blink, which was considerably below the mean of the group (22). This variation represents an 83% decrease in the number of endogenous blinks when compared to the first attempt. Furthermore, the participant demonstrated a 0% decrease in the percentage and number of fixations that were above 600 ms. Participant A-20 did not collect as many gooballs in the second attempt at the game (from 10 to 7) and increased the number of moves from four to seven. The participant also took more time to complete the level in the second attempt at the game (15 seconds).

From the data collected and the video evidence presented (Figure 4.15 and Figure 4.16), it is clear that the participant did not completely learn to play the game. In the first attempt at the game (Figure 4.15), the participant tried to move the gooball directly to the pipe (Frames 7 to 10). The participant then starts to build a tower, but then tries to move the gooball directly to the pipe again (Frames 27 to 29). The participant then successfully builds a tower and completes the level (Frames 30 to 38). In the second attempt at the game (Figure 4.16), the participant appears to struggle with the construction process. In frames 6 to 10, the participant does not appear to have grasped the process of moving the gooball to a position that is not too far away from the tower. Furthermore, in frames 12 to 25, the participant appears to continue to struggle with this concept. In frame 29, it appears that the participant once again tries to place the gooball directly in the pipe. This appears to be repeated several times in Frames 31 to 45. The participant eventually completes the level (Frame 51) however, from the construction method and processes used, it is clear that the participant did not completely understand how to play this video game.

The data collected from the eye tracker suggests that the participant did not produce any fixations that were above 600 ms. While it is possible that the participant did not undertake any problem solving during both experiments, the evidence suggests that this is highly unlikely. It is possible that for the fixations that were longer than 600 ms, the participant was an outlier. The multiple sources of data controlled for this.

### 5.1.3   Conclusions on the adult performance

The study of adults identified that the first hypothesis is supported; playing these commercial video games did positively affect user learning. From the data collected, it is possible to identify that the participants that did not have prior experience in playing the video game, learnt to play the game. Furthermore, the results of the analysis of the data on blinking and eye fixation supports this proposition. Finally, while the results of the pre- and post- questions and magnetic toy construction experiment are less than convincing, there is evidence that some participants did learn. There was some support for the second hypothesis that playing video games positively affects problem-solving ability. However, more empirical data is needed before this can be fully tested.

Given the prior experience of the adults and based on the reflective feedback on the reasons why the tower fell, it is more than likely that they already had some prior knowledge of the concepts required for success in the game. As noted in section 3.5, the implicit concepts in the game included static equilibrium, the importance of building sound structures, concept of force (gravity, wind, or buoyancy), the importance of strong foundations, the importance of support structures, and/or the importance of level structures. Therefore, the lack of identifiable change in the understanding of these principles is not difficult to understand. However, what is evident in this study is that those participants that had not played the game before, the majority (54%) learnt to play it. Furthermore, the participants who had prior experience in playing the game, the majority (80%) demonstrated an improvement in performance, which is an indicator of some improvement in skill learning (perceptual, cognitive, or motor), procedural learning (Ackerman, 1988; Beaunieux et al., 2006; Jarvis, 2006) and cognitive problem solving (Beaunieux et al., 2006).

## 5.2 THE STUDY OF CHILDREN

The study of children identified that the first hypothesis is supported; playing video games does positively affect user learning. The study of the children also found some support for the second hypothesis; that playing video games positively affects problem solving ability.

The results that support the hypotheses will be presented in three sections. The first Section (5.2.1) will discuss the performance of the group. The second Section (5.2.2) will discuss the individual performance. The final Section (5.2.3) will discuss the implications of these findings.

### 5.2.1 The combined performance of the children

The results of the study of the children will be discussed by first reviewing the answers to the quantitative data, and then the combined quantitative and qualitative data will be reviewed. As detailed in Section 4.1.2 there were five out of the six participants (83%) in the treatment group that changed their incorrect answer to the question, "What shape do you think is the strongest for building a really tall tower?" into a correct answer (Triangle) (Table 5.5). One participant, changed their incorrect answer to a different incorrect answer.

| Participant alias | Group | Pre-exposure answer | Post-exposure answer | Change |
|---|---|---|---|---|
| K-01 | Treatment | Square | Triangle | Positive |
| K-02 | Treatment | Square | Circle | Both incorrect |
| K-03 | Treatment | Circle | Triangle | Positive |
| K-08 | Treatment | Square | Triangle | Positive |
| K-10 | Treatment | Square | Triangle | Positive |
| K-11 | Treatment | Square | Triangle | Positive |

*Table 5.5 The treatment group answers to the construction question*

The evidence of learning is also apparent in the gameplay data. From Table 5.6, it is possible to conclude that all the participants demonstrated that some learning had taken place through a reduction in the time it took to complete the level. Furthermore, participants K-02, K-08, and K-11 also improved on both the number of gooballs collected and reduced the number of moves. This suggests that these participants who had no prior exposure to playing the game, learnt how to play it. In this experiment, all of the participants in the treatment group indicated that they had not played the game before.

| | Attempt 1 | | | Attempt 2 | | | Variance | | |
|---|---|---|---|---|---|---|---|---|---|
| ID | Gooballs collected | Number of moves | Time taken | Gooballs collected | Number of moves | Time taken | Additional gooballs collected | Additional moves | Time saved |
| K-01 | DNC | | 4:35 | DNC | | 1:52 | 0 | | 2:43 |
| K-02 | 4 | 9 | 5:30 | 7 | 7 | 0:44 | 3 | -2 | 4:46 |
| K-03 | 5 | 9 | 0:48 | 5 | 8 | 0:46 | 0 | -1 | 0:02 |
| K-08 | 9 | 5 | 1:33 | 10 | 4 | 1:29 | 1 | -1 | 0:04 |
| K-10 | 8 | 6 | 3:20 | 7 | 7 | 2:34 | -1 | 1 | 0:46 |
| K-11 | 4 | 8 | 2:42 | 10 | 4 | 0:46 | 6 | -4 | 1:56 |

*Table 5.6 The treatment group game performance*

Although participant A-01 did not complete the level, the time taken to run out of gooballs does suggest that some learning transpired. The individual performance data will be discussed further in Section 5.2.2. Figure 5.3 presents the percentage of endogenous blinks for both attempts at the game.

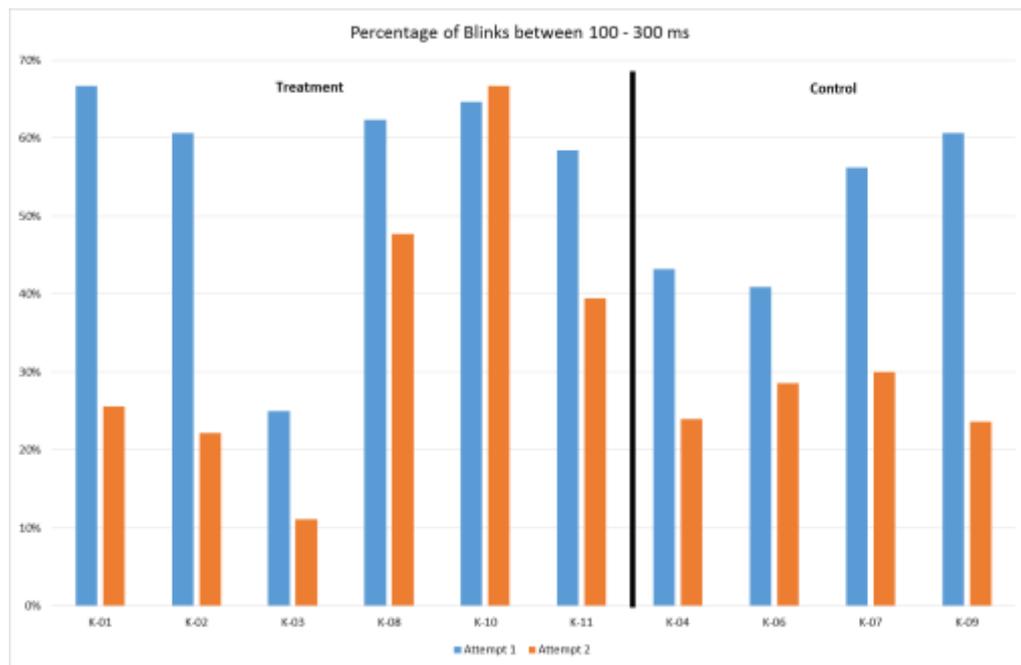

*Figure 5.3 Percentage of endogenous eye blinks*

From this data, it is possible to identify a reduction in the percentage the frequency of endogenous blinks in the second attempt at the game for participants K-01, K-02, K-03, K-04, K-06, K-07, K-08, K-09, and K-11. This suggests that there was less information processing in the second attempt at the game, which supports the findings from the data collected from the adults and the studies reviewed in the literature (Orchard & Stern, 1991; Stern, Walrath, & Goldstein, 1984; Ponder & Kennedy, 1927). Furthermore, participant K-10 performed more endogenous blinks in the second attempt at the game than in the first attempt.

The effect size for this analysis ($d = 1.20$) was found to exceed Cohen's (1988) convention for a large effect ($d = .80$). This indicates that the first attempt at the game had an effect on the second attempt at the game.

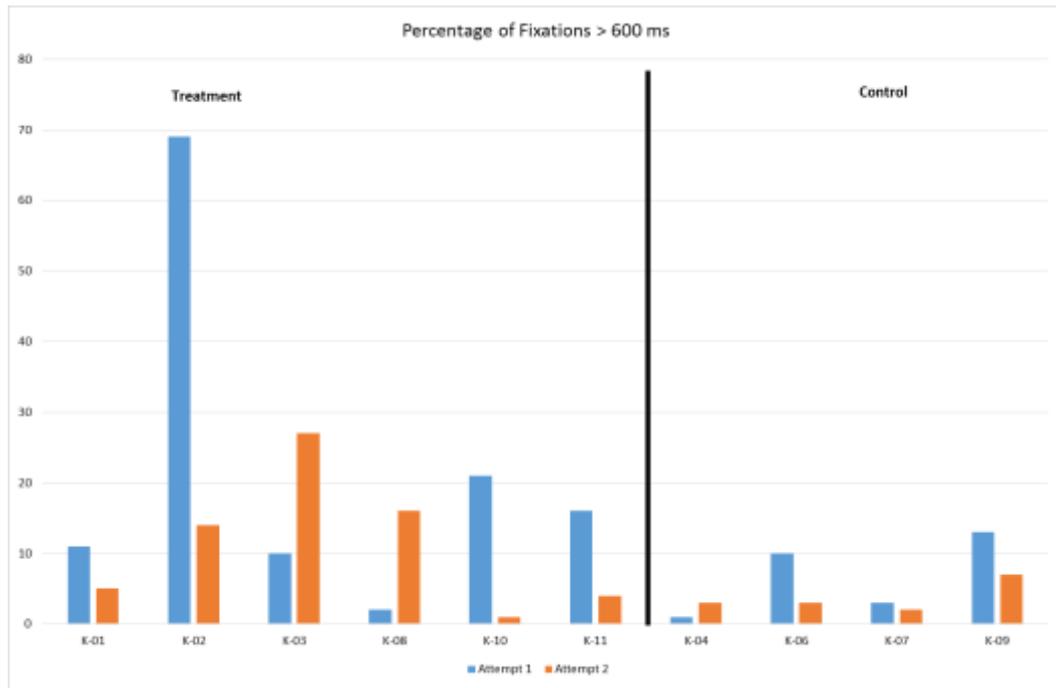

*Figure 5.4 The percentage of fixations above 600 ms*

Figure 5.4 presents the percentage of fixations that were above 600 ms. From this data, it is possible to identify that participants K-01, K-02, K-06, K-0, K-09, K-10, and K-11 all demonstrated a reduction in fixations that were above 600 ms in the second attempt at the game. However, participants K-03, K-04, and K-08 demonstrated increases in the percentage of fixations that were above 600 ms.

The effect size for this analysis ($d = 1.02$) was found to exceed Cohen's (1988) convention for a large effect ($d = .80$). This indicates that the first attempt at the game had a sizable effect on the second attempt at the game.

### 5.2.2    The individual performance of the children

As noted in Section 5.2.1, the in-game experience is dependent on the player's prior exposure to the game and/or games of a similar genre. It is worth considering the impact of the participants who did perform less endogenous blinks or reduce the number of fixations greater than 600 ms in the second attempt at the game. As identified in Section 5.2, there was one child (K-10) that performed more endogenous blinks and three children (K-03, K-04, and K-08) that exhibited more fixations that were longer than 600 ms in the second attempt at the game. It is, therefore, worth considering some of the underlying reasons for this variance. The prior gameplay experience and the in-game experience of each of these participants will be discussed to help understand the underlying causes. The performance of participant K-10 is not included as the variation in the percentage of endogenous blinks between the two exposures was minimal (3.12%). Participant K-01 did not complete the level and the reasons for this are worth considering.

### 5.2.2.1 Participant K-01

Participant K-01 was eight years old and reported that they had never played video games before. The participant was in the treatment group. In the first attempt, participant K-01 performed 4 endogenous blinks, which was below the mean of the group (62). In the second attempt participant K-01 blinked 3 endogenous blinks, which is also below the mean of the group (55). This reduction represents a 25% decrease in the number of endogenous blinks. Although this is what was expected, the percentage of decrease of endogenous blinks was considerably above the mean of the group. Participant K-01 did not complete the level on both attempts but reduced the time it took to build a tower in the video game by two minutes and thirty-one seconds.

From the video evidence, it is possible that the participant may not have learnt to play the game. In the first attempt at the game (Figure 4.25), the participant attempts to move the gooball directly in the pipe (Frame 65). After constructing a basic structure (Frame 117), the participant attempts to place another gooball directly in the pipe (Frames 118 to 120). The participant then attempts to place the gooball on the tree (Frames 165 to 166), which possibly looked like a pipe to the participant. The participant eventually uses all the available gooballs and is not able to complete the level. In the second attempt at the game (Figure 4.26), the participant attempts to place the gooball directly in the pipe (Frames 5 to 9) and appears not to have learnt that this was not an effective strategy in the first attempt at the game. Frames 23 to 26 indicate that the participant was struggling with the process of placing the gooball. This difficulty is further exhibited in Frames 27 to 32. In Frames 93 to 106, the participant attempts once again to place the gooball directly on the pipe. Eventually, the participant uses all the available gooballs and is not able to finish the level.

These results suggest that as the participant had never played a video game before, they may have been totally overloaded or overwhelmed by the information presented in the first attempt at the game. However, in the second attempt, they were able to make sense of this information and process it. This information combined with the reduction of fixations above 600 ms in the second attempt, suggests that there may have been less cognitive problem solving but a lot more information processing in the second attempt.

### 5.2.2.2 Participant K-03

Participant K-03 was eight years old and reported that they played video games more than six times a week. The participant was in the treatment group. In the first attempt, participant K-03 performed 6 endogenous blinks, which was below the mean of the group (62). In the second attempt the participant K-03 blinked 6 endogenous blinks, which also below the mean of the group (55), which represents a 0% increase in the number of endogenous blinks in the second attempt. The participant produced 10 fixations that lasted more than 600 ms in the first attempt at the game, which was slightly below the mean (15). However, in the second attempt at the game, the participant produced 27 fixations that were above 600 ms ($M = 8$), which represents a 170% increase. Participant K-03 collected no additional gooballs in the second attempt at the game and achieved this result by making one less move. The second attempt at the game took 2 seconds less than the first attempt at the game to complete.

From the video evidence, it is possible to identify that in the first attempt at the game (Figure 4.27), Participant K-03 attempted to place the gooball directly in the pipe (Frame 15). Frames 19 to 42 show the participant eventually makes the tower. However, in Frame 42, it is possible to identify that participant attempted to place the gooball directly in the pipe one more time. The participant eventually builds the tower high enough to make it to the pipe (Frames 45 to 51). However, in the second attempt at the game (Figure 4.28), the participant again appears to struggle with the tower construction process (Frame 8 to 30), but eventually makes the tower tall enough to get to the pipe (Frame 48 to 50). In the second attempt, Participant K-03 appears to have learnt not to place the gooball directly in the pipe. From the evidence collected it is difficult to explain the 170% increase in fixations that lasted over 600 ms and why there was no change in the number of endogenous eye blinks in the second attempt at the game. However, a possible explanation is that more cognitive problem solving did take place in the second attempt, and this is supported by the evidence that the participant did reduce the number of moves and the time to complete the level in the second game-play experience.

### 5.2.2.2 Participant K-04

Participant K-04 was eight years old and reported that they never had played video games before. This participant was in the control group. In the first attempt, K-04 performed 16 endogenous blinks, which was below the mean of the group (62). In the second attempt the participant K-04 blinked 14 endogenous blinks, which was also below the mean of the group (55), which represents a 12.5% decrease in the number of endogenous blinks in the second attempt. The participant produced 1 fixation that lasted more than 600 ms in the first attempt at the game, which was below the mean (16). However, in the second attempt at the game, the participant produced 3 fixations that were above 600 ms ($M = 8$), which represents a 200% increase.

In the first attempt at the game, the participant read the instruction book and needed to build three cars. The first two cars did not work (lack of wheels). The third car was successful and the participant achieved all the objectives of the level. In the second attempt at the game, the participant did not read the instruction book in the second attempt at the game. Furthermore, in the second attempt at the game, the participant only needed to build one car. The car was built with speed and precision. The level was completed with one attempt and the level objectives were met in the first attempt. This suggests that the participant had learnt the objectives of the level. Improvements in cognitive problem solving was evident through the number of cars built and the time it took to complete the level. Furthermore, although there were only two attempts at the level, the four methods used to construct the vehicle indicate improvements in procedural learning. Therefore, the 200% increase in the number of eye fixations that were longer than 600 ms is difficult to explain. However, when the normalised data is analysed, the percentage of fixations that lasted more than 600 ms, decreased from 1.23% to 0.8%. The normalised eye fixation data supports the performance data. Furthermore, the participant produced less endogenous eye blinks in the second attempt. This combined evidence does indicate that the participant did learn to play the game. However, as the game had no embedded transferable content to building magnetic towers, as expected the participant showed no improvements in building the magnetic tower (flat sticks flat on the table in both attempts) and the structured questionnaire (a correct answer first and then an incorrect answer).

### 5.2.2.3 Participant K-08

Participant K-08 was six years old and reported that they had never played video games before. The participant was in the treatment group. In the first attempt, participant K-08 performed 283 endogenous blinks, which well above the mean of the group (62). In the second attempt the participant K-08 blinked 253 endogenous blinks, which also well above the mean of the group (55), which represents a 10% decrease in the number of endogenous blinks in the second attempt. Although this reduction in endogenous blinks was expected, the percentage of frequency of endogenous blinks was considerably above the mean of the group. This participant produced 2 fixations that lasted more than 600 ms in the first attempt at the game, which was below the mean of the group (15.60). However, in the second attempt at the game, participant K-8 produced 16 fixations that lasted more than 600 ms, which was considerably higher than the mean of the group (8.20), and this represented a 700% increase when compared to the first attempt.

Participant K-08 collected one less gooball in the second attempt at the game and took 1 additional move to achieve this outcome. However, the participant reduced the amount of time to complete the level by 46 seconds.

From the video evidence (Figure 4.31), it is possible to identify that Participant K-08 attempts to place the gooball directly in the pipe (Frames 43 to 44). In frames 51 and 52, the participant appears to place the gooball directly on the tree (which possibly looked like a pipe to the participant). In frames 51 to 68, it is evident that the participant is struggling with the tower construction process. The participant then attempts to place the goo ball directly in the pipe (Frame 73) and eventually builds a tower tall enough (Frames 74 to 88). In the second attempt at the game (Figure 4.32), the participant attempted to move the gooball directly to the pipe (Frames 43 to 50). In frames 51 to 56, this is repeated. In frame 77, it appears that the participant attempts to place the gooball directly in the pipe, one more time. In frames 78 to 84, it is possible to identify the participant K-08 successfully completed the level.

From this evidence, it is possible to conclude that the participant may not have completely learnt how to play this video game. Moreover, the increase in fixations that lasted longer than 600 ms in the second attempt at the game may suggest that the participant spent a lot more time problem solving in this attempt.

### 5.2.3 Conclusions on the performance of the children

From the combined data collected, it is possible to accept that the first hypothesis is supported; that playing video games may positively affect user learning. The transfer of learning of the implicit content from the treatment to the construction of the magnetic towers was very apparent. There was some support for the second hypothesis that playing video games positively affects problem-solving ability. However, more empirical data is needed before this can be fully tested.

### 5.3 ANSWERS TO THE RESEARCH QUESTIONS

In this section, the research questions raised in Chapter 3 will be answered. Because this study involved two very divergent groups of participants, the answers to the research questions will be answered for each group separately.

### 5.3.1 Research Question 1 - *What learning takes place when playing the video game World of Goo?*

For both groups, there was evidence of skill learning through both attempts at the game in both the treatment and control group. This was evident through improvements in conceptual learning and conceptual problem solving. Furthermore, due to the repeated task procedural learning is evident in the improvements in performance between the two exposures to the game.

#### 5.3.1.1 The study of adults

As identified in Chapter Two according to Klabbers (2009), the types of knowledge a player gains and has to acquire for a successful game is explicit and tacit. Although this was not the focus of the study, the existing tacit knowledge was identifiable through the interaction with the technology. Although some participants had not played the games before, these participants spent minimal time learning to play the game. Although an instruction book was available in the control group, only one adult participant took the time to read it. This observation suggests that the prior experience helped these participants in completing an unknown task.

The influence of prior experience on performance was also evident. The participant in the treatment group that had played World of Goo previously (A-03), demonstrated the benefits of this experience through both the in-game performance and the percentage of fixations that were above 600 ms. Furthermore, three of the four participants that reported playing video games more than six times per week, produced less fixations that were above 600 ms than the other participants.

The explicit knowledge was easier to identify and measure, and this was identified through observing that the majority of the participants learnt to play the game (that is obtaining an understanding of the core game mechanics and being able to improve on the key performance metrics). From the evidence of the transfer of knowledge from the video game to the magnetic toy, it was not surprising to see limited improvements in the adults, and it was not possible to find any identifiable transfer from the treatment. The normalised blink and fixation data does suggest that there was a reduction in information processing problem solving and, therefore, when combined with the in-game performance metrics, it is possible to conclude that there were improvements in conceptual learning.

### 5.3.1.2 The study of children

The existing tacit knowledge was identifiable in the children. This was evident in the use of the devices provided. The children did not need any instruction on how to use a PC. Furthermore, they did not need any instruction on how to use the magnetic toy (although it was observed that some of the children struggled initially working out the positive and negative polarity of the magnets).

The influence of prior game play experience was evident for the majority (89%) of participants. As identified in Table 4.2, the participants that played video games more than five to six times a week produced eye fixations that were above 600 ms that were in the lower percentile of the group. Furthermore, the participants that had not played video games or did not play very often generally produced more fixations that were above 600 ms.

The explicit knowledge was easier to identify and measure. The transfer of knowledge from the treatment to the magnetic toy was visible in the study of the children. This evidence alone suggests that some learning did transpire. Furthermore, the normalised data on endogenous blinks and the fixations that were above 600 ms, does suggest that there was a reduction in cognition and problem solving. Therefore, when combined with the in-game performance metrics, it is possible to conclude that learning did transpire.

### 5.3.3 Research Question 2 - *Does problem solving ability improve through playing video games??*

This research explored the potential to extend the theory that endogenous blinks are an indicator of information processing (Fogerty & Stern, 1989; Orchard & Stern, 1991; Volkmann, 1986). Furthermore, it explored the potential to further the research that the number of endogenous blinks and the frequency of long eye fixations is an indicator of problem solving behaviour. While it is evident from the research that changes in both endogenous eye blinks and fixations above 600 ms did vary between the two exposures to the game, more empirical data is needed before this question can be fully answered.

### 5.3.3 Research Question 3 - *Do the participants that played the video game World of Goo learn tower construction from playing the game?*

The interest in the answer to this question is based on the literature that suggests that the skills learnt while playing a commercial (non-educational) video game are transferable to an external context (for example, Gee, 2003). However, there has not been much research to test the transfer of the knowledge acquired in commercial video games (apart from action video games) to an external context. The anecdotal evidence observed by this researcher suggests that this may be the case; however, this area needs further investigation. The major challenge in testing this assertion has been finding appropriate methods for testing the skills acquired in the game to an external context. Some authors (for example, Nardi 2010; Steinkuehler, 2008) have suggested that MMORPGs have the potential to teach leadership, an understanding of economics, and teamwork. However, concepts like leadership and teamwork are very challenging concepts to define and measure.

#### 5.3.3.1 The study of adults

The construction of the magnetic towers did not vary considerably between the participants of the treatment group and the control group. Furthermore, there were no notable differences to the answers to the questions about tower construction in the structured questionnaire. While the answer to the research question is 'No,' there are possibly mitigating factors that may have influenced this result. For example, it is possible that the pre-existing knowledge of basic construction principles had an impact on the result.

### 5.3.3.2  The study of children

The construction of the magnetic towers varied considerably between the control group and the treatment group. This result supports the research question. In the study of the children there was evidence that those participants who played the video game World of Goo (2D Boy, 2008) improved in magnetic tower building ability when compared to those participants who played the video game Bad Piggies (Rovio Entertainment, 2013). To ensure this result was not based on prior knowledge, a baseline measure of the understanding of tower construction methods was also taken. The result of this test demonstrated that an understanding of basic tower construction methods was not high and in the post-test assessment, the participants in the treatment group improved in their understanding of these principles. This result confirms that the intervention had a positive affect on the treatment group and minimal affect on the control group.

### 5.4    THE IMPLICATIONS FOR VIDEO GAME DESIGNERS

In this section, the observation noted in Chapter 3 that this study could benefit video game developers will be discussed. As this study researched the use of two commercial video games, the initial discussion will focus on identified usability issues in each of these games. From the literature reviewed, some recommendations will be made for both developers. However, while the focus of this study was on commercial video games, this research identified usability issues that also could be applied to the broader commercial video game development industry and the educational video game industry.

### 5.4.1   Recommendations for 2D Boy

The results of the eye tracking data and video evidence provide the basis for a number of recommendations for the developers of World of Goo, 2D Boy (2008). The research involved both groups of participants playing the first level of World of Goo. Because this game does not include a tutorial or help file, this first level appears to be designed to provide the functions of a tutorial through the following features. The image on the sign (left by the sign painter) provides very specific guidelines for playing the game. First, the text reads "drag n' drop to BUILD to the PIPE". The emphasis on the words build and pipe appears to be deliberate. Further, the sign painter also included an image with a hand that appears to move a gooball away from the starting frame. Fourth, the sign also includes an arrow that points toward the pipe (Figure 5.5). The first level also includes a starting frame that provides a vital clue as to the ideal structural design.

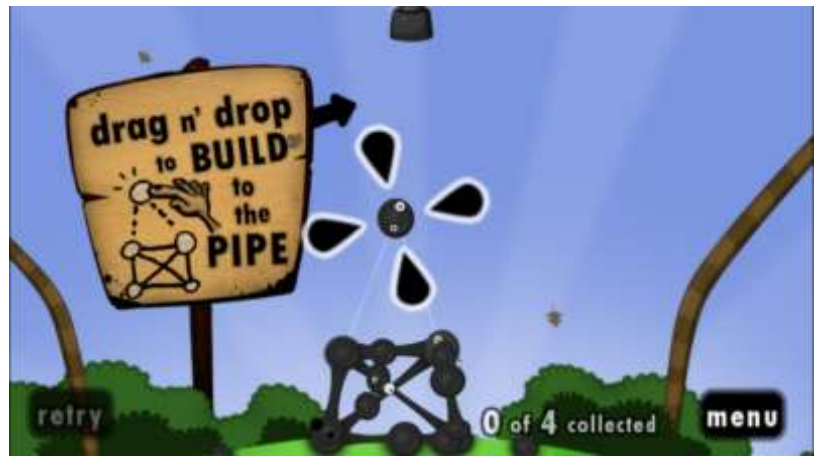

*Figure. 5.5. World of Goo – Level 1. (2D Boy, 2008)*

However, based on the observations and eye tracking data, very few participants actually looked at this sign. Furthermore, those that did read the sign, several participants clearly did not perceive that the instruction was related to the task. As detailed in the literature review (section 2.8.1), Mayer (2009) states that the multimedia principle indicates that students learn better from words and pictures than from words alone. While the developers of World of Goo (2D Boy, 2008) provide both an image and words, it is evident that this is was not clearly read or observed. This is possibly because the participant was preoccupied with the other content on the screen. Mayer (2009) states that the coherence principle suggests that students learn better, when extraneous words, pictures, and sounds are excluded rather than included. Therefore, the recommendation is that the sign needs to be presented to the participant while no other content is visible, or semi-visible. Furthermore, if the developers of World of Goo included a short video sequence at the beginning of the level that demonstrated the correct process, this would potentially provide the necessary scaffolding for successful completion of the level.

The game play video demonstrated that several participants did not realize that they needed to build the tower to get to the extraction pipe (instead, they moved the gooball there directly). From the evidence of the eye tracking data and the video of the game play, very few participants saw the semi-transparent triangle that indicated the direction of the extraction pipe. The second recommendation is 2D Boy need to make the semi-transparent triangle more explicit (less transparent) and possible more explicit that this is the direction of play (for example an arrow would provide a more explicit indication of the direction of play).

As detailed in section 4.1, several participants moved the gooball to the tree, which suggests that they mistook this for the pipe. It is recommended that the tree is either removed or made more like a tree (for example, made thicker, more irregularity in the trunk design, and include some foliage).

Finally, the one participant that did not complete the level probably did not realise this. Therefore, when the player does not complete the level, a visual and/or auditory cue that indicates that this was the case, should be provided (for example, try again, or better luck next time, and/or audio of the crowd expressing their disappointment).

### 5.4.2   Recommendations for Rovio Entertainment

The results of the eye tracking data and video evidence provide the basis for a number of recommendations for the developers of Rovio Entertainment (2013). The research involved both groups of participants playing a level of Bad Piggies. Although this game included a help facility, only one participant actually looked at this. From the analysis of the eye tracking data, none of the participants noticed the animated pointy hand pointing to the (tutorial) instruction book. Moreover, none of the participants noticed this animation pointing at the starting grid. According to the temporal contiguity principle (Mayer, 2009), students learn better when text and images are presented concurrently. Moreover, according to the coherence principle (Mayer, 2009), extraneous words, pictures and sounds should be excluded. Therefore, the first recommendation is that the instruction book icon should be more explicit in what its function is. The word Help should be included on the icon or the icon should be changed to a question mark with the word help on it. The pointy hand icon animation should be replaced with a bigger icon. It is also recommended that this be supplemented with video and audio indicating what each function does. Although many users will discover these elements through the assistance of a peer or through the Internet, by providing these basic cues, developers could reduce the user's initial frustration.

One participant was able to complete the level although the car broke and the pig rolled to the egg nest. Although this an acceptable feature of the game, it is recommended that in the beginning levels of the game, each element required to get the three stars is introduced individually. Therefore, the first game should not include the krack-snap objective. This would provide scaffolding for each core objective.

All of the participants involved in the pilot study did not realise what krack-snap actually meant. Moreover, none of the participants could read the text on this icon. The first recommendation is that an image and text should be included. The second recommendation is that the text on this icon needs to be made clearer.

### 5.4.3   Recommendations for other commercial and educational video game developers

The recommendations for other commercial and educational video game developers are based on Mayer (2009). To effectively play the game the users must learn how to play it. Therefore, the first recommendation is that both commercial and educational video game developers read and implement the principles that Mayer (2009) provides. The second recommendation based on this research is that these developers get a diverse group of participants to fully test their video games. Eye tracking technology is now a lot more affordable. Furthermore, there are more companies that specialise in video game usability studies that have the expertise and experience in performing this type of research that many video game developers lack.

### 5.4   CONCLUSION

This chapter discussed the findings that were the result of the data collected and has made a contribution to the research on using video games for learning. Other researchers have speculated about the link between commercial video games and learning, hence the research combined several tools to test and evaluate the link. A triangulation of methods was utilised to uncover the potential of video games for learning.

From the study of adults, the prior learning experience of the participants impacted any demonstrable evidence of improvements in skill learning and procedural learning through this experience. Furthermore, it was possible to identify indicators of cognition and problem solving behaviour.

Based on the game statistics and video evidence, it was possible to conclude that the hypothesis is supported; playing video games positively affects user learning. The adult participants that had no prior experience in playing the game demonstrated improvements in in-game achievements and reduced the time it took to complete the level.

In the experiment of the children, the cumulative percentages of fixations that were above 600 ms show that the long fixations in the first attempt have an effect on the long fixations in second attempt. The results of the out-of-game experiment were also positive. The majority of participants demonstrated an improvement in the understanding of basic construction principles, without intervention. Furthermore, the in-game performance of the both the treatment group and control group improved after the second attempt at the game.

One of the primary motivations for recruiting the young children is that the acquisition of new knowledge would be more apparent. From this study, this observation was correct. The change of the process of building the tower demonstrate that the treatment group learnt transferable skills from the game. As noted in section 4.1.2, the majority of the children placed the magnetic sticks flat on the ground during the pre-exposure test (Figure 5.6). However, in the post-exposure test, 83% of the children in the treatment group made or attempted to make a three-dimensional tower (Figure 5.7). The children did not receive any additional intervention or outside assistance in building the magnetic tower; they were provided the same equipment and asked to achieve the same output. The only delay between the two construction tests was the time it took to play the game.

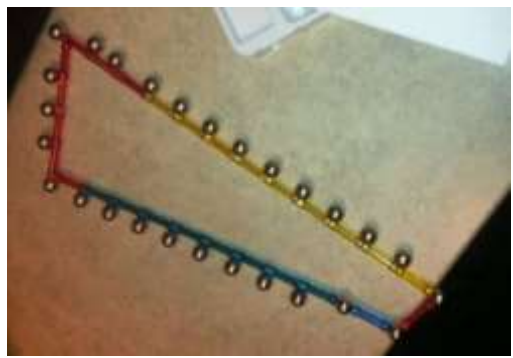

*Figure 5.6 Participant K-10 Pre-exposure tower*

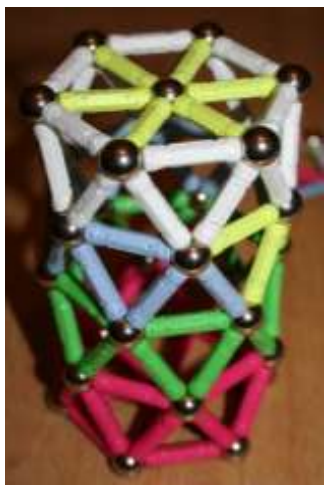

*Figure 5.7 Post exposure tower built in the pilot study*

It is apparent that the frequency of endogenous blinks and fixations that lasted longer than 600 ms reduced for the majority of participants. Those participants that did not reduce the frequency of endogenous blinks or fixations that lasted more than 600 ms, possibly needed another attempt at the game or facilitated instruction. As noted in the literature review, when humans learn a new task or skill, it is not unusual to need more than one or two attempts to be successful to retain the knowledge needed should this task or skill be required again in the future. Fundamental human motor skills can take several years to learn or master. For example, language learning usually requires more than two exposures. Therefore, it is not surprising that some of the participants in this study did not totally acquire the implicit concepts in the game after two exposures. As identified in the literature review, the process of education and learning usually requires more than one exposure (Dosher, 1984; Senechal, 1997; Stadler, & Frensch, 1998). Gibson and Gibson (1955), suggest that in humans, perceptual learning takes place when a person is repeatedly exposed to specific stimuli. Through repeated exposure to the same problem, humans can acquire knowledge. It is through repeated experience that advanced skills are obtained and maintained (Norman, 1993). Perceptual learning creates potentially long lasting changes to the human perceptual system that can improve the ability to respond to the environment.

The major finding of the research has shown a positive relationship between playing commercial video games and learning. Furthermore, it has shown a potential relationship between playing video games and cognitive problem solving, although, this was more evident in the participants that had minimal exposure to these learning concepts than those that had considerable exposure. However, there was evidence that those participants that had more exposure to playing video games, generally performed better than those participants that did not.

This research adds to the body of knowledge on the value of commercial video games for learning. It provides both qualitative and quantitative data to support this theory. Furthermore, the research provided evidence that some skills learnt in a commercial video game are transferable to an external context. The implications of these findings will be summarised further in Chapter Six.

CONCLUSIONS

## 6.0 INTRODUCTION

In Chapter One, the primary reasons and motivations for this research were provided. Furthermore, Chapter One also provided a summary of some of the associated publications that the author published that contributed to the development of this thesis. Chapter Two provided a review of that literature. Chapter Three included a review of research methods implemented in previous studies and proposed the chosen research methodology that would be used. Chapter Four provided the results of the data that was collected and Chapter Five provided a discussion of the research findings. This chapter summarised the findings in Section 6.1. In Section 6.2, the contribution to the field of knowledge is discussed. In Section 6.3, the limitations of this study will be reviewed. Section 6.4 will explore the areas for further research and Section 6.5 will conclude this chapter.

## 6.1 SUMMARY OF THE FINDINGS

The primary research question concerned the effectiveness of playing commercial video games to learn. Furthermore, this research also questioned the efficacy of using commercial video games as a tool to improve cognitive problem-solving. The research investigated the use of video games with two very different groups of participants. The first group were students enrolled at a tertiary education institution in the United States of America. These students had already experienced 12 to 14 years of formal education. As a result, the expected learning acquired by playing the video game would primarily be evident in learning how to play that game. This research found that the commercial video games studied did result in improvements in the conceptual skills needed to complete the task. The methods employed in this study found identifiable changes in indicators of information processing and cognitive problem solving. The second group were young children who were either just about to start their formal education or had experienced one or two years. The children in the treatment group demonstrated clear advances in the learning of how to build a magnetic tower. This was evident in the game and through the methods used to understand if any learning could be transferred to an out-of-game context.

The methods used found that the indicators of information processing and cognitive problem-solving decreased in the second attempt at the game. This evidence suggests that the participants did learn to play the game. Furthermore, the results of the answers to the pre- and post- questionnaires further suggest that the participants in the treatment group obtained an understanding of basic tower construction. The results of the construction methods used in building the magnetic towers further suggests that the learning that was obtained by playing the video game was identifiable and transferable to an out-of-game environment. The experimental design ensured that there was no additional intervention. This experiment involved the participants answering certain questions prior to playing the video game, then they were asked to build the magnetic tower and were subsequently asked to play the video game. Immediately after the first attempt at the game, the participants were given the magnetic toy to play with a second time. Upon completing this task, the participants were then asked the post-exposure questions. This entire process took on average 45 minutes to complete. Therefore, it is possible to conclude that the treatment had a direct and positive effect on the out-of-game tests.

### 6.1.2   Learning through mistakes

As identified in Chapter Two, one of the cited benefits of playing video games is learning through making mistakes. Perkinson (1979) suggests that learning from mistakes is potentially an evolutionary requirement. This Darwinian view of the necessity of learning for the survival of the species may be valid for learning basic survival skills, but it may not be necessarily valid for learning mathematics, science, or the arts.

This research validated what Piaget (1952) observed in his own child's behaviour. Piaget noted that the child reached for the watch; the child learnt from their own initial mistaken belief that the watch was no longer present after it had been hidden under the blanket. Similarly, in this research, when the children played the video game, the participants in the treatment group soon learnt that the tree to the right of the screen was not the pipe they were looking for. For some children, this required more than one attempt to learn this. However, it was eventually learnt.

This researcher has observed that in many formal education assessment tasks (tests, or exams), there are not many opportunities to learn from mistakes. If a student fails an end of year examination, there are usually not many second chances. However, in the experience and observations of the researcher, many of life's valuable lessons are the ones that were learnt by making mistakes. The researcher observed the results of his own son playing with a hot glue gun at kindergarten. After the first exposure to playing with a hot glue gun, the child came home with a burnt finger. After the second exposure, the child also came home with a burnt finger, but from that day on, the child never burnt his finger again. Video games provide a unique opportunity for children to learn from their own mistakes, without serious consequences (Gee, 2003; Grammenos, 2008; Juul, 2009; Prensky, 2003). The anecdote that the researcher provided in Chapter One about learning to race a production race car though playing a video game provides a useful reference. In the video game, it was safe to crash the car. It was safe to learn the importance of braking into a corner and not through a corner. Although this could have been learnt through reading a book, the video game provided immediate tangible feedback on the practical application of the theory. Although the video game may not have provided a complete learning experience (due to the lack of gravity, or g-forces), it did help the researcher learn some of the fundamental aspects of racing a car.

In the video games studied in this research, the beginning levels were chosen as these levels provided the scaffolding that would be valuable later in the game. The benefits of this scaffolding were apparent in the second attempt at the game, as the majority of participants demonstrated acquired learning from the first attempt. Moreover, the anecdotal evidence suggests that those participants that played the more advanced levels of the game directly benefited from the scaffolding that was provided in the beginning level of the game.

### 6.1.3   Video games as tools for learning

The experience of participant K-01, does suggest that not all users can or will learn from playing a commercial video game. From the evidence presented and the observations of this participant, it is highly likely that this participant did not realise that they had not successfully completed the level. If a peer, sibling, parent or teacher was present, this could have been explained to them. Furthermore, the participant would have benefited from informed advice in the first attempt at the game. This would have reduced the chances of the participant making the same mistakes in the second attempt at the game.

Gee (2004) asserts that parents and/or siblings or peers are an important element of the learning process and that video games **are not the only way to learn** (see also Clark, 1994). The process of learning any skill or concept generally requires multiple reference points before new knowledge can be acquired (Gee, 1997). These reference points can be tools (such as books, video, video games, simulations, calculators, and so on), people (such as family, teachers, friends, social groups, and so on), or communities (such as schools, religious organisations, sports clubs, and so on). However, as video games have the capability to present learning constructs in a truly interactive and immersive environment that provides immediate feedback and scaffolding, video game playing has the potential to provide a compelling, cost effective and cognitively efficient learning environment (Clark, 1994). When supported with directed instruction or support (through teachers, parents, or peers), video games can provide an effective and cost efficient learning experience.

## 6.2 CONTRIBUTION TO THE FIELD OF KNOWLEDGE

The most important contribution to the field of knowledge from the research is the confirmation that a commercial video game can facilitate skill learning that is identifiable. When implemented on learners with no or minimal exposure to the learning concepts, the learning is measurable and transferable. This contribution supports the assertions of Gee (2003), Shaffer et al. (2005), and Steinkuehler (2008).

Another important contribution is that this study has validated a method of measurement for obtaining quantitative data on information processing and cognitive problem solving. The evidence collected suggests that endogenous blinks and fixations provide reliable indicators of information processing and/or problem solving. Prior studies on the endogenous eye blinks used eye tracking devices that were intrusive for the participant. These devices were either a head-mounted camera or a coil device that was placed on the eyelid. By using a high-frequency desktop eye gaze camera, this study was able to obtain detailed data on the number and frequency of eye blinks and fixations with minimal configuration, calibration, or inconvenience to the participant. This enabled the researcher to process a reasonable number of participants in a relatively short timeframe.

The study found that the video game World of Goo was an effective tool for introducing young children to the concepts of basic tower construction.

Another contribution to the field of knowledge is that this research validated an effective method for testing the usability of a video game. The methods that were used clearly identified which parts of the game users struggled to comprehend. It further identified if and when specific users read the instructions and if the instructions provided sufficient details for the user to understand (or learn) how to successfully play the game. The research method could be implemented by video game developers and industry associations to provide objective baseline benchmarks as to which age group is most suited to playing these video games.

The method also supported Klabber's (2009) theory of the taxonomy of game learning. The variables of capability (dependent) and time (independent) are added to the conceptualisation of learning from playing a video game to capture the notion that the game players learn multi-faceted skills through interaction (experience in action). This is the notion that players have skill capability and can improve that capability (skill level) over time. This was evident in the reduction in time to complete the level, the reduction of endogenous blinks, and a reduction in fixations that lasted more than 600 ms.

The research also found that the proposed method for monitoring eye gaze and blinks on a tablet device might be better conducted with a head-mounted eye gaze camera or eye gaze glasses. Although, these tools were available to the researcher, the models that were available for the purpose of this research did not provide the required resolution of data needed for this study. However, these technologies are evolving at a rapid pace, and it might be possible to conduct this research on a tablet device with a pair of eye gaze glasses that will support the resolution needed in the not too distant future.

## 6.3 LIMITATIONS

The empirical study had inherent limitations from the individual methods chosen. The focus of the research was on two samples from a specific region in one particular country. However, without further research in other parts of the country or the broader populace, it is not possible to generalise these findings. Further, the research focused on two commercial video games, which were chosen for their similarities (they were both problem-solving games) and their unique differences (one game involved making a tower and the other game involved making a vehicle). However, although these video games provided the basis for a reasonably robust experimental design, it is difficult to generalise the findings to any other types of video games.

Although every effort was made to recruit as many participants as possible, the sample size in both studies further limits the generalisability of these results. As identified in the literature review, a considerable amount of the research that has been conducted using physiological measurements have typically involved sample sizes that are smaller than the sample size obtained in this study. Furthermore, the identified studies that used physiological measurements involving children had less than six participants.

Another limitation of the study is that it did not provide an opportunity for the participants to have a third attempt at the game. As noted in the literature review, many new concepts take more than one exposure before knowledge is obtained (Dosher, 1984; Senechal, 1997; Stadler, & Frensch, 1998). Furthermore, the research tested the immediate effects of the treatment. While this was valuable to identify the immediate effectiveness of the treatment, no efforts were made to test the temporal effects of the treatment. That is, the research did not seek to measure how long any of the new knowledge that was acquired would be sustained.

## 6.4 THE AREAS FOR FURTHER RESEARCH

The research has provided a foundation for additional studies that will address the noted limitations. Further research is planned that will take place in at least two other cities to improve the generalisability of the results. It is planned to conduct the study at a time of year that will increase the potential number of participants. The benefit of undertaking the research during the summer vacation is that there will potentially be more participants interested in participating. Furthermore, it was observed that as Boston has a high percentage of research institutions, another location with a lower density of research institutes should provide an improved participation rate.

While the research was able to provide data from two exposures to the video game, it would be beneficial to obtain data from additional exposures. Furthermore, it would be valuable to assess the effectiveness of a commercial video game like World of Goo (2DBoy, 2008) in a classroom environment and compare the effectiveness of this technology with traditional methods (such as books, video, or teachers). If this study could be conducted over a term, quarter, or semester, then this could measure the long-term effectiveness of using commercial video games for learning.

The experimental design used in the study of adults led to some data not being acquired or extraneous data being captured. To measure fixation and blinks while playing a tablet device, a desktop-based eye tracker may not currently be the best option. If the participant used a head-mounted or glasses-based eye tracker then the limitation may be addressed. However, as noted, the current eye tracking systems do not support resolutions that will provide the data needed for analysing endogenous eye blinks. Furthermore, the head-mounted devices are still very intrusive for the participant. When the technology in the eye tracking glasses advances to a stage where they can provide this information, then the study of the use of tablet devices could be investigated further.

The research investigated the benefits of using two specific video games. While this provided some very valuable results, before these findings can be generalised, it would be necessary to investigate the educational benefits of other commercial video games. Furthermore, the research specifically focused on commercial video games of a specific genre. It would also be necessary to investigate the effectiveness of using commercial video games from other genres as well.

## 6.5 CONCLUSIONS

In this chapter, the key findings of this research were identified. The contribution to the field of knowledge was discussed, and it identified the limitations of this study. Finally, the areas for further research were identified. The empirical research added to the body of knowledge of the value of using a commercial video game for learning. This research could be valuable to educators who are contemplating the benefits of using commercial video games for educational purposes.

**Pre-exposure Participants Questions**
Researcher to collect.

Participant number: _____________________

Start Time: _______________________

Begin the session by saying: Thank you, for agreeing to participate in this user study. My name is [name], and I'll be working with you today. In this study, we're exploring the learning process when playing video games. We'd like you to play this video game and we will watch while you build a really tall tower. We may ask you some questions while you work. I will try to be as unnoticed as I can sometimes, and at other times, I might ask you about what you are doing.

1. Personal Info
    a) Age
        1. 18-29 years old
        2. 30-49 years old
        3. 50-64 years old
        4. 65 years and over

    b) Gender
        1. Male
        2. Female

    c) What is the highest level of education you have completed?
        1. high school graduate
        2. college student
        3. college graduate
        4. some postgraduate study
        5. post graduate

    d) What is your major?

3. How frequently do you play computer games?
        1. Never
        2. Once or twice a week
        3. Three to four times a week
        4. Five times to six times a week
        5. More than six times a week

4. Do you play games on the following?
        1. Xbox, Wii, PlayStation,
        2. PSP, Nintendo DS, Game Boy, iPod, etc.
        3. iPhone, iPad, or Mobile Phone,
        4. Personal computer (PC)

5. Have you played this game (While showing them World of Goo)
   a) Yes
   b) No (Go to question 6)
   c) Not sure (Go to question 6)

5.1. (If yes) How many times have you played this game (While showing them World of Goo)
   1. Once
   2. Twice
   3. Three times
   4. Four times
   5. More than four times
   6. Not sure

6. What shape do you think is the strongest for building a really tall tower?
   a) Square
   b) Circle
   c) Triangle
   d) Not sure

**Post-Exposure Participants Questions**
Researcher to collect.

1. Did you have fun?
    a) Yes
    b) No
    c) Not sure

2. What do you think was the most fun part?

________________________________________________
________________________________________________
________________________________________________
________________________________________________

3. What do you think is important when building a tower?

________________________________________________
________________________________________________
________________________________________________
________________________________________________

4. Why did the tower break or fall down?

________________________________________________
________________________________________________
________________________________________________
________________________________________________

5. What shape do you think is the strongest for building a really tall tower?
    a) Square
    b) Circle
    c) Triangle
    d) Not sure

**Thank you for your time today. We really appreciate you helping out with this study.**

Measuring learning in commercial video games in young children: A proposed method

**Pre-exposure Participants Questions**
Researcher to collect.

Participant number: ____________________

Start Time: ______________________

1. Personal Info
   e) How old are you? __________

   f) Gender
      1. Male
      2. Female

   g) What grade are you in ______________?

3. How frequently do you play computer games?
   a) Never
   b) Once or twice a week
   c) Three to four times a week
   d) Five times to six times a week
   e) More than six times a week
   f) Not sure

4. Do you play games on the following?
   a) Xbox, Wii, PlayStation,
   b) PSP, Nintendo DS, Game Boy, iPod, etc.
   c) iPhone, iPad, or cell phone,
   d) Personal computer (PC)

5. What video games do you play?

_______________________________________________
_______________________________________________
_______________________________________________
_______________________________________________

5. Have you played this game? (While showing them World of Goo)
   d) Yes
   e) No (Go to question 6)
   f) Not sure (Go to question 6)

5.1. (If yes) How many times have you played this game? (While showing them World of Goo)
   a)  Once
   b)  Twice
   c)  Three times
   d)  Four times
   e)  More than four times
   f)  Not sure

6. What shape do you think is the strongest for building a really tall tower?
   e)  Square
   f)  Circle
   g)  Triangle
   h)  Not sure

1. Did you have fun?
    d) Yes
    e) No (go to question 2)
    f) Not sure (go to question 2)

2. What do you think was the most fun part?

    _______________________________________________
    _______________________________________________
    _______________________________________________
    _______________________________________________

3. What do you think is important when building a tower?

    _______________________________________________
    _______________________________________________
    _______________________________________________
    _______________________________________________

4. Why did the tower break or fall down?

    _______________________________________________
    _______________________________________________
    _______________________________________________
    _______________________________________________

5. What shape do you think is the strongest for building a really tall tower?
    e) Square
    f) Circle
    g) Triangle
    h) Not sure

**Thank you for your time today. We really appreciate you helping out with this study.**

# APPENDIX 2 – Recruitment Documents

## A2.1   Adult Recruitment email

Research Study Invitation

Dear «student»

We are inviting you to take part in a research study on video games. We are asking you to participate in this study because we are interested in finding out what people can learn through playing video games. You must be an NU undergraduate or graduate student and be at least 18 years of age to be in this research project.

The purpose of this research is to gain an understanding of the learning that takes place when people play a commercial video game. We will use a number of measurement techniques that are recognized to provide indicators of cognition.

The decision to participate in this research project is voluntary. Neither your decision whether to participate nor the content of your answers will affect in any way your relationship with anyone at Northeastern University or your status or standing at the university. The information gathered during this study is for research purposes only.

If you decide to take part in this study, you will be asked to play a non-violent video game. We think this will take you about 45 minutes. You will also be asked to play with a construction set. Your reactions will be recorded using a number of techniques. We will use a video recorder to record your on-screen actions and decisions. We will also use a special camera that will record where your eyes are focused. This camera uses infrared light, which is not visible to the user to track the movements of the cornea. We will ask you a few questions about the amount of time you spend playing video games and the types of video games you play. We will also ask you about the video game played during the study. The study will take place at Northeastern University's Playable Innovative Technologies (PLAIT) Lab.

You will be able to choose; a $20 gift card from Amazon or the iTunes *Store,* or a copy of the video game World of Goo as a token of appreciation for your participation. The information learned from this study may help us to make better informed decisions about the use of video games in education.

This study was approved by Northeastern University's Institutional Review Board (#13-09-10). If you are interested in participating or require any further information, please check out our website http://www.gurl.co.nz/college.html or contact me at allan.fowler@gurl.co.nz or Dr. Magy Seif EI-Nasr, College of Computer and Information Science, Northeastern University at m.seifel-nasr@neu.edu.

Kind regards

Allan Fowler

**A2.2    Recruitment Poster (for children)**

# APPENDIX 3 – Ethics Approval

## A3.1 Adult Study

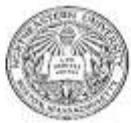

# Northeastern

## NOTIFICATION OF IRB ACTION

| | |
|---|---|
| Date: October 11, 2013 | IRB #: 13-09-10 |
| Principal Investigator(s): | Magy Seif El-Nasr |
| Department: | Game Design<br>College of Computer and Information Science<br>College of Arts, Media and Design |
| Address: | 100 Meserve Hall<br>Northeastern University |
| Title of Project: | Measuring Learning in Commercial Video Games: A Proposed method |
| Participating Sites: | N/A |
| DHHS Review Category: | Expedited #4, #6, #7 |
| Informed Consents: | One (1) signed consent form<br>One (1) debriefing statement |

Human Subject Research
Protection

960 Renaissance Park
360 Huntington Avenue
Boston, MA 02115

617.373.7570
fax 617.373.4595
northeastern.edu/hsrp

As per 45 CFR 46.116(d): (1) The research involves no more than minimal risk to the subjects; (2) The waiver or alteration of the consent process will not adversely affect the rights and welfare of the subjects; (3) The research could not practicably be carried out without the waiver or alteration; and (4) Whenever appropriate, the subjects will be provided with additional pertinent information after participation.

Monitoring Interval:     12 months

## APPROVAL EXPIRATION DATE: OCTOBER 10, 2014

### Investigator's Responsibilities:

1. The informed consent form bearing the IRB approval stamp must be used when recruiting participants into the study.
2. The investigator must notify IRB **immediately** of unexpected adverse reactions, or new information that may alter our perception of the benefit-risk ratio.
3. Study procedures and files are subject to audit any time.
4. Any modifications of the protocol or the informed consent as the study progresses must be reviewed and approved by this committee **prior to being instituted.**
5. Continuing Review Approval for the proposal should be requested at least one month prior to the expiration date above.
6. This approval applies to the protection of human subjects only. It does not apply to any other university approvals that may be necessary.

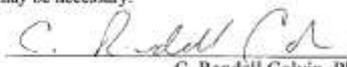

C. Randall Colvin, Ph.D., Chair
Northeastern University Institutional Review Board

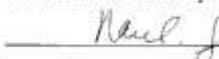

Nan C. Regina, Director
Human Subject Research Protection

Northeastern University FWA #4630

Research Study Invitation

Dear <<student>>

We are inviting you to take part in a research study on video games. We are asking you to participate in this study because we are interested in finding out what people can learn through playing video games. You must be an NU undergraduate or graduate student and be at least 18 years of age to be in this research project.

The purpose of this research is to gain an understanding of the learning that takes place when people play a commercial video game. We will use a number of measurement techniques that are recognized to provide indicators of cognition.

The decision to participate in this research project is voluntary. Neither your decision whether to participate nor the content of your answers will affect in any way your relationship with anyone at Northeastern University or your status or standing at the university. The information gathered during this study is for research purposes only.

If you decide to take part in this study, you will be asked to play a non-violent video game. We think this will take you about 45 minutes. You will also be asked to play with a construction set. Your reactions will be recorded using a number of techniques. We will use a video recorder to record your on-screen actions and decisions. We will also use a special camera that will record where your eyes are focused. This camera uses infrared light, which is not visible to the user to track the movements of the cornea. We will ask you a few questions about the amount of time you spend playing video games and the types of video games you play. We will also ask you about the video game played during the study. The study will take place at Northeastern University's Playable Innovative Technologies (PLAIT) Lab.

You will be able to choose; a $20 gift card from Amazon or the iTunes Store, or a copy of the video game World of Goo as a token of appreciation for your participation. The information learned from this study may help us to make better informed decisions about the use of video games in education.

This study was approved by Northeastern University's Institutional Review Board (#13-09-10).

If you are interested in participating or require any further information, please check out our website http://www.gurl.co.nz/college.html or contact me at allan.fowler@gurl.co.nz or Dr. Magy Seif El-Nasr, College of Computer and Information Science, Northeastern University at m.seifel-nasr@neu.edu.

Kind regards

Allan Fowler

Ver. 2013-10-10

**Game User Research Limited**

**College Students Study – Northeastern University, Boston (NU IRB# 13-09-10)**

**FAQ**

We have provided a list of some frequently asked questions below. If you have any additional questions or would like to send us comments please do not hesitate to contact us.

### What does participation in our studies involve?

Our study involves adults playing a non-violent video game that will tell us a bit about what people can learn from this experience. Participants will play a video game and play with a game. We will use a video recorder to record your on-screen actions and decisions. We will also use a special camera that will record where a participant's eyes are focused. This camera uses infrared light to track the movements of the cornea. Infrared light is invisible to people and is not harmful to the eyes. It is our intention you will have a great time!

### How long does a visit last?

Visits generally take 30 to 45 minutes. This gives us plenty of time to complete the activity and answer any of your questions.

### Where will the study take place?

The study will take place at Northeastern University's Playable Innovative Technologies (PLAIT) Lab.

### Will you need to come back in again?

Our studies involve a one-time visit only. If you are interested, you can always ask to be contacted when another study comes along.

### How old do you need to be?

We are currently interested in adults aged eighteen and over. Individuals with amblyopia (lazy eye) are not eligible for the study as the data tracking of one cornea may not be accurate. If you have questions or concerns please contact the researchers.

### Will you be paid?

You will be able to choose; a $20 gift card from Amazon or the iTunes Store, or a copy of the video game World of Goo as a token of appreciation for your participation. The information learned from this study may help us to make better informed decisions about the use of video games in education.

### Can you bring a friend along to appointments?

Sorry, we need you focus on the tasks that we give you and would like to reduce as many distractions as possible. Therefore, we would appreciate it if you were the only person that came along. However, your friends can sign up and join the fun.

### When can you come in?

We'll find a time that works best for you! We believe that attending classes is really important and therefore would prefer that the study is done outside of class hours.

### Interested?

If so, please give us a call or drop us an email, we will provide a detailed guide about the study.

### Contact Information

Allan Fowler, Phone: (646) 3091702, Email: allan.fowler@gurl.co.nz
Dr. Magy Seif El-Nasr: Email: m.seifel-nasr@neu.edu

APPROVED

Northeastern University. *450 Rennisance Pars Northeastern University Boston, MA 02115-5005 Tel 617.373.2872 Fax 617.373.4598*

Northeastern University *Human Subject Research Protection*

ver. 2013-10-10

**Northeastern University,** Department of Computer and Information Science
**Name of Investigator(s):** Dr. Magy Seif El-Nasr & Allan Fowler
**Title of Project:** Measuring learning in commercial video games

### Informed Consent to Participate in a Research Study

We are inviting you to take part in a research study. This form will tell you about the study, but the researcher will explain it to you first. You may ask this person any questions that you have. When you are ready to make a decision, you may tell the researcher if you want to participate or not. You do not have to participate if you do not want to. If you decide to participate, the researcher will ask you to sign this statement and will give you a copy to keep.

### Why am I being asked to take part in this research study?

We are asking you to participate in this study because we are interested in finding out what people can learn through playing video games.

### Why is this research study being done?

The purpose of this research is to gain an understanding of the learning that takes place when people play a commercial video game. We will use a number of measurement techniques that are recognized to provide indicators of cognition (the act or process of knowing).

### What will I be asked to do?

If you decide to take part in this study, you will be asked to play a non-violent video game. We think this will take you about 45 minutes. You will also be asked to play with a construction set. Your reactions will be recorded using a number of techniques. We will use a video recorder to record your on-screen actions and decisions. We will also use a special camera that will record where your eyes are focused. This camera uses infrared light to track the movements of the cornea. Infrared light is invisible to people and is not harmful to the eyes. We will ask you a few questions about the amount of time you spend playing video games and the types of video games you play. We will also ask you about the video game played during the study.

We believe that there is a relationship between playing video games and an increase in problem solving ability and this study is trying to investigate this link.

The investigators may stop the study at any time they judge it is in your best interest. They may also remove you from the study for various other reasons. They can do this without your consent. You can stop participating at any time. We have taken a random sample of undergraduate and post graduate students from Northeastern University.

### Where will this take place and how much of my time will it take?

The research will take place at the Playable Innovative Technologies (PLAIT) Lab at a time that is convenient for you. This will take about 45 minutes to one hour.

### Will there be any risk or discomfort to me?

There are no foreseeable risks or discomforts for participants. The study may include risks that are unknown at this time.

### Will I benefit by being in this research?

There are no direct benefits to you for participating in the study. The information learned from this study may help us to make better informed decisions about the use of video games in Education.

### Who will see the information about me?

Your part in this study will be handled in a confidential manner. Only the researchers will know that you participated in this study. Any reports or publications based on this research will use only group data and will not identify you or any individual as being of this project.

APPROVED
NU IRB# _______ 13-09-10
VALID _______ 10/11/13
THROUGH _______ 10/10/14

The data will be maintained on a secure server. This data will be only accessed by the Principal Investigator and the research assistant.

The data collected will be used solely for the purpose of this study.

In rare instances, authorized people may request to see research information about you and other people in this study. This is done only to be sure that the research is done properly. We would only permit people who are authorized by organizations such as the Northeastern University Institutional Review Board to see this information.

## What will happen if I suffer any harm from this research?

No special arrangements will be made for compensation or for payment for treatment solely because of your participation in this research.

## Can I stop my participation in this study?

Your participation in this research is completely voluntary. You do not have to participate if you do not want to and you can refuse to answer any question. Even if you begin the study, you may quit at any time. If you do not participate or if you decide to quit, you will not lose any rights, benefits, or services that you would otherwise have as a student.

## Who can I contact if I have questions or problems?

If you have any questions about this study, please feel free to contact Allan Fowler (646) 3091702 or allan.fowler@gurl.co.nz. You can also contact Dr. Magy Seif El-Nasr, College of Computer and Information Sciences, Northeastern University, 360 Huntington Avenue, Boston, MA 02115, m.seifelnasr@neu.edu, the Principal Investigator.

## Who can I contact about my rights as a participant?

If you have any questions about your rights in this research, you may contact Nan C. Regina, Director, Human Subject Research Protection, 960 Renaissance Park, Northeastern University, Boston, MA 02115. Tel: 617.373.4588, Email: n.regina@neu.edu. You may call anonymously if you wish.

## Will I be paid for my participation?

You will be able to choose; a $20 gift card from Amazon or the iTunes Store, or a copy of the video game World of Goo as a token of appreciation for your participation

## Will it cost me anything to participate?

There are no direct costs that will be incurred by the participant for the study.

## Is there anything else I need to know?

- You must be at least eighteen years old.
- Individuals with amblyopia (lazy eye) are not eligible for the study as the data tracking of one cornea may not be accurate. If you have questions or concerns please speak with the researcher.

## I agree to take part in this research.

Signature of person agreeing to take part                          Date

Printed name of person above

Signature of person who explained the study to the                 Date
participant above and obtained consent

APPROVED
NU IRB#
VALID
THROUGH

Printed name of person above

## Study Debriefing Document

This study is concerned with what learning takes place within a video game. Previous studies have found that that there is some link and we are looking to extend this research.

#### How was this tested?

In this study, you were asked to perform two tasks—play with a video game and play with a toy. All participants performed these same tasks for 15 minutes each, and your onscreen actions were recorded by the video camera. We also recorded the movements of your eyes and how many times you blinked. One group played The World of Goo whereas the other group played Bad Piggies. We also gave you a questionnaire to find what (if any) learning took place.

#### Hypotheses and main questions:

We expect to find that playing the World of Goo had a positive impact on playing with the magnetic toy.

We are also interested in if there are any improvements in problem solving after each time you replay a particular level of the game.

#### Why is this important to study?

There is a lot of debate about the value of video games in the learning process. A lot of this debate is speculative. We intend to provide quantitate data to substantiate these claims.

#### What if I want to know more?

If you are interested in learning more about video games for learning, you may want to consult:

Gee, J. P. (2003). *What video games have to teach us about learning and literacy*. New York, NY: Palgrave Macmillan. If you would like to receive a report of this research when it is completed (or a summary of the findings), please contact Allan Fowler at allan.fowler@gurl.co.nz

If you have concerns about your rights as a participant in this experiment, you may contact Nan C. Regina, Director, Human Subject Research Protection, 960 Renaissance Park, Northeastern University, Boston, MA 02115. Tel: 617.373.4588, Email: n.regina@ncu.edu. You may call anonymously if you wish.

Thank you again for your participation.

APPROVED
NU IRB# ___ 12-09-10
VALID ___ 10/1/13
THROUGH ___ 10/30/14

## A3.2 Study of children IRB approval

Northeastern

**NOTIFICATION OF IRB ACTION**

| | |
|---|---|
| Date: November 5, 2013 | IRB #: 13-10-13 |
| Principal Investigator(s): | Rupal Patel |
| Department: | Speech-Language Pathology and Audiology College of Science |
| Address: | 106A Forsyth Building Northeastern University |
| Title of Project: | Measuring Learning in Commercial Video Games in Young Children: A Proposed Method |
| Participating Sites: | N/A |
| DHHS Review Category: | Expedited #4, #6, #7 |
| Informed Consents: | One (1) signed parent/guardian consent form One (1) verbal child assent form |

This project is approved under 45CFR46.404 which applies to children on research subjects and involves research not involving greater than minimal risk. Adequate provisions are made for soliciting the assent of the children and the permission of their parents or guardians, as set forth in 45CFR46.408.

Monitoring Interval: 12 months

## APPROVAL EXPIRATION DATE: NOVEMBER 4, 2014

**Investigator's Responsibilities:**

1. The informed consent form bearing the IRB approval stamp must be used when recruiting participants into the study.
2. The investigator must notify IRB **immediately** of unexpected adverse reactions, or new information that may alter our perception of the benefit-risk ratio.
3. Study procedures and files are subject to audit any time.
4. Any modifications of the protocol or the informed consent as the study progresses must be reviewed and approved by this committee **prior to being instituted.**
5. Continuing Review Approval for the proposal should be requested at least one month prior to the expiration date above.
6. This approval applies to the protection of human subjects only. It does not apply to any other university approvals that may be necessary.

C. Randall Colvin, Ph.D., Chair
Northeastern University Institutional Review Board

Nan C. Regina, Director
Human Subject Research Protection

Northeastern University FWA #4630

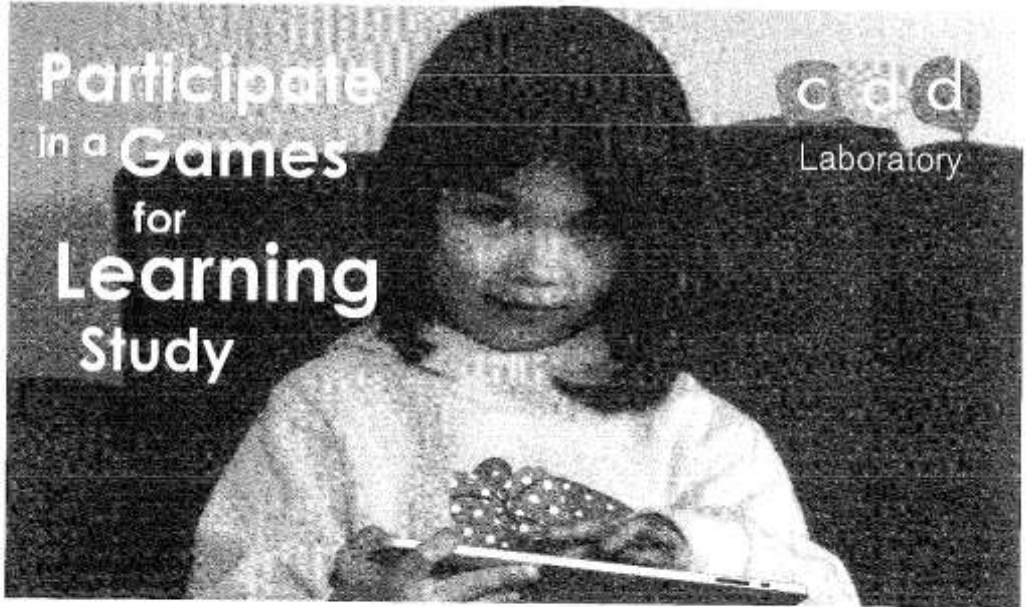

Participate in a **Games** for **Learning Study**

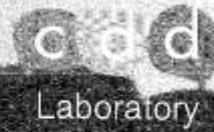
Laboratory

To be eligible for this study, your child should: 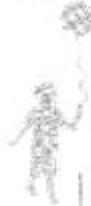

Be 7-8 years old

Have adequate
- (corrected) vision,
- hearing,
- manual dexterity

to play a building and computer game

**What** will my child do?

Play with a video game and building toy

70 Forsyth street, room 100 (CadLab)

**How long** will it take?

One session for a total of 45-60 minutes

APPROVED

Please Call: **617-373-6974** cadlab.neu@gmail.com

**Receive a $30 Amazon gift certificate**
& Free Parking

Northeastern University

www.cadlab.neu.edu

**Talking Points to share with potential participants**

Hi Mr/Mrs. ________ ,

How are you doing today? Before we get started, I wanted to see if you (your child) may be interested in participating in a study. Dr. Patel is a professor at Northeastern University in Boston and she is doing the research study. She and her research assistants would like to study your child playing a video game and playing with a magnet building toy. The session will take approximately 45-60 minutes. Your child's on screen actions and hand movements will be video recorded, his/her speech will be audio recorded, and his/her eye movements will be recorded with an eye-tracker. If you are interested in having your child participate, please let me know and we can schedule a time to meet. This study is completely voluntary, and you (your child) do not (does not) have to participate if you (your child) do not (does not) want to. Think about it and please let me know.

**Measuring learning in commercial video games in young children: A proposed method**

Rupal Patel, Ph.D.
Allan Fowler
Northeastern University
Department of Speech-Language Pathology and Audiology

Hello, my name is ___________, I am an investigator who works with Professor Rupal Patel at Northeastern University. We are looking at how video games may help children learn and are asking for your help with a research study. Research studies are done to find ways to help people or understand things better. I will tell you about the study we are doing so that you can decide whether or not you want to participate. Please ask any questions if you don't understand. It is okay if you don't want to be in the study. No one will be upset. Even if you agree to help us, you can stop at anytime. That is okay too.

- We will have you play with a building toy and play a video game. We will videorecord how your hands move, audiorecord your voice, and also record how your eyes move. Before and after you have played with the game and the toy, we will ask you some questions.

- If you get tired at any time, just let me know so we can take a short break.

- We have already asked your parent(s)/caregiver(s) to give their permission for you to help us. Even if your parent(s)/caregiver(s) said "yes," you can still say "no" and decide not to participate. It is your choice and no one will be upset with you. Even if you agree to help us, you can stop at anytime. That is okay too.

Do you have any questions? If you have a question later that you don't think of now, you can ask it at any time. You can also call me or ask your parent(s)/caregiver(s) to call me or Rupal Patel at (617) 373-5842.

...................................................................................

_______________________________________          ___________________
Printed name of child/adult                                             Date

I certify that I have explained the study to this child/adult and that the child/adult has assented to be a participant. A copy of this form has been given to the child/adult or his/her representative.

_______________________________________          ___________________
Signature of person obtaining assent                           Date

_______________________________________
Printed name of person above

APPROVED
NU IRB#        13-10-18
VALID          11/8/13
THROUGH        11/9/14

**Measuring learning in commercial video games in young children: A proposed method**

Rupal Patel, Ph.D.
Allan Fowler
Northeastern University
Department of Speech-Language Pathology and Audiology

We are inviting your child to take part in a research study. This form will tell you about the study, but the researcher will explain it to you in person if you require clarification or have questions. When you are ready to make a decision, you may tell the researcher if you want your child to participate or not. Your child does not have to participate if you do not want or if he/she does not wish to. If you decide to have your child participate, the researcher will ask you to sign this form and will give you a copy to keep for your records.

**Why is my child being asked to take part in this research study?** We are asking your child to participate in this study because we are interested in finding out what people can learn through playing video games.

**Why are we doing this research study?** The purpose of this research is to gain an understanding of the learning that takes place when children play a commercial video game. In addition, we want to see if there is a difference in the problem solving ability of those who played an educational video game (the World of Goo) vs. those who played a standard (non-educational) video game (Bad Piggies).

**What will my child be asked to do?** Your child will play with a building toy and be randomly assigned to play one of two non-violent video games (World of Goo or Bad Piggies). During the play, we will videorecord your child's hand movements to measure screen actions and decisions. We will audiorecord his/her voice to determine speech changes before and after the games. We will also use a special camera that will record his/her eye movements. This camera uses infrared light to track the movements of the cornea. Infrared light is invisible to people and is not harmful to the eyes. We will also ask your child some questions about building and video games at the beginning and end of the session.

**Where will this take place and how much time will it take?** All data collection will take place at the Communication Analysis and Design Laboratory at Northeastern University for a single session lasting between 45 and 60 minutes.

**Will there be any risk or discomfort to my child?** While any experimental procedure may carry with it some unforeseen risk, the chance of any harm from participating in this study is highly unlikely. The study does not involve any harmful or painful procedures. We only ask that your child play with a video game and a toy. If your child gets tired during the session, breaks will be provided.

APPROVED
IRB NEU     15-10-13
VALID
THROUGH

**Will my child benefit by being in this research?** There are no direct or immediate benefits to this project. Your child's participation may lead to findings that advance the knowledge base of the potential use of video games in education.

**Who will see the information about my child?** No one, except the researchers will know your child's name and identity. All records of identity and all recordings will be kept under lock and key in the office of Dr. Rupal Patel. Only the investigator and her research assistants will have access to the computer files and the video-, audio-, and eye tracker recordings. At the end of the study, the recordings will be archived for future research studies and educational purposes. The identity of participants will not be revealed in any publications or presentations, however, descriptions of participants will be provided. While it is not likely that someone might recognize your child from those descriptions, it is not impossible. In rare instances, authorized people may request to see research information about your child and other people in this study. This is done only to be sure that the research is done properly. We would only permit people who are authorized by organizations such as Northeastern University to see this information.

**Can I or my child stop his/her participation in this study?** Your child's participation in this research is completely voluntary. Your child does not have to participate if you or he/she does not want to. Even if he/she begins the study, he/she may quit at any time. If your child does not participate or decides to quit, your child will not lose any rights, benefits, or services that he/she would otherwise have. In any case, we are grateful that you are considering having your child participate.

**Who can I contact if I have questions or problems?** If you have any questions or problems, you should contact Dr. Rupal Patel at 617-373-5842 or at r.patel@neu.edu.

**Who can I contact about my rights as a participant?** If you have any questions about your rights as a participant, you may contact Nan Clark Regina, Director, Human Subject Research Protection, 960 Renaissance Park, Northeastern University Boston, MA 02115, Tel. 617-373-4588, Email: irb@neu.edu. You may call anonymously if you wish.

**Will my child be paid for his/her participation?** You/your child will receive a $30 Amazon gift card for participating.

**Will it cost me anything for my child to participate?** No.

**Is there anything else I need to know?** To participate in this study, your child must be between the ages of 7 to 8 years old and have adequate (corrected) vision, hearing, and manual dexterity to complete the experimental tasks.

**I have read the above explanation and I agree for my child to take part in this research.**

---

Signature of Parent/Guardian taking part      Date

---

Printed name of person above      Name of Child

---

Signature of person who explained the study to the      Date
participant above and obtained consent

Printed name of person above

APPROVED

NU IRB _______
VALID _______
THROUGH _______

# APPENDIX 4 – Informed Consent/ Ascent Forms

## A4.1 Adult Study

Northeastern University *Human Subject Research Protection*  960 Renaissance Park Northeastern University Boston, MA 02115-5005 Tel 617.373.7570 Fax 617.373.4588

ver. 2013-10-10

**Northeastern University,** Department of Computer and Information Science
**Name of Investigator(s):** Dr. Magy Seif El-Nasr & Allan Fowler
**Title of Project:** Measuring learning in commercial video games

### Informed Consent to Participate in a Research Study

We are inviting you to take part in a research study. This form will tell you about the study, but the researcher will explain it to you first. You may ask this person any questions that you have. When you are ready to make a decision, you may tell the researcher if you want to participate or not. You do not have to participate if you do not want to. If you decide to participate, the researcher will ask you to sign this statement and will give you a copy to keep.

### Why am I being asked to take part in this research study?

We are asking you to participate in this study because we are interested in finding out what people can learn through playing video games.

### Why is this research study being done?

The purpose of this research is to gain an understanding of the learning that takes place when people play a commercial video game. We will use a number of measurement techniques that are recognized to provide indicators of cognition (the act or process of knowing).

### What will I be asked to do?

If you decide to take part in this study, you will be asked to play a non-violent video game. We think this will take you about 45 minutes. You will also be asked to play with a construction set. Your reactions will be recorded using a number of techniques. We will use a video recorder to record your on-screen actions and decisions. We will also use a special camera that will record where your eyes are focused. This camera uses infrared light to track the movements of the cornea. Infrared light is invisible to people and is not harmful to the eyes. We will ask you a few questions about the amount of time you spend playing video games and the types of video games you play. We will also ask you about the video game played during the study.

We believe that there is a relationship between playing video games and an increase in problem solving ability and this study is trying to investigate this link.

The investigators may stop the study at any time they judge it is in your best interest. They may also remove you from the study for various other reasons. They can do this without your consent. You can stop participating at any time. We have taken a random sample of undergraduate and post graduate students from Northeastern University.

### Where will this take place and how much of my time will it take?

The research will take place at the Playable Innovative Technologies (PLAIT) Lab at a time that is convenient for you. This will take about 45 minutes to one hour.

### Will there be any risk or discomfort to me?

There are no foreseeable risks or discomforts for participants. The study may include risks that are unknown at this time.

### Will I benefit by being in this research?

There are no direct benefits to you for participating in the study. The information learned from this study may help us to make better informed decisions about the use of video games in Education.

### Who will see the information about me?

Your part in this study will be handled in a confidential manner. Only the researchers will know that you participated in this study. Any reports or publications based on this research will use only group data and will not identify you or any individual as being of this project.

APPROVED
NU IRB# _13-09-10_
VALID _10/11/13_
THROUGH _10/10/14_

Northeastern University  360 Huntington Ave.  
Human Subject Research Protection  Boston, MA 02115-5000  
Tel 617 373 7570, Fax 617 3734568

ver. 2013-10-10

The data will be maintained on a secure server. This data will be only accessed by the Principal Investigator and the research assistant.

The data collected will be used solely for the purpose of this study.

In rare instances, authorized people may request to see research information about you and other people in this study. This is done only to be sure that the research is done properly. We would only permit people who are authorized by organizations such as the Northeastern University Institutional Review Board to see this information.

## What will happen if I suffer any harm from this research?

No special arrangements will be made for compensation or for payment for treatment solely because of your participation in this research.

## Can I stop my participation in this study?

Your participation in this research is completely voluntary. You do not have to participate if you do not want to and you can refuse to answer any question. Even if you begin the study, you may quit at any time. If you do not participate or if you decide to quit, you will not lose any rights, benefits, or services that you would otherwise have as a student.

## Who can I contact if I have questions or problems?

If you have any questions about this study, please feel free to contact Allan Fowler (646) 3091702 or allan.fowler@gurl.co.nz. You can also contact Dr. Magy Seif El-Nasr, College of Computer and Information Sciences, Northeastern University, 360 Huntington Avenue, Boston, MA 02115, m.seifelnasr@neu.edu, the Principal Investigator.

## Who can I contact about my rights as a participant?

If you have any questions about your rights in this research, you may contact Nan C. Regina, Director, Human Subject Research Protection, 960 Renaissance Park, Northeastern University, Boston, MA 02115. Tel: 617.373.4588, Email: n.regina@neu.edu. You may call anonymously if you wish.

## Will I be paid for my participation?

You will be able to choose; a $20 gift card from Amazon or the iTunes Store, or a copy of the video game World of Goo as a token of appreciation for your participation

## Will it cost me anything to participate?

There are no direct costs that will be incurred by the participant for the study.

## Is there anything else I need to know?

- You must be at least eighteen years old.
- Individuals with amblyopia (lazy eye) are not eligible for the study as the data tracking of one cornea may not be accurate. If you have questions or concerns please speak with the researcher.

## I agree to take part in this research.

________________________________________          ________________
Signature of person agreeing to take part            Date

________________________________________
Printed name of person above

________________________________________          ________________
Signature of person who explained the study to the    Date
participant above and obtained consent

________________________________________
Printed name of person above

APPROVED

## A4.2 Study of Children Informed Consent

**Consent to Participate in a Research Study**

**Measuring learning in commercial video games in young children: A proposed method**

**Rupal Patel, Ph.D.**
**Allan Fowler**
**Northeastern University**
Department of Speech-Language Pathology and Audiology

We are inviting your child to take part in a research study. This form will tell you about the study, but the researcher will explain it to you in person if you require clarification or have questions. When you are ready to make a decision, you may tell the researcher if you want your child to participate or not. Your child does not have to participate if you do not want or if he/she does not wish to. If you decide to have your child participate, the researcher will ask you to sign this form and will give you a copy to keep for your records.

**Why is my child being asked to take part in this research study?** We are asking your child to participate in this study because we are interested in finding out what people can learn through playing video games.

**Why are we doing this research study?** The purpose of this research is to gain an understanding of the learning that takes place when children play a commercial video game. In addition, we want to see if there is a difference in the problem solving ability of those who played an educational video game (the World of Goo) vs. those who played a standard (non-educational) video game (Bad Piggies).

**What will my child be asked to do?** Your child will play with a building toy and be randomly assigned to play one of two non-violent video games (World of Goo or Bad Piggies). During the play, we will videorecord your child's hand movements to measure screen actions and decisions. We will audiorecord his/her voice to determine speech changes before and after the games. We will also use a special camera that will record his/her eye movements. This camera uses infrared light to track the movements of the cornea. Infrared light is invisible to people and is not harmful to the eyes. We will also ask your child some questions about building and video games at the beginning and end of the session.

**Where will this take place and how much time will it take?** All data collection will take place at the Communication Analysis and Design Laboratory at Northeastern University for a single session lasting between 45 and 60 minutes.

**Will there be any risk or discomfort to my child?** While any experimental procedure may carry with it some unforeseen risk, the chance of any harm from participating in this study is highly unlikely. The study does not involve any harmful or painful procedures. We only ask that your child play with a video game and a toy. If your child gets tired during the session, breaks will be provided.

APPROVED
NU IRB#  15-10-13
VALID
THROUGH

**Will my child benefit by being in this research?** There are no direct or immediate benefits to your child's participation in this project. Your child's participation may lead to findings that advance the knowledge base of the potential use of video games in education.

**Who will see the information about my child?** No one, except the researchers will know your child's name and identity. All records of identity and all recordings will be kept under lock and key in the office of Dr. Rupal Patel. Only the investigator and her research assistants will have access to the computer files and the video-, audio-, and eye tracker recordings. At the end of the study, the recordings will be archived for future research studies and educational purposes. The identity of participants will not be revealed in any publications or presentations, however, descriptions of participants will be provided. While it is not likely that someone might recognize your child from those descriptions, it is not impossible. In rare instances, authorized people may request to see research information about your child and other people in this study. This is done only to be sure that the research is done properly. We would only permit people who are authorized by organizations such as Northeastern University to see this information.

**Can I or my child stop his/her participation in this study?** Your child's participation in this research is completely voluntary. Your child does not have to participate if you or he/she does not want to. Even if he/she begins the study, he/she may quit at any time. If your child does not participate or decides to quit, your child will not lose any rights, benefits, or services that he/she would otherwise have. In any case, we are grateful that you are considering having your child participate.

**Who can I contact if I have questions or problems?** If you have any questions or problems, you should contact Dr. Rupal Patel at 617-373-5842 or at r.patel@neu.edu.

**Who can I contact about my rights as a participant?** If you have any questions about your rights as a participant, you may contact Nan Clark Regina, Director, Human Subject Research Protection, 960 Renaissance Park, Northeastern University Boston, MA 02115, Tel. 617-373-4588, Email: irb@neu.edu. You may call anonymously if you wish.

**Will my child be paid for his/her participation?** You/your child will receive a $30 Amazon gift card for participating.

**Will it cost me anything for my child to participate?** No.

**Is there anything else I need to know?** To participate in this study, your child must be between the ages of 7 to 8 years old and have adequate (corrected) vision, hearing, and manual dexterity to complete the experimental tasks.

**I have read the above explanation and I agree for my child to take part in this research.**

| | |
|---|---|
| Signature of Parent/Guardian taking part | Date |
| Printed name of person above | Name of Child |
| Signature of person who explained the study to the participant above and obtained consent | Date |

Printed name of person above

APPROVED

NU IRB ____________
VALID ____________
THROUGH ____________

## A4.2    Assent Form

Assent Form

**Measuring learning in commercial video games in young children: A proposed method**

Rupal Patel, Ph.D.
Allan Fowler
Northeastern University
Department of Speech-Language Pathology and Audiology

Hello, my name is ___________, I am an investigator who works with Professor Rupal Patel at Northeastern University. We are looking at how video games may help children learn and are asking for your help with a research study.  Research studies are done to find ways to help people or understand things better. I will tell you about the study we are doing so that you can decide whether or not you want to participate. Please ask any questions if you don't understand.  It is okay if you don't want to be in the study. No one will be upset. Even if you agree to help us, you can stop at anytime. That is okay too.

- We will have you play with a building toy and play a video game.  We will videorecord how your hands move, audiorecord your voice, and also record how your eyes move. Before and after you have played with the game and the toy, we will ask you some questions.

- If you get tired at any time, just let me know so we can take a short break.

- We have already asked your parent(s)/caregiver(s) to give their permission for you to help us. Even if your parent(s)/caregiver(s) said "yes," you can still say "no" and decide not to participate. It is your choice and no one will be upset with you.  Even if you agree to help us, you can stop at anytime. That is okay too.

Do you have any questions? If you have a question later that you don't think of now, you can ask it at any time. You can also call me or ask your parent(s)/caregiver(s) to call me or Rupal Patel at (617) 373-5842.

••••••••••••••••••••••••••••••••••••••••••••••••••••••••••••••••••••••••••••••••••••••••

_______________________________________          ___________________
Printed name of child/adult                                          Date

I certify that I have explained the study to this child/adult and that the child/adult has assented to be a participant.  A copy of this form has been given to the child/adult or his/her representative.

_______________________________________          ___________________
Signature of person obtaining assent                          Date

_______________________________________
Printed name of person above

APPROVED
NU IRB#        $13-10-18$
VAI ID        _______
THROUGH    11/4/14

# APPENDIX 5 Debriefing Document

## Study Debriefing Document

This study is concerned with what learning takes place within a video game. Previous studies have found that that there is some link and we are looking to extend this research.

How was this tested?
In this study, you were asked to perform two tasks—play with a video game and play with a toy. All participants performed these same tasks for 15 minutes each, and your onscreen actions were recorded by the video camera. We also recorded the movements of your eyes and how many times you blinked. One group played The World of Goo whereas the other group played Bad Piggies. We also gave you a questionnaire to find what (if any) learning took place.

Hypotheses and main questions:
We expect to find that playing the World of Goo had a positive impact on playing with the magnetic toy.

We are also interested in if there are any improvements in problem solving after each time you replay a particular level of the game.

Why is this important to study?
There is a lot of debate about the value of video games in the learning process. A lot of this debate is speculative. We intend to provide quantitate data to substantiate these claims.

What if I want to know more?
If you are interested in learning more about video games for learning, you may want to consult:

Gee, J. P. (2003). *What video games have to teach us about learning and literacy*. New York, NY: Palgrave Macmillan. If you would like to receive a report of this research when it is completed (or a summary of the findings), please contact Allan Fowler at allan.fowler@gurl.co.nz

If you have concerns about your rights as a participant in this experiment, you may contact Nan C. Regina, Director, Human Subject Research Protection, 960 Renaissance Park, Northeastern University, Boston, MA 02115. Tel: 617.373.4588, Email: n.regina@neu.edu. You may call anonymously if you wish.

Thank you again for your participation.

# APPENDIX 6 AUT Ethics Approval

## A6.1    AUT Ethics Approval – Study of Adults

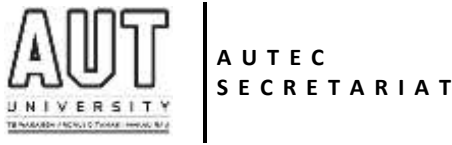

21 August 2014

Brian Cusack
Faculty of Design and Creative Technologies

Dear Brian

Ethics Application:             **14/270 Understanding learning within a commercial video game: A case study with adults**

Thank you for submitting your application for ethical review to the Auckland University of Technology Ethics Committee (AUTEC). I am pleased to confirm that the Chair and I have approved your ethics application for three years until 20 August 2017

As part of the ethics approval process, you are required to submit the following to AUTEC:

- A brief annual progress report using form EA2, which is available online through http://www.aut.ac.nz/researchethics. When necessary this form may also be used to request an extension of the approval at least one month prior to its expiry on 20 August 2017;
- A brief report on the status of the project using form EA3, which is available online through http://www.aut.ac.nz/researchethics. This report is to be submitted either when the approval expires on 20 August 2017 or on completion of the project;

It is a condition of approval that AUTEC is notified of any adverse events or if the research does not commence.  AUTEC approval needs to be sought for any alteration to the research, including any alteration of or addition to any documents that are provided to participants.  You are responsible for ensuring that research undertaken under this approval occurs within the parameters outlined in the approved application.

AUTEC grants ethical approval only.  If you require management approval from an institution or organisation for your research, then you will need to obtain this.  If your research is undertaken within a jurisdiction outside New Zealand, you will need to make the arrangements necessary to meet the legal and ethical requirements that apply there.

To enable us to provide you with efficient service, we ask that you use the application number and study title in all correspondence with us.  If you have any enquiries about this application, or anything else, please do contact us at ethics@aut.ac.nz.

All the very best with your research,

Kate O'Connor
Executive Secretary
**Auckland University of Technology Ethics Committee**

Cc:          Allan Fowler allan_fowler@hotmail.com

## A 6.2 AUT Ethics Approval – Study of Children

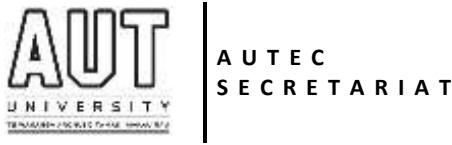

21 August 2014

Brian Cusack
Faculty of Design and Creative Technologies

Dear Brian

Ethics Application: **14/271 Understanding learning within a commercial video game: A case study with children**

Thank you for submitting your application for ethical review to the Auckland University of Technology Ethics Committee (AUTEC). I am pleased to confirm that the Chair and I have approved your ethics application for three years until 20 August 2017

As part of the ethics approval process, you are required to submit the following to AUTEC:

- A brief annual progress report using form EA2, which is available online through http://www.aut.ac.nz/researchethics. When necessary this form may also be used to request an extension of the approval at least one month prior to its expiry on 20 August 2017;

- A brief report on the status of the project using form EA3, which is available online through http://www.aut.ac.nz/researchethics. This report is to be submitted either when the approval expires on 20 August 2017 or on completion of the project;

It is a condition of approval that AUTEC is notified of any adverse events or if the research does not commence. AUTEC approval needs to be sought for any alteration to the research, including any alteration of or addition to any documents that are provided to participants. You are responsible for ensuring that research undertaken under this approval occurs within the parameters outlined in the approved application.

AUTEC grants ethical approval only. If you require management approval from an institution or organisation for your research, then you will need to obtain this. If your research is undertaken within a jurisdiction outside New Zealand, you will need to make the arrangements necessary to meet the legal and ethical requirements that apply there.

To enable us to provide you with efficient service, we ask that you use the application number and study title in all correspondence with us. If you have any enquiries about this application, or anything else, please do not contact us at ethics@aut.ac.nz.

All the very best with your research,

Kate O'Connor
Executive Secretary
**Auckland University of Technology Ethics Committee**

Cc: Allan Fowler allan_fowler@hotmail.com